\documentclass[10pt]{iopartmod}
\usepackage{amssymb}
\usepackage{amsmath}
\usepackage{makeidx}
\usepackage[dvips]{graphicx}

\input{tcilatex}

\begin{document}

\title{Nonlinear Photonic Crystals: IV. Nonlinear Schrodinger Equation Regime%
}
\author{Anatoli Babin and Alexander Figotin}

\begin{abstract}
We study here the nonlinear Schrodinger Equation (NLS) as the first term in
a sequence of approximations for an electromagnetic (EM) wave propagating
according to the nonlinear Maxwell equations (NLM). The dielectric medium is
assumed to be periodic, with a cubic nonlinearity, and with its linear
background possessing inversion symmetric dispersion relations. The medium
is excited by a current $\mathbf{J}$ producing an EM wave. The wave
nonlinear evolution is analyzed based on the modal decomposition and an
expansion of the exact solution to the NLM into an asymptotic series with
respect to three small parameters $\alpha $, $\beta $ and $\varrho $. These
parameters are introduced through the excitation current $\mathbf{J}$ to
scale respectively (i) its amplitude and consequently the magnitude of the
nonlinearity; (ii) the range of wavevectors involved in its modal
composition, with $\beta ^{-1}$ scaling its spatial extension; (iii) its
frequency bandwidth, with $\varrho ^{-1}$ scaling its time extension. We
develop a consistent theory of approximations of increasing accuracy for the
NLM with its first term governed by the NLS. We show that such NLS regime is
the medium response to an almost monochromatic excitation current $\mathbf{J}
$. The developed approach not only provides rigorous estimates of the
approximation accuracy of the NLM with the NLS in terms of powers of $\alpha 
$, $\beta $ and $\varrho $, but it also produces new extended NLS (ENLS)
equations providing better approximations. Remarkably, quantitative
estimates\ show that properly tailored ENLS can significantly improve the
approximation accuracy of the NLM compared with the classical NLS.
\end{abstract}

\pacs{42.65.-k, 42.65.Ky, 42.65.Hw, 0.3.50.De, 42.70.Qs}


\address{Department of Mathematics, University of California at Irvine, CA
92697}

\section{Introduction}

The subject of this work is the accuracy of approximation of solutions to
the nonlinear Maxwell equations (NLM) for periodic dielectric media, i.e.
photonic crystals, by solutions to the Nonlinear Schrodinger equations (NLS)
or, more broadly, by similar to the NLS equations. The both NLM and NLS
equations are widely used in the nonlinear optics, and there are many
derivations of the NLS in different situations and of different level of
rigor in physical literature. There is extensive literature devoted to
studies of solutions to the NLS\ (see \cite{Abl}, \cite{Cai02}, \cite{Dodd}, 
\cite{Drazin}, \cite{Lamb}, \cite{Mikh}, \cite{Sulem},\ \cite{TrZP},\ \cite%
{Zakh} and references therein). If the evolultion of an electromagnetic (EM)
wave is governed by the NLM equations and it can be approximated by an NLS
equation or, may be, by a slightly more general extended NLS equation, we
refer to it as \emph{NLS regime of propagation or just NLS regime}.

The NLS describes a universal wave propagation regime occurring in a
dispersive medium with a dispersive relation $\omega \left( \mathbf{k}%
\right) $ for its linear background. A derivation\ of the NLS emphasizing
its universal nature can be obtained by introducing an amplitude-dependent
dispersion relation (see \cite[p.4-5]{Sulem}, \cite[p. 50-51]{NewM}, \cite{W}%
)) of the form $\omega \left( \mathbf{k}\right) +\delta \left\vert
Z\right\vert ^{2}$ formally implying the following NLS evolution equation: 
\begin{equation}
\partial _{t}Z=-\mathrm{i}\left[ \omega \left( -\mathrm{i}\vec{\nabla}_{%
\mathbf{r}}\right) Z+\delta \left\vert Z\right\vert ^{2}Z\right] .
\end{equation}%
More elaborate derivations of the NLS based on the NLM evolution equation
make use of the Fourier expansions of the envolved fields in the infinite
space, \cite[p.6-7]{Sulem}, \cite[p. 67-71 and 83-104]{NewM}. A similar
analysis of the NLM for periodic dielectric media, i.e. photonic crystals,
was carried out based on Bloch expansions in \cite{BhatSipe}. In a number of
mathematical studies the NLS are derived based on equations other than the
nonlinear Maxwell equations (see \cite{Bambusi}, \cite{Colin}, \cite%
{Schneider98}, \cite{Schneider98a}).

Looking at different NLS derivations one can see that they are based on the
following fundamental assumptions:

\begin{itemize}
\item the nonlinear component of the wave is relatively small (the
nonlinearity is weak);

\item the wave is defined as a real-valued function;

\item the wavevectors (quasimomenta) $\mathbf{k}$ involved in the wave
composition are close to a certain $\mathbf{k}_{\ast }$;

\item the time evolution of the wave envelope is slow compared to the
typical carrier wave frequency;

\item the dispersion relation $\omega \left( \mathbf{k}\right) =\omega
\left( \mathbf{k}_{\ast }+\mathbf{\eta }\right) $ in a vicinity of $\mathbf{k%
}_{\ast }$ is approximated by its second-order Taylor polynomial $\gamma
_{\left( 2\right) }\left( \mathbf{\eta }\right) $;

\item the non-frequency-matched wave interactions (in particular the third
harmonic generation) are neglected;

\item the frequency dependence of the susceptibility tensor is neglected and
its value at $\mathbf{k}_{\ast }$ \ is used.
\end{itemize}

All the above factors are presented and to some degree are refined in our
quantitative approach to the approximation of solutions to the NLM by the
NLS. The approach is based on the framework described in \cite{BF1}-\cite%
{BF3}, and its outline is as follows. A wave propagating in the nonlinear
media is generated by an excitation current $\mathbf{J}$ which is turned on
at time $t=0$ and is turned off at a later time $t=t_{0}$. Hence for $%
t>t_{0} $ there are no external currents and the wave dynamics is determined
entirely by the medium.

Suppose that the excitation current $\mathbf{J}$ has the form of a
wavepacket with the carrier frequency $\omega =\omega _{n_{0}}\left( \mathbf{%
k}_{\ast }\right) $ where $\omega _{n}\left( \mathbf{k}\right) $ is the
dispersion relation of the underlying linear medium with the band number $n$
and the wave numbers (quasimomenta) $\mathbf{k}$, and $n_{0}$ and $\mathbf{k}%
_{\ast }$ are chosen. The envelope amplitude of the excitation current $%
\mathbf{J}$ is supposed to vary slowly in space and time. The current $%
\mathbf{J}$ and the resulting wave evolution are determinded by three
dimensionless small parameters $\alpha $, $\beta $ and $\varrho $. The first
small parameter $\alpha $ scales the relative magnitude of the wave
nonlinear component and is related to the amplitude of the excitation
current. The second parameter $\beta $ scales the range of the wavevectors $%
\mathbf{k}$ in a vicinity $\mathbf{k}_{\ast }$ involved in the modal
composition $\mathbf{J}$, and, consequently, $\beta ^{-1}$ scales the
spatial extension of $\mathbf{J}$. Finally, the parameter $\varrho $ scales
\ the frequency bandwidth of $\mathbf{J}$, and, consequently, $\varrho ^{-1}$
scales the time extension of $\mathbf{J}$. \ It turns out, \cite{BF1}, that,
in particular, $\varrho $ determines the slow time $\tau =\varrho t$ related
to the nonlinear evolution.

Supposing that there is an excitation current $\mathbf{J}=\mathbf{J}\left(
\alpha ,\varrho ,\beta \right) $ as described above we consider the
resulting wave $\mathbf{U}=\mathbf{U}\left( \alpha ,\varrho ,\beta \right) $
which is a solution to the NLM. The NLM is a rather complicated nonlinear
evolution equation for electromagnetic vector fields varying in time and
space, and, naturally, we are interested in simpler scalar equations
approximating the NLM. It is well known that the NLS is one of such
approximations and we are interested in finding how the exact solutions $%
\mathbf{U}\left( \alpha ,\varrho ,\beta \right) $ of the NLM\ for small $%
\alpha $, $\beta $ and $\varrho $ for $t>t_{0}$ are approximated by
solutions to an NLS equation. In our analysis \emph{we take into account all
the modes and all possible interactions} as functions of the parameters $%
\alpha $, $\beta $ and $\varrho $. Using relevant series expansions
rigorously justified in \cite{BF4} we study the exact solution of the NLM
for small but still finite values of all three parameters $\alpha $, $\beta $
and $\varrho $, and relate this solution to a solution of a properly
tailored NLS. In particular, we show that the scalar amplitudes of the Bloch
modes in the modal composition of the solution $\mathbf{U}$ can be
approximated by amplitudes of the Fourier modes in the Fourier compostion of
the solution $Z$ of the relevant NLS with high precision, providing also
error estimates. Having a good control over all the steps of the
approximation, we identify all additional terms which should be added to the
classical NLS to improve the approximation accuracy. Those more accurate
equations are reffered to as extended NLS equations (ENLS). We provide
explicit expressions for those additional terms in ENLS which represent\ the
dominant discrepancy between the exact NLM\ equation and its classical NLS\
approximation. Consequently, the derived ENLS are intimately related to the
NLM. We provide here some analysis of the ENLS, for more information on the
subject see \cite{Sulem} and references therein.

One of interesting results of our quantitative analysis of the NLS regimes
for the NLM is their remarkable accuracy for small $\alpha $, $\beta $ and $%
\varrho $. Namely, quantitative estimates\ of nonlinear wave interactions
show that a properly tailored ENLS can be far more accurate than the
classical NLS. \emph{In particular, for the classical NLS characterized by
scaling }$\alpha \sim \varrho \sim \beta ^{2}$\emph{\ its approximation
accuracy of the NLM is proportional to }$\beta $\emph{\ whereas a properly
tailored ENLS of third and fourth order have the approximation accuracy
proportional, respectively, to }$\beta ^{2}$\emph{\ and }$\beta ^{3}$. An
explanation to this this phenomenon is based on the analysis of nonlinear
wave interactions, \cite{BF1}, \cite{BF2}, \cite{BF3}. Namely, we show in
following sections that under the condition $\alpha \sim \varrho \sim \beta
^{2}$ the nonlinear wave interactions that lead to the NLS-like regimes and
are described by diffrenent ENLS essentially exhaust all significant
interactions up to the order $\beta ^{4}$ whereas other nonlinear
interactions under same conditions are of the order not greater than $\beta
^{5}$. \emph{In other words, just by using ENLS, which are only a little
more complex than the classical NLS, we can improve the total approximation
accuracy of the NLM from }$\beta $\emph{\ to }$\beta ^{3}$\emph{.}

Complete analysis of the accuracy of the approximation of the NLM with the
NLS is laborious, and it is helpful to keep in mind the following key
elements of that analysis.

\begin{itemize}
\item The \emph{dispersion relations} $\omega _{n}\left( \mathbf{k}\right) $
of the underlying linear periodic medium, with $n$ and $\mathbf{k}$ being
respectively the band number and the quasimomentum, \emph{are inversion
symmetric}, i.e.%
\begin{equation}
\omega _{n}\left( -\mathbf{k}\right) =\omega _{n}\left( \mathbf{k}\right) ,\
n=1,2,\ldots \ .  \label{invsym}
\end{equation}%
The inversion symmetry condition (\ref{invsym}) is an important factor for
NLS regimes in dielectric media with cubic nonlinearities.

\item We use modal decompositions of all involved fields with respect to the
related Bloch modes of the underlying linear medium. We consider only weakly
nonlinear regimes for which, as it turns out, the modal decomposion is
instrumental to the analysis of the wave propagation. The physical and
mathematical significence of the spectral decomposition with respect to the
Bloch modes lies in the fact that they don't exhange the energy under the
linear evolution.

\item \emph{The NLS regime as a phenomenon of nonlinear wave interactions is
characterized by lack of significant nonlinear interactions and energy
exchanges between different spectral bands and different quasimomenta}. More
exactly, if the wave is initially composed of eigenmodes characterized by a
single band\ number $n_{0}$ and chosen quasimomentum $\pm \mathbf{k}_{\ast }$
then under the NLS regime its modal composition remains confined to this
band, and its quasimomenta remain close to $\pm \mathbf{k}_{\ast }$ for long
times with the nonlinear interactions essentially occuring only between this
narrow group of quasimomenta, whereas nonlinear interactions with all other
bands and quasimomenta being negligibly small.

\item The NLS describes approximately the evolution of the modal coefficient
of the solution of the NLM\ generated by a real-valued almost time-harmonic
excitation current composed of eigenmodes with a single band number $n_{0}$
and the quasimomentum $\mathbf{k}$ from a small vicinity of a chosen point $%
\mathbf{k}_{\ast }$. \emph{The NLS regime is a dielectric medium response to
almost time-harmonic excitations.}

\item The linear part of the NLS\ is determined by the second order (or
higher order for the ENLS) Taylor polynomial $\gamma _{\left( 2\right)
}\left( \mathbf{\eta }\right) $ of $\omega _{n_{0}}\left( \mathbf{k}_{\ast }+%
\mathbf{\eta }\right) $ at $\mathbf{k}_{\ast }$. It turns out that an exact,
one-to-one correspondence can be established between the modal amplitudes of
the linear NLM and the NLS.

\item To relate the NLM and the NLS we introduce spatial and time scales
through the excitation currents in the NLM, and then study\ its exact
solutions and their asymptotic expansions with respect the parameters $%
\alpha $, $\beta $ and $\varrho $ assuming that they are small. After that
we taylor the parameters of the NLS\ or an ENLS\ so that their solutions
have the same asymptotic expansions up to a prescribed accuracy. The
solutions comparison is carried out after the excitation currents are turned
off. \emph{We do not make any a-priori assumptions on the form of solutions
to the NLM}, and our analysis of the solutions is not based on any specific
ansatz. This allows us not to impose strict \ functional dependence between
the parameters $\alpha $, $\beta $ and $\varrho $, and, consequently, the
significance of different terms in the NLS and ENLS\ and their relation with
the NLM can be studied for different ranges of parameters.

\item The analysis of involved fields and equations is based on asymptotic
series expansions of interaction integrals with respect to small $\alpha $, $%
\varrho $ and $\beta $ and the fourth small parameter which equals either $%
\frac{\varrho }{\beta ^{2}}$ or $\frac{\beta ^{3}}{\varrho }$. In other
words, we consider two cases: $\frac{\varrho }{\beta ^{2}}$ is small or $%
\frac{\beta ^{3}}{\varrho }$ is small. \emph{The asymptotic expansions
involving }$\beta $\emph{\ and }$\varrho $\emph{\ stem from oscillatory
interaction integrals and they are not Taylor series expansions.}
\end{itemize}

Following \cite{BF1}-\cite{BF3} we recast the classical nonlinear Maxwell
equations in the following non-dimensional operator form 
\begin{equation}
\partial _{t}\mathbf{U}\left( \mathbf{r},t\right) =\mathbf{-}\mathrm{i}%
\mathbf{MU}\left( \mathbf{r},t\right) +\alpha \mathcal{F}_{\text{NL}}\left( 
\mathbf{U}\left( \mathbf{r},t\right) \right) -\mathbf{J};\ \mathbf{U}\left( 
\mathbf{r},t\right) =\mathbf{J}\left( \mathbf{r},t\right) =0\;\text{for }%
t\leq 0,  \label{MXshort}
\end{equation}%
\begin{gather}
\mathbf{U}\left( \mathbf{r},t\right) =\left[ 
\begin{array}{c}
\mathbf{D}\left( \mathbf{r},t\right) \\ 
\mathbf{B}\left( \mathbf{r},t\right)%
\end{array}%
\right] ,\ \mathbf{MU}\left( \mathbf{r},t\right) =\mathrm{i}\left[ 
\begin{array}{c}
\nabla \times \mathbf{B}\left( \mathbf{r},t\right) \\ 
-\nabla \times \left( \mathbf{\varepsilon }^{-1}\left( \mathbf{r}\right) 
\mathbf{D}\left( \mathbf{r},t\right) \right)%
\end{array}%
\right] ,  \label{ML1} \\
\mathbf{J}\left( \mathbf{r},t\right) =4\pi \left[ 
\begin{array}{c}
\mathbf{J}_{D}\left( \mathbf{r},t\right) \\ 
\mathbf{J}_{B}\left( \mathbf{r},t\right)%
\end{array}%
\right] ,  \notag
\end{gather}%
where $\mathcal{F}_{\text{NL}}$ is a nonlinearity with a cubic principal
part which may have a general tensorial form, and $\mathbf{\varepsilon }%
\left( \mathbf{r}\right) $ is the electric permittivity tensor depending on
the three-dimensional spatial variable $\mathbf{r}=\left(
r_{1},r_{2},r_{3}\right) $. We consider in this article the case of a
lossles medium, i.e. $\mathbf{\varepsilon }\left( \mathbf{r}\right) $ is a
Hermitian matrix satisfying%
\begin{equation}
\mathbf{\varepsilon }\left( \mathbf{r}\right) =\left[ \mathbf{\varepsilon }%
\left( \mathbf{r}\right) \right] ^{\ast },\ \mathbf{r}=\left(
r_{1},r_{2},r_{3}\right) ,  \label{eer1}
\end{equation}%
and our special interest is in the case when the permittivity tensor $%
\mathbf{\varepsilon }\left( \mathbf{r}\right) $ is also a real symmetric
matrix, i.e.%
\begin{equation}
\mathbf{\varepsilon }\left( \mathbf{r}\right) =\left\{ \varepsilon
_{jm}\left( \mathbf{r}\right) \right\} _{j,m=1}^{3}\text{ where all }%
\varepsilon _{jm}\left( \mathbf{r}\right) =\varepsilon _{mj}\left( \mathbf{r}%
\right) \text{ are real-valued.}  \label{eer2}
\end{equation}%
Notice that the condition (\ref{eer2}) implies the inversion symmetry
property (\ref{invsym}) as well as the complex conjugation property of the
eigenmodes (see (\ref{Gstar})). Though almost all our constructions assume
only the inversion symmetry property (\ref{invsym}), the dielecric media for
which the condition (\ref{eer2}) is satisfied get our special attention
since they can support real-valued waves described very accurately by the
classical NLS. Without the condition (\ref{eer2}) but still under the
inversion symmetry condition (\ref{invsym}) we obtain instead complex-valued
waves described by a system of two coupled NLS (see Section 1.4.5 below).

The cases when $\mathbf{\varepsilon }\left( \mathbf{r}\right) $, $\mathbf{J}%
\left( \mathbf{r}\right) $ and $\mathbf{U}\left( \mathbf{r}\right) $ depend
only on $r_{1}$ or on $r_{1},r_{2}$ are called, respectively,
one-dimensional and two-dimensional. All the fields $\mathbf{D}$,\textbf{\ }$%
\mathbf{B}$, $\mathbf{J}_{D}$ and $\mathbf{J}_{B}$ are assumed to be
divergence free. The dielectric permittivity $\mathbf{\varepsilon }\left( 
\mathbf{r}\right) $ and the nonlinear polarization $\mathbf{P}_{\text{NL}%
}\left( \mathbf{r}\right) $ involved in $\mathcal{F}_{\text{NL}}$ are
assumed to be periodic with respect to every $r_{i}$, $i=1,2,3$ with the
period one for simplicity. The nonlinearity $\mathcal{F}_{\text{NL}}$
originates from the nonlinear polarization which can be written in the
following canonical form (see \cite{BC}):%
\begin{equation}
\mathbf{P}_{\text{NL}}\left( \mathbf{r},t;\mathbf{E}\left( \cdot \right)
\right) =\mathbf{P}^{\left( 3\right) }\left( \mathbf{r},t;\mathbf{E}\left(
\cdot \right) \right) +\mathbf{P}^{\left( 5\right) }\left( \mathbf{r},t;%
\mathbf{E}\left( \cdot \right) \right) +\ldots ,  \label{Pser}
\end{equation}%
where $\mathbf{P}^{\left( m\right) }$ is $m$-homogeneous operator of the
form 
\begin{equation}
\mathbf{P}^{\left( m\right) }\left( \mathbf{r},t;\mathbf{E}\left( \cdot
\right) \right) =\int_{-\infty }^{t}\ldots \int_{-\infty }^{t}\mathbf{R}%
^{\left( m\right) }\left( \mathbf{r};t-t_{1},\ldots ,t-t_{h}\right) \mathbf{%
\vdots \,\,}\prod_{j=1}^{m}\mathbf{E}\left( \mathbf{r},t_{j}\right) \,%
\mathrm{d}t_{j},\   \label{mx7}
\end{equation}%
with $\mathbf{R}^{\left( m\right) }$, $m=3,5,\ldots $, describing the medium
response. The convergence of the series and the reduction of the nonlinear
Maxwell equations to the operator form (\ref{MXshort}) are discussed in
detail in \cite{BF4}. We consider here the case when the series (\ref{Pser})
has only odd order terms that is typical for a medium with central symmetry
allowing though the dependence on $\mathbf{r}$ which may be not central
symmetric. The parameter $\alpha $ in (\ref{MXshort}) evidently determines
the relative magnitude of the nonlinearity.

We assume that $\alpha \ll 1$ considering consequently weakly nonlinear
phenomena. Note that if we rescale $\mathbf{U}$ and $\mathbf{J}$ in (\ref%
{MXshort}) by replacing $\mathbf{U}$ by $\xi \mathbf{U}$\ and $\mathbf{J}$
by $\xi \mathbf{J}$ \ with a scaling parameter $\xi $, we obtain the same
equation (\ref{MXshort}) with $\alpha $ replaced by $\xi ^{2}\alpha $.

Hence, \emph{taking small values for }$\alpha $\emph{\ is equivalent to
taking small amplitudes for the excitation current }$\mathbf{J}$ \ and all
three small parameters $\alpha $, $\beta $, and $\varrho $ are ultimately
introduced into the NLM\ through the choice of the excitation current $%
\mathbf{J}$.

As it was already mentioned, we assume that the excitation current $\mathbf{J%
}\left( \mathbf{r},t\right) $ is nonzero only in a finite time interval $%
\left[ 0,\frac{\tau _{0}}{\varrho }\right] $, i.e. 
\begin{equation}
\mathbf{J}\left( \mathbf{r},t\right) =0\ \text{if }t\leq 0\ \text{or }t\geq 
\frac{\tau _{0}}{\varrho }\text{, where }\tau _{0}>0\text{ is a small
constant,}  \label{Jeq0}
\end{equation}%
and consider the NLS regime after the current is switched off, i.e. for $%
t\geq \frac{\tau _{0}}{\varrho }$. We also assume that the time dependence
of the modal coefficients of the currents $\mathbf{J}\left( \mathbf{r}%
,t\right) $ is described by \emph{almost time-harmonic} functions, that is
functions of the following form 
\begin{equation}
a\left( t\right) =a_{\varrho }\left( t\right) =\mathrm{e}^{-\mathrm{i}\omega
_{0}t}\psi \left( \varrho t\right) \text{ where }\psi \left( \tau \right) =0%
\text{ for }\tau \leq 0\text{ and }\tau \geq \tau _{0}.  \label{aa}
\end{equation}%
This type of time dependence corresponds to the well-known
slowly-varying-amplitude approximation, \cite{Bo}. Note that since we
prescribe this form to the excitation currents which are at our disposal and
not to the solutions, no approximation is made yet at this state. It turns
out that such currents in both linear and weakly nonlinear regimes generate
waves which also have almost time-harmonic amplitudes.

It was shown in \cite{BF1} and \cite{BF4} that the exact solution of (\ref%
{MXshort}) can be written in the form%
\begin{equation}
\mathbf{U}\left( \mathbf{r},t\right) =\mathbf{U}^{\left( 0\right) }\left( 
\mathbf{r},t\right) +\alpha \mathbf{U}^{\left( 1\right) }\left( \mathbf{r}%
,t\right) +O\left( \alpha ^{2}\right) ,\ 0\leq t\leq \frac{\tau _{\ast }}{%
\varrho },\text{ }\tau _{\ast }\gg \tau _{0}\text{ is a constant.}
\label{uMv1}
\end{equation}%
We remind that for any quantity $\xi $ the notation $O\left( \xi \right) $
stands for any quantity such that%
\begin{equation}
\left\vert O\left( \xi \right) \right\vert \leq C\left\vert \xi \right\vert 
\text{ where }C\text{ is a constant.}  \label{OO}
\end{equation}%
In (\ref{uMv1}) the term $\mathbf{U}^{\left( 0\right) }\left( t\right) $ is
the solution to the linear equation 
\begin{equation}
\partial _{t}\mathbf{U}^{\left( 0\right) }=\mathbf{-}\mathrm{i}\mathbf{MU}%
^{\left( 0\right) }-\mathbf{J}^{\left( 0\right) };\ \mathbf{U}^{\left(
0\right) }\left( t\right) =0\;\text{for }t\leq 0,  \label{MXlin}
\end{equation}%
obtained from (\ref{MXshort}) by setting there $\alpha =0$, and we refer to
this term as to the medium \emph{linear response}. The component $\mathbf{U}%
^{\left( 1\right) }\left( t\right) $ in (\ref{uMv1}), called the\ medium 
\emph{first nonlinear response (FNLR)}, is a solution of the linear equation
obtained by substitution of (\ref{uMv1}) into (\ref{MXshort}) with
consequent collection of terms proportional to $\alpha $, namely 
\begin{equation}
\partial _{t}\mathbf{U}^{\left( 1\right) }=\mathbf{-}\mathrm{i}\mathbf{MU}%
^{\left( 1\right) }+\mathcal{F}_{\text{NL}}\left( \mathbf{U}^{\left(
0\right) }\right) -\mathbf{J}^{\left( 1\right) };\ \mathbf{U}^{\left(
1\right) }\left( t\right) =0\;\text{for }t\leq 0.  \label{FNLR}
\end{equation}%
For the reader's convenience we collect in Table \ref{tabNLM} basic
quantities essential for the analysis of the NLM. 
\begin{table}[tbp] \centering%
\begin{tabular}{|l|r|}
\hline
\multicolumn{2}{|c|}{\textbf{Baisic quantities related to the NLM}} \\ \hline
EM wave, a solution to the NLM: 6-component vector field & $\mathbf{U}\left( 
\mathbf{r},t\right) $ \\ \hline
Excitation currents & $\mathbf{J}\left( \mathbf{r},t\right) $ \\ \hline
\multicolumn{2}{|c|}{\textbf{Linear part of the NLM}} \\ \hline
\begin{tabular}{l}
First-order Hermitian differential operator \\ 
with $1$-periodic coefficients%
\end{tabular}
& $\mathbf{M}$ \\ \hline
EM wave, a solution to the linear part & $\mathbf{U}^{\left( 0\right)
}\left( \mathbf{r},t\right) $ \\ \hline
Dispersion relations of $\mathbf{M}$, generic $2\pi $ -periodic functions & $%
\omega _{n}\left( \mathbf{\mathbf{k}}\right) $ \\ \hline
Floquet-Bloch eigenmodes of $\mathbf{M}$ & $\mathbf{\tilde{G}}_{\zeta
,n}\left( \mathbf{r},\mathbf{k}\right) $ \\ \hline
Modal coefficients of $\mathbf{U}\left( \mathbf{r},t\right) $ with respect
to $\mathbf{\tilde{G}}_{\zeta ,n}\left( \mathbf{r},\mathbf{k}\right) $ & $%
\tilde{U}_{\zeta ,n}\left( \mathbf{k},t\right) $ \\ \hline
Phase of the linear wave $\mathbf{U}^{\left( 0\right) }\left( \mathbf{r}%
,t\right) $ & $\zeta \omega _{n}\left( \mathbf{\mathbf{k}}\right) \frac{\tau 
}{\varrho }$ \\ \hline
\multicolumn{2}{|c|}{\textbf{Nonlinear part on the NLM}} \\ \hline
Tensorial nonlinearity & $\alpha \mathcal{F}_{\text{NL}}\left( \mathbf{U}%
\right) $ \\ \hline
\multicolumn{2}{|l|}{%
\begin{tabular}{l}
Cubic susceptibility tensor \\ 
$\mathbf{\chi }^{\left( 3\right) }\left( \mathbf{r};\zeta ^{\prime }\omega
_{n^{\prime }}\left( \mathbf{k}^{\prime }\right) ,\zeta ^{\prime \prime
}\omega _{n^{\prime \prime }}\left( \mathbf{k}^{\prime \prime }\right)
,\zeta ^{\prime \prime \prime }\omega _{n^{\prime \prime \prime }}\left( 
\mathbf{k}^{\prime \prime \prime }\right) \right) $%
\end{tabular}%
} \\ \hline
\multicolumn{2}{|l|}{%
\begin{tabular}{l}
Phase of cubic nonlinear interactions \\ 
$\left[ \zeta \omega _{n}\left( \mathbf{k}\right) -\zeta ^{\prime }\omega
_{n^{\prime }}\left( \mathbf{k}^{\prime }\right) -\zeta ^{\prime \prime
}\omega _{n^{\prime \prime }}\left( \mathbf{k}^{\prime \prime }\right)
-\zeta ^{\prime \prime \prime }\omega _{n^{\prime \prime \prime }}\left( 
\mathbf{k}^{\prime \prime \prime }\right) \right] \frac{\tau }{\varrho }$%
\end{tabular}%
} \\ \hline
\end{tabular}

\caption{The basic quantities needed for the NLM analysis. \label{tabNLM}}%
\end{table}%

\subsection{Sketch of nonlinear evolution essentials}

\subsubsection{\textbf{Magnitude, space and time scales}}

We study NLM-NLS approximations for the following time range 
\begin{equation}
\frac{\tau _{0}}{\varrho }<t<\frac{\tau _{\ast }}{\varrho },\;\frac{\tau
_{\ast }}{\varrho }\leq \frac{\alpha _{0}}{\alpha }\text{, where }\alpha
_{0},\ \tau _{0},\ \tau _{\ast }\text{ are constants.}  \label{rhoalph}
\end{equation}%
The constant $\alpha _{0}$ is related to the convergence of the series (\ref%
{uMv1}), and it is independent of the small parameters $\varrho $ and $\beta 
$. Observe that the relations (\ref{rhoalph}) imply that 
\begin{equation}
\alpha \leq \varrho \frac{\alpha _{0}}{\tau _{\ast }},\text{ in particular }%
\alpha \sim \varrho ^{\varkappa _{0}},\ \varkappa _{0}\geq 1.  \label{rhoal}
\end{equation}%
Our primary focus is on an important particular case of (\ref{rhoal}) when 
\begin{equation}
\alpha \sim \varrho ,\;\varkappa _{0}=1,  \label{aeqr}
\end{equation}%
and in Section 7 we discuss the wave evolution for longer time intervals.

The nonlinear evolution governed by the NLM naturally involves two time
scales related to $t$ and $\tau =\varrho t$ (see for details Subsection
5.2). The time (\emph{fast time})\emph{\ }$t$ is just the "real" time,
whereas the \emph{slow time} $\tau =\varrho t$ describes a typical time
scale for a noticeble nonlinear evolution as in the rescaled NLS (\ref{NLSy}%
) below. In other words, $\frac{1}{\varrho }$ is the time for which a
noticeble nonlinear evolution can occur. Recasting (\ref{rhoalph}) in terms
of the slow time $\tau $ we obtain%
\begin{equation}
\tau _{0}\leq \tau <\tau _{\ast }\text{, where }\tau _{0}<\tau _{\ast }\text{
are constants.}  \label{taurhoalph}
\end{equation}%
We study the nonlinear evolution of a wavepacket within a time interval $%
\tau _{0}\leq \tau =\varrho t\leq \tau _{\ast }$. We remind that the
excitation current $\mathbf{J}\left( \mathbf{r},t\right) $ vanishes outside
the intermal $0\leq t\leq \frac{\tau _{0}}{\varrho }$, and the focus is on
the wavepacket produced by $\mathbf{J}\left( \mathbf{r},t\right) $ for times 
$t>\frac{\tau _{0}}{\varrho }$. The lesser times $t\leq \frac{\tau _{0}}{%
\varrho }$, corresponding to transient regimes, are beyond the scope of our
studies.

Now we give a preliminary sketch of the NLS which approximates the NLM in
the one-dimensional case $d=1$. The NLS equation has the form 
\begin{equation}
\partial _{t}Z=-\mathrm{i}\gamma _{0}Z-\gamma _{1}\partial _{x}Z+\mathrm{i}%
\gamma _{2}\partial _{x}^{2}Z+\mathrm{i}\alpha q\left\vert Z\right\vert
^{2}Z,\;Z\left( x,t\right) |_{t=0}=h\left( \beta x\right)  \label{NLS1}
\end{equation}%
where 
\begin{equation}
\gamma _{0}=\omega _{n_{0}}\left( \mathbf{k}_{\ast }\right) ,\ \gamma
_{1}=\omega _{n_{0}}^{\prime }\left( \mathbf{k}_{\ast }\right) ,\ \gamma
_{2}=\frac{1}{2}\omega _{n_{0}}^{\prime \prime }\left( \mathbf{k}_{\ast
}\right) .  \label{gami}
\end{equation}%
Note that the spatial scale $\frac{1}{\beta }$ is explicitly introduced in
the initial condition for $Z$ in (\ref{NLS1}). In the rescaled variables 
\begin{equation}
\tau =\varrho t,\ y=\beta x,\ Z\left( x,t\right) =z\left( y,\tau \right) ,
\label{rescale}
\end{equation}%
the equation (\ref{NLS1}) turns into%
\begin{equation}
\partial _{\tau }z=-\mathrm{i}\frac{\gamma _{0}}{\varrho }z-\frac{\beta
\gamma _{1}}{\varrho }\partial _{y}z+\mathrm{i}\gamma _{2}\frac{\beta ^{2}}{%
\varrho }\partial _{y}^{2}z+\mathrm{i}\frac{\alpha }{\varrho }q\left\vert
z\right\vert ^{2}z,\;z\left( y,t\right) |_{\tau =0}=h\left( y\right) .
\label{NLSy}
\end{equation}%
Evidently, the coefficients of the NLS equation (\ref{NLSy}) explicitly
depend on the small parameters $\alpha $, $\varrho $ and $\beta $ whereas
the initial condition does not depend on them. Note that the terms $\frac{%
\beta \gamma _{1}}{\varrho }\partial _{y}z$ and $\mathrm{i}\frac{\gamma _{0}%
}{\varrho }z$, describing\ respectively the propagation of the wavepacket
with the group velocity $\frac{\beta \gamma _{1}}{\varrho }$ and time
oscillations at the frequency $\frac{\gamma _{0}}{\varrho }$, can be
eliminated by a standard change of variables yielding the following reduced
classical NLS%
\begin{equation}
\partial _{\tau }z=\mathrm{i}\gamma _{2}\frac{\beta ^{2}}{\varrho }\partial
_{y}^{2}z+\mathrm{i}\frac{\alpha }{\varrho }q\left\vert z\right\vert ^{2}z.
\label{NLSclas}
\end{equation}

\begin{figure}[tbph]
\scalebox{0.5}{\includegraphics[viewport=100 50 700 500,clip]{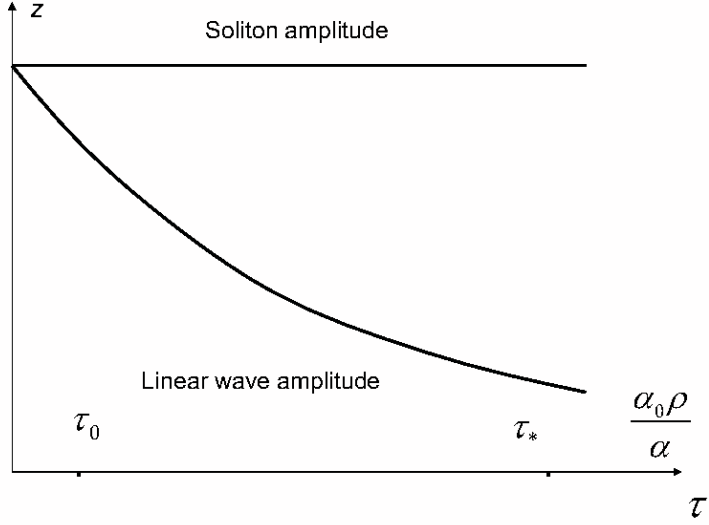}}
\caption{This plot shows a qualitative comparative picture of the wave
amplitude evolution for a linear medium versus a medium with a cubic
nonlinearity. If $\protect\varrho \sim \protect\alpha \sim \protect\beta %
^{2} $ the linear dispersion is exactly balanced by the nonlinearity in a
relevant soliton wave.}
\label{figsolt}
\end{figure}

Let us look now at the term $\mathrm{i}\gamma _{2}\frac{\beta ^{2}}{\varrho }%
\partial _{y}^{2}z$ in (\ref{NLSclas}) describing linear dispersive effect
and introduce the following parameter 
\begin{equation}
\theta =\frac{\varrho }{\beta ^{2}}  \label{theta}
\end{equation}%
to which we refer as the \emph{inverse dispersion parameter} since it
determines the magnitude of the linear dispersion effects. \emph{It is well
known that the ultimate magnitude of nonlinear effects is essentially
determined by an interplay between nonlinearity caused by sufficiently large
wave amplitudes and the linear wave dispersion causing a reduction of the
wave amplitude.} In particular,%
\begin{eqnarray}
\text{if }\theta ^{-1} &\ll &1\text{ the dispersive effects are weaker, }
\label{thedis} \\
\text{if }\theta ^{-1} &\gg &1\text{ the dispersive effects are stronger.} 
\notag
\end{eqnarray}%
The significance of the inverse dispersion parameter $\theta $ is also
supported by our analysis of the error of the NLM-NLS approximation. The
dispersive effects already show themselves when $\theta $ is fixed and
bounded uniformly in $\beta $ and $\varrho $. Indeeed, in the linear case $%
\alpha =0$, the dispersion causes a reduction of the wave amplitude
approximaely at the rate $\left( 1+\frac{\gamma _{2}\tau }{\theta }\right)
^{-\frac{1}{2}}$ as the slow time $\tau $ increases. In contrast, in the
nonlinear case $\alpha \neq 0$ the wave amplude does not fall with time as
in the linear case under assumption that $\frac{\varrho }{\alpha }$ is
bounded, indicating a signiciant nonlinear effect on the wave evolution. In
particular, if 
\begin{equation}
\alpha \sim \varrho \sim \beta ^{2},  \label{thet1alph}
\end{equation}%
the NLS (\ref{NLSclas}) has soliton solutions with amplitudes that do not
fall as $\tau $ increases. A qualitative comparative picture of the wave
amplitude evolutions for a linear medium versus a nonlinear one is shown in
Fig. \ref{figsolt}, which indicates, in particular, that for \emph{for time
ranges as in (\ref{taurhoalph}) and under conditions (\ref{thet1alph}) the
wave evolution shows significant nonlinear effects.}

Note that the effect of the nonlinearity is already significant when the
fraction $\frac{\tau _{0}}{\tau _{\ast }}<1$ in (\ref{rhoalph}) is a fixed
number and it does not have to be infinitesimally small.

A closer look at the classical NLS\ equation (\ref{NLSclas}) shows \emph{%
that if the small parameters vary so that} 
\begin{equation}
\frac{\beta ^{2}}{\varrho }=\theta ^{-1}=\limfunc{Const},\ \frac{\alpha }{%
\varrho }=\limfunc{Const}  \label{NLSinv}
\end{equation}%
its form is essentially preserved. We refer to the relations (\ref{NLSinv})
as \emph{classical NLS scaling}. Notice that the condition (\ref{thet1alph})
is an equivalent form of the classical NLS scaling. In particular, from the
linear wave dispersion point of view the classical scaling is the marginal
case when the inverse dispersion parameter $\theta $\ is neither
infinitesimally small nor large but rather it is finite. \emph{Existence of
the solitons manifests the balance between dispersion and nonlinearity }%
reached at the classical NLS scaling (\ref{thet1alph}).

As to the further analysis of the interplay of the linear dispersion with
the nonlinearity we consider two cases: (i) $\theta \rightarrow 0$; (ii) $%
\theta \geq \theta _{0}>0$. The first case as $\theta \rightarrow 0$,
corresponds to stronger dispersion effects, and it can be characterised more
accurately by the inequality 
\begin{equation}
\left\Vert \omega _{n_{0}}^{\prime \prime }\left( \mathbf{k}_{\ast }\right)
^{-1}\right\Vert \theta =\left\Vert \omega _{n_{0}}^{\prime \prime }\left( 
\mathbf{k}_{\ast }\right) ^{-1}\right\Vert \frac{\varrho }{\beta ^{2}}\ll 1,
\label{nonM}
\end{equation}%
where $\omega _{n_{0}}^{\prime \prime }\left( \mathbf{k}_{\ast }\right) $ in
the multidimensional case $d>1$ is the matrix of the second differential of $%
\omega _{n_{0}}\left( \mathbf{k}\right) $ at $\mathbf{k}_{\ast }$ and the
symbol $\left\Vert \cdot \right\Vert $ stands for the matrix norm. Notice
that in the case $d=1$ the expression $\omega _{n_{0}}^{\prime \prime
}\left( \mathbf{k}_{\ast }\right) $ is just the second derivative and $%
\left\Vert \omega _{n_{0}}^{\prime \prime }\left( \mathbf{k}_{\ast }\right)
\right\Vert $ is its absolute value, implying $\left\Vert \omega
_{n_{0}}^{\prime \prime }\left( \mathbf{k}_{\ast }\right) ^{-1}\right\Vert
=\left\Vert \omega _{n_{0}}^{\prime \prime }\left( \mathbf{k}_{\ast }\right)
\right\Vert ^{-1}$. In view of (\ref{thedis}), we refer to the case
described by (\ref{nonM}) as to \emph{dispersive case}. In particular, the
dispersive case takes place if 
\begin{equation}
\varrho \sim \beta ^{\varkappa _{1}},\ \varkappa _{1}>2.  \label{kap1}
\end{equation}%
The other case, $\theta \geq \theta _{0}>0$, occurs if%
\begin{equation}
\frac{\theta }{\left\Vert \omega _{n_{0}}^{\prime \prime }\left( \mathbf{k}%
_{\ast }\right) \right\Vert }=\frac{\varrho }{\left\Vert \omega
_{n_{0}}^{\prime \prime }\left( \mathbf{k}_{\ast }\right) \right\Vert \beta
^{2}}\gg 1\text{\ or\ }\frac{\varrho }{\left\Vert \omega _{n_{0}}^{\prime
\prime }\left( \mathbf{k}_{\ast }\right) \right\Vert \beta ^{2}}\sim 1.
\label{nonM1}
\end{equation}%
Again, in view of (\ref{thedis}), we refer to the case described by (\ref%
{nonM1}) as to \emph{weakly dispersive}. In particular, the weakly
dispersive case takes place if%
\begin{equation}
\varrho \sim \beta ^{\varkappa _{1}},\ 2\geq \varkappa _{1}>0.  \label{kap11}
\end{equation}%
Notice that the classical NLS scaling is covered by the second alternative
condition of the weakly dispersive case (\ref{nonM1}) and (\ref{kap11}),
namely 
\begin{equation}
\frac{\varrho }{\left\vert \omega _{n_{0}}^{\prime \prime }\left( \mathbf{k}%
_{\ast }\right) \right\vert \beta ^{2}}\sim 1,\;\left\Vert \omega
_{n_{0}}^{\prime \prime }\left( \mathbf{k}_{\ast }\right) ^{-1}\right\Vert 
\frac{\varrho }{\beta ^{2}}\sim 1,\;\varkappa _{1}=2.  \label{border}
\end{equation}%
In the case (\ref{border}) the dispersive effects are already significant as
one can see from Fig. \ref{figsolt}.

For the both dispersive and weakly dispersive cases we get the same set of
NLS or ENLS\ equations, but the properties of the equations are different in
different ranges of the parameters. Mathematical techniques used to study
them are different as well. Namely, in the dispersive case of (\ref{nonM})
we apply the Stationary Phase Method to the FNLR. The weakly dispersive case
(\ref{nonM1}) is technically simpler and is studied based on the Taylor
expansion of relevant oscillatory phases. Remarkably, in both cases the
dynamics of the directly excited modes is explicitly expressed in terms of a
solution of the same NLS.

In addition to the above conditions, we assume that $\varrho $ is small
enough to provide%
\begin{equation}
\frac{\omega _{n_{0}}\left( \mathbf{k}_{\ast }\right) }{\varrho }\gg 1,
\label{FrM}
\end{equation}%
where factor $\omega _{n_{0}}\left( \mathbf{k}_{\ast }\right) $ signifies
the relevance of the Frequency Matching condition. We also assume the
following condition 
\begin{equation}
\frac{\left\vert \omega _{n_{0}}^{\prime }\left( \mathbf{k}_{\ast }\right)
\right\vert \beta }{\varrho }\gg 1,  \label{rho/beta}
\end{equation}%
which allows to use the group velocity for the analysis of wave
interactions. Note though that the conditions (\ref{FrM}) and (\ref{rho/beta}%
) are not always necessary.

\subsubsection{Relation between the NLM and the NLS}

Observe that the excitation current $\mathbf{J}$ determines uniquely the
solution of the NLM equation (\ref{MXshort}) whereas the initial data $h$
determines the solution of the NLS (\ref{NLS1}). Consequently, if we want to
select regimes of the NLM which are well approximated by solutions of the
NLS we have (i) to construct the NLS, in other words to determine its
coefficients, based on the NLM; (ii) \ to describe the correspondence
between $\mathbf{J}$ and $h$. It turns out that the current $\mathbf{J}$
should be of the form 
\begin{equation}
\mathbf{J}\left( \mathbf{r},t\right) =\mathbf{J}^{\left( 0\right) }\left( 
\mathbf{r},t\right) +\alpha \mathbf{J}^{\left( 1\right) }\left( \mathbf{r}%
,t\right) ,\ \mathbf{J}^{\left( j\right) }\left( \mathbf{r},t\right) =0\ 
\text{if }t\leq 0\ \text{or }t\geq \frac{\tau _{0}}{\varrho },\ j=0,1,
\label{J01}
\end{equation}%
where the principal part $\mathbf{J}^{\left( 0\right) }$ and the corrective
part $\mathbf{J}^{\left( 1\right) }$ of the current $\mathbf{J}$ are
properly selected to produce a NLS-type regime (see Subsections 2.1 and 5.2
for details). Notice that the current $\mathbf{J}$ substitutes for the
initial data for the NLM and is based on the initial data of the NLS. As it
was explained in \cite{BF1}, the introduction of the excitation current $%
\mathbf{J}$ is both mathematically and physically a more suitable\ option
for the NLM having nonlinear polarization of the form (\ref{mx7}), since a
prescription of instantaneous initial data for $t=0$ is inconsistent with
the form of the nonlinearity (\ref{mx7}) which requires to know the fields
at all times. The standard classical NLS though assumes the prescription of
an instantaneous initial data at $t=0$. We overcome this difference by
setting a proper form for the current $\mathbf{J}\left( \mathbf{r},t\right) $
including, in particular, its composition of the form (\ref{J01}). To
produce NLS-type regimes the current $\mathbf{J}$ has to possess two
properties. Firstly, it should be almost time-harmonic, as in (\ref{aa})
with a deviation from time-harmonicity measured by a small parameter $%
\varrho $ which consequently determines the ratio of the slow time and the
fast time scales. Secondly, following the framework described in \cite{BF1}--%
\cite{BF4} we choose the excitation current composed of Bloch modes from a
single spectral band, described by an index $n=n_{0}$, and with the
quasimomenta $\mathbf{k}$ from a small $\beta $-vicinity $\left\vert \mathbf{%
k}\pm \mathbf{k}_{\ast }\right\vert =O\left( \beta \right) $ of a fixed
quasimomentum $\pm \mathbf{k}_{\ast }$ in the Brillouin zone. In the case of
the NLS the parameter $\beta $ is introduced through its initial data $%
h\left( \beta \mathbf{r}\right) $ at $t=0$. Then we provide an explicit
construction of the excitation current $\mathbf{J}\left( \mathbf{r},t\right) 
$ based on the prescribed initial data $h\left( \beta \mathbf{r}\right) $
for the NLS. Notice that $\beta ^{-1}$ determines the length scale for $%
\mathbf{J}\left( \mathbf{r},t\right) $. Thus, the both parameters $\varrho $
and $\beta $ are introduced into the NLM\ via the excitation current $%
\mathbf{J}\left( \mathbf{r},t\right) $. Then we study the solution $\mathbf{U%
}\left( \mathbf{r},t\right) $ for times $t\geq \frac{\tau _{0}}{\varrho }$
when $\mathbf{J}\left( \mathbf{r},t\right) =0$. 
\begin{table}[tbp] \centering%
\begin{tabular}{|l|r|}
\hline
\multicolumn{2}{|c|}{\textbf{Basic quantities related to the NLS}} \\ \hline
Wave, a solution to the NLS: scalar function & $Z_{\zeta }\left( \mathbf{r}%
,t\right) $ \\ \hline
Initial data & $h\left( \beta r\right) $ \\ \hline
\multicolumn{2}{|c|}{\textbf{Linear part of the NLS}} \\ \hline
Second-order Hermitian differential operator & $\gamma _{\left( 2\right)
}\left( -\mathrm{i}\partial _{r}\right) $ \\ \hline
Wave, a solution to the linear part & $Z_{\zeta }^{\left( 0\right) }\left( 
\mathbf{r},t\right) $ \\ \hline
Dispersion relations of $\gamma _{\left( 2\right) }\left( -\mathrm{i}%
\partial _{r}\right) $, a polynomial & $\gamma _{\left( 2\right) }\left( 
\mathbf{\xi }\right) $ \\ \hline
Fourier \ eigenmodes of $\gamma _{\left( 2\right) }\left( -\mathrm{i}%
\partial _{r}\right) $: exponentials & $e^{i\mathbf{r\cdot \xi }}$ \\ \hline
Modal coefficients of $Z_{\zeta }\left( \mathbf{r},t\right) $ with respect
to $e^{i\mathbf{r\cdot \xi }}$ & $\hat{Z}_{\zeta }\left( \mathbf{\xi }%
,t\right) $ \\ \hline
Phase of the linear wave $Z_{\zeta }^{\left( 0\right) }\left( \mathbf{r}%
,t\right) $ & $\gamma _{\left( 2\right) }\left( \mathbf{\xi }\right) \frac{%
\tau }{\varrho }$ \\ \hline
\multicolumn{2}{|c|}{\textbf{Nonlinear part on the NLS}} \\ \hline
Scalar nonlinearity & $\alpha Q_{\zeta }\left\vert Z\right\vert ^{2}Z$ \\ 
\hline
\multicolumn{2}{|l|}{%
\begin{tabular}{l}
Phase of nonlinear interactions \\ 
$\left[ \gamma _{\left( 2\right) }\left( \zeta \beta \mathbf{\xi }\right)
-\gamma _{\left( 2\right) }\left( \zeta \beta \mathbf{\xi }^{\prime }\right)
-\gamma _{\left( 2\right) }\left( \zeta \beta \mathbf{\xi }^{\prime \prime
}\right) +\gamma _{\left( 2\right) }\left( -\zeta \beta \mathbf{\xi }%
^{\prime \prime \prime }\right) \right] \frac{\tau }{\varrho }$%
\end{tabular}%
} \\ \hline
\end{tabular}

\caption{The basic quantities needed for the NLS analysis. \label{tabNLS}}%
\end{table}
Let us consider now an outline of the analysis of the approximation\ of the
NLM by the NLS in the simpler one-dimensional case $d=1$, i.e. when the
medium coefficients and solutions of (\ref{MXshort}) depend only on the
coordinate $r_{1}=x$ and do not depend on the remaining coordinates $r_{2}$
and $r_{3}$. The first important observation is that the NLM evolution
reduces approximately to an NLS regime if the excitation current $\mathbf{J}%
\left( \mathbf{r},t\right) $ and, consequently, the wavepacket $\mathbf{U}%
\left( \mathbf{r},t\right) $ are composed of eigenmodes from a single band $%
n_{0}$ with $\mathbf{k}$ from a small vicinity of quasimomenta $\pm \mathbf{k%
}_{\ast }$. The reason for having two quasimomenta $\pm \mathbf{k}_{\ast }$
(a doublet) rather than just one $\mathbf{k}_{\ast }$ is that it is the
minimal set of quasimomenta producing a\ real-valued $\mathbf{U}\left( 
\mathbf{r},t\right) $. The next logical step is to introduce two scalar
functions $Z_{\pm }\left( \mathbf{r},t\right) $ approximating the two
relevant modal coefficients in the compostion of the solution $\mathbf{U}%
\left( \mathbf{r},t\right) $ to the NLM. These two intimately related scalar
functions $Z_{\pm }\left( \mathbf{r},t\right) $ satisfy two related NLS
equations \ 
\begin{gather}
\partial _{t}Z_{+}=-\mathrm{i}\gamma _{0}Z_{+}-\gamma _{1}\partial _{x}Z_{+}+%
\mathrm{i}\gamma _{2}\partial _{x}^{2}Z_{+}+\alpha _{\pi
}Q_{+}Z_{-}Z_{+}^{2},  \label{Si} \\
Z_{+}\left( x,t\right) |_{t=0}=h_{+}\left( \beta x\right) ,\ \alpha _{\pi
}=3\alpha \left( 2\pi \right) ^{2},  \notag
\end{gather}%
\begin{gather}
\partial _{t}Z_{-}=\mathrm{i}\gamma _{0}Z_{-}-\gamma _{1}\partial _{x}Z_{-}-%
\mathrm{i}\gamma _{2}\partial _{x}^{2}Z_{-}+\alpha _{\pi
}Q_{-}Z_{-}^{2}Z_{+},  \label{Si1} \\
Z_{-}\left( x,t\right) |_{t=0}=h_{-}\left( \beta x\right) ,\ h_{-}\left(
\beta x\right) =h_{+}^{\ast }\left( \beta x\right) ,  \notag
\end{gather}%
with the asterisk denoting the complex conjugation. In (\ref{Si}) and (\ref%
{Si1}) the coefficeints $\gamma _{0},\gamma _{1},\gamma _{2}$ satisfy (\ref%
{gami}) and $h_{+}\left( x\right) $ is a smooth function decaying
sufficiently fast as $x\rightarrow \infty $. The function $h_{+}\left(
x\right) $ can be chosen as we please. The coefficients $Q_{\pm }$ in (\ref%
{Si}) and (\ref{Si1}) are certain complex valued numbers related to the the
third-order susceptibility tensor associated with the cubic nonlinearity $%
\mathcal{F}_{\text{NL}}\left( \mathbf{U}\right) $. We do not impose any
conditions on the structure of the cubic tensor in $\mathcal{F}_{\text{NL}%
}\left( \mathbf{U}\right) $, which affects only the values of the
coefficients $Q_{\pm }$ in (\ref{Si}) and (\ref{Si1}). With no structural
conditions imposed on the tensors related to the nonlinearity, the complex
coefficients $Q_{\pm }$ may be such that $Q_{+}\neq Q_{-}^{\ast }$. In the
latter case $Z_{-}\left( x,t\right) $ might be different from $Z_{+}^{\ast
}\left( x,t\right) $. \emph{Though in the case when the nonlinearity maps
real-valued fields into real-valued and (\ref{eer2}) holds we always have }%
\begin{equation}
Z_{-}\left( x,t\right) =Z_{+}^{\ast }\left( x,t\right) ,\emph{\ }%
Z_{-}Z_{+}^{2}=\left\vert Z_{+}\right\vert ^{2}Z_{+}  \label{conjug}
\end{equation}%
\emph{\ and the system (\ref{Si}) and (\ref{Si1}) effectively is reduced to
a single scalar equation (\ref{Si}):}%
\begin{equation}
\partial _{t}Z=-\mathrm{i}\gamma _{0}Z-\gamma _{1}\partial _{x}Z+\mathrm{i}%
\gamma _{2}\partial _{x}^{2}Z+\alpha _{\pi }Q_{+}\left\vert Z\right\vert
^{2}Z,\emph{\ }Z\left( x,t\right) |_{t=0}=h_{+}\left( \beta x\right) .
\label{Sireal}
\end{equation}%
The two functions $Z_{\pm }\left( x,t\right) $ satisfying the NLS equations (%
\ref{Si}) and (\ref{Si1}) yield an approximation $\mathbf{U}_{Z}\left( 
\mathbf{r},t\right) $ to the exact solution $\mathbf{U}\left( \mathbf{r}%
,t\right) $ of the NLM. An analysis of the approximate solution $\mathbf{U}%
_{Z}\left( \mathbf{r},t\right) $ leads to a natural partition of modes
involved in its composition into two groups: "directly" and "indirectly"
excited modes, and it suggests splitting of the approximate solution into
two parts 
\begin{equation}
\mathbf{U}_{Z}\left( \mathbf{r},t\right) =\mathbf{U}_{Z}^{\text{dir}}\left( 
\mathbf{r},t\right) +\mathbf{U}_{Z}^{\text{ind}}\left( \mathbf{r},t\right) .
\label{UZdirind}
\end{equation}%
The \emph{directly excited modes} which contribute to $\mathbf{U}_{Z}^{\text{%
dir}}$ are the ones presented in the excitation current $\mathbf{J}$ and
excited through the linear mechanism, i.e. when $\alpha =0$, whereas \emph{%
indirectly excited modes} which form $\mathbf{U}_{Z}^{\text{ind}}$ are
excited only through the nonlinearity and, consequently, for $\alpha =0$
their amplitudes are zero (see Section 3 for details). The modal
coefficients of the indirectly excited modes are much smaller compared to
ones related to the directly excited modes, therefore $\mathbf{U}_{Z}^{\text{%
ind}}$ is much smaller than $\mathbf{U}_{Z}^{\text{dir}}$, namely%
\begin{equation}
\left\vert \mathbf{U}_{Z}^{\text{ind}}\right\vert =O\left( \alpha \right)
O\left( \left\vert \mathbf{U}_{Z}^{\text{dir}}\right\vert \right) .
\end{equation}%
It turns out that high precision approximations for the modal coefficients
of the indirectly excited modes are based only the FNLR, and, consequently,
are expressed in terms of the excitation currents and do need the NLS. \emph{%
In contrast, approximations of the same accuracy for directly excited modes
are ultimately reduced to relevant NLS's which account for nonlinear
self-interactions of these modes.}

The directly excited part of the approximate solution $\mathbf{U}_{Z}$ has
the following form in the space domain:%
\begin{gather}
\mathbf{U}_{Z}^{\text{dir}}\left( \mathbf{r},t\right) =  \label{Uphys} \\
\mathbf{\tilde{G}}_{+,n_{0}}\left( \mathbf{r},\mathbf{k}_{\ast }\right)
Z_{+}\left( \mathbf{r},t\right) +\mathbf{\tilde{G}}_{-,n_{0}}\left( \mathbf{r%
},\mathbf{k}_{\ast }\right) Z_{-}\left( \mathbf{r},t\right) +\beta \mathbf{U}%
_{Z,n_{0}}^{1}\left( \mathbf{r},t\right) +O\left( \beta ^{2}\right) O\left(
\left\vert \mathbf{U}_{Z}\right\vert \right) ,  \notag
\end{gather}%
with $\mathbf{\tilde{G}}_{\zeta ,n_{0}}\left( \mathbf{r},\mathbf{k}_{\ast
}\right) $, $\zeta =\pm $,$\mathbf{\ }$ being Bloch eigenmodes of the linear
Maxwell operator $\mathbf{M}$. We remind that%
\begin{equation}
\mathbf{\tilde{G}}_{\zeta ,n_{0}}\left( \mathbf{r},\mathbf{k}_{\ast }\right)
=\mathrm{e}^{\mathrm{i}\mathbf{k}_{\ast }\cdot \mathbf{r}}\mathbf{\hat{G}}%
_{\zeta ,n_{0}}\left( \mathbf{r},\mathbf{k}_{\ast }\right) \text{ where }%
\mathbf{\hat{G}}_{\zeta ,n_{0}}\left( \mathbf{r},\mathbf{k}_{\ast }\right) 
\text{ is periodic in }\mathbf{r}\text{.}  \label{Ghat}
\end{equation}%
The next order correction $\mathbf{U}_{Z}^{1}\left( \mathbf{r},t\right) $ in
this representation in the one-dimensional case when $\mathbf{r}=x$ is given
by the following formula (see Subsection 5.5 for the general case of the
space dimesions 2 and 3) 
\begin{equation}
\mathbf{U}_{Z}^{1}\left( \mathbf{r},t\right) =\mathbf{U}_{Z_{+}}^{1}\left(
x,t\right) +\mathbf{U}_{Z_{-}}^{1}\left( x,t\right) ,
\end{equation}%
\begin{eqnarray}
\mathbf{U}_{Z_{+}}^{1}\left( x,t\right) &=&-\mathrm{ie}^{\mathrm{i}k_{\ast
}\cdot x}\beta ^{-1}\partial _{x}Z_{+}\left( x,t\right) \partial _{k}\mathbf{%
\hat{G}}_{+,n_{0}}\left( x,k_{\ast }\right) ,  \label{UZ11} \\
\mathbf{U}_{Z_{-}}^{1}\left( x,t\right) &=&\mathrm{ie}^{-\mathrm{i}k_{\ast
}\cdot x}\beta ^{-1}\partial _{x}Z_{-}\left( x,t\right) \partial _{k}\mathbf{%
\hat{G}}_{-,n_{0}}\left( x,k_{\ast }\right) .  \notag
\end{eqnarray}%
This correction reflects finer effects of the periodicity of the medium and
it is present even in the linear case when $\alpha =0$. The terms $\beta
^{-1}\partial _{x}Z_{\pm }\left( x,t\right) $ in (\ref{UZ11}) are bounded
for small $\beta $ because \ after the rescaling (\ref{rescale}) $\beta
^{-1}\partial _{x}Z_{\pm }\left( x,t\right) $ equals $\partial _{y}z_{\pm
}\left( y,t\right) $ \ where $z_{\pm }\left( y,t\right) $ \ solve equations
of the form (\ref{NLSy}). Note also that the approximate expression (\ref%
{Uphys}) is not an ansatz, it is a consequence of the exact formula (\ref%
{UNLS}) written in the next subsection in terms of the Floquet-Bloch
transform. Note that according to (\ref{Uphys}) and (\ref{Ghat}) the
quasimomentum $\mathbf{k}_{\ast }$ describes the phase shift of the carrier
wave over the period cell.

The difference between the approximate solution $\mathbf{U}_{Z}\left( 
\mathbf{r},t\right) $, based on the NLS equations (\ref{Si}), (\ref{Si1})
and the exact solution $\mathbf{U}\left( \mathbf{r},t\right) $ is called the 
\emph{approximation error}. Using the modal decomposition and analytic
methods developed in \cite{BF1}- \cite{BF4} we proved the following estimate
for the approximation error:%
\begin{equation}
\mathbf{U}\left( \mathbf{r},t\right) -\mathbf{U}_{Z}\left( \mathbf{r}%
,t\right) =\left[ O\left( \alpha ^{2}\right) +O\left( \alpha \beta \right)
+O\left( \alpha \varrho \right) \right] O\left( \left\vert \mathbf{U}%
^{\left( 1\right) }\right\vert \right) ,  \label{UUZ}
\end{equation}%
with the symbol $O\left( \xi \right) $ defined by (\ref{OO}). When $\mathbf{U%
}=\mathbf{U}\left( \mathbf{r},t\right) $ is a function of $\mathbf{r},t$ we
write $O\left( \left\vert \mathbf{U}\right\vert \right) $ for a function of $%
\mathbf{r},t$ such that it is bounded in some sense when $\mathbf{U}$ is
bounded, assuming that $O\left( \left\vert \mathbf{U}\right\vert \right) $
is homogenious in $\mathbf{U}$\textbf{\ (}in particular $O\left( \left\vert
\beta ^{q}\mathbf{U}\right\vert \right) =\beta ^{q}O\left( \left\vert 
\mathbf{U}\right\vert \right) $). We do not want to elaborate and get more
specific on the definition of $O\left( \left\vert \mathbf{U}\right\vert
\right) $\ since a mathematically rigorous discussion of this subject would
require to introduce concepts and technicalities that though are important
for a mathematical justification, but are not essential for presenting the
results of our analysis.

The approximation error $\mathbf{U}\left( \mathbf{r},t\right) -\mathbf{U}%
_{Z}\left( \mathbf{r},t\right) $ can be reduced by adding certain corrective
terms to the NLS (\ref{Si}), (\ref{Si1}). We call such equations with added
corrective terms \emph{extended NLS equations (ENLS)} (see Subsection 1.3
for details). The simplest extended NLS have corrective terms of the form $%
\partial _{x}^{3}Z_{+}$, $Z_{+}^{\ast }\partial _{x}Z_{+}^{2}$ and $%
Z_{+}^{2}\partial _{x}Z_{+}^{\ast }$ with calculable coefficients $Q_{1,\pm
} $, and they are as follows 
\begin{gather}
\partial _{t}Z_{+}=-\mathrm{i}\gamma _{0}Z_{+}-\gamma _{1}\partial _{x}Z_{+}+%
\mathrm{i}\gamma _{2}\partial _{x}^{2}Z_{+}+\gamma _{3}\partial
_{x}^{3}Z_{+}+  \label{Six} \\
\alpha _{\pi }\left[ Q_{+}Z_{+}^{2}Z_{-}+Q_{1,+}Z_{+}Z_{-}\partial
_{x}Z_{+}+Q_{1,\ast ,+}Z_{+}^{2}\partial _{x}Z_{-}\right] ,\;Z_{+}\left(
x,t\right) |_{t=0}=h_{+}\left( \beta x\right) ,  \notag
\end{gather}%
\begin{gather}
\partial _{t}Z_{-}=\mathrm{i}\gamma _{0}Z_{-}-\gamma _{1}\partial _{x}Z_{-}-%
\mathrm{i}\gamma _{2}\partial _{x}^{2}Z_{-}+\gamma _{3}\partial
_{x}^{3}Z_{-}+  \label{Six1} \\
\alpha _{\pi }\left[ Q_{-}Z_{+}Z_{-}^{2}+Q_{1,-}Z_{+}Z_{-}\partial
_{x}Z_{-}+Q_{1,\ast ,-}Z_{-}^{2}\partial _{x}Z_{+}\right] ,\;Z_{-}\left(
x,t\right) |_{t=0}=h_{+}^{\ast }\left( \beta x\right) .  \notag
\end{gather}%
The coefficients $Q_{1,\pm }$, $Q_{1,\ast ,\pm }$ in (\ref{Six}), (\ref{Six1}%
) take into account the dependence of the susceptibility and Bloch
eigenfunctions on the Bloch spectral variable $\mathbf{k}$ (the
quasimomentum) which are neglected in the standard NLS (\ref{Si}), (\ref{Si1}%
). In the real-valued case we use (\ref{conjug}) to reduce two equations (%
\ref{Six}), (\ref{Six1}) to one equation (\ref{Six}). In Table \ref{tabAdd1}
we list additional terms of the order of $\beta $ showing their relations to
the NLM. 
\begin{table}[tbp] \centering%
\begin{tabular}{|c|c|}
\hline
\multicolumn{2}{|c|}{%
\begin{tabular}{c}
Additional terms of the order $\beta $ in the ENLS improving the accuracy \\ 
of the NLM-NLS approximation%
\end{tabular}%
} \\ \hline
Source in the NLM & Term in the ENLS \\ \hline
\multicolumn{1}{|l|}{Dispersion relation} & $\beta \gamma _{3}\partial
_{y}^{3}z$ \\ \hline
\multicolumn{1}{|l|}{Susceptibility} & $\beta q_{1}\left\vert z\right\vert
^{2}\partial _{y}z+\beta q_{1,\ast ,+}z^{2}\partial _{y}z^{\ast }$ \\ \hline
\end{tabular}%
\newline
\caption{The list of additional terms in the ENLS improving the accuracy of
the NLM-NLS approximation and showing their origin from the NLM.
\label{tabAdd1}}%
\end{table}
Note that change of variables $\beta x=y$ transforms (\ref{Six})\ into a
form similar to (\ref{NLSy}), namely%
\begin{gather}
\partial _{\tau }z=-\mathrm{i}\frac{\gamma _{0}}{\varrho }z-\frac{\beta
\gamma _{1}}{\varrho }\partial _{y}z+\frac{\beta ^{2}}{\varrho }\left[ 
\mathrm{i}\gamma _{2}\partial _{y}^{2}z+\beta \gamma _{3}\partial _{y}^{3}z%
\right] +  \label{Sixy} \\
\frac{\alpha }{\varrho }\left[ \mathrm{i}q_{0}\left\vert z\right\vert
^{2}z+\beta q_{1}\left\vert z\right\vert ^{2}\partial _{y}z+\beta q_{1,\ast
,+}z^{2}\partial _{y}z^{\ast }\right] ,\;z\left( y,t\right) |_{\tau
=0}=h\left( y\right) .  \notag
\end{gather}%
If $Z_{\pm }$ are solutions to the ENLS (\ref{Six}), (\ref{Six1}), then the
approximate solution $\mathbf{U}_{Z}$ of the NLM\ given by (\ref{Uphys})
provides a better approximation of $\mathbf{U}$, than $\mathbf{U}_{Z}$ based
on $Z_{\pm }$ which are solutions to the standard NLS (\ref{Si}), (\ref{Si1}%
), namely 
\begin{equation}
\mathbf{U}\left( \mathbf{r},t\right) -\mathbf{U}_{Z}\left( \mathbf{r}%
,t\right) =\left[ O\left( \alpha ^{2}\right) +O\left( \alpha \beta
^{2}\right) +O\left( \alpha \varrho \right) \right] O\left( \left\vert 
\mathbf{U}^{\left( 1\right) }\right\vert \right) .  \label{UUZx}
\end{equation}%
A comparison of the estimates (\ref{UUZx}) and (\ref{UUZ}) indicates that
the introduction of the corrective terms into the ENLS improves the accuracy
of the aproximation, namely the error term $O\left( \alpha \beta \right) $
is replaced by a smaller $O\left( \alpha \beta ^{2}\right) $. Evidently that
is a significant improvement in the dispersive case $\varrho \ll \beta ^{2}$
or $\varrho \sim \beta ^{2}$. Such a refinement of the approximation is
possible due to the specific form of matching between solutions of the NLS
and the NLM which is described in the next subsection, see (\ref{UNLS}). In
Subsection 1.3 we consider ENLS\ having more corrective terms and yielding
even better approximations.

Another way to construct approximate solutions of NLM is by using not the
differential equations of the form of NLS or ENLS, but rather
finite-difference lattice equations, see Section 9 for details. In the
one-dimensional case the lattice counterpart of the equation (\ref{Sireal})
is as follows 
\begin{gather}
\partial _{t}Z_{+}\left( m\right) =-\mathrm{i}\left( \gamma _{0}+\gamma
_{2}\right) Z_{+}\left( m\right) -\gamma _{1}\left( \frac{1}{2}\left[
Z_{+}\left( m+1\right) -Z_{+}\left( m-1\right) \right] \right)
\label{eqLatIntr} \\
+\mathrm{i}\frac{\gamma _{2}}{2}\left[ Z_{+}\left( m+1\right) +Z_{+}\left(
m-1\right) \right] +\alpha _{\pi }Q_{+}\left\vert Z_{+}\left( m\right)
\right\vert ^{2}Z_{+}\left( m\right) ,  \notag \\
Z_{+}\left( m,t\right) |_{t=0}=h_{+}\left( \beta m\right) ,\ \alpha _{\pi
}=3\alpha \left( 2\pi \right) ^{2},\;m=\ldots -1,0,1,2,\ldots .  \notag
\end{gather}%
The lattice equation (\ref{eqLatIntr}) is obtained by a direct approximation
of the NLM, and it is not a finite-difference approximation of the NLS
equation (\ref{Si}). Technically, approximations of dispersion relations by
algebraic polynomials yield differential operators whereas approximations by
trigonometric polynomials yield finite-difference lattice operators. Instead
of (\ref{Uphys}) a similar formula holds with the same leading term, see (%
\ref{UZLr})-(\ref{UZLr2}) for details. \emph{The accuracy of the
approximation of the NLM\ in terms of the lattice NLS is the same}, it is
given by (\ref{UUZ}).

Summarizing we single out the following factors essential for forming
NLS-type regimes of the NLM and for determing the coefficients of the
relevant NLS or ENLS:

\begin{itemize}
\item dispersion relations $\omega _{n}\left( \mathbf{k}\right) $;

\item band number $n_{0}$, quasimomentum $\mathbf{k}_{\ast }$ and the
dispersion relation $\omega _{n_{0}}\left( \mathbf{k}\right) $\ which
determine, in parituclar, the wave carrier frequency $\omega _{n_{0}}\left( 
\mathbf{k}_{\ast }\right) $;

\item the susceptibility tensor $\chi ^{\left( 3\right) }$;

\item the Bloch mode $\mathbf{\tilde{G}}_{\zeta ,n_{0}}\left( \mathbf{r},%
\mathbf{k}_{\ast }\right) $ corresponding to the band $n_{0}$ and
quasimomentum $\mathbf{k}_{\ast }$;

\item the chosen order $\nu $ of the NLS which is often equals $2$.
\end{itemize}

The Table \ref{tabNLMS1} shows elements of the construction of the classical
second-order NLS, for the order $\nu =2$. 
\begin{table}[tbp] \centering%
\begin{tabular}{|c|c|c|}
\hline
\multicolumn{3}{|c|}{\textbf{Construction of NLS based on NLM}} \\ \hline
\begin{tabular}{l}
\textbf{NLM} \\ 
\textbf{Characteristics}%
\end{tabular}%
\textbf{\ } & 
\begin{tabular}{l}
\textbf{Mechanism} \\ 
\textbf{of correspondence}%
\end{tabular}%
\textbf{\ } & 
\begin{tabular}{l}
\textbf{NLS} \\ 
\textbf{characteristics}%
\end{tabular}%
\textbf{\ } \\ \hline
\begin{tabular}{l}
Dispersion relation \\ 
$\omega _{n_{0}}\left( \mathbf{\mathbf{k}}\right) $%
\end{tabular}
& 
\begin{tabular}{l}
$\gamma _{\left( 2\right) }\left( \mathbf{\xi }\right) $ is Taylor polynomial
\\ 
of $\omega _{n_{0}}\left( \mathbf{\mathbf{k}}\right) $ for $\mathbf{\mathbf{%
k=k}}_{\ast }+\mathbf{\xi }$%
\end{tabular}
& 
\begin{tabular}{l}
Dispersion relation \\ 
$\gamma _{\left( 2\right) }\left( \mathbf{\xi }\right) $%
\end{tabular}
\\ \hline
\begin{tabular}{l}
Nonlinearity $\mathcal{F}_{\text{NL}}\left( \mathbf{U}\right) $ \\ 
and the susceptibility \\ 
$\chi ^{\left( 3\right) }$%
\end{tabular}
& 
\begin{tabular}{l}
the susceptibility $\chi ^{\left( 3\right) }$ and \\ 
modes $\mathbf{\tilde{G}}_{\zeta ,n_{0}}\left( \mathbf{r},\mathbf{k}\right) $
at $\mathbf{\mathbf{k=k}}_{\ast }$ \\ 
determine $Q_{+}$%
\end{tabular}
& 
\begin{tabular}{l}
Coefficient $Q_{+}$ \\ 
at the nonlinearity \\ 
$\left\vert Z_{+}\right\vert ^{2}Z_{+}$%
\end{tabular}
\\ \hline
\end{tabular}%
\caption{The table shows the origin of terms in the classical second-order
NLS as an approximation the NLM.\label{tabNLMS1}}%
\end{table}
After the value of $\nu $ is chosen and the NLS equation is constructed we
move to the construction \ of NLS-type solutions for the NLM based on the
initial data $h\left( \beta \mathbf{r}\right) $. Such NLS-type solutions are
constucted by setting a proper expression for the excitation currents $%
\mathbf{\mathbf{J}}$ in terms of the initial data $h\left( \beta \mathbf{r}%
\right) $. 
\begin{table}[tbp] \centering%
\begin{tabular}{|c|c|c|}
\hline
\multicolumn{3}{|c|}{\textbf{Relation between solutions and data of NLS and
NLM}} \\ \hline
\textbf{NLM} & 
\begin{tabular}{l}
\textbf{Mechanism} \\ 
\textbf{of correspondence}%
\end{tabular}
& \textbf{NLS} \\ \hline
solution $\mathbf{U}\left( \mathbf{r},t\right) $ & matching modal
coefficients & solution $Z\left( \mathbf{r},t\right) $ \\ \hline
phase $\omega _{n}\left( \mathbf{k}_{\ast }+\mathbf{\xi }\right) \frac{\tau 
}{\varrho }$ & Taylor polynomial & phase $\gamma _{\left( 2\right) }\left( 
\mathbf{\xi }\right) \frac{\tau }{\varrho }$ \\ \hline
$%
\begin{tabular}{l}
modal coefficient \\ 
$\tilde{U}_{\zeta ,n_{0}}\left( \mathbf{k},t\right) $%
\end{tabular}%
$ & $\tilde{U}_{\zeta ,n_{0}}\left( \mathbf{k}_{\ast }+\mathbf{\eta }%
,t\right) =\hat{Z}_{\zeta }\left( \mathbf{\eta },t\right) $ & $%
\begin{tabular}{l}
Fourier coefficient \\ 
$\hat{Z}_{\zeta }\left( \mathbf{\xi },t\right) $%
\end{tabular}%
$ \\ \hline
$%
\begin{tabular}{l}
Excitation current \\ 
$\mathbf{J}\left( \mathbf{r},t\right) $%
\end{tabular}%
$ & 
\begin{tabular}{l}
$\mathbf{J}$ is of the form $\psi \left( \varrho t\right) \mathbf{\Psi }%
\left( \mathbf{r},t\right) $, \\ 
where $\psi $ is a cutoff function, \\ 
and $\mathbf{\Psi }\left( \mathbf{r},t\right) $ is determined \\ 
by the initial data $h\left( \beta r\right) $%
\end{tabular}
& Initial data $h\left( \beta r\right) $ \\ \hline
\end{tabular}%
\caption{The table shows in a simplified form the relation between the NLS
as it approximates the NLM. \label{tabNLMS2}}%
\end{table}
In Table \ref{tabNLMS2} we give a simplified form of the relation between $%
\tilde{U}_{\zeta ,n_{0}}\left( \mathbf{k},t\right) $ and $\hat{Z}_{\zeta
}\left( \mathbf{\xi },t\right) $. That simplified relation applies only for
some scalings for $\alpha $, $\varrho $ and $\beta $ which incude the
important classical NLS scaling $\alpha \sim \varrho \sim \beta ^{2}$. For
more general scalings for $\alpha $, $\varrho $ and $\beta $ \ desired
accuracy is obtained by the following more complicated relation%
\begin{equation}
\tilde{U}_{\zeta ,n_{0}}\left( \zeta \mathbf{k}_{\ast }+\mathbf{\eta }%
,t\right) =\hat{Z}_{\zeta }\left( Y_{\zeta }^{-1}\left( \mathbf{\eta }%
\right) ,t\right)  \label{uzz1}
\end{equation}%
\ where $\mathbf{\xi }=Y^{-1}\left( \mathbf{\eta }\right) $ is a rectifying
change of variables defined by%
\begin{equation}
\omega _{n_{0}}\left( \zeta \mathbf{k}_{\ast }+\mathbf{\eta }\right) =\gamma
_{\left( 2\right) }\left( Y_{\zeta }^{-1}\left( \mathbf{\eta }\right)
\right) \text{ in a vicinity of }\mathbf{k}_{\ast }  \label{uzz2}
\end{equation}%
\emph{This rectifying change of variables (\ref{uzz2}) exactly reduces the
linear part of the NLM to the linear part of the NLS}.

\subsection{Basics of the modal analysis}

Following \cite{BF1}-\cite{BF4} \ we study the nonlinear Maxwell equations
in periodic media based on the Floquet-Bloch modal decomposition. The
importance and even necessity of such a decomposition is based on the
absence of the energy transfer between Bloch modes in the linear
approximation which is instrumental for the construction of the perturbation
theory of the nonlinear evolution. As long as the amplitude of the wave
component due to the nonlinearity does not exceed the amplitude of its
linear component the Floquet-Bloch modal expansions continue to be an
excellent framework capturing well the nonlinear evolution. The
Floquet-Bloch expansion of the exact solution of (\ref{MXshort}) has the form%
\begin{equation}
\mathbf{U}\left( \mathbf{r},t\right) =\frac{1}{\left( 2\pi \right) ^{d}}%
\sum_{\bar{n}}\int_{\left[ -\pi ,\pi \right] ^{d}}\tilde{U}_{\bar{n}}\left( 
\mathbf{k},t\right) \mathbf{\tilde{G}}_{\bar{n}}\left( \mathbf{r},\mathbf{k}%
\right) \,\mathrm{d}\mathbf{k},  \label{scpr3}
\end{equation}%
where $\mathbf{\tilde{G}}_{\bar{n}}\left( \mathbf{r},\mathbf{k}\right) $ are
the Bloch eigenfunctions corresponding to the eigenvalues $\omega _{\bar{n}%
}\left( \mathbf{k}\right) $ of the Maxwell operator $\mathbf{M}$ and \textbf{%
\ }$\mathbf{k}$ is the quasimomentum with values in the Brillouin zone $%
\left[ -\pi ,\pi \right] ^{d}$. The scalar functions $\tilde{U}_{\bar{n}%
}\left( \mathbf{k},t\right) $ in (\ref{scpr3}) are the \emph{modal
coefficients} of $\mathbf{U}\left( \mathbf{r},t\right) $ corresponding to
the mode $\left( \bar{n},\mathbf{k}\right) $. In the combined index $\bar{n}%
=\left( \zeta ,n\right) $, the integer index $n=1,2,\ldots $ is the band
number and the binary index $\zeta =\pm 1$ labels two conjugate
eigenfunctions of the Maxwell operator $\mathbf{M}$ with opposite
eigenvalues $\omega _{\bar{n}}\left( \mathbf{k}\right) =\omega _{\zeta
,n}\left( \mathbf{k}\right) =\zeta \omega _{n}\left( \mathbf{k}\right) $.
The field 
\begin{equation}
\mathbf{\tilde{U}}\left( \mathbf{k},\mathbf{r},t\right) =\sum_{\bar{n}}%
\tilde{U}_{\bar{n}}\left( \mathbf{k},t\right) \mathbf{\tilde{G}}_{\bar{n}%
}\left( \mathbf{r},\mathbf{k}\right) =\sum_{\zeta =\pm 1}\sum_{n=1}^{\infty }%
\tilde{U}_{\zeta ,n}\left( \mathbf{k},t\right) \mathbf{\tilde{G}}_{\zeta
,n}\left( \mathbf{r},\mathbf{k}\right) ,  \label{Utild}
\end{equation}%
which is the integrand of the integral in the right-hand side of (\ref{scpr3}%
), is called the Floquet-Bloch transform of $\mathbf{U}\left( \mathbf{r}%
,t\right) $, see \cite{BF1} for details. By setting $\alpha =0,$ $\mathbf{J}=%
\mathbf{0}$ in (\ref{MXshort}) we obtain the linear homogenious Maxwell
equation%
\begin{equation}
\partial _{t}\mathbf{U}\left( \mathbf{r},t\right) =\mathbf{-}\mathrm{i}%
\mathbf{MU}\left( \mathbf{r},t\right) ,  \label{Maxlin}
\end{equation}%
its general solution has the following Floquet-Bloch transform 
\begin{equation}
\mathbf{\tilde{U}}\left( \mathbf{k},\mathbf{r},t\right) =\sum_{\bar{n}}%
\tilde{u}_{\bar{n}}\left( \mathbf{k}\right) \mathrm{e}^{-\mathrm{i}\omega _{%
\bar{n}}\left( \mathbf{k}\right) t}\mathbf{\tilde{G}}_{\bar{n}}\left( 
\mathbf{r},\mathbf{k}\right) .  \label{Ulin}
\end{equation}

If we ask now what kind of current $\mathbf{J}\left( \mathbf{r},t\right) $
can produce a wave that evolves essentially according to an NLS equation the
answer is as follows. We set $\mathbf{J}\left( \mathbf{r},\mathbf{t}\right) $%
, firstly, to be of the form (\ref{J01}) and composed of eigenmodes with a
single band number $n_{0}$, and, secondly, we set the modal form of its
principal part $\mathbf{J}^{\left( 0\right) }\left( \mathbf{r},\mathbf{t}%
\right) $ to be as follows 
\begin{gather}
\mathbf{\tilde{J}}_{n_{0}}^{\left( 0\right) }\left( \mathbf{r},\mathbf{k}%
,t\right) =\tilde{j}_{+,n_{0}}^{\left( 0\right) }\left( \mathbf{k},\tau
\right) \mathbf{\tilde{G}}_{+,n_{0}}\left( \mathbf{r},\mathbf{k}\right) 
\mathrm{e}^{-\mathrm{i}\omega _{n_{0}}\left( \mathbf{k}\right) t}+\tilde{j}%
_{-,n_{0}}^{\left( 0\right) }\left( \mathbf{k},\tau \right) \mathbf{\tilde{G}%
}_{-,n_{0}}\left( \mathbf{r},\mathbf{k}\right) \mathrm{e}^{\mathrm{i}\omega
_{n_{0}}\left( \mathbf{k}\right) t},  \label{Jzin} \\
\tilde{j}_{\zeta ,n_{0}}^{\left( 0\right) }\left( \mathbf{k},\tau \right)
=-\varrho \beta ^{-d}\psi _{0}\left( \tau \right) \Psi _{0}\left( \mathbf{k}-%
\mathbf{k}_{\ast }\right) \hat{h}_{\zeta }\left( \frac{1}{\beta }Y_{\zeta
}^{-1}\left( \mathbf{k}-\mathbf{k}_{\ast }\right) \right) ,\ \tau =\varrho t,
\notag \\
\mathbf{\tilde{J}}_{n}^{\left( 0\right) }\left( \mathbf{r},\mathbf{k}%
,t\right) =0,\ n\neq n_{0},\;\zeta =\pm .  \notag
\end{gather}%
We call such an excitation current \emph{almost single-mode excitation}.
Evidently, the current $\mathbf{J}^{\left( 0\right) }\left( \mathbf{r},%
\mathbf{k},t\right) $ defined by (\ref{Jzin}) is an almost time-harmonic
function of the time $t$ as in (\ref{aa}) for every $\mathbf{k}$. Observe
also, that in (\ref{Jzin}) $\mathbf{k}_{\ast }$ is a chosen quasimomentum in
the Brillouin zone $\left[ -\pi ,\pi \right] ^{d}$. \ The currents of the
above form are determined by the choice of the function $\beta ^{-d}\hat{h}%
_{\pm }\left( \frac{\mathbf{\xi }}{\beta }\right) $, which is the Fourier
transform of the function $h_{\pm }\left( \beta \mathbf{r}\right) $, which,
in turn, corresponds to the initial data of the NLS. The rectifying change
of variables $\mathbf{\xi }=Y_{\zeta }^{-1}\left( \mathbf{\eta }\right) $ is
very close to the identity, and its purpose is to provide exact matching
between the linear NLM and NLS when $\alpha =0$. Therefore, $h_{\pm }\left(
\beta \mathbf{r}\right) \mathbf{\tilde{G}}_{\pm ,n_{0}}\left( \mathbf{r},%
\mathbf{k}_{\ast }\right) $ is a proper substitute for the initial data for
the NLM. Note that for small $\beta $ the spread of the function $h\left(
\beta \mathbf{r}\right) $ is large and proportional to $\frac{1}{\beta }$,
whereas the spread of its Fourier transform $\beta ^{-d}\hat{h}_{\pm }\left( 
\frac{\mathbf{\xi }}{\beta }\right) $ is small and propotional to $\beta $.
The cut-off function $\Psi _{0}$ in (\ref{Jzin}) is introduced to restrict $%
\hat{h}_{\pm }\left( \frac{\mathbf{\xi }}{\beta }\right) $ from the entire
space to the Brillouin zone $\left[ -\pi ,\pi \right] ^{d}$ and its
properties are listed in (\ref{j0}). The slowly varying function $\varrho
\psi _{0}\left( \varrho t\right) $ is set to be non-zero only for $0\leq
\tau =\varrho t\leq \tau _{0}$. Its purpose is to provide a transition from
the rest solution to a nonzero solution of the NLM, and also to introduce a
finite, proportinal to $\varrho $ frequency bandwidth, and, consequently,
the slow time scale $\tau =\varrho t$, into the excitation current. We refer
to currents and waves of the form similar to (\ref{Jzin}) as \emph{almost
single-mode waves}. The concept of almost single-mode wave is instrumental
for studies on nonlinear wave interactions and the NLS regimes.

To explain the construction of an NLS corresponding to the NLM we introduce
first an abstract nonlinear equation for a 2-component vector valued
amplitude $V$%
\begin{equation}
\partial _{t}V=-\mathrm{i}\mathcal{L}V+\alpha F^{\left( 3\right) }\left(
V\right) -f  \label{Veq}
\end{equation}%
where $\mathcal{L}$ is a linear differential operator with constant
coefficients, $F^{\left( 3\right) }\left( V\right) $ is a cubic nonlinearity
with a simplest possible structure and $f=0$ when $t<0$ and $t>\tau
_{0}/\varrho $. The vector $V$ \ in (\ref{Veq}) includes two components
which correspond to two modes $\pm \mathbf{k}_{\ast }$ excited by a
real-valued almost single-mode current regime. Our goal is to construct $%
\mathcal{L}$ and $F^{\left( 3\right) }$and choose $f$ so that the sum of the
linear and the first nonlinear responses associated with (\ref{Veq}) would
approximate well the directly excited modal coefficients $\tilde{U}_{\zeta
,n_{0}}\left( \mathbf{k},t\right) $ when $t>\tau _{0}/\varrho $. We
rigorously show that equations providing accurate approximations to the NLM
are of the form (\ref{Veq}), in particular, they are the classical NLS or
ENLS for higher order approximations. An analysis shows that for excitation
currents as in (\ref{Jzin}) only the modes close to $\pm \mathbf{k}_{\ast }$
interact nonlinearly with themselves strongly enough to determine the
nonlinear evolution, whereas all other nonlinear interactions are
generically negligible. An important element in the construction of accurate
approximations is a rectifying change of variables which recasts the linear
Maxwell equation into the corresponding linear Schrodinger equation in the
quasimomentum domain. Note that the difference between the NLS and NLM is
obvious even when the nonlinearity is absent, since the NLM is an equation
with variable coefficients for 6-component vector fields which includes only
first-order spatial derivatives whereas the NLS has two components
(reducible to one by complex conjugation) with constant coefficients and
with second-order spatial derivatives.

The relation between the NLM and corresponding NLS is as follows. The
coefficients of the NLS\ can be explicitly written in terms of the Bloch
dispersion relations, the eigenfunctions and the cubic susceptibility. Then
the approximate solution $\mathbf{U}_{Z}\left( \mathbf{r},t\right) $ of the
NLM\ is constructed based on solutions $Z_{\pm }\left( \mathbf{r},t\right) $
to the NLS (\ref{Si}), (\ref{Si1})\ by the formula (\ref{UZdirind}) where
the leading, directly excited component does not include modes which are not
present in the excitation current 
\begin{equation}
\mathbf{U}_{Z,n}^{\text{dir}}\left( \mathbf{r},t\right) =0,\ n\neq n_{0},
\label{Udirn}
\end{equation}%
and the component in the excited band is given by the following fundamental
formula%
\begin{gather}
\mathbf{U}_{Z,n_{0}}^{\text{dir}}\left( \mathbf{r},t\right) =\frac{1}{\left(
2\pi \right) ^{d}}\int_{\left[ -\pi ,\pi \right] ^{d}}\Psi _{0}\left( 
\mathbf{\eta }\right)  \label{UNLS} \\
\left[ \hat{Z}_{+}\left( Y_{+}^{-1}\left( \mathbf{\eta }\right) ,t\right) 
\mathbf{\tilde{G}}_{+,n_{0}}\left( \mathbf{r},\mathbf{k}_{\ast }+\mathbf{%
\eta }\right) +\hat{Z}_{-}\left( Y_{-}^{-1}\left( \mathbf{\eta }\right)
,t\right) \mathbf{\tilde{G}}_{-,n_{0}}\left( \mathbf{r},-\mathbf{k}_{\ast }+%
\mathbf{\eta }\right) \right] \,\mathrm{d}\mathbf{\eta }  \notag
\end{gather}%
where $\hat{Z}_{\pm }\left( \mathbf{k},t\right) $ is the Fourier transform
of $Z_{\pm }\left( \mathbf{r},t\right) $, $t\geq \frac{\tau _{0}}{\varrho }$%
. The expression $Y^{-1}\left( \mathbf{\eta }\right) $, which is very close
to $\mathbf{\eta }$, is the inverse to the rectifying change of variables
which reduces the dispersion relation of the NLM to that of the NLS, namely 
\begin{equation}
\omega _{n_{0}}\left( \zeta \mathbf{k}_{\ast }+\mathbf{\eta }\right) =\gamma
_{\left( \nu \right) }\left( Y_{\zeta }^{-1}\left( \mathbf{\eta }\right)
\right) ,\ \text{where }Y_{\zeta }^{-1}\left( \mathbf{\eta }\right) =\mathbf{%
\eta }+O\left( \left\vert \mathbf{\eta }\right\vert ^{\nu +1}\right) ,\ 
\text{for small }\left\vert \mathbf{\eta }\right\vert .
\end{equation}%
The power $\nu =2$ \ for the classical NLS and $\nu =3,4$ for the ENLS (in
special cases $Y_{\zeta }^{-1}\left( \mathbf{\eta }\right) $ may be replaced
by $\mathbf{\eta }$ without loss of accuracy). The cuttoff function $\Psi
_{0}\left( \mathbf{\eta }\right) $ is introduced to select only $\mathbf{%
\eta }$ from a fixed small vicinity of $\mathbf{k}_{\ast }$ in the Brillouin
zone. The components $\mathbf{U}_{Z,n}\left( \mathbf{r},t\right) $ with $%
n\neq n_{0}$ of $\mathbf{U}_{Z}\left( \mathbf{r},t\right) $ are included in
the indirectly excited part $\mathbf{U}_{Z}^{\text{ind}}\left( \mathbf{r}%
,t\right) $, they are much smaller and are described in the end of this
subsection. Formula (\ref{UNLS}) shows that the dynamics of the solution of
the Maxwell equation on time intervals of order $\frac{1}{\varrho }$ is
reduced to the dynamics of solutions to the NLS or extended NLS. Formula (%
\ref{UNLS}) also shows that the time evolution of the pair of the
coefficients $\tilde{U}_{\zeta ,n_{0}}\left( \mathbf{k},t\right) $, $\zeta
=\pm 1$, determinded by the almost single-mode excitations, is described by
solutions $\hat{Z}_{\pm }$ to the NLS. The rectifying change of variables $%
Y_{\zeta }^{-1}\left( \mathbf{\eta }\right) $ allows to establish an \emph{%
exact} equivalence between the dynamics of the NLM\ and the NLS\ in the
linear approximation ($\alpha =0$) \emph{for arbitrary long times}. It is
important to note that this rectifying change of variables does not affect
neither dynamics of the NLM, nor dynamics of the NLS, but rather establishes
a relation between solutions of the two equations.

The relation between the modal coefficients of the approximate solution $%
\mathbf{U}_{Z}\left( \mathbf{r},t\right) $ of the NLS and the exact solution 
$\mathbf{U}\left( \mathbf{r},t\right) $ of the nonlinear Maxwell equation is
represented by the formula 
\begin{eqnarray}
\fl\;\tilde{U}_{+,n_{0}}\left( \mathbf{k}_{\ast }+\mathbf{\eta },t\right) &=&%
\hat{Z}_{+}\left( Y_{+}^{-1}\left( \mathbf{\eta }\right) ,t\right) +\left[
O\left( \alpha ^{2}\right) +O\left( \alpha \beta \right) +O\left( \alpha
\varrho \right) \right] O\left( \left\vert \mathbf{U}^{\left( 1\right)
}\right\vert \right) ,  \label{UNLS1} \\
\fl\;\tilde{U}_{-,n_{0}}\left( -\mathbf{k}_{\ast }+\mathbf{\eta },t\right)
&=&\hat{Z}_{-}\left( Y_{-}^{-1}\left( \mathbf{\eta }\right) ,t\right) +\left[
O\left( \alpha ^{2}\right) +O\left( \alpha \beta \right) +O\left( \alpha
\varrho \right) \right] O\left( \left\vert \mathbf{U}^{\left( 1\right)
}\right\vert \right) .  \notag
\end{eqnarray}%
The equalities (\ref{UNLS1}) hold for%
\begin{equation}
\frac{\tau _{0}}{\varrho }\leq t\leq \frac{\tau _{\ast }}{\varrho },\text{
where }\tau _{0}\ll 1,\ \tau _{\ast }\gg 1\text{ are fixed numbers, and }%
\left\vert \mathbf{\eta }\right\vert \leq \pi _{0}.
\end{equation}%
The magnitude of the FNLR $\mathbf{U}^{\left( 1\right) }$ on time intervals
which satisfy (\ref{rhoalph}) is estimated as follows: 
\begin{eqnarray}
\fl\;O\left( \left\vert \mathbf{U}^{\left( 1\right) }\right\vert \right)
&=&O\left( \varrho ^{d-1}\right) O\left( \left\vert \mathbf{J}\right\vert
^{3}\right) \text{ in the dispersive case }\theta ^{-1}\gg \left\Vert \omega
_{n_{0}}^{\prime \prime }\left( \mathbf{k}_{\ast }\right) ^{-1}\right\Vert 
\text{,}  \label{OU1} \\
\fl\;O\left( \left\vert \mathbf{U}^{\left( 1\right) }\right\vert \right)
&=&O\left( \varrho ^{-1}\right) O\left( \left\vert \mathbf{J}\right\vert
^{3}\right) \text{ in the weakly dispersive case }\theta ^{-1}\ll \left\Vert
\omega _{n_{0}}^{\prime \prime }\left( \mathbf{k}_{\ast }\right) \right\Vert
^{-1}\text{.}  \notag
\end{eqnarray}

The coefficients of the NLS can be found as follows. Using the analytic
expansion (\ref{uMv1}) of the solution of (\ref{MXshort}) we obtain the
following representation for the modal coefficients 
\begin{equation}
\tilde{U}_{\zeta ,n}\left( \mathbf{k},t\right) =\tilde{U}_{\zeta ,n}^{\left(
0\right) }\left( \mathbf{k},t\right) +\alpha \tilde{U}_{\zeta ,n}^{\left(
1\right) }\left( \mathbf{k},t\right) +O\left( \alpha ^{2}\right) O\left(
\left\vert \mathbf{U}^{\left( 1\right) }\right\vert \right) ,\;\frac{\tau
_{0}}{\varrho }\leq t<\frac{\tau _{\ast }}{\varrho }.  \label{Unexp}
\end{equation}%
The first order term of the power expansion (\ref{Unexp}) of $\tilde{U}_{%
\bar{n}}\left( \mathbf{k},t\right) $ is given by the modal coefficient $%
\tilde{U}_{\bar{n}}^{\left( 1\right) }\left( \mathbf{k},t\right) $ of the
first nonlinear response (FNLR)\ determined by (\ref{FNLR}). We also have a
similar expansion for $Z_{\zeta }\left( \mathbf{r},t\right) $ and its
Fourier transform $\hat{Z}_{\zeta }\left( \mathbf{\xi },t\right) $,%
\begin{equation}
\hat{Z}_{\zeta }\left( \mathbf{\xi },t\right) =\hat{Z}_{\zeta }^{\left(
0\right) }\left( \mathbf{\xi },t\right) +\alpha \hat{Z}_{\zeta }^{\left(
1\right) }\left( \mathbf{\xi },t\right) +O\left( \alpha ^{2}\right) O\left( 
\hat{Z}_{\zeta }^{\left( 1\right) }\right) ,\ \zeta =\pm .
\end{equation}%
Note that for regular initial data $h_{\zeta }$ we have $\hat{Z}_{\zeta
}^{\left( 1\right) }=O\left( \left\vert \mathbf{U}^{\left( 1\right)
}\right\vert \right) $. \emph{Then the coefficients to the NLS are
determined from the following requirement.} The FNLR of the NLS must
approximate the FNLR of the NLM\ with an error $O_{\text{FNLR}}$ so that 
\emph{for all initial data }$h_{\zeta }$\emph{\ }the following two
identities hold: 
\begin{equation}
\hat{Z}_{\zeta }^{\left( 0\right) }\left( Y^{-1}\left( \mathbf{\eta }\right)
,t\right) =\tilde{U}_{\zeta ,n_{0}}^{\left( 0\right) }\left( \zeta \mathbf{k}%
_{\ast }+\mathbf{\eta },t\right) \text{ if }\left\vert \mathbf{\eta }%
\right\vert \leq \pi _{0},\ 0\leq t<\infty ,  \label{Eq0}
\end{equation}%
\begin{equation}
\hat{Z}_{\zeta }^{\left( 1\right) }\left( Y^{-1}\left( \mathbf{\eta }\right)
,t\right) =\tilde{U}_{\zeta ,n_{0}}^{\left( 1\right) }\left( \zeta \mathbf{k}%
_{\ast }+\mathbf{\eta },t\right) +O_{\text{FNLR}}\text{ if }\left\vert 
\mathbf{\eta }\right\vert \leq \pi _{0},\ \frac{\tau _{0}}{\varrho }\leq t<%
\frac{\tau _{\ast }}{\varrho }.  \label{Eq1}
\end{equation}%
Note that (\ref{Eq0}) and (\ref{Eq1}) imply 
\begin{equation}
\tilde{U}_{\zeta ,n_{0}}^{\text{dir}}\left( \zeta \mathbf{k}_{\ast }+\mathbf{%
\eta },t\right) =\hat{Z}_{\zeta }\left( Y_{\zeta }^{-1}\left( \mathbf{\eta }%
\right) ,t\right) +\alpha O_{\text{FNLR}}+O\left( \alpha ^{2}\right) O\left(
\left\vert \mathbf{U}^{\left( 1\right) }\right\vert ^{2}\right) .
\label{UZO}
\end{equation}%
For instance, if the NLS is given by (\ref{Si}) (\ref{Si1}) the error is 
\begin{equation}
O_{\text{FNLR}}=\left[ O\left( \beta \right) +O\left( \varrho \right) \right]
O\left( \left\vert \mathbf{U}^{\left( 1\right) }\right\vert \right) .
\label{OFNLR}
\end{equation}%
If we use solutions of the extended NLS that involve additional terms the
error becomes smaller (see Subsection 1.3 for details) 
\begin{equation}
O_{\text{FNLR}}=\left[ O\left( \beta ^{3}\right) +O\left( \beta \varrho
\right) \right] O\left( \left\vert \mathbf{U}^{\left( 1\right) }\right\vert
\right) .  \label{OFNLR1}
\end{equation}

All remaining, indirectly excited modes of the approximate solution are
given in terms of the FNLR: 
\begin{equation}
\tilde{U}_{Z,\zeta ,n}^{\text{ind}}\left( \mathbf{k},t\right) =\tilde{U}%
_{\zeta ,n}^{\left( 1\right) }\left( \mathbf{k},t\right)  \label{Uind1}
\end{equation}%
and the approximation error 
\begin{equation}
\tilde{U}_{\bar{n}}\left( \mathbf{k},t\right) -\tilde{U}_{Z,\bar{n}}^{\text{%
ind}}\left( \mathbf{k},t\right) =O\left( \varrho \alpha ^{2}\right) O\left(
\left\vert \mathbf{U}^{\left( 1\right) }\right\vert \right) \;\text{when}\
n\neq n_{0}\ \text{or }\left\vert \mathbf{k}-\mathbf{k}_{\ast }\right\vert
>\pi _{0}.  \label{Unne0}
\end{equation}%
Note that indirectly excited modes are much smaller than directly excited,
namely%
\begin{equation}
\tilde{U}_{Z,\zeta ,n}^{\text{ind}}\left( \mathbf{k},t\right) =O\left(
\varrho \alpha \right) O\left( \left\vert \mathbf{U}^{\left( 1\right)
}\right\vert \right)  \label{UindO}
\end{equation}%
compared with $\tilde{U}_{Z,n_{0}}^{\text{dir}}=O\left( 1\right) $. Note
that (\ref{UindO}) implies that \emph{the indirectly excited modes can be
neglected in the cases (\ref{UUZ}) and (\ref{UUZx}) but have to be taken
into account when higher precision approximation is used}. An analysis given
in Section 7 shows that though we determine the coefficients of the NLS
based on the FNLR of the NLM, using \emph{exact} solution $Z_{\zeta }$ of
the NLS in (\ref{UNLS}) allows to obtain estimates (\ref{UNLS1}). The
indirectly excited part of the approximate solution is given by the formula 
\begin{equation}
\mathbf{U}_{Z}^{\text{ind}}\left( \mathbf{r},t\right) =\sum_{n=1}^{\infty }%
\mathbf{U}_{Z,n}^{\text{ind}}\left( \mathbf{r},t\right)  \label{Uindsum}
\end{equation}%
where%
\begin{gather}
\mathbf{U}_{Z,n}^{\text{ind}}\left( \mathbf{r},t\right) =\frac{\alpha }{%
\left( 2\pi \right) ^{d}}\int_{\left[ -\pi ,\pi \right] ^{d}}  \label{Uindn}
\\
\left[ \tilde{U}_{+,n}^{\left( 1\right) }\left( \mathbf{k}_{\ast }+\mathbf{%
\eta },t\right) \mathbf{\tilde{G}}_{+,n}\left( \mathbf{r},\mathbf{k}_{\ast }+%
\mathbf{\eta }\right) +\tilde{U}_{-,n}^{\left( 1\right) }\left( -\mathbf{k}%
_{\ast }+\mathbf{\eta },t\right) \mathbf{\tilde{G}}_{-,n}\left( \mathbf{r},-%
\mathbf{k}_{\ast }+\mathbf{\eta }\right) \right] \,\mathrm{d}\mathbf{\eta }%
,\ n\neq n_{0};  \notag
\end{gather}%
\begin{gather}
\mathbf{U}_{Z,n_{0}}^{\text{ind}}\left( \mathbf{r},t\right) =\frac{\alpha }{%
\left( 2\pi \right) ^{d}}\int_{\left[ -\pi ,\pi \right] ^{d}}\left( 1-\Psi
_{0}\left( \mathbf{\eta }\right) \right)  \label{Uindn0} \\
\left[ \tilde{U}_{+,n_{0}}^{\left( 1\right) }\left( \mathbf{k}_{\ast }+%
\mathbf{\eta },t\right) \mathbf{\tilde{G}}_{+,n_{0}}\left( \mathbf{r},%
\mathbf{k}_{\ast }+\mathbf{\eta }\right) +\tilde{U}_{-,n_{0}}^{\left(
1\right) }\left( -\mathbf{k}_{\ast }+\mathbf{\eta },t\right) \mathbf{\tilde{G%
}}_{-,n_{0}}\left( \mathbf{r},-\mathbf{k}_{\ast }+\mathbf{\eta }\right) %
\right] \,\mathrm{d}\mathbf{\eta },  \notag
\end{gather}%
with $\tilde{U}_{\bar{n}}^{\left( 1\right) }\left( \mathbf{k},t\right) $
being the modal coefficient of the solution $\mathbf{U}^{\left( 1\right)
}\left( \mathbf{r},t\right) $ of the linear equation (\ref{FNLR}), for an
explicit formula see (\ref{u1rho}), (\ref{Vn}), (\ref{V0}).

\begin{table}[tbp] \centering%
\begin{tabular}{|l|r|}
\hline
\multicolumn{2}{|l|}{%
\begin{tabular}{c}
\textbf{Order of magnitude of fields and its components for}$\;\frac{\tau
_{0}}{\varrho }\leq t\leq \frac{\tau _{\ast }}{\varrho }$ \\ 
\textbf{under the classical NLS scaling} $\varrho \sim \alpha \sim \beta
^{2} $ \textbf{in the one-dimensional case}%
\end{tabular}%
} \\ \hline
Excitation current $\mathbf{J}\left( \mathbf{r},t\right) $ for $t\leq \frac{%
\tau _{0}}{\varrho }$ & $\varrho \sim \beta ^{2}$ \\ \hline
Linear response $\mathbf{U}^{\left( 0\right) }\left( \mathbf{r},t\right) ,$
& $1$ \\ \hline
Directly excited part of the FNLR $\alpha \mathbf{U}^{\left( 1\right) \text{%
dir}}\left( \mathbf{r},t\right) ,$ & $\alpha \varrho ^{-1}\sim 1$ \\ \hline
Indirectly excited part of the FNLR $\alpha \mathbf{U}^{\left( 1\right) 
\text{ind}}\left( \mathbf{r},t\right) $ & $\alpha \sim \beta ^{2}$ \\ \hline
Exact solution of the NLM\ $\mathbf{U}\left( \mathbf{r},t\right) $ & $1$ \\ 
\hline
\end{tabular}%
\caption{The entries show the order of magnitude of excitation current $\mathbf{J}\left( \mathbf{r},t\right) $ before it vanishes for $t\leq \frac{\tau _{0}}{\varrho }$, the field $\mathbf{U}\left( \mathbf{r},t\right) $, which in an
exact solution to the NLM, and its components during the time period $\frac{\tau _{0}}{\varrho }\leq t\leq \frac{\tau _{\ast }}{\varrho }$. Notice that
the indirectly excited part of the FNLR is far smaller than the directly
excited one. \label{tabmagJ}}%
\end{table}%

In conclusion, the developed method allows to find higher order
approximations of the solutions of the NLM by solutions of NLS-type
equations with a rigorous control of errors on time intervals consistent
with the FNLR, i.e. if the relations (\ref{rhoalph}) hold. We would to point
out that remarkably though the formula (\ref{UNLS}) is derived based on the
analysis of the FNLR which is applicable on time intervals of order $\frac{1%
}{\alpha }$, it turns out that the formula may still be valid for larger
times as long as the solution $Z\left( \mathbf{r},t\right) $ of the NLS has
"good" properties. In the latter case, since we use an exact solution of the
NLS in (\ref{UNLS}), $\mathbf{U}_{Z}\left( \mathbf{r},t\right) $ still
solves the NLM with a higher precision, see Section 7 for details. If
additional information on the solution of the NLS is available, in
particular, if appropriate stability conditions are fulfilled, $\mathbf{U}%
_{Z}\left( \mathbf{r},t\right) $ approximates the exact solution $\mathbf{U}%
\left( \mathbf{r},t\right) $ well on the longer time intervals.

\subsection{Wave interactions and multimode NLS regimes}

It is interesting and instructive to look at NLS regimes of nonlinear wave
propagation in periodic dielectric media in the context of nonlinear
interactions between the eigenmodes of the underlying linear medium. From
that perspecitve an \emph{NLS regime can be characterized as such a regime
of nonlinear mode interactions when for a generic mode its self-interaction
(that is interaction with the conjugate mode) significantly dominates the
nonlinear interactions with all other modes}. More accurate description of
an NLS regime is based on finer estimations of magnitudes of nonlinear
interactions between different modes and their dependence on values of the
small parameters $\alpha $, $\varrho $ and $\beta $. In turns out, that in
the case of an NLS regime when a generic mode, described by a quasimomentum $%
\mathbf{k}_{\ast }$ and a band index $\bar{n}=\left( \zeta ,n_{0}\right) $
is excited, it interacts significanlty stronger with modes from the same
band $n_{0}$ and with quasimomenta located about $\mathbf{k}_{\ast }$ than
with all other modes. In addition to that, nonlinear interactions between
mode $\left( \left( \zeta ,n_{0}\right) ,\zeta \mathbf{k}_{\ast }\right) $
and its conjugate mode $\left( \left( -\zeta ,n_{0}\right) ,-\zeta \mathbf{k}%
_{\ast }\right) $ are much stronger compared to other mode interactions in
this band. We call such a modal pair, occuring often in our analysis, a 
\emph{doublet} and denote it by%
\begin{equation}
\left\uparrow n_{0},\mathbf{k}_{\ast }\right\downarrow =\left\{ \left(
+,n_{0},\mathbf{k}_{\ast }\right) ,\left( -,n_{0},-\mathbf{k}_{\ast }\right)
\right\} =\left\{ \left( \zeta ,n_{0},\zeta \mathbf{k}_{\ast }\right) :\zeta
=\pm \right\} .  \label{doublet}
\end{equation}

In this article (excluding this subsection) we consider primarily almost
single-mode current excitations based on a single doublet $\left\uparrow
n_{0},\mathbf{k}_{\ast }\right\downarrow $ formed by a mode $\left( \left(
+,n_{0}\right) ,\mathbf{k}_{\ast }\right) $ together with its conjugate
counterpart $\left( \left( -,n_{0}\right) ,-\mathbf{k}_{\ast }\right) $ that
would allow to produce a real-valued field. The dynamics of a doublet is
described by the NLS (\ref{Si}), (\ref{Si1}) or with a higher precision by
the ENLS\ (\ref{Six}), (\ref{Six1}). A more detailed investigation of
nonlinear mode interactions would naturally require the introduction of
multimode current excitation involving small vicinities of several doublets $%
\left\uparrow n_{l},\mathbf{k}_{\ast l}\right\downarrow $, $l=1,\ldots ,N$,
rather than just an almost single mode excitation and leading to groups of
excited modes.

An analysis below suggests a view on the NLS and ENLS as regimes of
nonlinear wave propagation when wave modal components admit a decomposition
into essentially noninteracting groups. Consequently, the existence,
conditions and accuracy of such a decomposition as well as the derivation of
relevant simplified evolution equations of smaller modal groups become a
subject of the theory of ENLS equations. \emph{In other words, a "big
picture" characterizing an NLS regime for the electromagnetic wave
propagation is that the evolution of components of its modal composition
occurs essentially independently for groups of modes with separated carrier
frequences and quasimomenta whereas the interactions inside every single
group occur according to a rather universal scenario described by NLS-type
equations.}

The first and fundamental step in the analysis is to find and classify all
the interactions between the modal groups as well as with the rest of modes
with estimations of their relative magnitudes. We do this based on the
quantitative theory of nonlinear mode interactions and, in particular, with
the help of selection rules for stronger interactions studied in \cite{BF1}-%
\cite{BF3}. The essentials of the analysis are provided below.

\subsubsection{Selection rules for stronger wave interactions and NLS regimes%
}

To find the wave decomposition into almost independent components we use the
selection rules for stronger interactions, \cite{BF1}-\cite{BF3}, which are
as follows. Consider the modal coefficients $\tilde{U}_{\bar{n}}\left( 
\mathbf{k},t\right) $ of the wave goverened by the NLM. Notice that if $%
\alpha =0$ the NLM turns into a linear equation and according to the
classical spectral theory the modal coefficients $\tilde{U}_{\bar{n}}\left( 
\mathbf{k},t\right) $ for\ different $\bar{n}$ and $\mathbf{k}$ evolve
independently one from another as in (\ref{Ulin}). For $\alpha \neq 0$ the
cubic nonlinearity introduces interactions between all the modes. In the
case when the nonlinear term of the electric polarization has the same
spatial period as the underlying linear medium, the first fundamental
restriction on any quadruplet of interacting Bloch modes is given by the 
\emph{phase matching condition} 
\begin{equation}
\mathbf{k}=\mathbf{\mathbf{k}^{\prime }}+\mathbf{k}^{\prime \prime }+\mathbf{%
k}^{\prime \prime \prime }\quad \func{mod}\left( 2\pi \right) ,  \label{PM0}
\end{equation}%
where $\mathbf{k}$ is the quasimomentum of the mode which is affected by a
triad of modes with the quasimomenta $\mathbf{\mathbf{k}^{\prime }},\mathbf{k%
}^{\prime \prime },\mathbf{k}^{\prime \prime \prime }$. We call the triad%
\begin{equation}
\left( \left( \zeta ^{\prime },n^{\prime }\right) ,\mathbf{k}^{\prime
}\right) ,\left( \left( \zeta ^{\prime \prime },n^{\prime \prime }\right) ,%
\mathbf{k}^{\prime \prime }\right) ,\left( \left( \zeta ^{\prime \prime
\prime },n^{\prime \prime \prime }\right) ,\mathbf{k}^{\prime \prime \prime
}\right)  \label{triad}
\end{equation}%
the \emph{origin triad} or \emph{origin modes} of the interaction quadruplet
and $\left( \left( \zeta ,n\right) ,\mathbf{k}\right) $ the \emph{end mode}
of the quadruplet. The \emph{interaction quadruplet} is completely defined
by its origin triad and its end mode.

If in the excitation current $\mathbf{J}$ of the form (\ref{J01}), (\ref%
{Jzin}), or of a more general form decribed in \cite{BF1}-\cite{BF3}, the
both parameters $\alpha $ and $\varrho $ are small, and (\ref{rho/beta}) is
fulfilled(or, more precisely, (\ref{GVMC}) holds) then stronger interacting
modal quadruplets satisfy also the \emph{group velocity matching} condition 
\begin{equation}
\nabla \omega _{\bar{n}^{\prime }}\left( \mathbf{k}^{\prime }\right) =\nabla
\omega _{\bar{n}^{\prime \prime }}\left( \mathbf{k}^{\prime \prime }\right)
=\nabla \omega _{\bar{n}^{\prime \prime \prime }}\left( \mathbf{k}^{\prime
\prime \prime }\right) .  \label{GVM0}
\end{equation}%
Note that (\ref{GVM0}) is a constraint only on the origin triad of the
quadruplet. The selection rule (\ref{GVM0})\ is the most important one,
since if it is not fulfilled, the magnitude of the interaction is estimated
by $O\left( \left( \frac{\varrho }{\beta }\right) ^{\kappa }\right) $ with
arbitrarily large $\kappa $,\ and, in view of (\ref{rho/beta}), is not a
strong interaction.

Finally, a modal quadruplet would have even stronger nonlinear interactions
if in addition to the phase and group velocity matching it satisfies the 
\emph{frequency matching} condition 
\begin{equation}
\omega _{\bar{n}^{\prime }}\left( \mathbf{k}^{\prime }\right) +\omega _{\bar{%
n}^{\prime \prime }}\left( \mathbf{k}^{\prime \prime }\right) +\omega _{\bar{%
n}^{\prime \prime \prime }}\left( \mathbf{k}^{\prime \prime \prime }\right)
=\omega _{\bar{n}}\left( \mathbf{k}\right) .  \label{FMC0}
\end{equation}%
For many cases of interest there are modal quadruplets satisfying all three
conditions of (\ref{PM0}), (\ref{GVM0}) and (\ref{FMC0}), \cite{BF3}. In any
case, the selection rules (\ref{PM0}), (\ref{GVM0}), (\ref{FMC0}) determine
stronger interacting quadruplets of modes with a detailed classification of
generic mode interactions provided in \cite{BF1}-\cite{BF3}.

In the present article we primarily focus on the case of waves excited by
almost single-mode excitation currents given by (\ref{J01}), (\ref{Jzin}),
that is only the modes in $\beta $-vicinity of a fixed quasimomentum $\pm 
\mathbf{k}_{\ast }$ are directly excited, creating a directly excited
doublet $\left\uparrow n_{0},\mathbf{k}_{\ast }\right\downarrow $. \ One can
form $2^{4}$ different interaction quadruplets based on two modes from a
single doublet. \ We have shown in \cite{BF3} that if the inversion symmetry
condition (\ref{invsym}) holds then every such quadruplet formed based on a
given doublet always satisfy the group velocity matching condition (\ref{PM0}%
), see for details a discussion below, see also Section 3. In addition to
that, for the phase matching condition (\ref{PM0}) to hold for a quadruplet
with modes from a doublet $\left\uparrow n_{0},\mathbf{k}_{\ast
}\right\downarrow $ the following relation must hold: 
\begin{equation}
\zeta \mathbf{k}_{\ast }=\zeta ^{\prime }\mathbf{k}_{\ast }+\zeta ^{\prime
\prime }\mathbf{k}_{\ast }+\zeta ^{\prime \prime \prime }\mathbf{k}_{\ast
}+O\left( \beta \right) .  \label{zzkk}
\end{equation}%
Since $\zeta =\pm 1$ (\ref{zzkk}) holds if and only if $\zeta =\zeta
^{\prime }+\zeta ^{\prime \prime }+\zeta ^{\prime \prime \prime }$.
Therefore some two of the three binary indices $\zeta ^{\prime },\zeta
^{\prime \prime },\zeta ^{\prime \prime \prime }$ must coincide with $\zeta $%
, implying that any strongly interacting quadruplet that contains the mode$\
\left( \left( \zeta ,n_{0}\right) ,\zeta \mathbf{k}_{\ast }\right) $ has
also to contain two more copies of the very same $\left( \left( \zeta
,n_{0}\right) ,\zeta \mathbf{k}_{\ast }\right) $ and one mode $\left( \left(
-\zeta ,n_{0}\right) ,-\zeta \mathbf{k}_{\ast }\right) $. Since the
interaction is trilinear and there are two copies of $\left( \left( \zeta
,n_{0}\right) ,\zeta \mathbf{k}_{\ast }\right) $ and one copy of $\left(
\left( -\zeta ,n_{0}\right) ,-\zeta \mathbf{k}_{\ast }\right) $ the
magnitude of this interaction is proportional to the product $\tilde{U}%
_{\zeta ,n_{0}}\left( \zeta \mathbf{k}_{\ast }\right) \tilde{U}_{\zeta
,n_{0}}\left( \zeta \mathbf{k}_{\ast }\right) \tilde{U}_{-\zeta
,n_{0}}\left( -\zeta \mathbf{k}_{\ast }\right) $ where $\tilde{U}_{\zeta
,n_{0}}\left( \zeta \mathbf{k}_{\ast }\right) $ and $\tilde{U}_{-\zeta
,n_{0}}\left( -\zeta \mathbf{k}_{\ast }\right) $ are the corresponding modal
coefficients. Recall now that for real-valued fields their Bloch coefficient 
$\tilde{U}_{-\zeta ,n_{0}}\left( -\zeta \mathbf{k}_{\ast }\right) $ equals
the Bloch coefficient of the complex conjugate field, and, consequently, the
nonlinear interaction magnitude is proportional to $\left\vert U_{\zeta
,n_{0}}\right\vert ^{2}U_{\zeta ,n_{0}}$ leading to the NLS\ with a
nonlinearity of the form $\left\vert Z\right\vert ^{2}Z$. In the following
sections we provide rigorous and detailed derivation of the two coupled NLS
equations of the form (\ref{Si}), (\ref{Si1}) or (\ref{Six}), (\ref{Six1})
describing the mode interaction for a doublet as well as the error
estimates. Conversely, an analysis of \cite{BF3} shows that in a medium with
the inversion symmetry any generic quadruplet of strongly interacting modes
must have all its four modes from a single doublet.

In this article we advance the analysis of mode interactions further,
showing that an almost time-harmonic excitation based on a single doublet
yields a wave described approximately by the Nonlinear Schrodinger equations
with the approximation accuracy depending on the three small parameters $%
\alpha $, $\varrho \ $and $\beta $. As we have already pointed out, the
solution of the NLM\ depends on every one of these parameters in a different
way. We can add to the said that the dependence on $\alpha $ is relatively
simple, the analytic expansion with respect to $\alpha $ regularly converges
uniformly on time intervals of order $\frac{1}{\alpha }$ as long as the
relevant fields remain bounded. This allows to effectively reduce the
analysis to the zero and first order (or, in some special cases , zero,
first and second order) terms in the power expansion of the solution with
respect to $\alpha $. The zero and first order terms are explicitly given by
the linear response and the FNLR respectively, \ and we can explicitly
estimate the contribution given by higher order terms in $\alpha $. The
analysis of the dependence on the parameter $\varrho $, which describes the
slow modulation of the excitation currents, allows to recast the FNLR, given
in the form of causal integrals, in terms of simpler expressions involving
the frequency-dependent susceptibility. These two steps can be done for
general excitation currents and solutions which are not necesserily
localized in the quasimomentum domain, see \cite{BF1}-\cite{BF3}, see also
Section 6. The third step which introduces the NLS regimes is based on
almost time-harmonic excitation currents in a fixed band labeled by $n_{0}$
with quasimomenta from a small vicinity of a fixed quasimomentum $\mathbf{k}%
_{\ast }$, with the linear dimensions of the vicinity described by a small
parameter $\beta $. The modes are separated into two classes: (i) directly
excited modes, for which the linear response is not zero, these modes must
have quasimomenta about $\pm \mathbf{k}_{\ast }$; (ii) indirectly excited
ones, for which the linear response is zero. These two classes obviously
differ by the magnitude of their modal coefficients. Since there is no
exchange of energy between Bloch modes for the linear Maxwell equations, the
indirectly excited modes are excited only through nonlinear interactions and
their principal part is explicitly given by the FNLR. Thanks to the
inversion symmetry (\ref{invsym}) the strongly excited doublets interact
with themselves much stronger than with indirectly excited modes. In fact,
their self-interactions are described with a high precision by a sistem of
two (or four when the backward propagating mode is excited) NLS or extended
NLS equations.

We also would like to remark that in the analysis of NLS regimes it is
rather common to introduce a single small parameter. In our framework it can
be achieved by setting, for instance, $\alpha =\varrho =\beta ^{2}$. We find
that such a reduction to a single small parameter not only does not simplify
the analysis, but, on the contrary, it entangles needlessly in bundles
different interaction terms obscuring roles played by different parameters
in nonlinear interactions. Moreover, the approximation of the NLM by the NLS
is valid for any power dependence $\varrho =\beta ^{q}$ independently on the
particular value of $q$, thus allowing to use the analysis \ of the
dependence of solutions of the NLS on these parameters to study solutions of
the original NLM. Of course, after the basic analysis is done one can choose
a fixed dependence between the parameters, for example the classical NLS\
scaling $\alpha =\varrho =\beta ^{2}$ and look at finer details under this
specific assumption; the different scalings may lead to \ different NLS-type
equations, see for example subsection 1.3.7.

\subsubsection{Multiple mode excitations and waves}

We have already pointed out that it is natural and useful to consider
multimode excitations and waves when studing the nonlinear wave evolution .
Such mulimode excitations can be introduced as follows. First we introduce
the excitation current of a more general form than in (\ref{Jzin}), namely%
\begin{equation}
\mathbf{J}=\sum_{l=1}^{N}\mathbf{J}_{l}  \label{JJ1}
\end{equation}%
with every of $\mathbf{J}_{l}$ being an almost single-mode excitation given
by (\ref{J01}), (\ref{Jzin}) with corresponding $\mathbf{k}_{\ast l}$ and $%
n_{0l}$,\ $l=1,\ldots ,N$. Consequently, every $\mathbf{J}_{l}$ excites the
corresponding doublet $\left\uparrow n_{l},\mathbf{k}_{\ast
l}\right\downarrow $. The modal components corresponding to the group $B_{l}$
of modes with $\left\vert \mathbf{k}-\mathbf{k}_{\ast l}\right\vert \precsim
\beta $ are directly excited through the linear process, and the amplitudes
of the directly excited modes are considerably higher (of the order $O\left(
\alpha ^{-1}\right) $ times) than the same for the indirectly excited modes.

We assume in this subsection that the ratio$\frac{\varrho }{\beta }$
satisfies the condition (\ref{rho/beta}), or, more precisely, that 
\begin{equation}
\frac{\varrho }{\beta }\ll \max_{l=1,\ldots ,N}\left\Vert \omega
_{n_{0}}^{\prime }\left( \mathbf{k}_{\ast l}\right) \right\Vert ,\;N\geq 2.
\label{GVMC}
\end{equation}%
The condition (\ref{GVMC}) evidently requires the group velocities to be
much larger compare to $\frac{\varrho }{\beta }$ (this condition is not
required in the single mode case $N=1$).

To determine finer features of the wave dynamics we pose the following
questions:

\begin{itemize}
\item Which modes are excited through nonlinear interactions and what are
the magnitudes of the amplitudes of such modes.

\item Which interactions determine the dynamics of the directly excited
modes with a given precision.

\item What are the equations which determine the dynamics of the directly
excited modes.

\item What is the influence of indirectly excited modes on the directly
excited modes.
\end{itemize}

The answers to the above questions depend on the choice of the quasimomenta $%
\mathbf{k}_{\ast l}$,\ $l=1,\ldots ,N$. It turns out that there are special
combinations of modes having the strongest interactions and playing the
dominant role for the wave nonlinear evolution. Such special combinations
involve exactly two ( $N=2$) special pairs of modes corresponding to the two
values of $\vartheta =\pm 1$ for a two doublets $\left\uparrow n_{0},\mathbf{%
k}_{\ast }\right\downarrow _{\vartheta }=\left\uparrow n_{0},\vartheta 
\mathbf{k}_{\ast }\right\downarrow $. The values $\pm \omega _{n}\left( 
\mathbf{k}_{\ast }\right) $ of the carrier frequencies of the excitation and
the quasimomenta $\pm \mathbf{k}_{\ast }$ are respectively the same for the
both doublets. The difference between the doublets is in the value of the
group velocity $\vartheta \omega _{n}^{\prime }\left( \mathbf{k}_{\ast
}\right) $ which is opposite for alternate doublets with $\vartheta =\pm 1$.
Such an excitation and the corresponding wave can be interpreted as \emph{%
bidirectional}, see also Subsection 1.3.6 and 5.4. In the case of a
bidirectional excitation the wave evolution can be approximated by a
four-component system of NLS equations which reduces to \ a two-component
system (\ref{Bi5}), (\ref{Bi51}) for the real-valued fields. Note that
relevant interactions between the four modes of the bi-directional
quadruplet are determined by the selection rules. Let us look briefly at the
interactions. The consideration will be useful for a more general case we
consider below. There are $4^{4}=16$ interacting quadruplets that can be
formed based on the described four modes. Taking into account that the
excitation currents are localized about $\vartheta \zeta \mathbf{k}_{\ast }$
we deduce from the selection rules (\ref{PM0}), (\ref{GVM0}), (\ref{FMC0}
the following approximate equalities becoming exact as $\beta \rightarrow 0$%
: 
\begin{equation}
\vartheta ^{\prime }\zeta ^{\prime }\mathbf{k}_{\ast }+\vartheta ^{\prime
\prime }\zeta ^{\prime \prime }\mathbf{k}_{\ast }+\vartheta ^{\prime \prime
\prime }\zeta ^{\prime \prime \prime }\mathbf{k}_{\ast }=\vartheta \zeta 
\mathbf{k}_{\ast }+O\left( \beta \right) \quad \func{mod}\left( 2\pi \right)
,  \label{PM0*}
\end{equation}%
\begin{equation}
\zeta ^{\prime }\omega _{n^{\prime }}\left( \vartheta ^{\prime }\zeta
^{\prime }\mathbf{k}_{\ast }\right) +\zeta ^{\prime \prime }\omega
_{n^{\prime \prime }}\left( \vartheta ^{\prime \prime }\zeta ^{\prime \prime
}\mathbf{k}_{\ast }\right) +\zeta ^{\prime \prime \prime }\omega _{n^{\prime
\prime \prime }}\left( \vartheta ^{\prime \prime \prime }\zeta ^{\prime
\prime \prime }\mathbf{k}_{\ast }\right) =\zeta \omega _{n}\left( \vartheta
\zeta \mathbf{k}_{\ast }\right) +O\left( \beta \right) ,  \label{FMC0*}
\end{equation}%
\begin{eqnarray}
\vartheta ^{\prime }\omega _{n^{\prime }}^{\prime }\left( \vartheta ^{\prime
}\zeta ^{\prime }\mathbf{k}_{\ast }\right) &=&\vartheta ^{\prime \prime
}\omega _{n^{\prime \prime }}^{\prime }\left( \vartheta ^{\prime \prime
}\zeta ^{\prime \prime }\mathbf{k}_{\ast }\right) +O\left( \beta \right) ,
\label{GVM0*} \\
\vartheta ^{\prime \prime }\omega _{n^{\prime \prime }}^{\prime }\left(
\vartheta ^{\prime \prime }\zeta ^{\prime \prime }\mathbf{k}_{\ast }\right)
&=&\vartheta ^{\prime \prime \prime }\omega _{n^{\prime \prime \prime
}}^{\prime }\left( \vartheta ^{\prime \prime \prime }\zeta ^{\prime \prime
\prime }\mathbf{k}_{\ast }\right) +O\left( \beta \right) .  \notag
\end{eqnarray}%
For small $\beta $ the above equations would hold if the binary variables $%
\vartheta =\pm 1$, $\zeta \pm 1$ satisfy exactly\ the following equations 
\begin{equation}
\vartheta ^{\prime }\zeta ^{\prime }+\vartheta ^{\prime \prime }\zeta
^{\prime \prime }+\vartheta ^{\prime \prime \prime }\zeta ^{\prime \prime
\prime }=\vartheta \zeta ,  \label{gamzetaph}
\end{equation}

\begin{equation}
\zeta ^{\prime }+\zeta ^{\prime \prime }+\zeta ^{\prime \prime \prime
}=\zeta ,  \label{gamzetafr}
\end{equation}%
\begin{equation}
\vartheta ^{\prime }=\vartheta ^{\prime \prime }=\vartheta ^{\prime \prime
\prime }.  \label{gamzetagv}
\end{equation}%
For example, the set 
\begin{equation}
\zeta ^{\prime }=1,\ \zeta ^{\prime \prime }=1,\ \zeta ^{\prime \prime
\prime }=-1,\ \zeta =1;\quad \vartheta ^{\prime }=1,\ \vartheta ^{\prime
\prime }=1,\ \vartheta ^{\prime \prime \prime }=1,\ \vartheta =1,
\label{gamz1}
\end{equation}%
satisfies all three conditions (\ref{gamzetaph}), (\ref{gamzetafr}),\ (\ref%
{gamzetagv}). The related solution corresponds to the interaction inside one
doublet with $\vartheta =1$, and the corresponding interaction is well
approximated by the NLS nonlinearity. Note that when the origin modes are
all from the same doublet $\vartheta ^{\prime }=\vartheta ^{\prime \prime
}=\vartheta ^{\prime \prime \prime }$, the equations (\ref{gamzetaph}) and (%
\ref{gamzetafr}) necesserily imply that the end mode is too from the same
doublet. A solution of (\ref{gamzetaph}), (\ref{gamzetafr}) which has the
origin modes from both doublets 
\begin{equation}
\zeta ^{\prime }=1,\ \zeta ^{\prime \prime }=1,\ \zeta ^{\prime \prime
\prime }=-1,\ \zeta =1;\quad \vartheta ^{\prime }=-1,\ \vartheta ^{\prime
\prime }=1,\ \vartheta ^{\prime \prime \prime }=-1,\ \vartheta =1.
\label{gamz2}
\end{equation}%
corresponds to the interaction for which the phase matching and frequency
matching conditions are fulfilled, but this interaction does not satisfy the
group velocity condition (\ref{gamzetagv}). Hence, the magnitude\ of
interaction is of order $O\left( \left( \frac{\varrho }{\beta }\right)
^{\kappa }\right) $ with arbitrary large $\kappa $ and, consequently, it is
negligible. Now let us look at the solution of (\ref{gamzetaph}), (\ref%
{gamzetagv}) for which \ 
\begin{equation}
\zeta ^{\prime }=-1,\ \zeta ^{\prime \prime }=1,\ \zeta ^{\prime \prime
\prime }=1,\ \zeta =1;\quad \vartheta ^{\prime }=1,\ \vartheta ^{\prime
\prime }=1,\ \vartheta ^{\prime \prime \prime }=1,\ \vartheta =-1.
\label{gamz3}
\end{equation}%
This solution corresponds to the interaction with the origin triad taken
from the doublet with $\vartheta =1$\ and the end mode from the second
doublet with $\vartheta =-1$. The frequency matching condition for this
interaction does not hold and its relative magnitude is $O\left( \varrho
\right) $ times of the magnitude of the intraduplet interaction. At the
lowest order of approximation this interaction can be neglected leading to
uncoupled NLS equations for every one of the two douplets. When we
approximate NLM\ with a higher accuracy, we have to take into account this
interaction, it is well approximated by the term with the coefficient $%
\delta _{\times ,+}^{+}$ in the bi-directional ENLS system (\ref{Bi1}), (\ref%
{Bi2}).

Now let us consider the case of a general multimodal excitation. Similarly
to (\ref{PM0*})-(\ref{GVM0*}) applying the selection rules to origin modes
from directly excited doublets $\left\uparrow n_{l},\mathbf{k}_{\ast
l}\right\downarrow $ and the end mode from an arbitrary doublet $%
\left\uparrow n,\mathbf{k}_{\ast \ast }\right\downarrow $ we get 
\begin{equation}
\zeta _{l_{1}}\mathbf{k}_{\ast l_{1}}+\zeta _{l_{2}}\mathbf{k}_{\ast
l_{2}}+\zeta _{l_{3}}\mathbf{k}_{\ast l_{3}}=\zeta \mathbf{k}_{\ast \ast
}+O\left( \beta \right) \quad \func{mod}\left( 2\pi \right) ,  \label{PM04}
\end{equation}%
\begin{equation}
\zeta _{l_{1}}\omega _{n_{l_{1}}}\left( \zeta _{l_{1}}\mathbf{k}_{\ast
l_{1}}\right) +\zeta _{l_{2}}\omega _{n_{l_{2}}}\left( \zeta _{l_{2}}\mathbf{%
k}_{\ast l_{2}}\right) +\zeta _{l_{3}}\omega _{n_{l_{3}}}\left( \zeta
_{l_{3}}\mathbf{k}_{\ast l_{3}}\right) =\zeta \omega _{n}\left( \zeta 
\mathbf{k}_{\ast \ast }\right) +O\left( \beta \right) ,  \label{FMC04}
\end{equation}%
\begin{equation}
\omega _{n_{l_{1}}}^{\prime }\left( \zeta _{l_{1}}\mathbf{k}_{\ast
l_{1}}\right) =\omega _{n_{l_{2}}}^{\prime }\left( \zeta _{l_{2}}\mathbf{k}%
_{\ast l_{2}}\right) +O\left( \beta \right) ,\omega _{n_{l_{2}}}^{\prime
}\left( \zeta _{l_{2}}\mathbf{k}_{\ast l_{2}}\right) =\omega
_{n_{l_{3}}}^{\prime }\left( \zeta _{l_{3}}\mathbf{k}_{\ast l_{3}}\right)
+O\left( \beta \right) .  \label{GVM04}
\end{equation}%
First we consider the case when the end mode belongs to a directly excited
doublet, i.e. 
\begin{equation}
n=n_{l_{4}},\quad \zeta =\zeta _{l_{4}},\quad \mathbf{k}_{\ast \ast }=%
\mathbf{k}_{\ast l_{4}}.  \label{dird}
\end{equation}%
In a generic case these equations have solutions only if $%
l_{1}=l_{2}=l_{3}=l_{4}$. This means that the evolution of the modal
components corresponding to a group $B_{l}$ of modes with $\left\vert 
\mathbf{k}-\mathbf{k}_{\ast l}\right\vert \precsim \beta $ is essentially
independent from similar components for $B_{m}$ where $m\neq l$. In addition
to that, the nonlinear evolution for the components from the same group $%
B_{l}$ is described by a Nonlinear Schrodinger equation denoted by $\limfunc{%
NLS}_{l}$, $l=1,\ldots ,N$. \emph{In other words, we have a system of }$N$%
\emph{\ completely decoupled two-component systems }$\limfunc{NLS}_{l}$\emph{%
.} In view of our general view on "almost independence" between different
groups of modes $B_{l}$ we notice that though the interactions between
different groups are not zero, in the generic case they are smaller (of
higher powers of the parameters $\varrho $ and $\beta $) compared with the
interactions inside of the doublet. In some special cases, for example, in
the bi-directional case (which is not, strictly speaking, generic) the
magnitude of the interactions between a two doublets is not negligible when
a higher order of precision is assumed, it can be estimated in terms of
positive powers of the small parameters $\alpha $, $\varrho $ and $\beta $. 
\emph{Therefore, the exact system of }$N$\emph{\ evolution equations for the
groups of modes }$B_{l}$\emph{,\ }$l=1,\ldots ,N$\emph{\ reduces to a system
of }$N$\emph{\ completely decoupled equations }$\left\{ \limfunc{NLS}%
_{l}\right\} $\emph{\ only at a certain level of accuracy.} To get more
accurate evolution equations one has to introduce new terms in the system $%
\left\{ \limfunc{NLS}_{l}\right\} $ which would couple the equations from
this system. The construction of such new most significant coupling terms
can be approached as follows. Note first that if (\ref{GVM04}) or (\ref{PM04}%
) is not satisfied the corresponding interaction is negligible at any level
of accuracy. This leads to the following conditions 
\begin{equation}
l_{1}=l_{2}=l_{3},  \label{triple}
\end{equation}%
\begin{equation}
\left( \zeta _{l_{1}}+\zeta _{l_{2}}+\zeta _{l_{3}}\right) \mathbf{k}_{\ast
\ast }=\zeta \mathbf{k}_{\ast \ast }+O\left( \beta \right) \quad \func{mod}%
\left( 2\pi \right) ,  \label{triple1}
\end{equation}%
Here $\zeta _{l_{1}},\zeta _{l_{2}},\zeta _{l_{3}}$ can independently take
values $\pm 1$. \ It follows from (\ref{triple}) that all the three modes of
the origin triplet of the interaction quadruplet has to be chosen from the
same doublet. The phase matching condition (\ref{triple1}) for the
quasimomentum $\mathbf{k=k}_{\ast \ast }=\mathbf{k}_{\ast l_{4}}\in
B_{l_{4}} $ from the end mode of the interaction quadruplet implies that%
\begin{eqnarray}
\mathbf{k}_{\ast \ast } &=&\pm \mathbf{k}_{\ast l}\quad \func{mod}\left(
2\pi \right) ,  \label{JJ2} \\
\text{ or }\mathbf{k}_{\ast \ast } &=&\pm 3\mathbf{k}_{\ast l}\quad \func{mod%
}\left( 2\pi \right) .  \label{JJ23}
\end{eqnarray}%
As to the band number $n=n_{l_{4}}$, it can be different from $%
n_{l_{1}}=n_{l_{2}}=n_{l_{3}}$. The case (\ref{JJ2}) corresponds to the
excitation of the second doublet of the bi-directional quadruplet, it can
also excite doublets $\left\uparrow n,\pm \mathbf{k}_{\ast
l}\right\downarrow $ in all bands with $n\neq n_{l_{1}}$. With the exclusion
of the bi-directional case $n=n_{l_{1}}$, in a generic situation all these
doublets are indirectly excited.\ Similarly, the case (\ref{JJ23})
corresponds to the indirect excitation of all doublets $\left\uparrow n,\pm 3%
\mathbf{k}_{\ast l}\right\downarrow $ in all bands. The amplitudes of the
indirectly excited doublets in this case are determined by the directly
excited modes solely with the principal part given by the first nonlinear
response (though when $\alpha \sim \varrho $ higher order responses have to
be taken into account, but the magnitude of the contribution can be
estimated by the same expression). In the generic case the frequency
matching condition (\ref{FMC04}) for the indirectly excited modes does not
hold, and, hence, the magnitude of these interactions estimated by $O\left(
\alpha \varrho \left\vert \mathbf{U}^{\left( 1\right) }\right\vert \right) $%
, $\varrho \ll 1$, it is evidently much smaller compared with $O\left(
\alpha \left\vert \mathbf{U}^{\left( 1\right) }\right\vert \right) $ which
is the magnitude of nonlinear interactions inside a doublet.

\subsubsection{Mode-to-mode coupling and almost independence}

It turns out that properly defined different types of mode combinations
evolve almost-independently for long times and high accuracy. In this
section we introduce concepts and give a sketch\ of constructions needed for
establishing that amost-indepedence and more.

As in previous subsesection we consider the current $\mathbf{J}%
=\sum_{l=1}^{N}\mathbf{J}_{l}$ with currents $\mathbf{J}_{l}$ described
there, and denote by $B_{l}$ of a set of directly excited modes by the
current $\mathbf{J}_{l}$, namely%
\begin{equation}
B_{l}=\left\{ \left( \zeta ,n,\mathbf{k}\right) :n=n_{0l},\;\left\vert 
\mathbf{k}-\zeta \mathbf{k}_{\ast l}\right\vert \leq \pi _{0},\;\zeta =+%
\text{ \ or \ }\zeta =-\right\} ,\ l=1,\ldots ,N.  \label{Bl}
\end{equation}%
We consider also: (i) the complement $B_{l}^{\limfunc{C}}$ for every $B_{l}$%
, (ii) $B$ as the union of all $B_{l}$; (iii) the complement $B^{\limfunc{C}%
} $, namely%
\begin{equation}
B=\bigcup_{l=1,\ldots ,N}B_{l},\ B^{\limfunc{C}}=\left( \bigcup_{l=1,\ldots
,N}B_{l}\right) ^{\limfunc{C}}.  \label{Bl2}
\end{equation}%
Then we introduce a decomposition of the wave $\mathbf{U}$ governed by the
NLM based on $B_{l}$, namely 
\begin{equation}
\mathbf{U}=\mathbf{U}_{B_{1}}+\ldots +\mathbf{U}_{B_{N}}+\mathbf{U}_{B^{C}}
\label{UB}
\end{equation}%
where $\mathbf{U}_{B_{l}}$ is composed of modes from $B_{l}$. Using such a
decomposition we recast NLM (\ref{MXshort}) in the form of the following
system of equations: 
\begin{gather}
\partial _{t}\mathbf{U}_{B_{1}}=\mathbf{-}\mathrm{i}\mathbf{MU}%
_{B_{1}}+\alpha \left. \mathcal{F}_{\text{NL}}\left( \mathbf{U}%
_{B_{1}}+\ldots +\mathbf{U}_{B_{N}}+\mathbf{U}_{B^{C}}\right) \right\vert
_{B_{1}}-\mathbf{J}_{1};  \label{UB1} \\
\ldots  \notag \\
\partial _{t}\mathbf{U}_{B_{N}}=\mathbf{-}\mathrm{i}\mathbf{MU}%
_{B_{N}}+\alpha \left. \mathcal{F}_{\text{NL}}\left( \mathbf{U}%
_{B_{1}}+\ldots +\mathbf{U}_{B_{N}}+\mathbf{U}_{B^{C}}\right) \right\vert
_{B_{N}}-\mathbf{J}_{N};  \label{UBN} \\
\partial _{t}\mathbf{U}_{B^{\limfunc{C}}}=\mathbf{-}\mathrm{i}\mathbf{MU}%
_{B^{\limfunc{C}}}+\alpha \left. \mathcal{F}_{\text{NL}}\left( \mathbf{U}%
_{B_{1}}+\ldots +\mathbf{U}_{B_{N}}+\mathbf{U}_{B^{\limfunc{C}}}\right)
\right\vert _{B^{\limfunc{C}}};\   \label{UBC0} \\
\mathbf{U}_{B_{1}}=0,...,\mathbf{U}_{B_{N}}=0;\;\mathbf{U}_{B^{\limfunc{C}%
}}=0\;\text{for }t\leq 0.  \label{UBC}
\end{gather}%
Now having the system (\ref{UB1})-(\ref{UBC}) we can give a precise meaning
to the almost-independence of different mode combinations such as $B_{l}$
for different $l$ and their independence of $B^{\limfunc{C}}$. Indeed, we
interpret and define the almost-independence of different $B_{l}$ as the
almost-independence of the components $\mathbf{U}_{B_{l}}$ and $\mathbf{U}%
_{B^{\limfunc{C}}}$ for different $l$ which satisfy the system (\ref{UB1})-(%
\ref{UBC}). It remains, of course, to define the almost-independence of the
components $\mathbf{U}_{B_{l}}$ and $\mathbf{U}_{B^{\limfunc{C}}}$ for
different $l$ which satisfy the system (\ref{UB1})-(\ref{UBC}), that we do
as follows.

If $\mathbf{U}_{B_{1}},$...,$\mathbf{U}_{B_{N}}$ solving the system (\ref%
{UB1})-(\ref{UBC}) were indepedent then we would be able to drop in every
right-hand side of every equation in (\ref{UB1})-(\ref{UBC}) everything but
the corresponding $\mathbf{U}_{B_{l}}$, and would get the following system 
\begin{gather}
\partial _{t}\mathbf{V}_{B_{1}}=\mathbf{-}\mathrm{i}\mathbf{MV}%
_{B_{1}}+\alpha \left. \mathcal{F}_{\text{NL}}\left( \mathbf{V}%
_{B_{1}}\right) \right\vert _{B_{1}}-\mathbf{J}_{1};\ \mathbf{V}_{B_{1}}=0\;%
\text{for }t\leq 0,  \label{VB1} \\
\ldots  \notag \\
\partial _{t}\mathbf{V}_{B_{N}}=\mathbf{-}\mathrm{i}\mathbf{MV}%
_{B_{N}}+\alpha \left. \mathcal{F}_{\text{NL}}\left( \mathbf{V}%
_{B_{N}}\right) \right\vert _{B_{N}}-\mathbf{J}_{N};\ \mathbf{V}_{B_{N}}=0\;%
\text{for }t\leq 0,  \label{VBN} \\
\partial _{t}\mathbf{V}_{B^{\limfunc{C}}}=\mathbf{-}\mathrm{i}\mathbf{MV}%
_{B^{\limfunc{C}}}+\alpha \left. \mathcal{F}_{\text{NL}}\left( \mathbf{V}%
_{B_{1}}+\ldots +\mathbf{V}_{B_{N}}\right) \right\vert _{B^{\limfunc{C}}};\ 
\mathbf{V}_{B^{\limfunc{C}}}=0\;\text{for }t\leq 0.  \label{VBC}
\end{gather}%
In other words, the $l$-th \ equation in (\ref{VB1})-(\ref{VBN}) for $%
\mathbf{V}_{B_{l}}$ is obtained from the $l$-th \ equation for $\mathbf{U}%
_{B_{l}}$\ by dropping the $\mathbf{U}_{B_{j}}$, $j\neq l$ and $\mathbf{U}%
_{B^{\limfunc{C}}}$ in the nonlinear term. Obviously, the first $N$
equations in (\ref{VB1})-(\ref{VBC}) can be solved independently, and the
very last equation (\ref{VBC}) is linear with respect to $\mathbf{V}_{B^{C}}$
and can be easily solved too.

To find the nonlinear influence of modes from $B_{l}$ onto themself we take
the $l$-th \ equation in (\ref{VB1})-(\ref{VBN}) and set $\mathbf{V}_{B_{l}}$
in the nonlinear term $\mathcal{F}_{\text{NL}}$ to be zero that leads to the
following linear equation 
\begin{equation}
\partial _{t}\mathbf{V}_{B_{l}}^{\left( 0\right) }=\mathbf{-}\mathrm{i}%
\mathbf{MV}_{B_{l}}^{\left( 0\right) }-\mathbf{J}_{l};\ \mathbf{V}%
_{B_{l}}^{\left( 0\right) }=0\;\text{for }t\leq 0.  \label{VBB1}
\end{equation}

Now we can assess the level of independence or coupling of different $%
\mathbf{U}_{B_{l}}$ and $\mathbf{U}_{B^{\limfunc{C}}}$ by comparing them
with the corresponding $\mathbf{V}_{B_{l}}$ and $\mathbf{V}_{B^{\limfunc{C}%
}} $, and similarly we can compare $\mathbf{V}_{B_{l}}$ with $\mathbf{V}%
_{B_{l}}^{\left( 0\right) }$ to assess the nonlinear influence of modes $%
B_{l}$ onto themself. Namely, we define the \emph{mode-to-mode coupling} as
follows%
\begin{gather}
\text{mode-to-mode coupling }B_{l}^{\limfunc{C}}\rightarrow B_{l}\equiv 
\mathbf{U}_{B_{l}}-\mathbf{V}_{B_{l}},  \label{VBB2} \\
\text{mode-to-mode coupling }B_{l}\rightarrow B_{l}\equiv \mathbf{V}_{B_{l}}-%
\mathbf{V}_{B_{l}}^{\left( 0\right) }.  \notag
\end{gather}%
We would like to underline that \emph{the definition of mode-to-mode
coupling includes the direction of influence} via the corresponding
evolution equations (\ref{UB1})-(\ref{UBC}) and (\ref{VB1})-(\ref{VBC}), and
the\emph{\ mode-to-mode coupling is not symmetric}. The analysis of
nonlinear evolution requires to introduce such a direction of influence for
nonlinearly interacting modes.

An additional analysis of the equations (\ref{VB1})-(\ref{VBC}) also shows
that $\mathbf{V}_{B_{l}}$ can be well approximated by a solution of a
corresponding NLS or ENLS systems. In addition to that, estimates similar to
(\ref{Unne0}) for indirectly excited modes $\mathbf{U}_{B^{C}}$ in one
dimensional case $d=1$ with the classical NLS scaling $\varrho \sim \alpha
\sim \beta ^{2}$ yield that 
\begin{equation}
\mathbf{U}_{B^{\limfunc{C}}}-\mathbf{V}_{B^{\limfunc{C}}}=O\left( \alpha
^{2}\right) =O\left( \beta ^{4}\right) .
\end{equation}

In Table \ref{tabmagintD} we have collected order of magnitude estimates of
the mode-to-mode interactions involving unidirectional excitations and
doublets ($\beta ^{\infty }$ in this table means arbitrarily large power of $%
\beta $).

\begin{table}[tbp] \centering%
\begin{tabular}{|l|r|}
\hline
\multicolumn{2}{|c|}{%
\begin{tabular}{c}
\textbf{Order of nonlinear mode-to-mode coupling for} \\ 
\textbf{unidirectional excitations under the classical NLS scaling } $%
\varrho \sim \alpha \sim \beta ^{2}$%
\end{tabular}%
} \\ \hline
Mode-to-mode coupling & Order of the mode-to-mode coupling for$\;\frac{\tau
_{0}}{\varrho }\leq t\leq \frac{\tau _{\ast }}{\varrho }$ \\ \hline
$\limfunc{doublet}_{l}\rightarrow \limfunc{doublet}_{l}$ & $\frac{\alpha }{%
\varrho }\sim 1$ \\ \hline
$\limfunc{doublet}_{l}\rightarrow \limfunc{doublet}_{l}^{\limfunc{C}}$ & $%
\alpha \sim \beta ^{2}$ \\ \hline
$\limfunc{doublet}_{l}^{\limfunc{C}}\rightarrow \limfunc{doublet}_{l}$ & $%
\beta ^{\infty }$ \\ \hline
\end{tabular}%
\caption{The entries show the magnitude of nonlinesr mode-to-mode coupling for unidirectional excitations 
in one-dimensional case under the classical NLS scaling.. \label{tabmagintD}}%
\end{table}%

When the excitations are bi-directional, quadruplets of modes are excited,
and in this case magnitudes of nonlinear interactions are as in Table \ref%
{tabmagintQ} which is similar to Table \ref{tabmagintD}.

\begin{table}[tbp] \centering%
\begin{tabular}{|l|r|}
\hline
\multicolumn{2}{|c|}{%
\begin{tabular}{c}
\textbf{Order of nonlinear mode-to-mode coupling for} \\ 
\textbf{bidirectional excitations under the classical NLS scaling } $\varrho
\sim \alpha \sim \beta ^{2}$%
\end{tabular}%
} \\ \hline
Mode-to-mode coupling & Order of the mode-to-mode coupling$\;$for$\;\frac{%
\tau _{0}}{\varrho }\leq t\leq \frac{\tau _{\ast }}{\varrho }$ \\ \hline
$\limfunc{quadruplet}_{l}\rightarrow \limfunc{quadruplet}_{l}$ & $\frac{%
\alpha }{\varrho }\sim 1$ \\ \hline
$\limfunc{quadruplet}_{l}\rightarrow \limfunc{quadruplet}_{l}^{\limfunc{C}}$
& $\alpha \sim \beta ^{2}$ \\ \hline
$\limfunc{quadruplet}_{l}^{\limfunc{C}}\rightarrow \limfunc{quadruplet}_{l}$
& $\beta ^{\infty }$ \\ \hline
\end{tabular}%
\caption{The entries show the magnitudes of nonlinesr  mode-to-mode coupling for bidirectional excitations in one-dimensional 
case under the classical NLS scaling. \label{tabmagintQ}}%
\end{table}%

The order of magnitude comparative estimates for the basic system (\ref{UB1}%
)-(\ref{UBC}) and its decoupled counterpart (\ref{VB1})-(\ref{VBC}) provide
additional facts on the interplay between dispersion and nonlinearity. These
estimates are collected in Table \ref{tabscal}, and they are based on the
analysis of the exact solution $\mathbf{U}\left( t\right) $ of the NLM
involving instrumentally: (i) the analytic expansion (\ref{uMv1}) for $%
\mathbf{U}\left( t\right) $; (ii) representation of the terms of that
expansion (\ref{uMv1}) by oscillatory integrals; (iii) computation of
asymptotic approximations and series for these oscillatory integrals as
powers of the small parameters $\alpha $, $\varrho $ and $\beta $. Observe
that for a generic $\mathbf{W}$, which can be expanded as in (\ref{UB}), the
value $\alpha \left. \mathcal{F}_{\text{NL}}\left( \mathbf{W}\right)
\right\vert _{B_{l}}\mathcal{\ }$differs noticeably from $\alpha \left. 
\mathcal{F}_{\text{NL}}\left( \mathbf{W}_{B_{l}}\right) \right\vert _{B_{l}}$
and the difference is of order $\beta ^{2}$. In contrast, in the case when $%
\mathbf{W}$ is the exact solution $\mathbf{U}\left( t\right) $ of the NLM
the same difference for $\frac{\tau _{0}}{\varrho }\leq t\leq \frac{\tau
_{\ast }}{\varrho }$ is of order $\beta ^{\infty }$, that is much smaller.
Such a difference is due to distructive wave interference and wave
dispersion for a wave governed exactly by the NLM.

\begin{table}[tbp] \centering%
\begin{tabular}{|l|r|}
\hline
\multicolumn{2}{|l|}{%
\begin{tabular}{c}
\textbf{Comparison of solutions to the basic system and its decoupled} \\ 
\textbf{\ counterpart under the classical\ NLS scaling} $\varrho \sim \alpha
\sim \beta ^{2}$\textbf{\ } \\ 
\textbf{for }$\frac{\tau _{0}}{\varrho }\leq t\leq \frac{\tau _{\ast }}{%
\varrho }$%
\end{tabular}%
} \\ \hline
Solutions $\mathbf{U}_{B_{l}},\mathbf{V}_{B_{l}}$ & $1$ \\ \hline
Nonlinearity $\alpha \left. \mathcal{F}_{\text{NL}}\left( \mathbf{U}\right)
\right\vert _{B_{l}}$, $\alpha \left. \mathcal{F}_{\text{NL}}\left( \mathbf{V%
}_{B_{l}}\right) \right\vert _{B_{l}}$ & $\alpha $ \\ \hline
\begin{tabular}{l}
Difference of the values of nonlinearity on generic test functions \\ 
$\alpha \left. \mathcal{F}_{\text{NL}}\left( \mathbf{W}_{B_{1}}+\ldots +%
\mathbf{W}_{B_{N}}+\mathbf{W}_{B^{C}}\right) \right\vert _{B_{l}}-\alpha
\left. \mathcal{F}_{\text{NL}}\left( \mathbf{W}_{B_{l}}\right) \right\vert
_{B_{l}}$%
\end{tabular}
& $\alpha \sim \beta ^{2}$ \\ \hline
\begin{tabular}{l}
Difference of the values of nonlinearity applied to solutions \\ 
$\alpha \left. \mathcal{F}_{\text{NL}}\left( \mathbf{U}_{B_{1}}\left(
t\right) +\ldots +\mathbf{U}_{B_{N}}\left( t\right) +\mathbf{U}%
_{B^{C}}\left( t\right) \right) \right\vert _{B_{l}}-\alpha \left. \mathcal{F%
}_{\text{NL}}\left( \mathbf{U}_{B_{l}}\left( t\right) \right) \right\vert
_{B_{l}}$%
\end{tabular}
& $\beta ^{\infty }$ \\ \hline
Difference of solutions $\mathbf{U}_{B_{l}}\left( t\right) -\mathbf{V}%
_{B_{l}}\left( t\right) $ & $\beta ^{\infty }$ \\ \hline
\end{tabular}%
\caption{
The order of magnitude estimates collected here are based on the analysis of the
exact solution $\mathbf{U}\left( t\right) $ of the NLM involving
instrumentally: (i) the analytic expansion (\ref{uMv1}) for $\mathbf{U}\left( t\right) $; (ii) representation of the terms of that expansion (\ref{uMv1}) by oscillatory integrals; (iii) computation of asymptotic
approximations and series for these oscillatory integrals as powers of the
small parameters $\alpha $, $\varrho $ and $\beta $. 
 \label{tabscal}}%
\end{table}%

We end the section by the following qualitative conclusions on the interplay
between dispersive and nonlinear effects:

\begin{itemize}
\item dispersive effects balance nonlinear effects when mode interact inside
one doublet leading to NLS/ENLS type dynamics;

\item dispersive effects are dominant in interactions between different
doublets, and nonlinear effects are less pronounced.
\end{itemize}

\subsubsection{Spectral theory of nonlinear wave propagation}

The above discussion suggests that the theory of NLS, ENLS and systems of
coupled ENLS equations can be viewed as the \emph{spectral theory of
nonlinear wave propagation}. The word "spectral" here refers to the property
of certain classes of waves to be decomposable into components evolving 
\emph{almost independently} for long times as described in the previous
section. The "almost independence", in turn, means that the coupling between
the components is small, and, more precisely, that the coupling terms\ in
the relevant exact evolution equations can be classified by powers $\alpha
^{l_{0}}\varrho ^{l_{1}}\beta ^{l_{2}}$. We remind that the small parameters 
$\alpha $, $\varrho $ and $\beta $ introduced in previous sections
characterize respectively the relative magnitude of nonlinearity, the time
and the space scales related to the nonlinear evolutions. The parameter $%
\alpha $ characterizing the magnitude of the nonlinearity plays the leading
role in ordering levels of different nonlinear mode interactions by the
scale of positive integer powers of $\alpha $. The next is the small
parameter $\varrho $, which characterizes the degree of time-harmonicity of
the excitation wave. Positive powers $\varrho $ provide another scale for
mode interactions. And, finally, the third small parameter $\beta $
characterizes the linear dimensions of a small vicinity of a single or
several quasimomenta $\mathbf{k}_{\ast j}$ involved in the modal
decomposition of the wave. The parameter $\beta $ refines further the above
classification. \emph{When accounting for different magnitudes of mode
interactions as powers }$\alpha ^{l_{0}}\varrho ^{l_{1}}\beta ^{l_{2}}$ 
\emph{we come to either the classical NLS, ENLS or a system of ENLS
equations.} The obtained so equations take into account at the prescribed
precision level all relevant nonlinear interactions and with that level of
accuracy describe the nonlinear wave evolution. In such a contstruction, the
linear spectral theory forms a fundamental basis for the nonlinear one. It
yields the system of eigenmodes which evolve indepenently and set a
framework for the nonlinear spectral theory. \ 

In this article we focus primarily at almost single-mode excitation currents
and only sketch the case of multimode excitations. More detailed studies of
waves generated by multimode excitation currents and, in particular, the
derivation of the corresponding systems of ENLS equations accounting for
smaller coupling between essentially nointeracting groups of modes are
naturally to be conducted as the next step.

The essence of above disscusion on nonlinear evolution and wave interactions
can be formulated in the form of the following \emph{principle of
approximate superposition}. Let us call a solution to the NLM \emph{a
multiple-mode solution} if it correpsonds to an excitation current which is
generic and is a sum of almost time harmonic single mode excitations. \emph{%
Then being given a level of accuracy and any multiple-mode solution we can
decompose it into the sum of certain single-mode solutions each of which is
goverened by NLS or ENLS (can be a system) with a prescribed accuracy.}

More accurate formulation of the principle of approximate superposition is
as follows. Let $\mathbf{U}_{l}$ \ be a solution of the NLM corresponding to
an almost single-mode excitation $\mathbf{J}_{l}$ around $\mathbf{k}_{\ast
l} $, i.e. 
\begin{equation}
\partial _{t}\mathbf{U}_{l}=-\mathrm{i}\mathbf{MU}_{l}+\alpha \mathcal{F}_{%
\text{NL}}\left( \mathbf{U}_{l}\right) -\mathbf{J}_{l},\;l=1,\ldots ,N.
\label{NLMl}
\end{equation}%
Then for a generic collection of $\mathbf{k}_{\ast l}$ the multiple-mode
solution $\mathbf{U}$ corresponding to sum of $\mathbf{J}_{l}$ satisfies 
\begin{equation}
\partial _{t}\mathbf{U}=-\mathrm{i}\mathbf{MU}+\alpha \mathcal{F}_{\text{NL}%
}\left( \mathbf{U}\right) -\mathbf{J},\mathbf{\;J}=\sum_{l=1}^{N}\mathbf{J}%
_{l}  \label{NLMsum}
\end{equation}%
and 
\begin{equation}
\mathbf{U}=\sum_{l=1}^{N}\mathbf{U}_{l}+O\left( \frac{\varrho }{\beta }%
\right) ^{N_{1}},\text{ where }N_{1}\text{ can be arbitrarily large.}
\label{superpos}
\end{equation}%
Observe a remarkable "\emph{superaccuracy}" of the superpostion formula in (%
\ref{superpos}). For the typical scaling $\varrho \sim \alpha \sim \beta
^{\varkappa _{1}},\ \varkappa _{1}\geq 2,$ as in (\ref{kap1}), the
approximation error is smaller than any power of $\alpha $ whereas the
nonlinearity itself is of order $\alpha $. The explanation of the
superaccuracy follows from an analysis of nonlinear wave interactions which
we present here in a concise form.

First, the linear response $\mathbf{U}^{\left( 0\right) }$, i.e. the
solution of the linear equation (\ref{NLMsum}) with $\alpha =0$, satisfies
exactly the superposition principle, i.e. 
\begin{equation}
\mathbf{U}^{\left( 0\right) }=\sum_{l=1}^{N}\mathbf{U}_{l}^{\left( 0\right)
}.
\end{equation}%
The first nonlinear responses $\mathbf{U}_{l}^{\left( 1\right) }$ and $%
\mathbf{U}^{\left( 1\right) }$ to respectively the almost single-mode
currents $\mathbf{J}_{l}$ and to the multimple-mode sum $\mathbf{J}$ are
defined as the solutions to 
\begin{eqnarray}
\partial _{t}\mathbf{U}_{l}^{\left( 1\right) } &=&-\mathrm{i}\mathbf{MU}%
_{l}^{\left( 1\right) }+\alpha \mathcal{F}_{\text{NL}}^{\left( 3\right)
}\left( \mathbf{U}_{l}^{\left( 0\right) }\right) ,  \label{U1l} \\
\partial _{t}\mathbf{U}^{\left( 1\right) } &=&-\mathrm{i}\mathbf{MU}^{\left(
1\right) }+\alpha \mathcal{F}_{\text{NL}}^{\left( 3\right) }\left( \mathbf{U}%
^{\left( 0\right) }\right) ,  \label{U1sum}
\end{eqnarray}%
with all the solutions vanishing for negative times. Note that the solution $%
\mathbf{U}_{l}$ of (\ref{NLMl}) involves all modes excited by $\mathbf{J}%
_{l} $ directly and indirectly. The evolution of the directly excited mode $%
\mathbf{U}_{l}$ is described by an $\limfunc{NLS}_{l}$ or an ENLS$_{l}$
equations, whereas the evolution of indirecly excited modes of $\mathbf{U}%
_{l}$ is essentially described by the FNLR $\mathbf{U}_{l}^{\left( 1\right)
} $ to the excitation $\mathbf{J}_{l}$ which is of order $O\left( \varrho
\alpha \left\vert \mathbf{U}^{\left( 1\right) }\right\vert \right) $. The
approximation of $\mathbf{U}_{l}$ by ENLS$_{l}$ and $\mathbf{U}_{l}^{\left(
1\right) }$, which is the main topic of this article, holds with a high
precision, but the accuracy of the superposition formula (\ref{superpos}) is
even higher. To see why the relation (\ref{superpos}) holds let us take a
closer look at $\mathbf{U}^{\left( 1\right) }$. The equation (\ref{U1sum})
involves the term $\alpha \mathcal{F}_{\text{NL}}\left( \mathbf{U}^{\left(
0\right) }\right) $, since it is trilinear it has the form 
\begin{equation}
\mathcal{F}_{\text{NL}}^{\left( 3\right) }\left( \mathbf{U}^{\left( 0\right)
}\right) =\mathcal{F}_{\text{NL}}^{\left( 3\right) }\vdots \left(
\sum_{l=1}^{N}\mathbf{U}_{l}^{\left( 0\right) }\right)
^{3}=\sum_{l_{1},l_{2},l_{3}=1}^{N}\mathcal{F}_{\text{NL}}^{\left( 3\right)
}\left( \mathbf{U}_{l_{1}}^{\left( 0\right) }\mathbf{U}_{l_{2}}^{\left(
0\right) }\mathbf{U}_{l_{3}}^{\left( 0\right) }\right) ,
\end{equation}%
where the sum in the right-hand side contains $N^{3}$ terms. Therefore the
solution $\mathbf{U}^{\left( 1\right) }$ consists of $N^{3}$ components $%
\mathbf{U}_{l_{1},l_{2},l_{3}}^{\left( 1\right) }$ labeled by $%
l_{1},l_{2},l_{3}$. Observe now that if $l_{1}\neq l_{2}$ or $l_{1}\neq
l_{3} $ or $l_{2}\neq l_{3}$ the group velocity matching condition does not
generically hold and as an analysis shows we have%
\begin{equation}
\mathbf{U}_{l_{1},l_{2},l_{3}}^{\left( 1\right) }=O\left( \left( \frac{%
\varrho }{\beta }\right) ^{N_{1}}\right) \text{ where }N_{1}\text{ can be
taken arbitrary large,}
\end{equation}%
and, consequently, $\mathbf{U}_{l_{1},l_{2},l_{3}}^{\left( 1\right) }$ is
negligible at any level of accuracy. Therefore only terms with $%
l_{1}=l_{2}=l_{3}$ \ are left and we obtain 
\begin{equation}
\mathbf{U}^{\left( 1\right) }=\sum_{l=1}^{N}\mathbf{U}_{l}^{\left( 1\right)
}+O\left( \left( \frac{\varrho }{\beta }\right) ^{N_{1}}\right) .
\end{equation}%
Similarly for higher order responses in the expansion (\ref{UU1}) we obtain 
\begin{equation}
\mathbf{U}^{\left( m\right) }=\sum_{l=1}^{N}\mathbf{U}_{l}^{\left( m\right)
}+O\left( \left( \frac{\varrho }{\beta }\right) ^{N_{1}}\right)
,m=1,2,\ldots ,  \label{Um}
\end{equation}%
which implies (\ref{superpos}). Now we explain why (\ref{Um}) holds for $%
m>1. $ In this article we discuss in detail the zero order and first order
terms in the expansion (\ref{UU1}), which is sufficient in many cases, in
particular when $\alpha \ll \varrho $. The expansion (\ref{UU1}) includes
higher order terms $\alpha ^{m}\mathbf{U}^{\left( m\right) }$ which have to
be taken into account in the case $\alpha \sim \varrho $ and which we
discuss now. The term $\mathbf{U}^{\left( m\right) }$ depends \ on the
excitation current $\mathbf{J}$, its dependence is given by a $2m+1$-linear
operator applied to the excitation current $\mathbf{J}$. \ In the case $m=1$
such cubic operator (FNLR)\ is described in detail in Subsection 6.1. Here
we use a less detailed description. First, we use (\ref{Utild}) \ to
determine the modal coefficients $\tilde{j}_{\bar{n}}\left( \mathbf{k}%
,t\right) $ of \ the excitation current $\mathbf{J}$. The term $\mathbf{U}%
^{\left( m\right) }$ can be written in terms of the modal coefficients \ of $%
\ \mathbf{J}$ as follows: 
\begin{gather}
\alpha ^{m}\tilde{U}_{\bar{n}}^{\left( m\right) }\left( \mathbf{k},t\right)
=\alpha ^{m}\sum_{\bar{n}^{\prime },\ldots ,\bar{n}^{\left( m\right) }}\int 
_{\substack{ \lbrack -\pi ,\pi ]^{2d}  \\ \mathbf{\mathbf{k}^{\prime }}+%
\mathbf{\ldots }+\mathbf{k}^{\left( m\right) }=\mathbf{k}}}  \label{UT} \\
T\left( \vec{k}\right) \left[ \tilde{j}_{\bar{n}^{\prime }}\left( \mathbf{k}%
^{\prime },\cdot \right) \ldots \tilde{j}_{\bar{n}^{\left( m\right) }}\left( 
\mathbf{k}^{\left( m\right) },\cdot \right) \right] \,\mathrm{d}\mathbf{k}%
^{\prime }\ldots \mathrm{d}\mathbf{k}^{\left( m\right) }-\frac{1}{\varrho }%
\int_{0}^{\tau }\tilde{j}_{\bar{n}}^{\left( m\right) }\left( \mathbf{k},\tau
_{1}\right) \,\mathrm{d}\tau _{1}.  \notag
\end{gather}%
Here $T_{\vec{n}}\left( \vec{k}\right) $ for every $\vec{n}=\left( \bar{n},%
\bar{n}^{\prime },\ldots ,\bar{n}^{\left( m\right) }\right) $ $\ \vec{k}%
=\left( \mathbf{k},\mathbf{k}^{\prime },\ldots ,\mathbf{k}^{\left( m\right)
}\right) $, \ is a $2m+1$- linear operator which linearly acts on every $%
\tilde{j}_{\bar{n}^{\prime }}\left( \mathbf{k}^{\prime },\cdot \right) $,
..., $\tilde{j}_{\bar{n}^{\left( m\right) }}\left( \mathbf{k}^{\left(
m\right) },\cdot \right) $. The operator $T_{\vec{n}}\left( \vec{k}\right) $
includes integration with respect to time variables. Formula (\ref{UT}) is
not very simple, but it is still possible to show that the terms with given
values of $\bar{n},\bar{n}^{\prime },\ldots ,\bar{n}^{\left( m\right) }$ and 
$\mathbf{k},\mathbf{k}^{\prime },\ldots ,\mathbf{k}^{\left( m\right) }$ \
which do not satisfy the Phase Matching, Group Velocity Matching and
Frequency Matching rules have a small magnitude. To this end we use almost
time-harmonic analysis as in Section 6 and after that an analysis similar to
the case of FNLR (that is $m=1$) which is considered above. \ Note that
according to (\ref{JJ1}) 
\begin{equation}
\tilde{j}_{\bar{n}}\left( \mathbf{k},t\right) =\tilde{j}_{1,\bar{n}}\left( 
\mathbf{k},t\right) +\ldots +\tilde{j}_{N,\bar{n}}\left( \mathbf{k},t\right)
\end{equation}%
Therefore, 
\begin{gather*}
\int_{\substack{ \lbrack -\pi ,\pi ]^{2d}  \\ \mathbf{\mathbf{k}^{\prime }}+%
\mathbf{\ldots }+\mathbf{k}^{\left( m\right) }=\mathbf{k}}}T\left( \vec{k}%
\right) \left[ \tilde{j}_{\bar{n}^{\prime }}\left( \mathbf{k}^{\prime
},\cdot \right) \ldots \tilde{j}_{\bar{n}^{\left( m\right) }}\left( \mathbf{k%
}^{\left( m\right) },\cdot \right) \right] \mathrm{d}\mathbf{k}^{\prime
}\ldots \mathrm{d}\mathbf{k}^{\left( m\right) }= \\
\sum_{l_{1},\ldots ,l_{m}}\int_{\substack{ \lbrack -\pi ,\pi ]^{2d}  \\ 
\mathbf{\mathbf{k}^{\prime }}+\mathbf{\ldots }+\mathbf{k}^{\left( m\right) }=%
\mathbf{k}}}T\left( \vec{k}\right) \left[ \tilde{j}_{l_{1},\bar{n}^{\prime
}}\left( \mathbf{k}^{\prime },\cdot \right) \ldots \tilde{j}_{l_{m},\bar{n}%
^{\left( m\right) }}\left( \mathbf{k}^{\left( m\right) },\cdot \right) %
\right] \mathrm{d}\mathbf{k}^{\prime }\ldots \mathrm{d}\mathbf{k}^{\left(
m\right) }.
\end{gather*}%
An analysis shows that only the terms where $l_{1}=l_{2}=\ldots =l_{m}=l$
satisfy the selection rules and are not small. Collecting such term we
obtain the formula for $\mathbf{U}_{l}^{\left( m\right) }$. Since we can
explicitly estimate smallness of the negligible terms for every $m$ and we
have uniform convergence of the series (\ref{UU1}) \ we can estimate in
terms of $\alpha ,\beta $ and $\varrho $ the order of total contribution of
all the \ thrown away terms to the exact solution. A separate analysis (see
Section 7) shows that the remaining, non-negligible terms in (\ref{UT}) are
in exact correspondence with the terms of a similar expansion of the exact
solution of the NLS/ENLS\ system. The principal parts of expansions in $%
\beta $ of non-negligible terms in (\ref{UT}) have the same form as
corresponding terms of the expansion of the exact solution of the NLS/ENLS\
system. Based on the above considerations and an estimate of form (\ref%
{difexp}) we can estimate the difference between the NLS/ENLS approximation $%
\mathbf{U}_{Z}$ given by (\ref{uzz1}), (\ref{UNLS}), (\ref{Uind1}) and the
exact solution $\mathbf{U}$ as in (\ref{UUZx}).

Notice that the principle of approximate superposition has its natural
limitations, and the condition $\left( n_{l_{1}},\mathbf{k}_{\ast
l_{1}}\right) \neq \left( n_{l_{1}},\mathbf{k}_{\ast l_{2}}\right) \mathbf{\ 
}$for $l_{1}\neq l_{2}$ in (\ref{JJ1}) is absolutely instrumental. For
instance, though evidently $2\mathbf{J}_{l}=\mathbf{J}_{l}+\mathbf{J}_{l}$
the solution for $2\mathbf{J}_{l}$ is evidently not $2\mathbf{U}_{l}=\mathbf{%
U}_{l}+\mathbf{U}_{l}$ since $\mathbf{U}_{l}$ is a solution of a nonlinear
equation which is well-approximated by the NLS. So, to have (\ref{superpos})
with arbitrary large $N_{1}$\ the proper genericity condition has to include%
\begin{equation}
\nabla \omega _{\bar{n}_{l_{2}}}\left( \mathbf{k}_{\ast l_{2}}\right) \neq
\nabla \omega _{\bar{n}_{l_{1}}}\left( \mathbf{k}_{\ast l_{1}}\right) \text{
for }l_{2}\neq l_{1}.  \label{genGVM}
\end{equation}

\subsection{Extended Nonlinear Schrodinger equations}

As it was discussed above, the NLS describes the evolution of a doublet of
directly excited modes of the NLM. More accurate higher order approximations
of solutions to the NLM\ can be obtained by constructing extended NLS (ENLS)
instead of the classical NLS. Extended NLS\ are widely used in nonlinear
optics (see, \cite{A} \ and recent publications \cite{Dudley02}, \cite%
{Hong02}, \cite{Kumagai03}). The corrective terms in the ENLS originate from
several sources, resulting in relatively smaller alterations of solutions
compared to the basic (classical) NLS. Note that the nonlinearity $\alpha
\left\vert Z\right\vert ^{2}Z$ in the classical NLS gives an alteration of
the linear Schrodinger equation of order $O\left( \alpha \right) $ and the
error of approximation (when time $t$ is $O\left( 1\right) $) of the NLM by
a linear Schrodinger equation is $O\left( \alpha \right) $ too. One though
has to take into account that an $\ O\left( \alpha \right) $ alteration of
the equation leads on time intervals of length $O\left( \varrho ^{-1}\right) 
$, which we consider here, to alterations of solutions of order $O\left(
\alpha \varrho ^{-1}\right) $. In the case of classical NLS scaling (\ref%
{thet1alph}) $O\left( \alpha \right) =O\left( \varrho \right) =O\left( \beta
^{2}\right) $ and solutions of \ the classical NLS \ give approximations of
solutions of the NLM with the error $O\left( \beta \right) $. We consider in
this article two types of the ENLS: third and fourth order. Using third
order ENLS \ improves the error estimate in the case of classical NLS
scaling (\ref{thet1alph}) from $O\left( \beta \right) $ \ to $O\left( \beta
^{2}\right) $ and solutions of fourth order ENLS\ approximate solutions of
the NLM\ with the accuracy $O\left( \beta ^{3}\right) $.

Here is a complete list of all sources of the additional corrective terms
that are required to be added to the NLS to improve the accuracy of
approximation with estimations of their magnitude:

\begin{itemize}
\item Cubic and the fourth order polynomial approximations of the dispersion
relation $\omega _{n_{0}}\left( \mathbf{k}\right) $ at $\mathbf{k}_{\ast }$
with the corrective terms magnitude $O\left( \beta ^{3}\right) $ and $%
O\left( \beta ^{4}\right) $ respectively;

\item The first-order approximation of the cubic susceptibility accounting
for its frequency dependence (see subsection 6.3) with the corrective terms
magnitude $O\left( \alpha \varrho \right) $;

\item Polynomial approximation of the modal susceptibility $\breve{Q}_{\vec{n%
}}\left( \vec{k}\right) $ in (\ref{Qn}) at $\vec{k}_{\ast }$ with the
corrective terms magnitude $O\left( \alpha \beta \right) $ or $O\left(
\alpha \beta ^{2}\right) $;

\item Non-frequency-matched interactions between waves propagating in
opposite directions (see Subsection 5.4) with the correction term magnitude $%
O\left( \alpha \varrho \right) $;

\item The fifth order nonlinear terms in the expansion of the nonlinearity
in the NLM with the corrective terms magnitude $O\left( \alpha ^{2}\right) $

\item The interband interaction terms with the corrective terms magnitude $%
O\left( \alpha \varrho \right) $
\end{itemize}

If all the corrective terms from the above list are taken into account then
the accuracy of the approximation by the fourth-order ENLS of the NLM is
estimated by the following expression%
\begin{equation}
\mathbf{U}-\mathbf{U}_{Z}=\left[ O\left( \alpha \beta ^{3}\right) +O\left(
\alpha ^{2}\beta \right) +O\left( \alpha ^{3}\right) +O\left( \alpha \varrho
^{2}\right) \right] O\left( \left\vert \mathbf{U}^{\left( 1\right)
}\right\vert \right) .  \label{UUZ0}
\end{equation}%
\ In the case of the classical NLS scaling $\varrho \sim \alpha \sim \beta
^{2}$ we find that \ the neglected terms in the right-hand side of (\ref%
{UUZ0}) are of order $\beta ^{5}$, whereas the classical NLS nonlinearity
itself is of order $\beta ^{2}$ and the introduced above additional
corrective terms in ENLS are of order $\beta ^{3}$ or $\beta ^{4}$.

Let us introduce the following short notation for the linear Schrodinger
operator of the order $\nu $: 
\begin{equation}
\mathcal{L}_{+}^{\left[ \nu \right] }Z=\gamma _{\left( \nu \right) }\left[ -%
\mathrm{i}\vec{\nabla}_{\mathbf{r}}\right] Z,\ \mathcal{L}_{-}^{\left[ \nu %
\right] }Z=-\gamma _{\left( \nu \right) }\left[ \mathrm{i}\nabla _{\mathbf{r}%
}\right] Z,  \label{LZ}
\end{equation}%
where the so-called symbol (characteristic polynomial) $\gamma _{\left( \nu
\right) }\left( \mathbf{\eta }\right) $ of the differential operator $\gamma
_{\left( \nu \right) }\left[ -\mathrm{i}\vec{\nabla}_{\mathbf{r}}\right] $
is the Taylor polynomial of the order $\nu $ of the dispersion relation $%
\omega _{n_{0}}\left( \mathbf{k}\right) $ at $\mathbf{k}=\mathbf{k}_{\ast }$%
. For instance, for $\nu =2$ 
\begin{equation}
\gamma _{\left( 2\right) }\left( \mathbf{\eta }\right) =\omega
_{n_{0}}\left( \mathbf{\mathbf{k}_{\ast }}\right) +\omega _{n_{0}}^{\prime
}\left( \mathbf{\mathbf{k}_{\ast }}\right) \left( \mathbf{\mathbf{\eta }}%
\right) +\frac{1}{2}\omega _{n_{0}}^{\prime \prime }\left( \mathbf{\mathbf{k}%
_{\ast }}\right) \left( \mathbf{\eta }^{2}\right) .  \label{g2}
\end{equation}%
We always consider the situation where the FNLR is applicable, that is the
time interval satisifies (\ref{rhoalph}), namely 
\begin{equation}
\frac{\tau _{0}}{\varrho }\leq t\leq \frac{\tau _{\ast }}{\varrho }\text{
where }\frac{\tau _{0}}{\tau _{\ast }}<1\;\text{is fixed.}  \label{t0tt2}
\end{equation}%
We consider the cases $\nu =2,$ $\nu =3$ and $\nu =4$. The resulting ENLS
equations and the approximation error estimates are the same in both the
dispersive and weakly dispersive cases (\ref{nonM}) and (\ref{nonM1}). In
the error estimates we assume that (\ref{kap1}) or (\ref{kap11}) holds with
some fixed value of $\varkappa _{1}$. Often from general error estimates
which include three parameters $\alpha ,\beta ,\varrho $ we deduce in the
case of the classical NLS scaling (\ref{thet1alph}) simpler estimates in
terms of single parameter $\beta $ \ as a consequence.

For illustration we give the form of a typical ENLS of order $\nu $ (for
simplicity skipping some corrective terms) 
\begin{equation}
\partial _{t}Z_{+}=-\mathrm{i}\mathcal{L}_{+}^{\left[ \nu \right]
}Z_{+}+\alpha _{\pi }p_{+}^{\left[ \nu -2\right] }\left[ -\mathrm{i}\vec{%
\nabla}_{\mathbf{r}}\right] \left( Z_{+}^{2}Z_{-}\right) ,\ \alpha _{\pi
}=3\alpha \left( 2\pi \right) ^{2d},  \label{Znu}
\end{equation}%
\begin{equation}
\partial _{t}Z_{-}=-\mathrm{i}\mathcal{L}_{-}^{\left[ \nu \right]
}Z_{+}+\alpha _{\pi }p_{-}^{\left[ \nu -2\right] }\left[ -\mathrm{i}\vec{%
\nabla}_{\mathbf{r}}\right] \left( Z_{+}^{2}Z_{-}\right) ,\   \label{Znu-}
\end{equation}%
where the nonlinearity $p_{+}^{\left[ \nu -2\right] }\left[ -\mathrm{i}\vec{%
\nabla}_{\mathbf{r}}\right] \left( Z_{+}^{2}Z_{-}\right) $ includes spatial
derivatives of $Z_{+}$ and $Z_{-}$ of order up to $\ \nu -2$. \ 

Note that if (i) the excitation currents are real valued; (ii) (\ref{eer2})
holds and (iii) the polarization tensors have real coefficients, then the
equation for $Z_{+}$ is obtained by the complex conjugation of the equation
for $Z_{-}$ and $Z_{-}=Z_{+}^{\ast }$. Moreover, we can use (\ref{conjug})
and reduce the system for two equations to one equation for $Z_{+}$.

The extended NLS we describe below are \emph{universal}, they do not depend
on the relation between $\varrho ,\alpha $ and $\beta $ in particular on the
exponents $\varkappa _{0}$ in (\ref{rhoal}) and $\varkappa _{1}$ in (\ref%
{kap1}) or (\ref{kap11}) . From the universal ENLS\ one may deduce reduced
ENLS for a particular scaling, see Subsection 1.4.7 for examples of such
reduction. The reduced ENLS\ may depend on the choice of $\varkappa _{0}$
and $\varkappa _{1}$.

\textbf{Remark.} When we discuss the magnitude of the terms in the NLS, one
has to take into account that we study the NLM and NLS on intervals of order 
$\frac{1}{\varrho }$. Integration of the equation with respect to time leads
to a factor $\frac{1}{\varrho }$ in the contribution of the corresponding
terms to the exact solution of the NLS. In our error estimates, for example (%
\ref{UUZ0}) the effects of integration are taken into account, namely the
factor $\frac{1}{\varrho }$ is included into $O\left( \left\vert \mathbf{U}%
^{\left( 1\right) }\right\vert \right) $. For example, adding the NLS
nonlinearity which has order $O\left( \alpha \right) $ leads to a change of
a solution of the linear equation of order $O\left( \alpha \right) O\left(
\left\vert \mathbf{U}^{\left( 1\right) }\right\vert \right) =O\left( \frac{%
\alpha }{\varrho }\right) $, and in the case of the classical NLS scaling
this change of the solution has a finite, non-vanishing magnitude $O\left( 
\frac{\alpha }{\varrho }\right) =O\left( 1\right) $ as can be seen from (\ref%
{NLSy}). \ Skipping $O\left( \left\vert \mathbf{U}^{\left( 1\right)
}\right\vert \right) $ in the estimates is consistent with taking estimates
of terms in the NLS and ENLS without integrating in time. In this section we
will systematically do that without further reference, we give some details
only when it is necessary, as in (\ref{Orho}). To simplify the discussion of
the magnitude of the nonlinear terms we everywhere in this section assume
that (\ref{aeqr}) holds, that is , $\alpha \sim \varrho $. Sometimes, for a
further simplification, we consider the classical NLS\ scaling, that is $%
\alpha \sim \varrho $, $\varrho \sim \beta ^{2}$.$\blacklozenge $

\subsubsection{The second-order ENLS}

If the order of the linear part $\nu =2$ then the extended Nonlinear
Schrodinger equations take the form 
\begin{equation}
\partial _{t}Z_{+}=-\mathrm{i}\mathcal{L}_{+}^{\left[ 2\right] }Z_{+}+\alpha
_{\pi }p_{+}^{\left[ 0\right] }\left[ -\mathrm{i}\vec{\nabla}_{\mathbf{r}}%
\right] \left( Z_{+}^{2}Z_{-}\right) ,\ \alpha _{\pi }=3\alpha \left( 2\pi
\right) ^{2d},  \label{GNLS+}
\end{equation}%
\begin{equation}
\partial _{t}Z_{-}=-\mathrm{i}\mathcal{L}_{-}^{\left[ 2\right] }Z_{-}+\alpha
_{\pi }p_{-}^{\left[ 0\right] }\left[ -\mathrm{i}\vec{\nabla}_{\mathbf{r}}%
\right] \left( Z_{-}^{2}Z_{+}\right) ,  \label{GNLS-}
\end{equation}%
with the initial conditions 
\begin{equation}
Z_{+}\left( \mathbf{r},t\right) |_{t=0}=h_{+}\left( \beta \mathbf{r}\right)
,\ Z_{-}\left( \mathbf{r},t\right) |_{t=0}=h_{-}\left( \beta \mathbf{r}%
\right) ,  \label{inZ}
\end{equation}%
where%
\begin{equation}
h_{-}\left( \beta \mathbf{r}\right) =h_{+}^{\ast }\left( \beta \mathbf{r}%
\right) .  \label{hreal1}
\end{equation}%
The linear operator%
\begin{equation}
\mathcal{L}_{\zeta }^{\left[ \nu \right] }=\zeta \gamma _{\left( \nu \right)
}\left[ -\mathrm{i}\zeta \vec{\nabla}_{\mathbf{r}}\right] ,\nu =2,\zeta =\pm
\end{equation}%
is the second order linear differential operator with constant coefficients
given by formulas involving $\omega _{n_{0}}\left( \mathbf{\mathbf{k}}%
\right) $ and its derivatives at $\mathbf{\mathbf{k}_{\ast }}$ which are
similar to (\ref{gami}) (see (\ref{Gam1}) for general case and details). The
action of $p_{\pm }^{\left[ 0\right] }=p_{\pm }^{\left[ \sigma \right] }$ \
with $\sigma =0$ is just the multiplication by a constant, that is%
\begin{equation}
p_{+}^{\left[ 0\right] }\left[ -\mathrm{i}\vec{\nabla}_{\mathbf{r}}\right]
\left( Z_{+}^{2}Z_{-}\right) =Q_{\pm }Z_{+}^{2}Z_{-},  \label{p0}
\end{equation}%
where $Q_{\pm }$ is determined by the modal susceptibility(\ref{Qn}), (see (%
\ref{Q0}) for details) and in this case we obtain the classical NLS (\ref{Si}%
). The order of approximation is given by the formula 
\begin{equation}
\mathbf{U}-\mathbf{U}_{Z}=\left[ O\left( \alpha \beta \right) +O\left(
\alpha \varrho \right) +O\left( \alpha ^{2}\right) \right] O\left(
\left\vert \mathbf{U}^{\left( 1\right) }\right\vert \right) ,  \label{O20}
\end{equation}%
which implies (\ref{UZO}).

\textbf{Remark.} \ When together with fulfillment of (\ref{hreal1}) and (\ref%
{eer2}) the nonlinearity in NLM is real-valued for real-valued vector
fields, we have 
\begin{equation}
Q_{-}=Q_{+}^{\ast },\;Z_{-}\left( \mathbf{r},t\right) =Z_{+}^{\ast }\left( 
\mathbf{r},t\right) ,  \label{Zreal}
\end{equation}%
therefore we can use (\ref{conjug}) and (\ref{GNLS+}), (\ref{GNLS-}) is
equivalent to the NLS in its classical form:%
\begin{equation}
\partial _{t}Z_{+}=-\mathrm{i}\mathcal{L}_{+}^{\left[ 2\right] }Z_{+}+\alpha
_{\pi }Q_{+}\left\vert Z_{+}\right\vert ^{2}Z_{+}.
\end{equation}

\subsubsection{The third-order ENLS}

For the classical NLS scaling (\ref{thet1alph}) the leading term in the
error estimate (\ref{O20}) is $O\left( \alpha \beta \right) =O\left( \beta
^{3}\right) $ compared with $O\left( \alpha \varrho \right) =O\left( \beta
^{4}\right) $ and $O\left( \alpha ^{2}\right) =O\left( \beta ^{4}\right) $.
To reduce this term to $O\left( \alpha \beta ^{2}\right) =O\left( \beta
^{4}\right) $ and to get higher accuracy of approximation we take $\nu =3$
and obtain the following system of two third-order equations similar to (\ref%
{GNLS+}), (\ref{GNLS-}): 
\begin{equation}
\partial _{t}Z_{+}+\mathrm{i}\mathcal{L}_{+}^{\left[ 3\right] }Z_{+}=\alpha
_{\pi }p_{+}^{\left[ 1\right] }\left[ -\mathrm{i}\vec{\nabla}_{\mathbf{r}}%
\right] \left( Z_{+}^{2}Z_{-}\right) ,  \label{G3+}
\end{equation}%
\begin{equation}
\partial _{t}Z_{-}+\mathrm{i}\mathcal{L}_{-}^{\left[ 3\right] }Z_{-}=\alpha
_{\pi }p_{-}^{\left[ 1\right] }\left[ -\mathrm{i}\vec{\nabla}_{\mathbf{r}}%
\right] \left( Z_{+}^{2}Z_{-}\right)  \label{G3-}
\end{equation}%
with the initial conditions (\ref{inZ}) (see also \cite[p.44-45]{A} and \cite%
{Hile96}, \cite{Karpman00}, \cite{Nakkeeran00}, \cite{Berge00}, \cite%
{Frantzeskakis96}, \cite{Solange91}, \cite{Hayashi02}, \cite{Potasek87},
where similar equations are studied). Now $\gamma _{\left( 3\right) }\left( -%
\mathrm{i}\partial _{1}\right) $ is the third-order linear operator with the
symbol which is the third degree Taylor polynomial of the dispersion
relation $\omega _{n_{0}}\left( \mathbf{k}\right) $ at $\mathbf{k}_{\ast }$.
For $\sigma =1$ the polynomial%
\begin{equation}
p_{\pm }^{\left[ 1\right] }\left( \mathbf{\eta }\right) =p_{\pm }^{\left[ 0%
\right] }\left( \mathbf{\eta }\right) +p_{1,\pm }\left( \mathbf{\eta }\right)
\end{equation}%
originates from the Taylor approximation of the order one for the modal
susceptibility $\breve{Q}_{\vec{n}}\left( \vec{k}\right) $ in (\ref{Qn}) at
the point $\vec{k}\mathbf{_{\ast }}$ determined by $\mathbf{\mathbf{k}_{\ast
}}$. The zero order term is given by (\ref{p0}). The action of $p_{1,\pm }%
\left[ -\mathrm{i}\vec{\nabla}_{\mathbf{r}}\right] $ on the product $%
Z_{+}^{2}Z_{+}^{\ast }$ is defined by the formula 
\begin{gather}
p_{1,+}\left[ -\mathrm{i}\vec{\nabla}_{\mathbf{r}}\right] \left(
Z_{+}^{2}Z_{-}\right) =  \label{p1} \\
-\mathrm{i}Z_{+}Z_{-}\left( a_{11,+}+a_{12,+}\right) \cdot \nabla _{\mathbf{r%
}}\left( Z_{+}\right) -\mathrm{i}Z_{+}^{2}a_{13,+}\cdot \nabla _{\mathbf{r}%
}\left( Z_{-}\right) +Q_{+}Z_{+}^{2}Z_{-},  \notag
\end{gather}%
where $a_{11,+}$, $a_{12,+}$ and $a_{13,+}$ are constant vectors explicitly
given in terms of the gradient of $\breve{Q}_{\vec{n}}$ at $\vec{k}\mathbf{%
_{\ast }}$ by formula (\ref{a11}) which also defines $p_{-}^{\left[ 1\right]
}\left[ \mathrm{i}\nabla _{\mathbf{r}}\right] $. Note that the order of the
factors $Z_{+}$ and $Z_{-}$ in the notation $p_{+}^{\left[ 1\right] }\left[ -%
\mathrm{i}\vec{\nabla}_{\mathbf{r}}\right] \left( Z_{+}^{2}Z_{-}\right) $ is
important, see (\ref{nonVVV}). The corrective terms $\alpha _{\pi }p_{+}^{%
\left[ 1\right] }\left[ -\mathrm{i}\vec{\nabla}_{\mathbf{r}}\right] \left(
Z_{+}^{2}Z_{-}\right) $ can be considered as nonlinear corrections to the
linear operator $\gamma _{\left( 3\right) }\left[ -\mathrm{i}\vec{\nabla}_{%
\mathbf{r}}\right] $. Note that in the case of real-valued fields using (\ref%
{conjug}) the first-order part of $p_{+}^{\left[ 1\right] }\left[ -\mathrm{i}%
\vec{\nabla}_{\mathbf{r}}\right] \left( Z_{+}^{2}Z_{-}\right) $ can be
rewritten in the following commonly used form (see \cite{A} p. 44-45):%
\begin{gather*}
-\mathrm{i}Z_{+}Z_{-}\left( a_{11,+}+a_{12,+}\right) \cdot \nabla _{\mathbf{r%
}}\left( Z_{+}\right) -\mathrm{i}Z_{+}^{2}a_{13,+}\cdot \nabla _{\mathbf{r}%
}\left( Z_{-}\right) \\
=-\mathrm{i}\left\vert Z_{+}\right\vert ^{2}\left(
a_{11,+}+a_{12,+}-a_{13,+}\right) \cdot \nabla _{\mathbf{r}}\left(
Z_{+}\right) -\mathrm{i}Z_{+}a_{13,+}\cdot \nabla _{\mathbf{r}}\left(
\left\vert Z_{+}\right\vert ^{2}\right) .
\end{gather*}%
The error of approximation in the case $\nu =3$ is 
\begin{equation}
\mathbf{U}-\mathbf{U}_{Z}=\left[ O\left( \alpha \beta ^{2}\right) +O\left(
\alpha \varrho \right) +O\left( \alpha ^{2}\right) \right] O\left(
\left\vert \mathbf{U}^{\left( 1\right) }\right\vert \right) .  \label{O3}
\end{equation}%
\ The improvement $O\left( \alpha \beta ^{2}\right) $ in (\ref{O3}) compared
with $O\left( \alpha \beta \right) $ in (\ref{O20}) is obtained by taking
the variability of $\breve{Q}_{\vec{n}}\left( \vec{k}\right) $ into account
and by more precise approximation of $\omega _{n_{0}}\left( \mathbf{k}%
\right) $. \ The terms with spatial derivatives of $Z_{+}$ and $Z_{+}^{\ast
} $ in (\ref{p1}) are computed in terms of the gradient of $\breve{Q}_{\vec{n%
}}\left( \vec{k}\right) $ at $\mathbf{k=k}_{\ast }$.

In particular, for the classical NLS scaling (\ref{thet1alph}) the error in (%
\ref{O3}) is $O\left( \beta ^{4}\right) $. \ According to (\ref{inZ}) 
\begin{equation}
\vec{\nabla}_{\mathbf{r}}Z_{+}=O\left( \beta \right) ,\vec{\nabla}_{\mathbf{r%
}}^{3}Z_{+}=O\left( \beta ^{3}\right)  \label{grbet1}
\end{equation}%
therefore the third order terms added in $\mathcal{L}_{+}^{\left[ 3\right] }$
and the first order terms added in (\ref{p1}) to the ENLS\ are respectively
of order $O\left( \beta ^{3}\right) $ and $\alpha O\left( \beta \right)
=O\left( \beta ^{3}\right) $, and they are generically non-zero and much
larger than the approximation error $O\left( \beta ^{4}\right) $. Note that
if the terms of order $O\left( \beta ^{3}\right) $ in (\ref{G3+}+, (\ref{G3-}%
) \ were thrown away, we would arrive to the classical second-order NLS.
Therefore, the difference of \ solutions of the NLM\ and the classical NLS
really is of order $O\left( \beta ^{3}\right) $ and is represented by the
additional terms in the ENLS (\ref{G3+}), (\ref{G3-}). Hence, \emph{the
corrections introduced into the nonlinear Schrodinger equations capture the
actual properties of solutions of the NLM\ and they are necessary if one
wants to approximate the solutions to the Nonlinear Maxwell equation with a
higher accuracy than the classical NLS}.

\subsubsection{The fourth-order ENLS}

>From the very form of the error estimate (\ref{O3}), one can see that to
improve the error estimate we have to make smaller every one of the three
terms $O\left( \alpha \beta ^{2}\right) $, $O\left( \alpha \varrho \right) $
\ and $O\left( \alpha ^{2}\right) $, \ which are of the same order under the
classical NLS scaling $\varrho \sim \alpha \sim \beta ^{2}$ (and may be of
different magnitude in a more general situation) . The third one comes
mainly from the next term of the fifth rank with the coefficient $\alpha
^{2} $ in the expansion of the nonlinearity in the NLM, and to make it
smaller we have to take the mentioned term into account. To reduce $O\left(
\alpha \beta ^{2}\right) $ we have to approximate better the modal
susceptibility $\breve{Q}_{\vec{n}}\left( \vec{k}\right) $ about $\mathbf{k=k%
}_{\ast }$, and to this end we use the second degree Taylor polynomial of $%
\breve{Q}_{\vec{n}}$ instead of the first degree one, that yields the
second-order linear operator $p_{\zeta }^{\left[ 2\right] }\left[ -\mathrm{i}%
\vec{\nabla}_{\mathbf{r}}\right] $, see (\ref{psig}) for an explicit
formula. To reduce $O\left( \alpha \varrho \right) $ we have to take into
account finer effects related to the convolution integrals in the
nonlinearity, which, in turn, are reflected in the frequency dependence of
the susceptibility.

\subsubsection{Corrections related to the frequency-dependence of the
susceptibility}

To take into account the first-order corrections due to the frequency
dependence of the susceptibility tensor (discussed in detail in Section 6),
some terms involving time derivatives must be added to the ENLS (\ref{G3+}),
(\ref{G3-}) yielding the following ENLS equations 
\begin{gather}
\left( \partial _{t}+\mathrm{i}\mathcal{L}_{+}^{\left[ 4\right] }\right)
Z_{+}=  \label{GNLS1+} \\
-\alpha _{\pi }\delta _{1,+}Z_{+}Z_{-}\left( \partial _{t}+\mathrm{i}%
\mathcal{L}_{+}^{\left[ 4\right] }\right) Z_{+}-\alpha _{\pi }\delta _{2,+\
}Z_{+}^{2}\left( \partial _{t}+\mathrm{i}\mathcal{L}_{-}^{\left[ 4\right]
}\right) Z_{-}+\alpha _{\pi }p_{+}^{\left[ 2\right] }\left[ -\mathrm{i}\vec{%
\nabla}_{\mathbf{r}}\right] \left( Z_{+}^{2}Z_{-}\right) ,  \notag
\end{gather}%
\begin{gather}
\left[ \partial _{t}+\mathrm{i}\mathcal{L}_{-}^{\left[ 4\right] }\right]
Z_{-}=  \label{GNLS1-} \\
-\alpha _{\pi }\delta _{1,-}Z_{+}Z_{-}\left( \partial _{t}+\mathrm{i}%
\mathcal{L}_{-}^{\left[ 4\right] }\right) Z_{-}-\alpha _{\pi }\delta
_{2,-}Z_{-}^{2}\left( \partial _{t}+\mathrm{i}\mathcal{L}_{+}^{\left[ 4%
\right] }\right) Z_{+}+\alpha _{\pi }p_{-}^{\left[ 2\right] }\left[ -\mathrm{%
i}\vec{\nabla}_{\mathbf{r}}\right] \left( Z_{-}^{2}Z_{+}\right) ,  \notag
\end{gather}%
with the initial conditions (\ref{inZ}) (here we consider the case when the
fifth order term in the nonlinearity in the NLM is absent; a general case is
considered a little later). The coefficients $\delta _{1,\pm }$, $\delta
_{2,\pm }$ in (\ref{GNLS1+}), (\ref{GNLS1-}) are proportional to the
derivatives of the susceptibility with respect to the frequency and defined
by (\ref{del+}), (\ref{Qnl}). The Fourier transforms of the new terms in (%
\ref{GNLS1+}), (\ref{GNLS1-}) create the same FNLR\ as the terms in (\ref%
{sus1}), see for details Subsections 6.2 and 8.7.2. The approximation error
is of order 
\begin{equation}
\mathbf{U}-\mathbf{U}_{Z}=\left[ O\left( \alpha \beta ^{3}\right) +O\left(
\alpha \beta \varrho \right) \right] O\left( \left\vert \mathbf{U}^{\left(
1\right) }\right\vert \right) +O\left( \alpha ^{2}\varrho \right) O\left(
\left\vert \mathbf{U}^{\left( 1\right) }\right\vert ^{2}\right) .
\label{ErM}
\end{equation}

The parameter $\varrho $ does not enter into (\ref{GNLS1+}), (\ref{GNLS1-}),
but it is important for the matching of the initial data for the NLS with
the excitation currents for the NLM \ (see Subsection 5.2 for details) and
determines the slow time scale in the NLM.

Now we briefly discuss the relative magnitude of terms in (\ref{GNLS1+}), (%
\ref{GNLS1-}). It follows from (\ref{GNLS1+}), (\ref{GNLS1-}) that 
\begin{equation}
\left( \partial _{t}+\mathrm{i}\mathcal{L}_{+}^{\left[ 4\right] }\right)
Z_{+}=O\left( \alpha \right) ,\ \left( \partial _{t}+\mathrm{i}\mathcal{L}%
_{-}^{\left[ 4\right] }\right) Z_{-}=O\left( \alpha \right) .
\end{equation}%
Hence, the correction \ terms due to the frequency dependence have magnitude 
$O\left( \alpha ^{2}\right) $. This agrees with the smallest value of $%
\varrho \sim \alpha $ allowed in (\ref{rhoalph}). Now we compare the
contribution of the corrective terms with the terms of order $O\left( \alpha
^{2}\right) $ that come from other sources. More detailed analysis (see
Subsection 7) shows that since (\ref{UNLS}) uses the exact solution of ENLS,
the corrective terms involved in (\ref{GNLS1+}), (\ref{GNLS1-}) are in many
cases more important than other terms that we neglected. In particular, when
the fifth and higher order terms in the expansion of $\mathcal{F}_{\text{NL}%
}\left( \mathbf{U}\right) $\ are much smaller than $O\left( \alpha
^{2}\right) $, namely (\ref{alph1}) holds, the neglected terms are much
smaller than $O\left( \alpha ^{2}\right) $. When the fifth order term in $%
\mathcal{F}_{\text{NL}}\left( \mathbf{U}\right) $ is exactly of order $%
O\left( \alpha ^{2}\right) $, it has to be taken into account by adding the
terms of the form $Q_{5,+}\left\vert Z_{+}\right\vert ^{4}Z_{+}$ to (\ref%
{GNLS1+}) and $Q_{5,-}\left\vert Z_{-}\right\vert ^{4}Z_{-}$ to (\ref{GNLS1-}%
), (an explicit formula for $Q_{5,\pm }$ is given in (\ref{Q5})) thus
reducing the error of the approximation from this source to $O\left( \beta
\alpha ^{2}\right) $.

\textbf{Simplification of the system.} Now we can simplify (\ref{GNLS1+})
and (\ref{GNLS1-}). We consider the case (\ref{conjug}). First, we write (%
\ref{GNLS1+}) in the form, 
\begin{gather}
\left( 1+\alpha _{\pi }\delta _{1,+}\left\vert Z_{+}\right\vert ^{2}\right) 
\left[ \partial _{t}+\mathrm{i}\mathcal{L}_{+}^{\left[ 4\right] }\right]
Z_{+}+\alpha _{\pi }\delta _{2,+}Z_{+}^{2}\left( \left( \partial _{t}+%
\mathrm{i}\mathcal{L}_{+}^{\left[ 4\right] }\right) Z_{+}\right) ^{\ast }= \\
\alpha _{\pi }p_{+}^{\left[ 2\right] }\left[ -\mathrm{i}\vec{\nabla}_{%
\mathbf{r}}\right] \left( Z_{+}^{2}Z_{+}^{\ast }\right)  \notag
\end{gather}%
\ and solve for $\left( \partial _{t}+\mathrm{i}\mathcal{L}_{+}^{\left[ 4%
\right] }\right) Z_{+}$ obtaining the equation equivalent to (\ref{GNLS1+}):
\ 
\begin{gather}
\left[ \partial _{t}+\mathrm{i}\mathcal{L}_{+}^{\left[ 4\right] }\right]
Z_{+}  \label{Zrat} \\
=\frac{\alpha _{\pi }\left( 1+\alpha _{\pi }\delta _{1,+}^{\ast }\left\vert
Z_{+}\right\vert ^{2}\right) \left( p_{+}^{\left[ 2\right] }\left[ -\mathrm{i%
}\vec{\nabla}_{\mathbf{r}}\right] \left( Z_{+}^{2}Z_{+}^{\ast }\right)
\right) -\alpha _{\pi }^{2}\delta _{2,+}Z_{+}^{2}\left( p_{+}^{\left[ 2%
\right] }\left[ -\mathrm{i}\vec{\nabla}_{\mathbf{r}}\right] \left(
Z_{+}^{2}Z_{+}^{\ast }\right) \right) ^{\ast }}{\left( \left\vert 1+\alpha
_{\pi }\delta _{1,+}\left\vert Z_{+}\right\vert ^{2}\right\vert ^{2}-\alpha
_{\pi }^{2}\left\vert \delta _{2,+\ }\right\vert ^{2}\left\vert
Z_{+}\right\vert ^{4}\right) }.  \notag
\end{gather}%
Then we expand (\ref{Zrat}) keeping terms of order $\alpha _{\pi }$ and $%
\alpha _{\pi }^{2}$, i.e.%
\begin{gather*}
\left[ \partial _{t}+\mathrm{i}\mathcal{L}_{+}^{\left[ 4\right] }\right]
Z_{+}=\alpha _{\pi }\left( 1+\alpha _{\pi }\delta _{1,+}^{\ast }\left\vert
Z_{+}\right\vert ^{2}\right) \left( p_{+}^{\left[ 2\right] }\left[ -\mathrm{i%
}\vec{\nabla}_{\mathbf{r}}\right] \left( Z_{+}^{2}Z_{+}^{\ast }\right)
\right) \\
-\alpha _{\pi }^{2}\delta _{2,+}Z_{+}^{2}\left( p_{+}^{\left[ 2\right] }%
\left[ -\mathrm{i}\vec{\nabla}_{\mathbf{r}}\right] \left(
Z_{+}^{2}Z_{+}^{\ast }\right) \right) ^{\ast }-2\alpha _{\pi }^{2}\func{Re}%
\delta _{1,+}\left\vert Z_{+}\right\vert ^{2}p_{+}^{\left[ 2\right] }\left[ -%
\mathrm{i}\vec{\nabla}_{\mathbf{r}}\right] \left( Z_{+}^{2}Z_{+}^{\ast
}\right) +O\left( \alpha ^{3}\right) .
\end{gather*}%
According to (\ref{inZ}) 
\begin{equation}
\vec{\nabla}_{\mathbf{r}}Z_{+}=O\left( \beta \right) ,\qquad \vec{\nabla}_{%
\mathbf{r}}^{2}Z_{+}=O\left( \beta ^{2}\right) ,  \label{grbet}
\end{equation}%
and, hence, we can neglect spatial derivatives in terms with the factor $%
\alpha _{\pi }^{2}$ (this requires, strictly speaking, some regularity of
solutions of the ENLS, see Section 7 for references) and obtain%
\begin{gather*}
\left[ \partial _{t}+\mathrm{i}\mathcal{L}_{+}^{\left[ 4\right] }\right]
Z_{+}=\alpha _{\pi }\left( p_{+}^{\left[ 2\right] }\left[ -\mathrm{i}\vec{%
\nabla}_{\mathbf{r}}\right] \left( Z_{+}^{2}Z_{+}^{\ast }\right) \right)
+\alpha _{\pi }^{2}\delta _{1,+}^{\ast }Q_{+}\left\vert Z_{+}\right\vert
^{4}Z_{+} \\
-\alpha _{\pi }^{2}\delta _{2,+}Q_{+}^{\ast }\left\vert Z_{+}\right\vert
^{4}Z_{+}-2\alpha _{\pi }^{2}\func{Re}\delta _{1,+}\left\vert
Z_{+}\right\vert ^{4}Q_{+}Z_{+}+O\left( \alpha ^{3}\right) +O\left( \beta
\alpha ^{2}\right) .
\end{gather*}%
Consequently, we can introduce the following equation with a quintic
nonlinearity 
\begin{gather}
\left[ \partial _{t}+\mathrm{i}\mathcal{L}_{+}^{\left[ 4\right] }\right]
Z_{+}=\alpha _{\pi }\left( p_{+}^{\left[ 2\right] }\left[ -\mathrm{i}\vec{%
\nabla}_{\mathbf{r}}\right] \left( Z_{+}^{2}Z_{+}^{\ast }\right) \right)
+\alpha _{\pi }^{2}\delta _{5,+}\left\vert Z_{+}\right\vert ^{4}Z_{+},
\label{ENL+} \\
\delta _{5,+}=-\delta _{2,+}Q_{+}^{\ast }-\delta _{1,+}Q_{+}.  \label{del5}
\end{gather}%
and with the initial condition (\ref{inZ}). The solution of this equation
approximates the solution of (\ref{GNLS1+}). The solution of the following
equation for $Z_{-}$ 
\begin{equation}
\left[ \partial _{t}+\mathrm{i}\mathcal{L}_{-}^{\left[ 4\right] }\right]
Z_{-}=\alpha _{\pi }\left( p_{-}^{\left[ 2\right] }\left[ -\mathrm{i}\vec{%
\nabla}_{\mathbf{r}}\right] \left( Z_{-}^{2}Z_{-}^{\ast }\right) \right)
+\alpha _{\pi }^{2}\delta _{5,-}\left\vert Z_{-}\right\vert ^{4}Z_{-}
\label{ENL-}
\end{equation}%
approximates the solution of (\ref{GNLS1-}) with $\delta _{5,-}=-\delta
_{2,-}Q_{-}^{\ast }-\delta _{1,-}Q_{-}$. Solutions of (\ref{ENL+}), \ (\ref%
{ENL-}) approximate solutions of the NLM\ with the same order of accuracy as
solutions of (\ref{GNLS1+}) and (\ref{GNLS1-}). Hence, (\ref{UNLS}) gives an
approximate solution to the NLM\ with the error estimate (\ref{ErM}). Note
that to take into account the fifth order term in the expansion of $\mathcal{%
F}_{\text{NL}}\left( \mathbf{U}\right) $\ we have to use the coefficients $%
Q_{5,\pm }$ \ defined by (\ref{Q5}), which effect the values of $\delta
_{5,+},$ $\delta _{5,-}$ in (\ref{ENL+}). Namely, the values of the
coefficients that take into account the fifth-order terms of the NLM are 
\begin{equation}
\delta _{5,+}=-\delta _{2,+}Q_{+}^{\ast }-\delta _{1,+}Q_{+}+Q_{5,+},\
\delta _{5,-}=-\delta _{2,-}Q_{-}^{\ast }-\delta _{1,-}Q_{-}+Q_{5,-},
\label{dl55}
\end{equation}%
with $Q_{5,\pm }$ and $\delta _{1,\pm }$, $\delta _{2,\pm }$ be respectively
defined the formulas (\ref{Q5}) and (\ref{del+}).

\emph{It is interesting that the both refinements coming from the frequency
dependence of the cubic susceptibility and the fifth order susceptibility
are taken care of by the same fifth-order term }$\delta _{5,\pm }\left\vert
Z_{\pm }\right\vert ^{4}Z_{\pm }$\emph{\ in the NLS.}\ \ In conclusion, to
take into account these effects we take in (\ref{UNLS}) the solution $Z_{\pm
}$ of \ (\ref{ENL+}), (\ref{ENL-}).\ 

If the quintic terms of the NLM\ are taken into account as in (\ref{dl55}),
the excitations currents of NLM\ are formed as in Section 7 and $\varrho
\sim \alpha $, the estimate of error of approximation by solutions of (\ref%
{ENL+}), (\ref{ENL-}) takes the form%
\begin{equation}
\mathbf{U}-\mathbf{U}_{Z}=\left[ O\left( \alpha \beta ^{3}\right) +O\left(
\alpha ^{2}\beta \right) +O\left( \alpha ^{3}\right) \right] O\left(
\left\vert \mathbf{U}^{\left( 1\right) }\right\vert \right) +O\left( \alpha
^{2}\varrho \right) O\left( \left\vert \mathbf{U}^{\left( 1\right)
}\right\vert ^{2}\right) .  \label{er5}
\end{equation}%
In particular, for the classical NLS scaling (\ref{thet1alph}) the error is $%
O\left( \beta ^{5}\right) O\left( \left\vert \mathbf{U}^{\left( 1\right)
}\right\vert \right) $. Let us compare the extended NLS with the classical
NLS. According to (\ref{grbet}) the corrective terms in%
\begin{equation}
\alpha _{\pi }p_{\zeta }^{\left[ 2\right] }\left[ -\mathrm{i}\vec{\nabla}_{%
\mathbf{r}}\right] \left( Z_{\zeta }^{2}Z_{\zeta }^{\ast }\right)
\end{equation}%
involving the second order derivatives which we added here are estimated by $%
O\left( \alpha \beta ^{2}\right) =O\left( \beta ^{4}\right) $. The
corrective term $\alpha _{\pi }^{2}\delta _{5,+}\left\vert Z_{+}\right\vert
^{4}Z_{+}$ is estimated by $O\left( \alpha ^{2}\right) =O\left( \beta
^{4}\right) $ too, which is larger than the first term in the difference of
NLM and ENLS solutions $O\left( \beta ^{5}\right) $ in (\ref{er5}). The
second term has a different nature, we discuss it in the following remark.

\textbf{Effect of interband interactions.} In the one-dimensional case $%
O\left( \left\vert \mathbf{U}^{\left( 1\right) }\right\vert \right) =O\left( 
\frac{1}{\varrho }\right) $ \ and when $\varrho \sim \alpha $ the term $%
O\left( \alpha ^{2}\varrho \right) O\left( \left\vert \mathbf{U}^{\left(
1\right) }\right\vert ^{2}\right) $ in (\ref{er5}) has magnitude $O\left(
\beta ^{2}\right) $. This term originates from the interband interactions,
that is non-frequency matched interactions which envolve indirectly excited
modes. The significance and exact contribution of these interactions to the
NLM\ can be found when all higher order terms in the analytic expansion (\ref%
{uap1}) are taken into account. We can construct the ENLS system which takes
the effect of such interactions into account and admits an improved error
estimate replacing $O\left( \alpha ^{2}\varrho \right) O\left( \left\vert 
\mathbf{U}^{\left( 1\right) }\right\vert ^{2}\right) $ in (\ref{er5}) by $%
O\left( \alpha ^{2}\varrho ^{2}\right) O\left( \left\vert \mathbf{U}^{\left(
1\right) }\right\vert ^{2}\right) $: 
\begin{equation}
\mathbf{U}-\mathbf{U}_{Z}=\left[ O\left( \alpha \beta ^{3}\right) +O\left(
\alpha ^{2}\beta \right) +O\left( \alpha ^{3}\right) \right] O\left(
\left\vert \mathbf{U}^{\left( 1\right) }\right\vert \right) +O\left( \alpha
^{2}\varrho ^{2}\right) O\left( \left\vert \mathbf{U}^{\left( 1\right)
}\right\vert ^{2}\right)  \label{er5a}
\end{equation}%
which implies (\ref{UUZ0}), that is reduces the total error from $O\left(
\beta ^{2}\right) $ to $O\left( \beta ^{3}\right) $. These ENLS\ in addition
to (\ref{ENL+}), \ (\ref{ENL-}) have to include another pair of scalar
NLS-type equations with zero initial data, the additional NLS-type nonlinear
terms couple \ these equations with (\ref{ENL+}), \ (\ref{ENL-}) forming a
four-component system. The coefficients at the coupling cubic terms describe
nonlinear interactions between spectral bands related to the third harmonic
generation. Since a detailed explanation and introduction of the coupling
coefficients would require new notations and techniques which are beyond the
scope of this paper we leave it for a future article.

\subsubsection{Complex initial data}

There are situations when complex electromagnetic vector fields are of
interest and useful, \cite{Kong}. In this case the excitation currents still
are given essentially by (\ref{Jzin}), namely 
\begin{equation}
\tilde{j}_{\bar{n}}^{\left( 0\right) }\left( \mathbf{k},\tau \right) =\tilde{%
j}_{\zeta ,n_{0}}^{\left( 0\right) }\left( \mathbf{k},\tau \right) =-\varrho
\psi _{0}\left( \tau \right) \Psi _{0}\left( \mathbf{k}-\zeta \mathbf{k}%
_{\ast }\right) \beta ^{-d}\mathring{h}_{\zeta }\left( \frac{\mathbf{k}%
-\zeta \mathbf{k}_{\ast }}{\beta }\right) ,\zeta =\pm 1,\ \tau =\varrho t 
\notag
\end{equation}%
but now $\mathring{h}_{\zeta }\left( \left( \mathbf{k-}\zeta \mathbf{k}%
_{\ast }\right) \mathbf{/}\beta \right) $ and $\mathring{h}_{-\zeta }\left(
\left( \mathbf{k-}\zeta \mathbf{k}_{\ast }\right) \mathbf{/}\beta \right) $
are unrelated, and consequently the current can be complex-valued. This also
may happen if (\ref{eer2}) does not hold. In this case, in contrast to (\ref%
{jzeta}), the conjugation property does not have to hold and generically we
may have%
\begin{equation}
\mathring{h}_{+}\left( \mathbf{s}\right) \neq \mathring{h}_{-}\left( -%
\mathbf{s}\right) ^{\ast }.  \label{hneqh}
\end{equation}%
In this case the initial condition \ (\ref{inZ}) does not involve the
restiction (\ref{hreal1}). Consequently, we cannot assume that $%
Z_{-}=Z_{+}^{\ast }$, in (\ref{GNLS1+}), (\ref{GNLS1-}). This system also
can be reduced to the system (\ref{ENL+}), (\ref{ENL-}) with a quintic
nonlinearity. The estimate (\ref{er5}) holds in the complex-valued case too.

Note that if 
\begin{equation}
\mathring{h}_{+}\left( \mathbf{s}\right) \neq 0,\mathring{h}_{-}\left( 
\mathbf{s}\right) =0
\end{equation}%
the solutions of (\ref{GNLS1-}) and (\ref{ENL-}) with $\zeta =-$ equal zero: 
\begin{equation}
Z_{-}\left( \mathbf{r},t\right) =0.
\end{equation}%
Substituting $Z_{-}=0$ into (\ref{GNLS1+}) \ we observe that $Z_{+}\left( 
\mathbf{r},t\right) $ becomes a solution of the \emph{linear} Schrodinger
equation. \emph{This fact shows that the nonlinearity in the classical NLS
equation stems from the interaction of two modes of the doublet }$\left\{
\left( +,n_{0},\mathbf{k}_{\ast }\right) ,\left( -,n_{0},-\mathbf{k}_{\ast
}\right) \right\} $\emph{\ and when one of the modes is not initially
excited the nonlinear interaction disappears at the prescribed accuracy
level.}

\subsubsection{Bi-directional waves}

If the linearly excited waves propagate in the both directions $\pm \nabla
\omega _{n_{0}}\left( \mathbf{k}_{\ast }\right) $ the non-FM interactions
between two wavepackets are of the same order as the first-order
susceptibility corrections in (\ref{GNLS1+}), (\ref{GNLS1-}). The
corresponding interactions involve four modes $\tilde{U}_{\zeta
,n_{0}}\left( \pm \mathbf{k}_{\ast }+\mathbf{\eta },t\right) $, $\zeta =\pm
1 $, and their dynamics is approximated by the ENLS solutions $Z_{\zeta
}^{\vartheta }$, $\vartheta =\pm $. The ENLS system in the general complex
currents case consists of four coupled equations. Let us consider here the
system in the simplest case when the excitation currents and the
nonlinearity are real and we use (\ref{conjug}) (for the general case see
Subsection 5.4, in particular (\ref{ENc+}), (\ref{ENcx+})). In this case $%
Z_{-\zeta }^{\pm }\left( \mathbf{r},t\right) =Z_{\zeta }^{\pm }\left( 
\mathbf{r},t\right) ^{\ast }$ and the system reduces to two equations 
\begin{gather}
\left[ \partial _{t}+\mathrm{i}\gamma _{\left( 4\right) }\left[ -\mathrm{i}%
\vec{\nabla}_{\mathbf{r}}\right] \right] Z_{+}^{+}+\alpha _{\pi }\delta
_{\times ,+}^{+}\left( \left\vert Z_{+}^{-}\right\vert ^{2}Z_{+}^{-\ast
}\right) =  \label{Bi1} \\
-\alpha _{\pi }\delta _{1,+}^{+}Z_{+}^{+}Z_{+}^{+\ast }\left[ \partial _{t}+%
\mathrm{i}\gamma _{\left( 4\right) }\left[ -\mathrm{i}\vec{\nabla}_{\mathbf{r%
}}\right] \right] Z_{+}^{+}-\alpha _{\pi }\delta _{2,+}^{+}\left(
Z_{+}^{+}\right) ^{2}\left[ \partial _{t}-\mathrm{i}\gamma _{\left( 4\right)
}\left[ \mathrm{i}\nabla _{\mathbf{r}}\right] \right] Z_{+}^{+\ast }  \notag
\\
+\alpha _{\pi }p_{+}^{+,\left[ 2\right] }\left[ -\mathrm{i}\vec{\nabla}_{%
\mathbf{r}}\right] \left( \left( Z_{+}^{+}\right) ^{2}Z_{+}^{+\ast }\right) ,
\notag
\end{gather}%
\begin{gather}
\left[ \partial _{t}+\mathrm{i}\gamma _{\left( 4\right) }\left[ \mathrm{i}%
\nabla _{\mathbf{r}}\right] \right] Z_{+}^{-}+\alpha _{\pi }\delta _{\times
,+}^{-}\left( \left\vert Z_{+}^{+}\right\vert ^{2}Z_{+}^{+\ast }\right) =
\label{Bi2} \\
-\alpha _{\pi }\delta _{1,+}^{-}\left\vert \left( Z_{+}^{-}\right)
\right\vert ^{2}\left[ \partial _{t}+\mathrm{i}\gamma _{\left( 4\right) }%
\left[ \mathrm{i}\nabla _{\mathbf{r}}\right] \right] Z_{+}^{-}-\alpha _{\pi
}\delta _{2,+}^{-}\left( Z_{+}^{-}\right) ^{2}\left[ \partial _{t}-\mathrm{i}%
\gamma _{\left( 4\right) }\left[ -\mathrm{i}\vec{\nabla}_{\mathbf{r}}\right] %
\right] \left( Z_{+}^{-}\right) ^{\ast }  \notag \\
+\alpha _{\pi }p_{+}^{-,\left[ 2\right] }\left[ -\mathrm{i}\vec{\nabla}_{%
\mathbf{r}}\right] \left( Z_{+}^{-2}Z_{+}^{-\ast }\right) .  \notag
\end{gather}%
We can approximate this system similarly to (\ref{ENL+}), (\ref{ENL-}) by
the system 
\begin{gather}
\left[ \partial _{t}+\mathrm{i}\gamma _{\left( 4\right) }\left[ -\mathrm{i}%
\vec{\nabla}_{\mathbf{r}}\right] \right] Z_{+}^{+}=\alpha _{\pi }p_{+}^{+,%
\left[ 2\right] }\left[ -\mathrm{i}\vec{\nabla}_{\mathbf{r}}\right] \left(
\left( Z_{+}^{+}\right) ^{2}Z_{+}^{+\ast }\right)  \label{Bi5+} \\
-\alpha _{\pi }\delta _{\times ,+}^{+}\left( \left\vert Z_{+}^{-}\right\vert
^{2}Z_{+}^{-\ast }\right) +\alpha _{\pi }^{2}\delta _{5,+}^{+}\left\vert
Z_{+}^{+}\right\vert ^{4}Z_{+}^{+},  \notag
\end{gather}%
\begin{gather}
\left[ \partial _{t}+\mathrm{i}\gamma _{\left( 4\right) }\left[ \mathrm{i}%
\nabla _{\mathbf{r}}\right] \right] Z_{+}^{-}=\alpha _{\pi }p_{+}^{-,\left[ 2%
\right] }\left[ -\mathrm{i}\vec{\nabla}_{\mathbf{r}}\right] \left( \left(
Z_{+}^{-}\right) ^{2}Z_{+}^{+\ast }\right)  \label{Bi5-} \\
-\alpha _{\pi }\delta _{\times ,+}^{-}\left( \left\vert Z_{+}^{+}\right\vert
^{2}Z_{+}^{+\ast }\right) +\alpha _{\pi }^{2}\delta _{5,+}^{-}\left\vert
Z_{+}^{-}\right\vert ^{4}Z_{+}^{-},  \notag
\end{gather}%
where 
\begin{equation}
\delta _{5,+}^{+}=-\delta _{2,+}^{+}Q_{+}^{+\ast }-\delta
_{1,+}^{+}Q_{+}^{+},\ \delta _{5,+}^{-}=-\delta _{2,+}^{-}Q_{+}^{-\ast
}-\delta _{1,+}^{-}Q_{+}^{-}.
\end{equation}%
and $\delta _{\times ,\zeta }^{\pm }$ are some coefficients defined by (\ref%
{dcross}), (\ref{z0bi}), (\ref{kstar1}).

If (i) the fifth-order nonlinear terms in the NLM are taken into account by (%
\ref{dl55}), (ii) (\ref{t0tt2}) holds, (iii) excitation currents are formed
as in Section 7 and (iv) $\varrho \sim \alpha $, then the\ approximation
error estimate (\ref{er5}) holds.

Note that the substitution $Z_{+}^{\pm }=\mathrm{e}^{-\mathrm{i}\gamma
_{0}\tau /\varrho }z_{+}^{\pm }$, where $\gamma _{0}=\omega _{n_{0}}\left( 
\mathbf{k}_{\ast }\right) $, $\tau =\varrho t$, transforms the system (\ref%
{Bi5+}), (\ref{Bi5-}) into the similar one, namely 
\begin{gather}
\left[ \partial _{\tau }+\frac{\mathrm{i}}{\varrho }\gamma _{\left( 4\right)
}^{0}\left[ -\mathrm{i}\vec{\nabla}_{\mathbf{r}}\right] \right] z_{+}^{+}=
\label{Bi5} \\
\frac{\alpha _{\pi }}{\varrho }\left[ p_{+}^{+,\left[ 2\right] }\left[ -%
\mathrm{i}\vec{\nabla}_{\mathbf{r}}\right] \left( \left( z_{+}^{+}\right)
^{2}z_{+}^{+\ast }\right) -\delta _{\times ,+}^{+}\mathrm{e}^{\mathrm{i}%
\gamma _{0}\tau /\varrho }\left( \left\vert z_{+}^{-}\right\vert
^{2}z_{+}^{-\ast }\right) +\alpha _{\pi }\delta _{5,+}^{+}\left\vert
z_{+}^{+}\right\vert ^{4}z_{+}^{+}\right] ,  \notag
\end{gather}%
\begin{gather}
\left[ \partial _{\tau }+\frac{\mathrm{i}}{\varrho }\gamma _{\left( 4\right)
}^{0}\left[ \mathrm{i}\nabla _{\mathbf{r}}\right] \right] z_{+}^{-}=
\label{Bi51} \\
\frac{\alpha _{\pi }}{\varrho }\left[ p_{+}^{-,\left[ 2\right] }\left[ -%
\mathrm{i}\vec{\nabla}_{\mathbf{r}}\right] \left( \left( z_{+}^{-}\right)
^{2}z_{+}^{-\ast }\right) -\delta _{\times ,+}^{-}\mathrm{e}^{\mathrm{i}%
\gamma _{0}\tau /\varrho }\left( \left\vert z_{+}^{+}\right\vert
^{2}z_{+}^{+\ast }\right) +\alpha _{\pi }\delta _{5,+}\left\vert
z_{+}^{-}\right\vert ^{4}z_{+}^{-}\right]  \notag
\end{gather}%
with the differential operator $\gamma _{\left( 3\right) }^{0}\left[ \mathrm{%
i}\nabla _{\mathbf{r}}\right] $ having no zero-order terms. This system has
the oscillatory coefficients $\alpha _{\pi }\delta _{\times ,+}^{\pm }%
\mathrm{e}^{\mathrm{i}\gamma _{0}\tau /\varrho }$ accounting for the effect
of the non-FM\ interactions. Integration over $\tau $ of these coefficients
produces the factor $\frac{\varrho }{\omega _{n_{0}}\left( \mathbf{k}_{\ast
}\right) }$ and we get for $\tau ^{\prime }\leq \tau _{\ast }$ 
\begin{equation}
\alpha _{\pi }\int_{0}^{\tau ^{\prime }}\delta _{\times ,+}^{-}\mathrm{e}^{%
\mathrm{i}\gamma _{0}\tau /\varrho }\left( \left\vert z_{+}^{+}\right\vert
^{2}z_{+}^{+\ast }\right) \,\mathrm{d}\tau =O\left( \alpha \varrho \right)
\label{Orho}
\end{equation}%
thus showing that the coupling interactions are suppressed due to the
frequency mismatch and $\alpha _{\pi }\delta _{\times ,+}^{+}\left(
\left\vert Z_{+}^{-}\right\vert ^{2}Z_{+}^{-\ast }\right) $ in the case of
the classical NLS scaling after the integration have the same order of
magnitude $O\left( \alpha \varrho \right) =O\left( \beta ^{4}\right) $ as $%
\alpha _{\pi }^{2}\delta _{5,+}^{+}\left\vert Z_{+}^{+}\right\vert
^{4}Z_{+}^{+}=O\left( \alpha ^{2}\right) =O\left( \beta ^{4}\right) $ and
fourth-order derivatives in $\gamma _{\left( 4\right) }^{0}\left[ -\mathrm{i}%
\vec{\nabla}_{\mathbf{r}}\right] Z_{+}^{+}$ which are also $O\left( \beta
^{4}\right) $. \ 

\textbf{Remark.} For a derivation based on anharmonic Maxwell-Lorenz system
of coupled-mode equations which describe bi-directional propagation of waves
in one-dimensional periodic structures see \cite{GoodmanWH01} and references
therein.$\blacklozenge $

\subsubsection{Other scalings and the reduction of ENLS}

We remind that when deriving the NLS\ and ENLS\ and\ providing the related
error estimates, we allow an arbitrary power dependence between the
parameters $\varrho $ and $\beta $. The condition (\ref{rhoalph}) on $%
\varrho $ and $\alpha $ also has the form (\ref{rhoal}), we take here $%
\varkappa _{0}=1$ that is $\alpha \sim \varrho $. The properties of the
ENLS\ in different ranges of the parameters imply corresponding properties
for the NLM. The ENLS\ themselves can be reduced to simpler equations by
formally throwing away higher order terms. Estimating the order of the terms
in the ENLS one has to take into account the Remark in the beginning of
Subsection 1.3

\textbf{Example of a strongly dispersive scaling.} For example, let us
consider the particular case when (\ref{nonM}) holds 
\begin{equation}
\alpha \sim \varrho \sim \beta ^{3},  \label{rb3}
\end{equation}%
implying strong dispersion 
\begin{equation}
\theta ^{-1}=\frac{\beta ^{2}}{\varrho }\gg 1.  \label{nonM11}
\end{equation}%
The term in the right-hand side of (\ref{er5a}) takes the form%
\begin{equation}
\left[ O\left( \alpha \beta ^{3}\right) +O\left( \alpha ^{2}\beta \right)
+O\left( \alpha ^{3}\right) \right] =O\left( \beta ^{6}\right) .
\end{equation}%
The terms of order $O\left( \alpha \varrho \right) =O\left( \beta
^{6}\right) $ and $\ O\left( \alpha ^{2}\right) =O\left( \beta ^{6}\right) $
in (\ref{Bi5+}), (\ref{Bi5-}) are now of the same order as the error and can
be neglected. After discarding the higher order terms in (\ref{Bi5+}), (\ref%
{Bi5-}) we get the reduced system%
\begin{equation}
\left[ \partial _{t}+\mathrm{i}\gamma _{\left( 4\right) }\left[ -\mathrm{i}%
\vec{\nabla}_{\mathbf{r}}\right] \right] Z_{+}^{+}=\alpha _{\pi }p_{+}^{+,%
\left[ 2\right] }\left[ -\mathrm{i}\vec{\nabla}_{\mathbf{r}}\right] \left(
\left( Z_{+}^{+}\right) ^{2}Z_{+}^{+\ast }\right) ,  \notag
\end{equation}%
\begin{equation}
\left[ \partial _{t}+\mathrm{i}\gamma _{\left( 4\right) }\left[ \mathrm{i}%
\nabla _{\mathbf{r}}\right] \right] Z_{+}^{-}=\alpha _{\pi }p_{+}^{-,\left[ 1%
\right] }\left[ -\mathrm{i}\vec{\nabla}_{\mathbf{r}}\right] \left( \left(
Z_{+}^{-}\right) ^{2}Z_{+}^{-\ast }\right) ,  \notag
\end{equation}%
and the equation for $Z_{+}^{+}$ is now decoupled from the equation for $%
Z_{+}^{-}$ at the level of accuracy \ $O\left( \beta ^{6}\right) O\left(
\left\vert \mathbf{U}^{\left( 1\right) }\right\vert \right) $ on the time
interval of length $\frac{\tau _{\ast }}{\beta ^{3}}$. Note that the
coupling terms with the interband component mentioned in the end of Section
1.4.4 also are of order $O\left( \alpha \varrho \right) $ and can be
neglected.

\textbf{Example of a weakly dispersive scaling.} Let us consider the
particular case when (\ref{nonM}) holds: 
\begin{equation}
\varrho \sim \beta ,\;\alpha \sim \varrho ,  \label{rb1}
\end{equation}%
implying weak dispersion 
\begin{equation}
\theta ^{-1}=\frac{\beta ^{2}}{\varrho }\ll 1.  \label{wM}
\end{equation}%
First, consider the third-order ENLS (\ref{G3+}). The error term in (\ref{O3}%
) now is of order 
\begin{equation}
O\left( \alpha ^{2}\right) +O\left( \alpha \beta ^{2}\right) +O\left( \alpha
\varrho \right) =O\left( \beta ^{2}\right)
\end{equation}%
Since 
\begin{equation}
\vec{\nabla}_{\mathbf{r}}Z_{+}=O\left( \beta \right) ,\;\vec{\nabla}_{%
\mathbf{r}}^{2}Z_{+}=O\left( \beta ^{2}\right) ,\;\vec{\nabla}_{\mathbf{r}%
}^{3}Z_{+}=O\left( \beta ^{3}\right)  \label{Ospat}
\end{equation}%
the second and third derivatives in $\ \mathcal{L}_{+}^{\left[ 3\right]
}Z_{+}$ have order $\beta ^{2}$ and $\ \beta ^{3}$ respectively, they are $%
O\left( \beta ^{2}\right) $ and have to be thrown away. The first derivative
in $\alpha _{\pi }p_{+}^{\left[ 1\right] }\left[ -\mathrm{i}\vec{\nabla}_{%
\mathbf{r}}\right] \left( Z_{+}^{2}Z_{-}\right) $ also is $O\left( \alpha
\beta \right) =O\left( \beta ^{2}\right) $. The reduced equation takes the
form of a first-order hyperbolic equation%
\begin{equation}
\partial _{t}Z_{+}+\mathrm{i}\mathcal{L}_{+}^{\left[ 1\right] }Z_{+}=\alpha
_{\pi }p_{+}^{\left[ 0\right] }\left[ -\mathrm{i}\vec{\nabla}_{\mathbf{r}}%
\right] \left( Z_{+}^{2}Z_{-}\right) .  \label{G3r}
\end{equation}%
In the case of the space dimension $d=1$, $\mathbf{r}=x$, the reduced
equation takes the form%
\begin{equation}
\partial _{t}Z_{+}+\mathrm{i}\gamma _{0}Z_{+}+\frac{\beta \gamma _{1}}{%
\varrho }\partial _{x}Z_{+}=\alpha _{\pi }Q_{+}\left\vert Z_{+}\right\vert
^{2}Z_{+}  \label{G3r1d}
\end{equation}%
and a similar equation for $Z_{-}=Z_{+}^{\ast }$. This system approximates
the NLM with the accuracy $O\left( \beta ^{2}\right) $ in the case of the
scaling (\ref{rb1}). \ As we have pointed out earlier, the condition (\ref%
{wM}) implies that the dispersive effects are small, which agrees with the
form of the equation (\ref{G3r}) which does not include dispersive terms.

\textbf{Remark.} The described reduction of the universal ENLS to reduced
ENLS in the case of particular scaling relations between parameters $\alpha $%
, $\beta $ and $\varrho $ is quite simple. The nontrivial part is the
validity of the error estimates in the whole range of parameters, which
guarantees that the reduced equations well approximate the exact solutions
of the NLM itself. The estimates imply that the differences between
different reduced ENLS correspond to actual differences between different
classes of solutions of the NLM which are generated by different initial
excitations.$\blacklozenge $

\section{Modal decompositions and power series expansions of the linear and
the first nonlinear responses}

The very form (\ref{UNLS}) of the approximation $\mathbf{U}_{Z,n_{0}}\left( 
\mathbf{r},t\right) $ is based on the modal decomposition. We remind that
one of our goals is the construction of excitation currents producing waves
governed essentially by NLS equations. This construction is carried out in
terms of the modal decomposition of all fields which is absolutely
instrumental to the analysis of nonlinear wave propagation, \cite{BF1}. We
are particularly interested in approximations for propagating waves as the
quantities $\alpha $, $\varrho $ and $\beta $ approach zero, and these
approximations are constructed based on relevant asymptotic expansions of
the involved fields.

\subsection{Bloch modal decomposition}

We systematically use modal decompositions based on the Bloch eigenmodes $%
\mathbf{\tilde{G}}_{\bar{n}}\left( \mathbf{r},\mathbf{k}\right) $ of the
linear Maxwell operator $\mathbf{M}$ in (\ref{MXshort}):%
\begin{equation}
\mathbf{M\tilde{G}}_{\bar{n}}\left( \mathbf{r},\mathbf{k}\right) =\omega _{%
\bar{n}}\left( \mathbf{k}\right) \mathbf{\tilde{G}}_{\bar{n}}\left( \mathbf{r%
},\mathbf{k}\right) ,  \label{eigen}
\end{equation}%
where 
\begin{equation}
\bar{n}=\left( \zeta ,n\right) ,\ n=1,2\ldots .,\ \zeta =\pm 1;\ \omega _{%
\bar{n}}\left( \mathbf{k}\right) =\omega _{\zeta ,n}\left( \mathbf{k}\right)
=\zeta \omega _{n}\left( \mathbf{k}\right) ,\ \omega _{n+1}\left( \mathbf{k}%
\right) \geq \omega _{n}\left( \mathbf{k}\right) \geq 0,  \label{omn}
\end{equation}%
with $n$ being the band number and $\mathbf{k}$ being the quasimomentum
taking values in the Brillouin zone. For notational simplicity we consider
the cubic lattice with the lattice constant $L=1$ and, consequently, with
the Brillouin zone being the cube $\left[ -\pi ,\pi \right] ^{3}$. Note that
for given $n$ and $\mathbf{k}$ there are exactly two eigenvalues $\omega _{%
\bar{n}}\left( \mathbf{k}\right) =\zeta \omega _{n}\left( \mathbf{k}\right) $%
,$\ \zeta =\pm 1$.


\begin{figure}[tbph]
\scalebox{0.5}{\includegraphics[viewport= 100 50 800
600,clip]{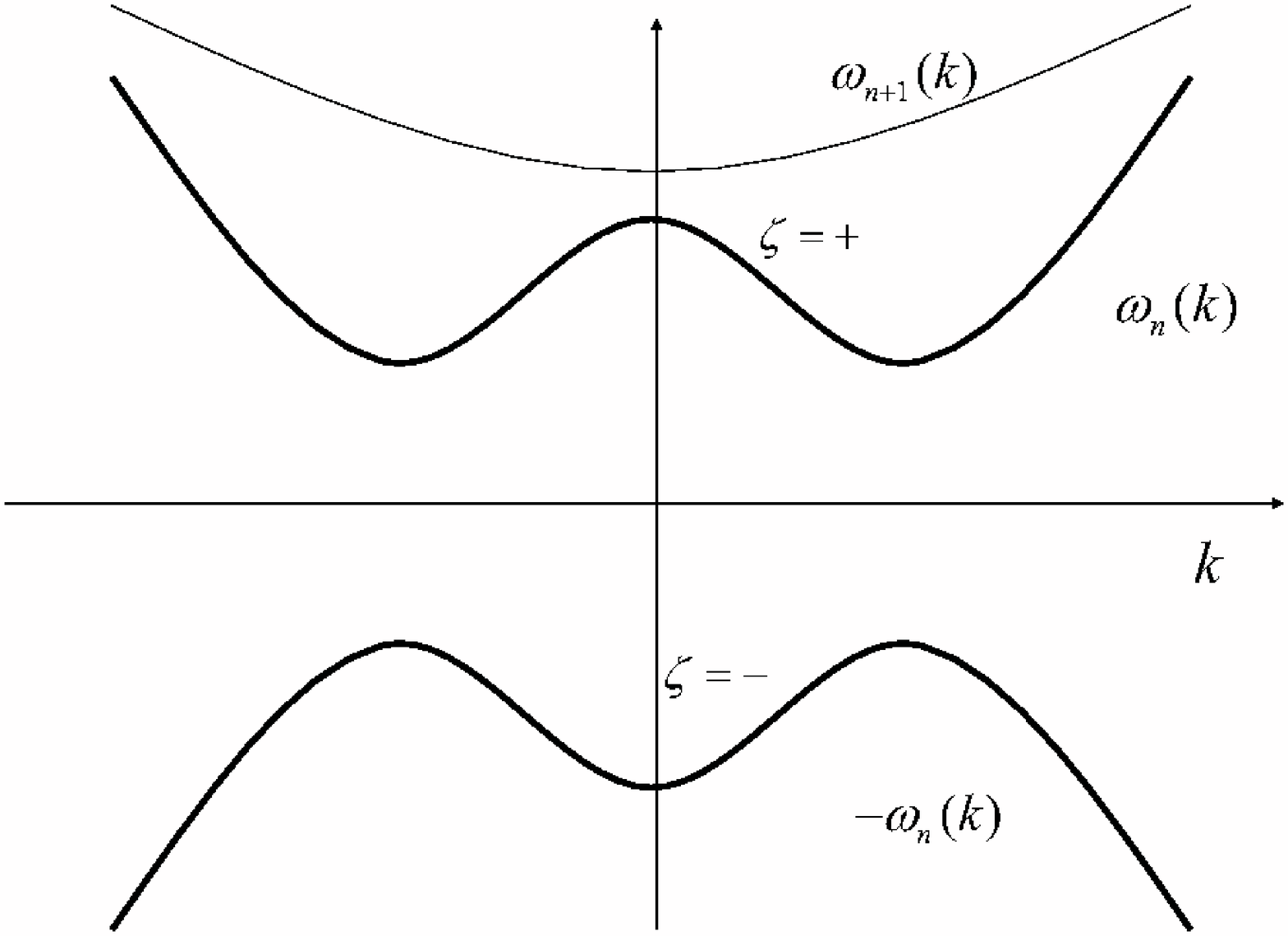}}
\caption{Schematic graphs of the dispersion relations corresponding to a
pair of conjugate bands $\protect\zeta \protect\omega _{n}\left( k\right) $, 
$\protect\zeta =\pm $ , which are inversion symmetric. }
\label{figband}
\end{figure}
The inversion symmetry condition (\ref{invsym}) for the dispersion
relations, which is%
\begin{equation}
\omega _{n}\left( -\mathbf{k}\right) =\omega _{n}\left( \mathbf{k}\right) ,\
n=1,2,\ldots \ ,
\end{equation}%
readily implies the following properties of its first, the second and higher
order differentials%
\begin{equation}
\omega _{n}^{\prime }\left( -\mathbf{k}\right) =-\omega _{n}^{\prime }\left( 
\mathbf{k}\right) ,\ \omega _{n}^{\prime \prime }\left( -\mathbf{k}\right)
=\omega _{n}^{\prime \prime }\left( \mathbf{k}\right) ,\ldots ,\ \omega
_{n}^{\left( j\right) }\left( -\mathbf{k}\right) =\left( -1\right)
^{j}\omega _{n}^{\left( j\right) }\left( \mathbf{k}\right) ,\ j,n=1,2,\ldots
.  \label{invsym2}
\end{equation}%
The eigenmodes $\mathbf{\tilde{G}}_{\bar{n}}\left( \mathbf{r},\mathbf{k}%
\right) $ are 6-component vector fields satsifying the following relations, 
\cite{BF1}, 
\begin{gather}
\mathbf{\tilde{G}}_{\bar{n}}\left( \mathbf{r},\mathbf{k}\right) =\left( 
\begin{array}{c}
\mathbf{\tilde{G}}_{D,\bar{n}}\left( \mathbf{r},\mathbf{k}\right) \\ 
\mathbf{\tilde{G}}_{B,\bar{n}}\left( \mathbf{r},\mathbf{k}\right)%
\end{array}%
\right) ,\ \nabla _{\mathbf{r}}\cdot \mathbf{\tilde{G}}_{D,\bar{n}}\left( 
\mathbf{r},\mathbf{k}\right) =0,\ \nabla _{\mathbf{r}}\cdot \mathbf{\tilde{G}%
}_{B,\bar{n}}\left( \mathbf{r},\mathbf{k}\right) =0,\ \mathbf{r}\text{ in }%
\left[ 0,1\right] ^{3},  \label{nz2} \\
\mathbf{\tilde{G}}_{\bar{n}}\left( \mathbf{r}+\mathbf{m},\mathbf{k}\right)
=\exp \left\{ \mathrm{i}\mathbf{k}\cdot \mathbf{m}\right\} \mathbf{\tilde{G}}%
_{\bar{n}}\left( \mathbf{r},\mathbf{k}\right) ,\ \mathbf{m}\text{ in }%
\mathbf{Z}^{3}.  \label{nz2a}
\end{gather}%
We also introduce the scalar product 
\begin{equation}
\left( \mathbf{U},\mathbf{V}\right) _{\mathcal{H}}=\int_{\left[ 0,1\right]
^{d}}\mathbf{U}\left( \mathbf{r}\right) \cdot \sigma _{\varepsilon }\left( 
\mathbf{r}\right) \mathbf{V}\left( \mathbf{r}\right) ^{\ast }\,\,\mathrm{d}%
\mathbf{r},\mathbf{\;}\sigma _{\varepsilon }\left( \mathbf{r}\right) =\left[ 
\begin{array}{cc}
\mathbf{\varepsilon }^{-1}\left( \mathbf{r}\right) & \mathbf{0} \\ 
\mathbf{0} & \mathbf{I}%
\end{array}%
\right] ,  \label{sigma}
\end{equation}%
and assume that $\mathbf{\tilde{G}}_{\zeta ,n}\left( \mathbf{r},\mathbf{k}%
\right) $\ are orthonormal in $\mathcal{H}$:%
\begin{equation}
\left\Vert \mathbf{\tilde{G}}_{\bar{n}}\left( \cdot ,\mathbf{k}\right)
\right\Vert _{\mathcal{H}}=\left( \mathbf{\tilde{G}}_{\bar{n}}\left( \cdot ,%
\mathbf{k}\right) ,\mathbf{\tilde{G}}_{\bar{n}}\left( \cdot ,\mathbf{k}%
\right) \right) _{\mathcal{H}}^{1/2}=1.
\end{equation}%
Notice that if the condition (\ref{eer2}) holds then the complex conjugate
of every eigenmode $\mathbf{\tilde{G}}_{\zeta ,n}\left( \mathbf{r},\mathbf{k}%
\right) $ coincides with the eigenmode $\mathbf{\tilde{G}}_{-\zeta ,n}\left( 
\mathbf{r},-\mathbf{k}\right) $, i.e. 
\begin{gather}
\left[ \mathbf{\tilde{G}}_{\zeta ,n}\left( \mathbf{r},\mathbf{k}\right) %
\right] ^{\ast }=\mathbf{\tilde{G}}_{-\zeta ,n}\left( \mathbf{r},-\mathbf{k}%
\right)  \label{Gstar} \\
\text{under the assumption }\func{Im}\mathbf{\varepsilon }\left( \mathbf{r}%
\right) =\left\{ \func{Im}\varepsilon _{jm}\left( \mathbf{r}\right) \right\}
_{j,m=1}^{3}=0.  \notag
\end{gather}

Let us consider now a solution $\mathbf{U}\left( \mathbf{r},t\right) $ to
the NLM (\ref{MXshort}) and its Floquet-Bloch modal decomposition, \cite{BF1}
\begin{equation}
\mathbf{U}\left( \mathbf{r},t\right) =\sum_{\bar{n}}\frac{1}{(2\pi )^{d}}%
\int_{\left[ -\pi ,\pi \right] ^{d}}\mathbf{\tilde{U}}_{\bar{n}}\left( 
\mathbf{r},\mathbf{k},t\right) \,\mathrm{d}\mathbf{k},\mathbf{\ \tilde{U}}_{%
\bar{n}}\left( \mathbf{r},\mathbf{k},t\right) =\tilde{U}_{\bar{n}}\left( 
\mathbf{k},t\right) \mathbf{\tilde{G}}_{\bar{n}}\left( \mathbf{r},\mathbf{k}%
\right) ,\   \label{UFB}
\end{equation}%
where $\mathbf{U}_{\bar{n}}\left( \mathbf{r},\mathbf{k},t\right) $ are the
modal components, and $\tilde{U}_{\bar{n}}\left( \mathbf{k},t\right) $ are
the (scalar) modal coefficients given by the formula%
\begin{equation}
\tilde{U}_{\bar{n}}\left( \mathbf{k},t\right) =\int_{\mathbf{R}^{d}}\mathbf{U%
}\left( \mathbf{r},t\right) \cdot \sigma _{\varepsilon }\left( \mathbf{r}%
\right) \mathbf{\tilde{G}}_{\bar{n}}^{\ast }\left( \mathbf{r},\mathbf{k}%
\right) \,\,\mathrm{d}\mathbf{r}.  \label{UG}
\end{equation}%
The property (\ref{Gstar}) implies the following relations for the modal
coefficients of the complex conjugate fields 
\begin{equation}
\left( \widetilde{U^{\ast }}\right) _{\zeta ,n}\left( \mathbf{k},t\right) =%
\left[ \tilde{U}_{-\zeta ,n}\left( -\mathbf{k},t\right) \right] ^{\ast }.
\label{Ustar}
\end{equation}%
The Floquet-Bloch transform $\mathbf{\tilde{U}}$ of $\mathbf{U}$ which
involves all modes is defined by the formula 
\begin{equation}
\mathbf{\tilde{U}}\left( \mathbf{r},\mathbf{k},t\right) =\sum_{\bar{n}}%
\mathbf{\tilde{U}}_{\bar{n}}\left( \mathbf{r},\mathbf{k},t\right) =\sum_{%
\bar{n}}\tilde{U}_{\bar{n}}\left( \mathbf{k},t\right) \mathbf{\tilde{G}}_{%
\bar{n}}\left( \mathbf{r},\mathbf{k}\right) ,  \label{Utilda}
\end{equation}%
with the properties of $\mathbf{\tilde{G}}_{\zeta ,n}\left( \mathbf{r},%
\mathbf{k}\right) $ and $\omega _{n}\left( \mathbf{k}\right) $ discussed in
detail in \cite{BF1}. It is often convenient to write the coefficients $%
\tilde{U}_{\bar{n}}\left( \mathbf{k},t\right) $ in a special form, namely 
\begin{equation}
\tilde{U}_{\bar{n}}\left( \mathbf{k},t\right) =\tilde{u}_{\bar{n}}\left( 
\mathbf{k},\tau \right) \mathrm{e}^{-\mathrm{i}\omega _{\bar{n}}\left( 
\mathbf{k}\right) t},\quad \tau =\varrho t,  \label{UFB1}
\end{equation}%
factoring out the carrier frequency $\omega _{\bar{n}}\left( \mathbf{k}%
\right) $. This equality defines the modal coefficient $\tilde{u}_{\bar{n}%
}\left( \mathbf{k},\tau \right) $ as a function of slow time $\tau $.
Similarly to (\ref{Utilda}) we define 
\begin{equation}
\mathbf{\tilde{u}}\left( \mathbf{r},\mathbf{k},\tau \right) =\sum_{\bar{n}}%
\mathbf{\tilde{u}}_{\bar{n}}\left( \mathbf{r},\mathbf{k},\tau \right) =\sum_{%
\bar{n}}\tilde{u}_{\bar{n}}\left( \mathbf{k},\tau \right) \mathbf{\tilde{G}}%
_{\bar{n}}\left( \mathbf{r},\mathbf{k}\right) .  \label{utilda}
\end{equation}

\subsection{Nonlinearity and related power expansions}

The nonlinear term $\mathcal{F}_{\text{NL}}\left( \mathbf{U}\right) $ in the
NLM equation (\ref{MXshort}) is given by the formula, \cite{BF1}-\cite{BF4}, 
\begin{equation}
\mathcal{F}_{\text{NL}}\left( \mathbf{U}\right) =\mathcal{F}_{\text{NL}%
}\left( \mathbf{U};\alpha \right) =\left[ 
\begin{array}{c}
\mathbf{0} \\ 
\nabla \times \mathbf{S}_{D}\left( \mathbf{r},t;\mathbf{D};\alpha \right)%
\end{array}%
\right] ,\ \mathbf{U}\left( \mathbf{r},t\right) =\left[ 
\begin{array}{c}
\mathbf{D}\left( \mathbf{r},t\right) \\ 
\mathbf{B}\left( \mathbf{r},t\right)%
\end{array}%
\right] ,  \label{FNL}
\end{equation}%
where 
\begin{equation}
\mathbf{S}_{D}\left( \mathbf{r},t;\mathbf{D}\right) =\mathbf{S}_{D}^{\left(
3\right) }\left( \mathbf{r},t;\mathbf{D}\right) +\alpha \mathbf{S}%
_{D}^{\left( 5\right) }\left( \mathbf{r},t;\mathbf{D}\right) +\alpha ^{2}%
\mathbf{S}_{D}^{\left( 7\right) }\left( \mathbf{r},t;\mathbf{D}\right)
+\ldots  \label{cd4}
\end{equation}%
is a series of causal integral operators $\mathbf{S}_{D}^{\left( 2n+1\right)
}$, which are determined based on the response functions from (\ref{mx7}).
Notice that the representation (\ref{cd4}) consists of only odd order terms
as it is typical for dielectric media allowing NLS regimes. The dominant
cubic nonlinearity is given by the causal integral 
\begin{equation}
\mathbf{S}_{D}^{\left( 3\right) }\left( \mathbf{r},t;\mathbf{D}\right)
=\int_{-\infty }^{t}\int_{-\infty }^{t}\int_{-\infty }^{t}\mathbf{R}%
_{D}^{\left( 3\right) }\left( \mathbf{r};t-t_{1},t-t_{2},t-t_{3}\right)
\vdots \,\prod_{j=1}^{3}\mathbf{D}\left( \mathbf{r},t_{j}\right) \,\mathrm{d}%
t_{j},  \label{caus}
\end{equation}%
where the trilinear tensorial operator $\mathbf{R}_{D}^{\left( 3\right) }$
is assumed to be symmetric: 
\begin{gather}
\mathbf{R}_{D}^{\left( 3\right) }\left( \mathbf{r};t-t_{1},t-t_{2},t-t_{3}%
\right) \vdots \,\mathbf{D}_{1}\mathbf{D}_{2}\mathbf{D}_{3}=\mathbf{R}%
_{D}^{\left( 3\right) }\left( \mathbf{r};t-t_{2},t-t_{1},t-t_{3}\right)
\vdots \,\mathbf{D}_{2}\mathbf{D}_{1}\mathbf{D}_{3}  \label{Rsym} \\
=\mathbf{R}_{D}^{\left( 3\right) }\left( \mathbf{r};t-t_{3},t-t_{2},t-t_{1}%
\right) \vdots \,\mathbf{D}_{3}\mathbf{D}_{1}\mathbf{D}_{2}.  \notag
\end{gather}%
An alternative and often used representation of the polarization is through
its frequency dependent susceptibility tensor $\mathbf{\chi }_{D}^{\left(
3\right) }$,%
\begin{equation}
\mathbf{\chi }_{D}^{\left( 3\right) }\left( \mathbf{r};\omega _{1},\omega
_{2},\omega _{3}\right) =\int_{0}^{\infty }\int_{0}^{\infty
}\int_{0}^{\infty }\mathbf{R}_{D}^{\left( 3\right) }\left( \mathbf{r}%
;t_{1},t_{2},t_{3}\right) \mathbf{\,}\mathrm{e}^{\left\{ \mathrm{i}\left(
\omega _{1}t_{1}+\omega _{2}t_{2}+\omega _{3}t_{3}\right) \right\} }\,%
\mathrm{d}t_{1}\mathrm{d}t_{2}\mathrm{d}t_{3}.  \label{cd4ab}
\end{equation}%
Note that the standard frequency dependent susceptibility tensor $\mathbf{%
\chi }^{\left( 3\right) }\left( \mathbf{r};\mathbf{\omega }\right) $ is \
determined in terms of the nonlinear polarization $\mathbf{P}_{\text{NL}%
}\left( \mathbf{r},t;\mathbf{E}\left( \cdot \right) \right) $ of the medium
by a formula similar to (\ref{cd4ab}) (see \cite{BC}). The tensor $\mathbf{%
\chi }_{D}^{\left( 3\right) }$ $\left( \mathbf{r};\mathbf{\omega }\right) $
(which acts on $\mathbf{D}$) is expressed in terms of $\ \mathbf{\chi }%
^{\left( 3\right) }\left( \mathbf{r};\mathbf{\omega }\right) $\ (which acts
on $\mathbf{E}$) and the dielectric tensor $\mathbf{\varepsilon }\left( 
\mathbf{r}\right) $ by the following formula (see \cite{BF1}, \cite{BF4} for
details):

\begin{equation}
\mathbf{\chi }_{D}^{\left( 3\right) }\left( \mathbf{r};\mathbf{\omega }%
\right) \vdots \,\prod_{j=1}^{3}\mathbf{D}_{j}=4\pi \mathbf{\varepsilon }%
^{-1}\left( \mathbf{r}\right) \mathbf{\chi }^{\left( 3\right) }\left( 
\mathbf{r};\mathbf{\omega }\right) \,\vdots \,\prod_{j=1}^{3}\left[ \mathbf{%
\varepsilon }^{-1}\left( \mathbf{r}\right) \mathbf{D}_{j}\right] .
\label{cd5b}
\end{equation}%
Let us consider the power series expansion (\ref{uMv1}) for the exact
solution $\mathbf{U}$ to the NML\ (\ref{MXshort}) with the current $\mathbf{J%
}$ satisfying the relations (\ref{Jeq0}) and (\ref{J01}), i.e.%
\begin{gather}
\mathbf{U}=\mathbf{U}^{\left( 0\right) }+\alpha \mathbf{U}^{\left( 1\right)
}+\alpha ^{2}\mathbf{U}^{\left( 2\right) }+\ldots ,\ \mathbf{J}=\mathbf{J}%
^{\left( 0\right) }+\alpha \mathbf{J}^{\left( 1\right) }+\ldots ,
\label{UU1} \\
\mathbf{J}^{\left( j\right) }\left( \mathbf{r},t\right) =0\ \text{if }t\leq
0\ \text{or }t\geq \frac{\tau _{0}}{\varrho },\ j=0,1.  \label{UU1a}
\end{gather}%
For every amplitude $\tilde{u}_{\bar{n}}\left( \mathbf{k},\tau \right) $
defined by (\ref{UFB}), (\ref{UFB1}) the series corresponding to (\ref{UU1})
becomes%
\begin{equation}
\tilde{u}_{\bar{n}}\left( \mathbf{k},\tau \right)
=\dsum\limits_{m=0}^{\infty }\tilde{u}_{\bar{n}}^{\left( m\right) }\left( 
\mathbf{k},\tau \right) \alpha ^{m}.  \label{uap1}
\end{equation}%
Power expansions for the amplitudes $\tilde{u}_{\bar{n}}\left( \mathbf{k}%
,\tau \right) $ as well as for other quantities of interest with respect to
the small parameter $\alpha $ are given by convergent Taylor series. The
expansions for the amplitudes $\tilde{u}_{\bar{n}}\left( \mathbf{k},\tau
\right) $ with respect to the small parameters $\varrho $ and $\beta $ are
of more complicated nature related to almost time-harmonic expansions and
asymptotic expansions for oscillatory integrals (see (\ref{asser}) in the
next subsection, see also Subsections 6.2, 8.1 and 8.3).

We remind that the current $\mathbf{J}^{\left( 1\right) }$ in (\ref{UU1}) is
introduced to provide proper transformation of the initial data for the NLS
into the excitation current (see Subsection 5.2 for details). The expansion (%
\ref{UU1}) determines the \emph{linear medium response} $\mathbf{U}^{\left(
0\right) }$ and \emph{the first nonlinear response} $\mathbf{U}^{\left(
1\right) }$ satisfying respectively the evolution equations (\ref{MXlin})
and (\ref{FNLR}), namely%
\begin{equation}
\partial _{t}\mathbf{U}^{\left( 0\right) }=\mathbf{-}\mathrm{i}\mathbf{MU}%
^{\left( 0\right) }-\mathbf{J}^{\left( 0\right) };\ \mathbf{U}^{\left(
0\right) }\left( t\right) =0\;\text{for }t\leq 0,  \label{UU2}
\end{equation}%
\begin{gather}
\partial _{t}\mathbf{U}^{\left( 1\right) }=\mathbf{-}\mathrm{i}\mathbf{MU}%
^{\left( 1\right) }+\mathcal{F}_{\text{NL}}^{\left( 1\right) }\left( \mathbf{%
U}^{\left( 0\right) }\right) -\mathbf{J}^{\left( 1\right) };\ \mathbf{U}%
^{\left( 1\right) }\left( t\right) =0\;\text{for }t\leq 0,  \label{UU3} \\
\mathcal{F}_{\text{NL}}^{\left( 1\right) }\left( \mathbf{U}^{\left( 0\right)
}\right) =\mathcal{F}_{\text{NL}}\left( \mathbf{U}^{\left( 0\right) };\alpha
\right) |_{\alpha =0}.  \notag
\end{gather}%
We introduce now the currents $\mathbf{J}^{\left( 0\right) }$ and $\mathbf{J}%
^{\left( 1\right) }$ satisfying the conditions (\ref{UU1a}) by their modal
coefficients as follows: 
\begin{gather}
\mathbf{\tilde{J}}_{\bar{n}}^{\left( j\right) }\left( \mathbf{r},\mathbf{k}%
,t\right) =\tilde{J}_{\bar{n}}^{\left( j\right) }\left( \mathbf{k},t\right) 
\mathbf{\tilde{G}}_{\bar{n}}\left( \mathbf{r},\mathbf{k}\right) ,\;\tilde{J}%
_{\bar{n}}^{\left( j\right) }\left( \mathbf{k},t\right) =\mathrm{e}_{\bar{n}%
}^{-\mathrm{i}\omega _{\bar{n}}\left( \mathbf{k}\right) t}\tilde{j}_{\bar{n}%
}^{\left( j\right) }\left( \mathbf{k},\tau \right) ,\;\tau =\varrho t,
\label{jjt2} \\
\tilde{j}_{\bar{n}}^{\left( j\right) }\left( \mathbf{k},\tau \right) =0\ 
\text{if }\tau \leq 0\text{ or }\tau \geq \tau _{0},\ j=0,1.  \notag
\end{gather}%
For the currents $\mathbf{J}^{\left( 0\right) }$ and $\mathbf{J}^{\left(
1\right) }$ to be real, in view of (\ref{Ustar}), their modal coefficients
should satisfy the relations 
\begin{equation}
\left[ j_{\zeta ,n}^{\left( j\right) }\left( \mathbf{k}\right) \right]
^{\ast }=j_{-\zeta ,n}^{\left( j\right) }\left( -\mathbf{k}\right) ,\ j=0,1.
\label{jstar}
\end{equation}%
>From (\ref{UU2}), (\ref{UU3}), (\ref{jjt2}) and (\ref{uap1}) we get the
following representation for the modal forms of the first two terms $\mathbf{%
U}^{\left( 0\right) }$ and $\mathbf{U}^{\left( 1\right) }$ of the power
expansion (\ref{UU1}): 
\begin{equation}
\mathbf{\tilde{U}}_{\bar{n}}^{\left( j\right) }\left( \mathbf{r},\mathbf{k}%
,t\right) =\tilde{U}_{\bar{n}}^{\left( j\right) }\left( \mathbf{k},t\right) 
\mathbf{\tilde{G}}_{\bar{n}}\left( \mathbf{r},\mathbf{k}\right) ,\;\tilde{U}%
_{\bar{n}}^{\left( j\right) }\left( \mathbf{k},t\right) =\tilde{u}_{\bar{n}%
}^{\left( j\right) }\left( \mathbf{k},\tau \right) \mathrm{e}^{-\mathrm{i}%
\omega _{\bar{n}}\left( \mathbf{k}\right) t},\ \tau =\varrho t,\ j=0,1,
\label{UlinBloch}
\end{equation}%
where 
\begin{eqnarray}
\tilde{u}_{\bar{n}}^{\left( 0\right) }\left( \mathbf{k},\tau \right)
&=&-\int_{0}^{\tau }\tilde{j}_{\bar{n}}^{\left( 0\right) }\left( \mathbf{k}%
,\tau \right) \,\mathrm{d}\tau _{1},  \label{V0} \\
\tilde{U}_{\bar{n}}^{\left( 1\right) }\left( \mathbf{k},\tau \right) &=&%
\frac{1}{\varrho }\int_{0}^{\tau }\mathrm{e}^{-\mathrm{i}\omega _{\bar{n}%
}\left( \mathbf{k}\right) \frac{\left( \tau -\tau _{1}\right) }{\varrho }%
}\left\{ \left[ \mathcal{F}_{\text{NL}}^{\left( 0\right) }\left( \mathbf{U}%
^{\left( 0\right) }\right) \right] _{\bar{n}}\left( \mathbf{k},\tau \right) -%
\tilde{J}_{\bar{n}}^{\left( 1\right) }\left( \mathbf{k},t\right) \right\} \,%
\mathrm{d}\tau _{1}.  \label{V01}
\end{eqnarray}%
Similarly to (\ref{utilda}) we introduce 
\begin{equation}
\mathbf{\tilde{u}}^{\left( 0\right) }\left( \mathbf{r},\mathbf{k},\tau
\right) =\sum_{\bar{n}}\mathbf{\tilde{u}}_{\bar{n}}^{\left( 0\right) }\left( 
\mathbf{r},\mathbf{k},\tau \right) =\sum_{\bar{n}}\tilde{u}_{\bar{n}%
}^{\left( 0\right) }\mathbf{\tilde{G}}_{\bar{n}}\left( \mathbf{r},\mathbf{k}%
\right) .  \label{u0tilda}
\end{equation}

\subsubsection{Structured asymptotic expansions}

We are interested in asymptotic approximations with respect to $\varrho $, $%
\beta $ of the coefficients $\tilde{u}_{\bar{n}}^{\left( j\right) }\left( 
\mathbf{k},\tau \right) =\tilde{u}_{\bar{n}}^{\left( j\right) }\left( 
\mathbf{k},\tau ;\varrho ,\beta \right) $ of the expansion (\ref{UU1}) given
by (\ref{UlinBloch}). We primarily study the modal amplitudes $\tilde{u}_{%
\bar{n}}^{\left( 1\right) }\left( \mathbf{k},\tau \right) =\tilde{u}_{\bar{n}%
}^{\left( 1\right) }\left( \mathbf{k},\tau ;\varrho ,\beta \right) $ of the
first nonlinear response $\mathbf{U}^{\left( 1\right) }$ for small $\varrho $
and $\beta $. Our analysis shows that the dependence on $\varrho $ and $%
\beta $ is more complicated than on $\alpha $. There are three different
types of asymptotic expansions which are envolved in the description of the
dependence of $\tilde{u}_{\bar{n}}^{\left( 1\right) }\left( \mathbf{k},\tau
;\varrho ,\beta \right) $ on $\varrho $, $\beta $ : the first type involves
powers of $\varrho $, the second one involves powers of $\beta $, and the
third type involves either powers of $\theta =\frac{\varrho }{\beta ^{2}}$
if $\theta =\frac{\varrho }{\beta ^{2}}\ll 1$, or powers of $\frac{\beta
^{\nu +1}}{\varrho }$ if $\theta ^{-1}=\frac{\beta ^{2}}{\varrho }\leq 1$,
where $\nu =2,3,4,\ldots $ \ is the order of the NLS or ENLS. The expansion
of the modal amplitudes of the NLM in the dispersive case $\theta =\frac{%
\varrho }{\beta ^{2}}\ll 1$ has the form of a \emph{structured power
asymptotic series} 
\begin{gather}
\tilde{u}_{\zeta ,n_{0}}^{\left( 1\right) }\left( \mathbf{k},\tau ;\varrho
,\beta \right) =\frac{\theta ^{d}}{\varrho }\sum_{l_{1}=0}^{N_{1}}%
\sum_{l_{2}=0}^{N_{2}}\sum_{l_{3}=0}^{N_{3}}C_{l_{1},l_{2},l_{3}}^{\text{NLM}%
}\left( \mathbf{k},\tau ;\zeta ,n_{0}\right) \varrho ^{l_{1}}\beta
^{l_{2}}\theta ^{l_{3}}  \label{asser} \\
+\frac{\theta ^{d}}{\varrho }\left[ O\left( \beta \varrho ^{N_{1}}\right)
+O\left( \beta ^{N_{2}+1}\right) +O\left( \theta ^{N_{3}+1}\right) \right]
,\ \theta =\frac{\varrho }{\beta ^{2}}.  \notag
\end{gather}%
\emph{We would like to emphasize that the form (\ref{asser}) for }$\tilde{u}%
_{\zeta ,n_{0}}^{\left( 1\right) }\left( \mathbf{k},\tau ;\varrho ,\beta
\right) $\emph{\ is not imposed as an ansatz, but it\ follows from the
analysis of the interaction integrals, and it describes properties of exact
solutions to the NLM.} Powers $\varrho ^{l_{1}}$ stem from the asymptotic
expansions for almost monochromatic waves, for details see Section 6 and
Subsection 8.3. Some expressions in the interaction integrals admit regular
Taylor expansions, see Subsection 4.1.1, that lead to the powers $\beta
^{l_{2}}$. More complicated terms with $\theta ^{l_{3}+d}=\left( \frac{%
\varrho }{\beta ^{2}}\right) ^{l_{3}+d}$come from a rapidly oscillating
phase function, and we use the Stationary Phase Method to take into account
its effect. Note that the expansion with respect to $\theta =\frac{\varrho }{%
\beta ^{2}}$ is mostly determined by (i) the phase function of the
interaction integral and (ii) by the rectifying change of variables which
exactly equates the phase functions for the NLS and the NLM. The complexity
of the expression (\ref{asser}) reflects the complexity of exact solutions
to the NLM. \emph{The type of dependence in (\ref{asser}) shows that formal
asymptotic expansions of solutions with respect to powers of independent
parameters }$\varrho $\emph{\ and }$\beta $\emph{\ are not very useful since
such expansions would involve negative powers of the small parameter }$\beta 
$\emph{.} In addition, this form of dependence implies that if one
prescribes the power relations of the form (\ref{rhoal}), (\ref{kap1}) or (%
\ref{kap11}) the resulting expansions in powers of one remaining parameter
would strongly depend on the choice of $\varkappa _{1}$ and $\varkappa _{0}$%
, whereas the higher order approximating ENLS, which we introduce, are \emph{%
universal}, they do not depend on $\varkappa _{1}$ and $\varkappa _{0}$.
Introduction of a specific power dependence $\varrho =\beta ^{\varkappa
_{1}} $ selects from the universal ENLS, which is described in Subsection
1.3, a specific reduced ENLS depending on the choice of $\varkappa _{1}$,
see Subsection 1.3.7 for examples. Note that the condition $\varrho \sim
\beta ^{\varkappa _{1}}$,$\ \varkappa _{1}>2$, in (\ref{kap1}) implies that
the ratio $\frac{\varrho }{\beta ^{2}}$ is small, but it may tend to zero
very slowly when $\beta \rightarrow 0$. To make $\left( \frac{\varrho }{%
\beta ^{2}}\right) ^{N_{3}+1}$ smaller than $\beta ^{N_{2}+1}$ or $\varrho
^{N_{1}+1}$ in (\ref{asser}) high values of $N_{3}$ may be required. Note
also that since $\frac{\varrho }{\beta ^{2}}\ll 1$ taking minimal $N_{1}$
which\ satisfies $N_{1}\geq N_{2}/2-1/2$ provides the same accuracy as
taking larger $N_{1}$.\ In this paper we consider in all details the case $%
N_{2}=\sigma =\nu -2\leq 2$, and $N_{1}\leq 1$, when the value of $N_{3}$ \
may be very large. The case of larger $N_{1}$, $N_{2}$ is similar.

Our strategy for approximating $\tilde{u}_{\zeta ,n_{0}}^{\left( 1\right) }$
by a solution of the NLS \ can be described \ as follows. We consider
solutions $V_{\zeta }\left( \mathbf{r},t\right) $ of the NLS, take their
Fourier transform $\hat{V}_{\zeta }\left( \mathbf{\xi },t\right) $ and
similarly to (\ref{UFB1}) \ introduce slowly varying coefficients $\hat{v}%
_{\zeta }\left( \mathbf{\xi },\tau \right) $%
\begin{equation}
\hat{V}_{\zeta }\left( \mathbf{\xi },t\right) =\hat{v}_{\zeta }\left( 
\mathbf{\xi },\tau \right) e^{-\mathrm{i}\zeta \gamma _{\left( \nu \right)
}\left( \zeta \mathbf{\xi }\right) t},\ \tau =\varrho t.\ 
\end{equation}%
Similarly to (\ref{uap1}) we introduce asymptotic expansions in $\alpha $%
\begin{equation}
\hat{v}_{\zeta }\left( \mathbf{\xi },\tau \right) =\hat{v}_{\zeta }^{\left(
0\right) }\left( \mathbf{\xi },\tau \right) +\alpha \hat{v}_{\zeta }^{\left(
1\right) }\left( \mathbf{\xi },\tau \right) +\ldots
\end{equation}%
(see Subsection 5.2 for details). We expand the Fourier transform of a
solution to the NLS similarly to (\ref{asser}): 
\begin{gather}
\hat{v}_{\zeta }^{\left( 1\right) }\left( \mathbf{\xi },\tau ;\varrho ,\beta
\right) =\frac{\theta ^{d}}{\varrho }\sum_{l_{1}=0}^{N_{1}}%
\sum_{l_{2}=0}^{N_{2}}\sum_{l_{3}=0}^{N_{3}}C_{l_{1},l_{2},l_{3}}^{\text{NLS}%
}\left( \mathbf{\xi },\tau ;\zeta \right) \varrho ^{l_{1}}\beta
^{l_{2}}\theta ^{l_{3}}  \label{asserNLS} \\
+\frac{\theta ^{d}}{\varrho }\left[ O\left( \varrho ^{N_{1}+1}\right)
+O\left( \beta ^{N_{2}+1}\right) +O\left( \left( \frac{\varrho }{\beta ^{2}}%
\right) ^{N_{3}+1}\right) \right] ,\ \theta =\frac{\varrho }{\beta ^{2}}. 
\notag
\end{gather}%
The coefficients $C_{l_{1},l_{2},l_{3}}^{\text{NLS}}$ of the expansion
depend on the choice of parameters $p_{\pm }^{\left[ \sigma \right] }$ and $%
\delta _{1,\pm }$ in the NLS (\ref{GNLS1+}), (\ref{GNLS1-}). These
parameters are chosen so that the following conditions are satisfied: 
\begin{equation}
C_{l_{1},l_{2},l_{3}}^{\text{NLS}}\left( \mathbf{\xi },\tau ;\zeta \right)
=C_{l_{1},l_{2},l_{3}}^{\text{NLM}}\left( \mathbf{k},\tau ;\zeta
,n_{0}\right) ,\quad l_{1}\leq N_{1},\ l_{2}\leq N_{2},\ l_{3}\leq N_{3}.
\label{match}
\end{equation}%
Recall that $N_{2}=\sigma \leq 2$ and $N_{1}\leq 1$ are relatively small,
and the value of $N_{3}$ \ may be large. Note that the number of additional
coefficients which are involved in the NLS to provide higher accuracy of the
approximation of the NLM depends only on $N_{1},N_{2}$ and does not depend
on $N_{3}$. \emph{Remarkably, the actual equations for coefficients which
follow from (\ref{match}) do not depend on }$l_{3}$\emph{, and this is the
reason we can satisfy all these conditions using a small number of
parameters that determine coefficients of the NLS.} To satisfy the
conditions we choose in a proper way the excitation currents for the NLM
based on initial data for the NLS. We get such independence of the equations
on $l_{3}$ through the use of the rectifying change of variables which
establishes direct correspondence between the NLM and the NLS. This change
of variables depends on $N_{2}$. The details of the related analysis are
rather technical and are considered in the subsequent sections. \emph{We
remind again that the form and coefficients of (\ref{asser}) are the result
of explicitly defined transformations of the interaction integral and should
be considered as a result of the analysis rather than a starting point.}
Similarly, fulfillment of (\ref{match}) follows from our choice of
excitation currents, coefficients of the NLS and rectifying change of
variables based on the analysis of the interaction integrals.

In the weakly dispersive case $\theta ^{-1}=\frac{\beta ^{2}}{\varrho }\leq
1 $ we have similar expansions for $\nu =4$, $N_{2}=2$ 
\begin{gather}
\tilde{u}_{\zeta ,n_{0}}^{\left( 1\right) }\left( \mathbf{k},\tau ;\varrho
,\beta \right) =\frac{1}{\varrho }\sum_{l_{1}=0}^{N_{1}}%
\sum_{l_{2}=0}^{N_{2}}\sum_{l_{3}=0}^{N_{3}}c_{l_{1},l_{2},l_{3}}^{\text{NLM}%
}\left( \mathbf{k},\tau ;\zeta ,n_{0}\right) \varrho ^{l_{1}}\beta
^{l_{2}}\left( \frac{\beta ^{5}}{\varrho }\right) ^{l_{3}}  \label{asserw} \\
+\frac{1}{\varrho }\left[ O\left( \beta \varrho ^{N_{1}}\right) +O\left(
\beta ^{N_{2}+1}\right) +O\left( \left( \frac{\beta ^{5}}{\varrho }\right)
^{N_{3}+1}\right) \right] .  \notag
\end{gather}%
and respective expansion for $\hat{v}_{\zeta }^{\left( 1\right) }$ 
\begin{gather}
\hat{v}_{\zeta }^{\left( 1\right) }\left( \mathbf{\xi },\tau ;\varrho ,\beta
\right) =\varrho
^{-1}\sum_{l_{1}=0}^{N_{1}}\sum_{l_{2}=0}^{N_{2}}%
\sum_{l_{3}=0}^{N_{3}}c_{l_{1},l_{2},l_{3}}^{\text{NLS}}\left( \mathbf{\xi }%
,\tau ;\zeta \right) \varrho ^{l_{1}}\beta ^{l_{2}}\left( \frac{\beta ^{5}}{%
\varrho }\right) ^{l_{3}} \\
+\frac{1}{\varrho }\left[ O\left( \beta \varrho ^{N_{1}}\right) +O\left(
\beta ^{N_{2}+1}\right) +O\left( \left( \frac{\beta ^{5}}{\varrho }\right)
^{N_{3}+1}\right) \right] .  \notag
\end{gather}%
The powers $\left( \frac{\beta ^{5}}{\varrho }\right) ^{l_{3}}$ come from
the expansion in (\ref{expbet}). Note that in this case $N_{3}$ does not
have to be large to acquire desired accuracy, and since (\ref{nonM1}) holds
one can take $N_{3}=0$ when $N_{2}=2$.

\subsubsection{First nonlinear response and modal susceptibility}

For the current $\mathbf{J}^{\left( 1\right) }$ of the form (\ref{jjt2}) and 
$\varrho \rightarrow 0$ the first nonlinear response $\mathbf{U}^{\left(
1\right) }$ determined by (\ref{V01}) can be represented as the following
series based on the time-harmonic expansion (see Sections 6.2 and 8.3 for
details)%
\begin{equation}
\mathbf{U}^{\left( 1\right) }=\mathbf{U}^{\left( 1,0\right) }+\varrho 
\mathbf{U}^{\left( 1,1\right) }+\varrho ^{2}\mathbf{U}^{\left( 1,2\right)
}+\ldots ,  \label{U1rho}
\end{equation}%
\begin{equation}
\tilde{u}_{\bar{n}}^{\left( 1\right) }=\tilde{u}_{\bar{n}}^{\left(
1,0\right) }+\varrho \tilde{u}_{\bar{n}}^{\left( 1,1\right) }+\varrho ^{2}%
\tilde{u}_{\bar{n}}^{\left( 1,2\right) }+\ldots .  \label{u1rho}
\end{equation}%
\emph{Notice that (\ref{U1rho}) and (\ref{u1rho}) are not the Taylor series}%
, and the quantities $\mathbf{U}^{\left( 1,s\right) }$ and $\tilde{u}_{\bar{n%
}}^{\left( 1,s\right) }$, $s=0,1,\ldots $ in (\ref{U1rho}) and (\ref{u1rho})
are represented as oscillatory integrals which depend on $\varrho $
themselves. As we will see in Section 4.1.3 and 4.1.2 respectively if $%
\varrho \rightarrow 0$ we have $\tilde{u}_{\bar{n}}^{\left( 1,s\right) }\sim
\varrho ^{d-1}$ for the dispersive case and $\tilde{u}_{\bar{n}}^{\left(
1,s\right) }\sim \varrho ^{-1}$ for the weakly dispersive case. Since we are
interested in the FNLR $\mathbf{U}^{\left( 1\right) }$ for small $\varrho $
we look first at the dominant term $\mathbf{U}^{\left( 1,0\right) }$ in the
series (\ref{U1rho}). We refer to $\mathbf{U}^{\left( 1,0\right) }$ as the 
\emph{time-harmonic FNLR. }Using the formula (\ref{V01}) and the
time-harmonic approximation of $\mathcal{F}_{\text{NL}}^{\left( 0\right) }$
together with the definitions of $\mathbf{\tilde{G}}_{\bar{n}}\left( \mathbf{%
r},\mathbf{k}\right) $, the susceptibility $\mathbf{\chi }_{D}^{\left(
3\right) }$, the inner product $\left( \cdot ,\cdot \right) _{\mathcal{H}}$
by respectively (\ref{nz2}), (\ref{cd4ab}) and (\ref{sigma}) we get the
following integral representation for $\tilde{u}_{\bar{n}}^{\left(
1,0\right) }$ (see \cite{BF1} and Subsection 6.2 for details) 
\begin{gather}
\tilde{u}_{\bar{n}}^{\left( 1,0\right) }\left( \mathbf{k},\tau \right) =%
\frac{1}{\varrho }\sum_{\bar{n}^{\prime },\bar{n}^{\prime \prime },\bar{n}%
^{\prime \prime \prime }}\int_{0}^{\tau }\int_{\substack{ \lbrack -\pi ,\pi
]^{2d}  \\ \mathbf{\mathbf{k}^{\prime }}+\mathbf{k}^{\prime \prime }+\mathbf{%
k}^{\prime \prime \prime }=\mathbf{k}}}\exp \left\{ \mathrm{i}\phi _{\vec{n}%
}\left( \vec{k}\right) \frac{\tau _{1}}{\varrho }\right\}  \label{Vn} \\
\breve{Q}_{\vec{n}}\left( \vec{k}\right) \tilde{u}_{\bar{n}^{\prime
}}^{\left( 0\right) }\left( \mathbf{k}^{\prime },\tau _{1}\right) \tilde{u}_{%
\bar{n}^{\prime \prime }}^{\left( 0\right) }\left( \mathbf{k}^{\prime \prime
},\tau _{1}\right) \tilde{u}_{\bar{n}^{\prime \prime \prime }}^{\left(
0\right) }\left( \mathbf{k}^{\prime \prime \prime },\tau _{1}\right) \,%
\mathrm{d}\mathbf{k}^{\prime }\mathrm{d}\mathbf{k}^{\prime \prime }\mathrm{d}%
\tau _{1}-\tilde{u}_{\bar{n}}^{\left( 1\right) }\left( \mathbf{J}_{1};%
\mathbf{k},\tau \right)  \notag
\end{gather}%
where 
\begin{eqnarray}
\vec{k} &=&\left( \mathbf{k},\mathbf{k}^{\prime },\mathbf{k}^{\prime \prime
},\mathbf{k}^{\prime \prime \prime }\right) ,\ \vec{\zeta}=\left( \zeta
,\zeta ^{\prime },\zeta ^{\prime \prime },\zeta ^{\prime \prime \prime
}\right) ,  \label{karrow} \\
\vec{n} &=&\left( \overline{n},\overline{n}^{\prime },\overline{n}^{\prime
\prime },\bar{n}^{\prime \prime \prime }\right) =\left( \left( \zeta
,n\right) ,\left( \zeta ^{\prime },n^{\prime }\right) ,\left( \zeta ^{\prime
\prime },n^{\prime \prime }\right) ,\left( \zeta ^{\prime \prime \prime
},n^{\prime \prime \prime }\right) \right)  \notag
\end{eqnarray}%
\begin{equation}
\phi _{\vec{n}}\left( \vec{k}\right) =\zeta \omega _{n}\left( \mathbf{k}%
\right) -\zeta ^{\prime }\omega _{n^{\prime }}\left( \mathbf{k}^{\prime
}\right) -\zeta ^{\prime \prime }\omega _{n^{\prime \prime }}\left( \mathbf{k%
}^{\prime \prime }\right) -\zeta ^{\prime \prime \prime }\omega _{n^{\prime
\prime \prime }}\left( \mathbf{k}^{\prime \prime \prime }\right) ,
\label{phink}
\end{equation}%
\begin{gather}
\breve{Q}_{\vec{n}}\left( \vec{k}\right) =\frac{1}{(2\pi )^{2d}}\left( \left[
\begin{array}{c}
\mathbf{0} \\ 
Q_{\mathbf{\chi }_{D}^{\left( 3\right) }}%
\end{array}%
\right] ,\mathbf{\tilde{G}}_{\bar{n}}\left( \mathbf{r},\mathbf{k}\right)
\right) _{\mathcal{H}},  \label{Qn} \\
Q_{\mathbf{\chi }_{D}^{\left( 3\right) }}=  \notag \\
\nabla \times \mathbf{\chi }_{D}^{\left( 3\right) }\left( \omega _{\bar{n}%
^{\prime }}\left( \mathbf{k}^{\prime }\right) ,\omega _{\bar{n}^{\prime
\prime }}\left( \mathbf{k}^{\prime \prime }\right) ,\omega _{\bar{n}^{\prime
\prime \prime }}\left( \mathbf{k}^{\prime \prime \prime }\right) \right)
\vdots \mathbf{\tilde{G}}_{D,\bar{n}^{\prime }}\left( \mathbf{r},\mathbf{k}%
^{\prime }\right) \mathbf{\tilde{G}}_{D,\bar{n}^{\prime \prime }}\left( 
\mathbf{r},\mathbf{k}^{\prime \prime }\right) \mathbf{\tilde{G}}_{D,\bar{n}%
^{\prime \prime \prime }}\left( \mathbf{r},\mathbf{k}^{\prime \prime \prime
}\right) ,  \notag
\end{gather}%
$\tilde{u}_{\bar{n}}^{\left( 0\right) }$ are defined in (\ref{V0}) and 
\begin{eqnarray}
\tilde{u}_{\zeta ,n_{0}}^{\left( 1\right) }\left( \mathbf{J}_{1};\mathbf{k}%
,\tau \right) &=&\frac{1}{\varrho }\int_{0}^{\tau }\tilde{j}_{\zeta
,n_{0}}^{\left( 1\right) }\left( \mathbf{k},\tau _{1}\right) \,\mathrm{d}%
\tau _{1},  \label{un1J} \\
\ \ \tilde{u}_{\bar{n}}^{\left( 1\right) }\left( \mathbf{J}_{1};\mathbf{k}%
,\tau \right) &=&0\text{ for }n\neq n_{0}.
\end{eqnarray}%
The quantatity $\breve{Q}_{\vec{n}}\left( \vec{k}\right) $ given by the
integral (\ref{Qn}) plays an important role in the approximation analysis
and we refer to it as the\emph{\ modal susceptibility}. An estimate for the
difference $\tilde{u}_{\bar{n}}^{\left( 1\right) }\left( \mathbf{k},\tau
\right) -\tilde{u}_{\bar{n}}^{\left( 1,0\right) }\left( \mathbf{k},\tau
\right) $ is given by (\ref{suserror}).

Note that though the formlula (\ref{Qn}) uses a specific form of $\mathcal{F}%
_{\text{NL}}$ in (\ref{FNL}), that particular form it is not essential for
our analysis. For example, if $\mathbf{D}$-component of $\mathcal{F}_{\text{%
NL}}$ would not be zero, or if $\mathbf{\chi }_{D}^{\left( 3\right) }$ acted
also on the $\mathbf{B}$-component of the vector $\mathbf{U}$, all steps and
conclusions of our analysis would remain the same.

We will also use the following notation which allows to rewrite (\ref{Vn})
in a shorter way: 
\begin{gather}
\tilde{F}_{\bar{n}}\left[ \left( \mathbf{u}^{\left( 0\right) }\right) ^{3}%
\right] \left( \mathbf{k},\tau \right) =  \label{FNR1} \\
\sum_{\bar{n}^{\prime },\bar{n}^{\prime \prime },\bar{n}^{\prime \prime
\prime }}\int_{0}^{\tau }\int_{\substack{ \lbrack -\pi ,\pi ]^{2d}  \\ 
\mathbf{\mathbf{k}^{\prime }}+\mathbf{k}^{\prime \prime }+\mathbf{k}^{\prime
\prime \prime }=\mathbf{k}}}\mathrm{e}^{\mathrm{i}\phi _{\vec{n}}\left( \vec{%
k}\right) \frac{\tau _{1}}{\varrho }}\frac{\left( \nabla \times I_{\mathbf{%
\chi ,}\vec{n}}\left( \mathbf{\tilde{u}}^{\left( 0\right) }\right) \,,%
\mathbf{\tilde{G}}_{\bar{n}}\left( \mathbf{r},\mathbf{k}\right) \right) _{%
\mathcal{H}}}{\varrho (2\pi )^{2d}}\mathrm{d}\mathbf{k}^{\prime }\mathrm{d}%
\mathbf{k}^{\prime \prime }\mathrm{d}\tau _{1},  \notag \\
I_{\mathbf{\chi ,}\vec{n}}\left( \mathbf{\tilde{u}}^{\left( 0\right)
}\right) =  \notag \\
\mathbf{\chi }_{D,B}^{\left( 3\right) }\left( \mathbf{r};\omega _{\bar{n}%
^{\prime }}\left( \mathbf{k}^{\prime }\right) ,\omega _{\bar{n}^{\prime
\prime }}\left( \mathbf{k}^{\prime \prime }\right) ,\omega _{\bar{n}^{\prime
\prime \prime }}\left( \mathbf{k}^{\prime \prime \prime }\right) \right)
\vdots \mathbf{\tilde{u}}_{\bar{n}^{\prime }}^{\left( 0\right) }\left( 
\mathbf{r},\mathbf{k}^{\prime },\tau _{1}\right) \mathbf{\tilde{u}}_{\bar{n}%
^{\prime \prime }}^{\left( 0\right) }\left( \mathbf{r},\mathbf{k}^{\prime
\prime },\tau _{1}\right) \mathbf{\tilde{u}}_{\bar{n}^{\prime \prime \prime
}}^{\left( 0\right) }\left( \mathbf{r},\mathbf{k}^{\prime \prime \prime
},\tau _{1}\right)  \notag
\end{gather}%
where $\mathbf{\tilde{u}}^{\left( 0\right) }$ is defined by (\ref{u0tilda}),%
\begin{equation}
\mathbf{\tilde{u}}_{\bar{n}}^{\left( 0\right) }\left( \mathbf{r},\mathbf{k}%
,\tau \right) =\tilde{u}_{\bar{n}}^{\left( 0\right) }\left( \mathbf{k},\tau
\right) \mathbf{\tilde{G}}_{\bar{n}}\left( \mathbf{r},\mathbf{k}\right) ,\ 
\mathbf{\tilde{u}}_{\bar{n}^{\left( i\right) }}=\left[ 
\begin{array}{c}
\mathbf{\tilde{u}}_{D,\bar{n}^{\left( i\right) }} \\ 
\mathbf{\tilde{u}}_{B,\bar{n}^{\left( i\right) }}%
\end{array}%
\right] ,
\end{equation}%
and%
\begin{equation}
\mathbf{\chi }_{D,B}^{\left( 3\right) }\left( \mathbf{r};\omega _{1},\omega
_{2},\omega _{3}\right) \vdots \mathbf{\tilde{u}}_{\bar{n}^{\prime }}\mathbf{%
\tilde{u}}_{\bar{n}^{\prime \prime }}\mathbf{\tilde{u}}_{\bar{n}^{\prime
\prime \prime }}=\left[ 
\begin{array}{c}
\mathbf{0} \\ 
\mathbf{\chi }_{D}^{\left( 3\right) }\left( \mathbf{r};\omega _{1},\omega
_{2},\omega _{3}\right) \vdots \mathbf{\tilde{u}}_{D,\bar{n}^{\prime }}%
\mathbf{\tilde{u}}_{D,\bar{n}^{\prime \prime }}\mathbf{\tilde{u}}_{D,\bar{n}%
^{\prime \prime \prime }}%
\end{array}%
\right]  \label{chiDB}
\end{equation}%
is a tensor obtained from $\mathbf{\chi }_{D}^{\left( 3\right) }$. This
tensor acts not in the 3-dimensional $\mathbf{D}$-space, but in the
6-dimensional $\left( \mathbf{D},\mathbf{B}\right) $-space. It acts on the $%
\mathbf{D}$-components of $\mathbf{\tilde{u}}_{\bar{n}^{\prime }}^{\left(
0\right) }$ taking values in the $\mathbf{B}$-component as in (\ref{Qn}).
Using \ (\ref{FNR1}) we can rewrite (\ref{Vn}) in the following form: 
\begin{equation}
\tilde{u}_{\bar{n}}^{\left( 1,0\right) }\left( \mathbf{k},\tau \right) =%
\tilde{F}_{\bar{n}}\left[ \left( \mathbf{u}^{\left( 0\right) }\right) ^{3}%
\right] \left( \mathbf{k},\tau \right) -\tilde{u}_{\bar{n}}^{\left( 1\right)
}\left( \mathbf{J}_{1};\mathbf{k},\tau \right) .  \label{uQ}
\end{equation}%
Below we analyze (\ref{Vn}) using the approach of \cite{BF1}-\cite{BF4} in
the case when the frequency of the excitation current (\ref{jjt2}) is in a
fixed band $n_{0}$, the quasimomentum $\mathbf{k}$ is in a vicinity of a
fixed quasimomentum $\mathbf{k}_{\ast }$, and the excitation current is
almost time-harmonic (such currents are described in detail in the
folllowing subsection). The term $\ \tilde{u}_{\bar{n}}^{\left( 1\right)
}\left( \mathbf{J}_{1};\mathbf{k},\tau \right) $ in (\ref{Vn}), as one can
see from (\ref{un1J}), is due to the excitation current $\mathbf{J}_{1}$
with amplitudes $\tilde{j}_{\bar{n}}^{\left( 1\right) }$. This current is
introduced to transform the initial data for the NLS into a proper
excitation current with maximal accuracy (see Subsection 5.2 for details).
The modal components of $j_{\bar{n}}^{\left( 1\right) }$ are defined by the
following formula%
\begin{equation}
\tilde{j}_{\zeta ,n_{0}}^{\left( 1\right) }\left( \zeta \mathbf{k}_{\ast
}+Y_{\zeta }\left( \beta \mathbf{q}\right) ,\tau \right) =\exp \left\{ 
\mathrm{i}\omega _{\zeta ,n_{0}}\left( \zeta \mathbf{k}_{\ast }+\beta 
\mathbf{q}\right) \frac{\tau }{\varrho }\right\} \hat{f}_{\zeta }^{\left(
1\right) }\left( \beta \mathbf{q},t\right) ,\ t=\frac{\tau }{\varrho },\
\zeta =\pm =\pm 1,  \label{jf}
\end{equation}%
with $\hat{f}_{+}^{\left( 1\right) }$ and $\ \hat{f}_{-}^{\left( 1\right) }$%
, in turn, being defined respectively by (\ref{f1}) and (\ref{f1m}).

\subsection{Almost time-harmonic excitations}

The concept of an almost time-harmonic excitation is central to the theory
of nonlinear mode interactions. An abstract form for an almost time-harmonic
function $a\left( t\right) $ is given by the formula (\ref{aa}) and the
basic properties of almost time-harmonic functions are considered in Section
8.3. In this section we give a precise definition of an almost time-harmonic
excitation current. Solutions to the NLM generated by almost time-harmonic
excitation currents are well approximated by solutions to properly
constructed NLS. The NLS, such as (\ref{Si}), (\ref{Si1}), are defined as
differential equations with the initial data $h_{\pm }\left( \mathbf{r}%
\right) $, whereas the NLM, with the nonlinear polarization (\ref{Pser}), (%
\ref{mx7}) defined by causal integrals, naturally involves the excitation
currents $\mathbf{J}\left( \mathbf{r},t\right) $ instead of the initial
data. To compare solutions to the NLM and NLS we have to resolve this
difference in settings. The difference is resolved by constructing a proper
form for the current $\mathbf{J}$ based on the initial data $h_{\pm }$ for
an NLS such as (\ref{Si}), (\ref{Si1}). It turns out that under the
assumption (\ref{eer2}) such a current $\mathbf{J}$ can be constructed as an
almost time-harmonic function as follows. The first step in setting up the
current $\mathbf{J}$ as defined by (\ref{UU1}) is to assume that its modal
composition (\ref{jjt2}), (\ref{jstar}) involves\ only a single spectral
band with the index $n_{0}$, i.e.%
\begin{gather}
\mathbf{\tilde{J}}_{n_{0}}^{\left( j\right) }\left( \mathbf{r},\mathbf{k}%
,t\right) =\tilde{j}_{+,n_{0}}^{\left( j\right) }\left( \mathbf{k},\tau
\right) \mathbf{\tilde{G}}_{+,n_{0}}\left( \mathbf{r},\mathbf{k}\right) 
\mathrm{e}^{-\mathrm{i}\omega _{n_{0}}\left( \mathbf{k}\right) t}+\tilde{j}%
_{-,n_{0}}^{\left( j\right) }\left( \mathbf{k},\tau \right) \mathbf{\tilde{G}%
}_{-,n_{0}}\left( \mathbf{r},\mathbf{k}\right) \mathrm{e}^{\mathrm{i}\omega
_{n_{0}}\left( \mathbf{k}\right) t},  \label{Jn} \\
\tau =\varrho t,\ \mathbf{\tilde{J}}_{n}^{\left( j\right) }\left( \mathbf{r},%
\mathbf{k},t\right) =0,\ n\neq n_{0},\ j=0,1;\;\mathbf{\tilde{J}}%
_{n}^{\left( j\right) }\left( \mathbf{r},\mathbf{k},t\right) =0\text{ for
all }n\text{ \ if }j>2.  \notag
\end{gather}%
The second step in the construction of the current $\mathbf{J}$ is to pick a
single quasimomentum $\mathbf{k}_{\ast }$ and to compose $\mathbf{J}$ of
only the modes with quasimomenta $\mathbf{k}$ in a $\beta $-vicinity of $\pm 
\mathbf{k}_{\ast }$. We can do that by picking a smooth function $\psi
_{0}\left( \tau \right) $ of the slow time $\tau $ and a smooth cutoff
function $\Psi _{0}\left( \mathbf{\eta }\right) $, $\mathbf{\eta }\in 
\mathbf{R}^{d}$, satisfying the following relations 
\begin{equation}
0\leq \psi _{0}\left( \tau \right) \leq 1,\ \psi _{0}\left( \tau \right)
=0,\ t\leq 0\text{ and }t\geq \tau _{0}>0,\ \dint\limits_{-\infty }^{\infty
}\psi _{0}\left( \tau \right) =1,  \label{psi0}
\end{equation}%
\begin{eqnarray}
0 &\leq &\Psi _{0}\left( \mathbf{\eta }\right) \leq 1,\ \Psi _{0}\left( -%
\mathbf{\eta }\right) =\Psi _{0}\left( \mathbf{\eta }\right) ,\   \label{j0}
\\
\Psi _{0}\left( \mathbf{\eta }\right) &=&1\text{ for }\left\vert \mathbf{%
\eta }\right\vert \leq \pi _{0}/2,\ \Psi _{0}\left( \mathbf{\eta }\right) =0%
\text{ for }\left\vert \mathbf{\eta }\right\vert \geq \pi _{0},  \notag
\end{eqnarray}%
where $\pi _{0}$ is a suffiently small constant which depends on $n_{0}$ and 
$\mathbf{k}_{\ast }$ and satisfies the inequalities $0<\pi _{0}<\pi /2$. The
function $\Psi _{0}\left( \mathbf{\eta }\right) $ is introduced to allow
useful local changes of variable $\mathbf{k}=\mathbf{k}_{\ast }+\mathbf{\eta 
}$\ in the $\pi _{0}$-vicinity of $\mathbf{k}_{\ast }$.

Suppose that we are given two scalar functions $h_{\pm }\left( \mathbf{r}%
\right) $ satisfying the relation 
\begin{equation}
h_{-}\left( \mathbf{r}\right) =h_{+}^{\ast }\left( \mathbf{r}\right) ,
\label{hhr}
\end{equation}%
and assume that these $h_{\pm }\left( \mathbf{r}\right) $ are the initial
data for NLS such as (\ref{Si}), (\ref{Si1}). Now we define the current
amplitudes $\tilde{j}_{\zeta ,n_{0}}^{\left( 0\right) }\left( \mathbf{k}%
,\tau \right) $ in (\ref{Jn}) by 
\begin{equation}
\tilde{j}_{\zeta ,n_{0}}^{\left( 0\right) }\left( \mathbf{k},\tau \right)
=-\varrho \psi _{0}\left( \tau \right) \Psi _{0}\left( \mathbf{k-}\zeta 
\mathbf{k}_{\ast }\right) \beta ^{-d}\mathring{h}_{\zeta }\left( \frac{%
\mathbf{k-}\zeta \mathbf{k}_{\ast }}{\beta }\right) ,\ \tau =\varrho t.
\label{jnn1}
\end{equation}%
We call an excitation current defined by (\ref{Jn})-(\ref{jnn1}) \emph{%
uni-directional} since the group velocities $\nabla \left( -\omega
_{n_{0}}\left( \mathbf{k}_{\ast }\right) \right) $ and $\nabla \left( \omega
_{n_{0}}\left( -\mathbf{k}_{\ast }\right) \right) $ corresponding to both
terms in (\ref{Jn}) coincide thanks to (\ref{invsym2}). The function $%
\mathring{h}_{\zeta }$ is related to the Fourier transform $\hat{h}_{\zeta }$
of the initial data $h_{\zeta }$ by the formula 
\begin{equation}
\mathring{h}_{\zeta }\left( \mathbf{s}\right) =\left\{ 
\begin{tabular}{lll}
$\hat{h}_{\zeta }\left( \frac{Y_{\zeta }^{-1}\left( \beta \mathbf{s}\right) 
}{\beta }\right) $ & for & $\left\vert \mathbf{s}\right\vert \leq \frac{3}{2}%
\pi _{0}$ \\ 
$0$ & for & $\left\vert \mathbf{s}\right\vert >\frac{3}{2}\pi _{0}$%
\end{tabular}%
\right. ,\ \hat{h}_{\zeta }\left( \mathbf{s}\right) =\frac{1}{\left( 2\pi
\right) ^{d}}\dint \mathrm{e}^{-\mathrm{i}\mathbf{s}\cdot \mathbf{r}%
}h_{\zeta }\left( \mathbf{r}\right) \,\mathrm{d}\mathbf{r},  \label{jnn2}
\end{equation}%
where $Y_{\zeta }$ is a \emph{rectifying change of variables} (See Section
2.2, Section 8.4 and (\ref{Y}), (\ref{redY}) for details.)

\textbf{Remark.} For small $\mathbf{\eta }$ the expression $Y_{\zeta
}^{-1}\left( \mathbf{\eta }\right) $ is very close to $\mathbf{\eta }$. In
the case of weak dispersion condition (\ref{nonM1}) and if $\alpha =O\left(
\varrho \right) $ one may assume for simplicity that $Y_{\zeta }^{-1}$ is
the identical change of variables and set in (\ref{jnn2}), (\ref{UNLS}), (%
\ref{UNLS1}) 
\begin{equation}
Y_{\zeta }^{-1}\left( \mathbf{\xi }\right) =\mathbf{\xi }\text{ \ and \ }%
\frac{Y_{\zeta }^{-1}\left( \beta \mathbf{s}\right) }{\beta }=\mathbf{s}.
\label{Yid}
\end{equation}%
A verification of all steps of estimates for the weakly dispersive case
shows that the estimates of the NLM-NLS\ approximation error still hold with
this simplification applied. Though to make all the results valid through
the whole range of parameters including the strongly dispersive case one has
to use the rectifying change of variables as it is defined in Section 2.2
and Section 8.4. As we already mentioned, the change of variables $Y$ does
not effect the NLS and their solutions, but makes the correspondence between
solutions of the NLS and the NLM more precise.$\blacklozenge $

Notice that (\ref{jnn2}) and (\ref{hhr}) imply that%
\begin{equation}
\mathring{h}_{-\zeta }\left( \mathbf{s}\right) =\mathring{h}_{\zeta }\left( -%
\mathbf{s}\right) ^{\ast },\ \zeta =\pm 1,  \label{jreal}
\end{equation}%
which together with (\ref{Jn}) \ and (\ref{jnn1}) yield 
\begin{equation}
\tilde{j}_{-\zeta ,n}^{\left( 0\right) }\left( -\mathbf{k},\tau \right) =%
\left[ \tilde{j}_{\zeta ,n}^{\left( 0\right) }\left( \mathbf{k},\tau \right) %
\right] ^{\ast }.  \label{jzeta}
\end{equation}%
In addition, (\ref{jnn2}), (\ref{hhr}) and (\ref{Gstar}) imply that $\mathbf{%
J}^{\left( 0\right) }\left( \mathbf{r},t\right) $ is real valued (notice
that (\ref{Gstar}) is satisfied due the condition (\ref{eer2})). Observe
also that it follows from (\ref{Jn}) and (\ref{jnn1}) that the current $%
\mathbf{J}^{\left( 0\right) }\left( \mathbf{r},t\right) $ is (i)
real-valued, (ii) almost time-harmonic and (iii) composed of modes from a
single band $n_{0}$ and quasaimomenta in a $\beta $-vicinity of $\pm \mathbf{%
k}_{\ast }$.

As to the corrective current $\mathbf{J}^{\left( 1\right) }$ in (\ref{Jn}),
its modal amplitudes are defined by the formula (\ref{jf})%
\begin{equation}
\tilde{j}_{\zeta ,n_{0}}^{\left( 1\right) }\left( \zeta \mathbf{k}_{\ast
}+Y_{\zeta }\left( \beta \mathbf{q}\right) ,\tau \right) =\hat{f}_{\zeta
}^{\left( 1\right) }\left( \beta \mathbf{q},\tau \right) ,  \label{jn1}
\end{equation}%
where $\hat{f}_{\zeta }^{\left( 1\right) }\left( \mathbf{q},\tau \right) $,
in turn, are defined by (\ref{f1F}), (\ref{f1m2}).


\begin{figure}[htbp]
\scalebox{0.5}{\includegraphics[viewport= 100 50 750
600,clip]{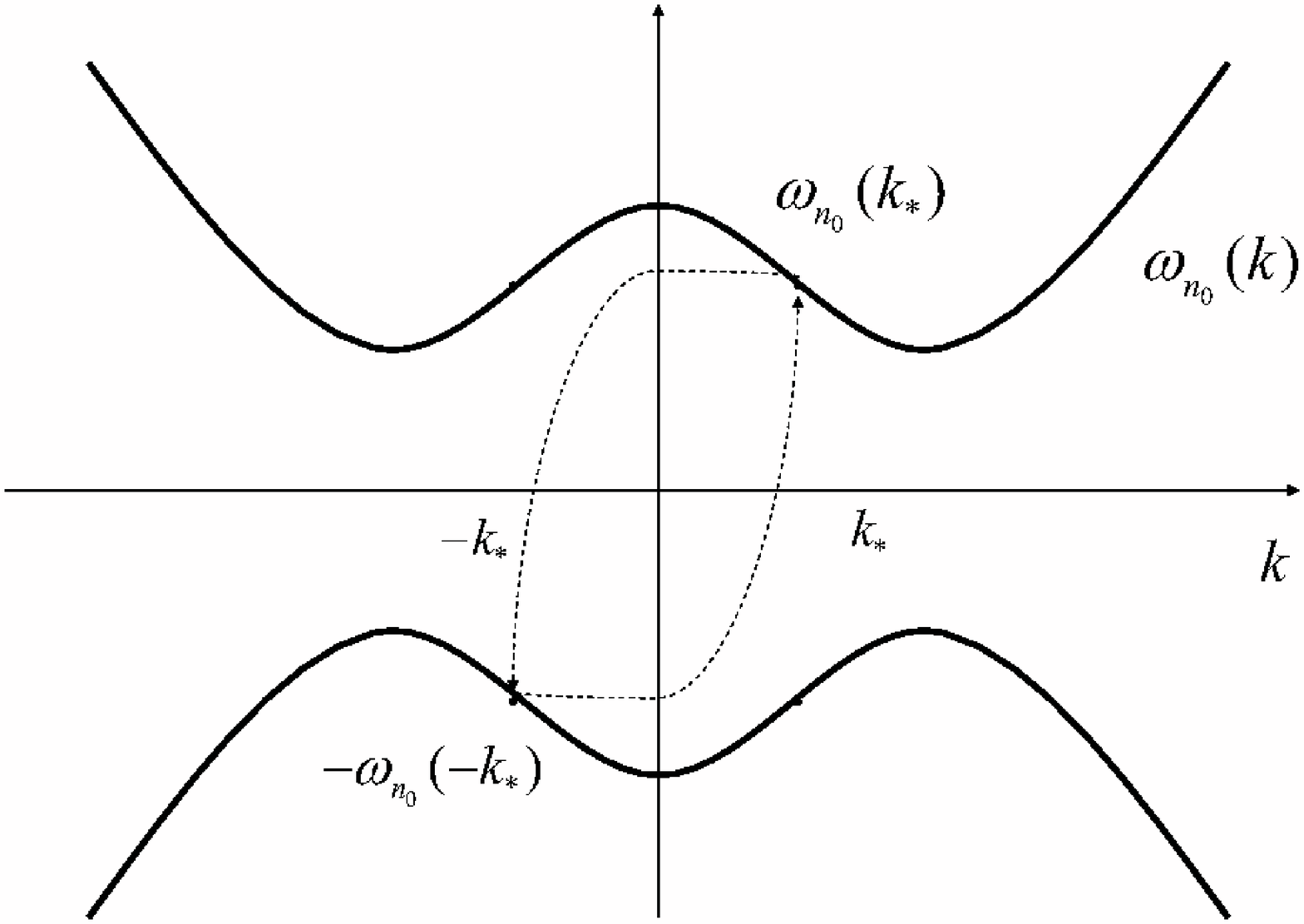}}
\caption{A real valued excitation current based on the band number $n_{0}$
and the quasimomentum $k_{\ast }$ directly excites a pair of modes with
quasimomenta $\protect\zeta k_{\ast }$ and frequencies $\protect\zeta 
\protect\omega _{n_{0}}\left( \protect\zeta k_{\ast }\right) $, $\protect%
\zeta =\pm $, forming a doublet. The modes in the doublet have a strong
nonlinear interaction.}
\label{figdoublet}
\end{figure}

\textbf{Remark.} In fact, our approach can be extended to excitation
currents involving several $\mathbf{k}_{\ast }$ and $n$. For such currents
the NLM generically can be reduced with high precision to several uncoupled
NLS, and we discuss this case in Subsection 1.2.$\blacklozenge $

\textbf{Remark.} Note that the magnitude of \ the inverse Fourier transform $%
h_{\zeta }\left( \beta \mathbf{r}\right) $ of $\beta ^{-d}\hat{h}_{\zeta
}\left( \mathbf{s/}\beta \right) $ in (\ref{jnn2}) does not depend on $\beta 
$, implying boundedness of the maximum of its the absolute value. To obtain
boundedness in a different norm one has to introduce an additional
dependence on $\beta $ into $h_{\zeta }\left( \beta \mathbf{r}\right) $. For
example, the integral of $\left\vert \beta ^{d/2}h_{\zeta }\left( \beta 
\mathbf{r}\right) \right\vert ^{2}$ is bounded uniformly in $\beta $.$%
\blacklozenge $

\subsection{Linear response and the rectifying variables for the NLM}

It follows from (\ref{UlinBloch}), (\ref{V0}) and (\ref{Jn}) 
\begin{eqnarray}
\mathbf{\tilde{U}}^{\left( 0\right) }\left( \mathbf{k},t\right) &=&\mathbf{%
\tilde{U}}_{+,n_{0}}^{\left( 0\right) }\left( \mathbf{k},t\right) +\mathbf{%
\tilde{U}}_{-,n_{0}}^{\left( 0\right) }\left( \mathbf{k},t\right) ,\quad
\label{U0n0} \\
\mathbf{\tilde{U}}_{\zeta ,n_{0}}^{\left( 0\right) }\left( \mathbf{k}%
,t\right) &=&\tilde{u}_{\zeta ,n_{0}}^{\left( 0\right) }\left( \mathbf{k}%
,\tau \right) \mathbf{\tilde{G}}_{\zeta ,n_{0}}\left( \mathbf{r},\mathbf{k}%
\right) \mathrm{e}^{-\mathrm{i}\zeta \omega _{n_{0}}\left( \mathbf{k}\right)
t},  \notag \\
\mathbf{\tilde{U}}_{\zeta ,n}^{\left( 0\right) }\left( \mathbf{k},t\right)
&=&0,\ n\neq n_{0},\ \tau =\varrho t.  \notag
\end{eqnarray}%
In addition to that, if we introduce%
\begin{equation}
\psi \left( \tau \right) =\int_{0}^{\tau }\psi _{0}\left( \tau \right) \,%
\mathrm{d}\tau _{1},  \label{psi1}
\end{equation}%
then using (\ref{V0}) and (\ref{jnn1}) we get 
\begin{equation}
\tilde{u}_{\zeta ,n_{0}}^{\left( 0\right) }\left( \mathbf{k},\tau \right)
=\psi \left( \tau \right) \Psi _{0}\left( \mathbf{k-}\zeta \mathbf{k}_{\ast
}\right) \beta ^{-d}\mathring{h}_{\zeta }\left( \frac{\mathbf{k-}\zeta 
\mathbf{k}_{\ast }}{\beta }\right) ,\ \tau =\varrho t,  \label{vz}
\end{equation}%
and, in particular, 
\begin{equation}
\tilde{u}_{\zeta ,n}^{\left( 0\right) }\left( \mathbf{k},\tau \right) =0%
\text{ if either }\left\vert \mathbf{k-}\zeta \mathbf{k}_{\ast }\right\vert
\geq \pi _{0},\ \text{or }\tau =\varrho t\leq 0,\ \text{or }n\neq n_{0}.
\label{un0}
\end{equation}%
By (\ref{UlinBloch}) and (\ref{vz}) 
\begin{equation}
\tilde{U}_{\bar{n}_{0}}^{\left( 0\right) }\left( \mathbf{k},t\right) =\tilde{%
u}_{\zeta ,n_{0}}^{\left( 0\right) }\left( \mathbf{k},\tau \right) \mathrm{e}%
^{-\mathrm{i}\zeta \omega _{n_{0}}\left( \mathbf{k}\right) t}=\psi \left(
\tau \right) \Psi _{0}\left( \mathbf{k-}\zeta \mathbf{k}_{\ast }\right)
\beta ^{-d}\mathring{h}_{\zeta }\left( \frac{\mathbf{k-}\zeta \mathbf{k}%
_{\ast }}{\beta }\right) \mathrm{e}^{-\mathrm{i}\zeta \omega _{n_{0}}\left( 
\mathbf{k}\right) t}.  \label{U0}
\end{equation}

\textbf{Remark.} It follows from (\ref{U0n0}), (\ref{vz}), (\ref{un0}) and (%
\ref{U0}) that for the chosen currents the linear medium response $\mathbf{U}%
^{\left( 0\right) }$ is composed of only the modes from a single band $n_{0}$
corresponding to the carrier wave frequency $\omega _{n_{0}}\left( \mathbf{k}%
\right) $ together with the opposite band corresponding to $-\omega
_{n_{0}}\left( \mathbf{k}\right) $ with wavenumberes in a $\beta $-vicinity
of respectively two wave vectors $\pm \mathbf{k}_{\ast }$. We call such an
excitaion in a vicinity of $\pm \omega _{n_{0}}\left( \pm \mathbf{k}\right) $
a \emph{doublet}, see Fig. \ref{figdoublet}. Note that the group velocities
of the two components of $\mathbf{U}^{\left( 0\right) }$ corresponding to
the points $\zeta \mathbf{k}_{\ast },$ $\zeta =\pm 1$\ and the bands $\zeta
\omega _{n_{0}}\left( \mathbf{k}\right) $ are equal to $\zeta \nabla \omega
_{n_{0}}\left( \zeta \mathbf{k}_{\ast }\right) $ for $\zeta =\pm 1$, and
these group velocities are the same in view of (\ref{invsym2}). Hence, a
doublet is a \emph{uni-directional}.excitation. Consequently, if $\beta $ is
small $\mathbf{U}^{\left( 0\right) }$ is a real-valued almost time-harmonic
wavepacket propagating in the direction of $\nabla \omega _{n_{0}}\left( 
\mathbf{k}_{\ast }\right) $ (see \cite{BF2} for a discussion of the group
velocity of wavepackets in photonic crystals).$\blacklozenge $

The formula (\ref{vz}) suggests to introduce a local variable $\mathbf{\eta }
$ and its scaled version $\mathbf{s}$ at $\zeta \mathbf{k}_{\ast }$ by the
following formulas 
\begin{equation}
\mathbf{\eta }=\beta \mathbf{s}=\mathbf{k}-\zeta \mathbf{k}_{\ast },
\label{eta}
\end{equation}%
allowing to recast (\ref{vz}) as 
\begin{equation}
\tilde{u}_{\zeta ,n_{0}}^{\left( 0\right) }\left( \mathbf{k},\tau \right) =%
\tilde{u}_{\zeta ,n_{0}}^{\left( 0\right) }\left( \zeta \mathbf{k}_{\ast
}+\beta \mathbf{s},\tau \right) =\psi \left( \tau \right) \Psi _{0}\left(
\beta \mathbf{s}\right) \beta ^{-d}\mathring{h}_{\zeta }\left( \mathbf{s}%
\right) .  \label{u0s0}
\end{equation}%
Let us approximate $\omega _{n_{0}}\left( \mathbf{\mathbf{k}_{\ast }}+%
\mathbf{\eta }\right) \ $in a vicinity of $\mathbf{\mathbf{\eta }}=\mathbf{%
\mathbf{0}}$ by its Taylor polynomial $\gamma _{\left( \nu \right) }\left( 
\mathbf{\eta }\right) $ of\ the degree $\nu $ (see Subsections 8.2, 8.3, 8.6
\ for notations) 
\begin{equation}
\gamma _{\left( \nu \right) }\left( \mathbf{\eta }\right) =\gamma _{\left(
\nu \right) }\left( \mathbf{\mathbf{k}_{\ast }};\mathbf{\eta }\right)
=\sum_{j=0}^{\nu }\frac{1}{j!}\omega _{n_{0}}^{\left( j\right) }\left( 
\mathbf{\mathbf{k}_{\ast }}\right) \left( \mathbf{\eta }^{j}\right) .
\label{Tayom}
\end{equation}%
In particular, for $\nu =2$ 
\begin{equation}
\gamma _{\left( 2\right) }\left( \mathbf{\eta }\right) =\omega
_{n_{0}}\left( \mathbf{\mathbf{k}_{\ast }}\right) +\omega _{n_{0}}^{\prime
}\left( \mathbf{\mathbf{k}_{\ast }}\right) \left( \mathbf{\mathbf{\eta }}%
\right) +\frac{1}{2}\omega _{n_{0}}^{\prime \prime }\left( \mathbf{\mathbf{k}%
_{\ast }}\right) \left( \mathbf{\eta }^{2}\right) \mathbf{,}  \label{defom2}
\end{equation}%
where $\omega _{n_{0}}^{\prime }\left( \mathbf{\mathbf{k}_{\ast }}\right) $
and $\omega _{n_{0}}^{\prime \prime }\left( \mathbf{\mathbf{k}_{\ast }}%
\right) $ are respectively linear and quadratic forms, i.e. a vector and a
matrix, i.e. 
\begin{equation}
\omega _{n_{0}}^{\prime }\left( \mathbf{\mathbf{k}_{\ast }}\right) \left( 
\mathbf{\mathbf{\eta }}\right) =\omega _{n_{0}}^{\prime }\left( \mathbf{%
\mathbf{k}_{\ast }}\right) \cdot \mathbf{\mathbf{\eta }},\mathbf{\mathbf{\ }}%
\omega _{n_{0}}^{\prime \prime }\left( \mathbf{\mathbf{k}_{\ast }}\right)
\left( \mathbf{\eta }^{2}\right) =\mathbf{\eta }\cdot \omega
_{n_{0}}^{\prime \prime }\left( \mathbf{\mathbf{k}_{\ast }}\right) \mathbf{%
\eta .}
\end{equation}%
Note that (\ref{invsym}), (\ref{invsym2}) and (\ref{Tayom}) imply that $%
\omega _{n_{0}}\left( \zeta \mathbf{\mathbf{k}_{\ast }}+\mathbf{\eta }%
\right) =\omega _{n_{0}}\left( \mathbf{\mathbf{k}_{\ast }}+\zeta \mathbf{%
\eta }\right) $ and that its Taylor polynomial coincides with $\gamma
_{\left( \nu \right) }\left( \zeta \mathbf{\eta }\right) $, namely%
\begin{equation}
\gamma _{\left( \nu \right) }\left( \zeta \mathbf{\mathbf{k}_{\ast }};%
\mathbf{\eta }\right) =\gamma _{\left( \nu \right) }\left( \mathbf{\mathbf{k}%
_{\ast }};\zeta \mathbf{\eta }\right) =\sum_{j=0}^{\nu }\frac{1}{j!}\omega
_{n_{0}}^{\left( j\right) }\left( \mathbf{\mathbf{k}_{\ast }}\right) \left(
\zeta \mathbf{\eta }^{j}\right) .  \label{Tayom2}
\end{equation}%
The following Taylor remainder estimation holds 
\begin{equation}
\left\vert \omega _{n_{0}}\left( \zeta \mathbf{\mathbf{k}_{\ast }}+\beta 
\mathbf{\mathbf{s}}\right) -\gamma _{\left( \nu \right) }\left( \zeta \beta 
\mathbf{\mathbf{s}}\right) \right\vert \leq C\beta ^{\nu +1}\left\vert 
\mathbf{\mathbf{s}}\right\vert ^{\nu +1},\ \mathbf{\mathbf{s}}\in \mathbf{%
\mathbf{R}}^{d}.  \label{estOm2}
\end{equation}

\subparagraph{The rectifying change of variables.}

The \emph{rectifying change of variables} $Y$ is a one-to-one mapping of a
small vicinity of a point $\mathbf{k}_{\ast }$ onto a similar vicinity, i.e.%
\begin{equation}
\mathbf{\xi }=Y^{-1}\left( \mathbf{\eta }\right) ,\ \mathbf{\eta }=Y\left( 
\mathbf{\xi }\right) \text{ if }\left\vert \mathbf{\eta }\right\vert \leq
2\pi _{0},\ \mathbf{\eta }=Y\left( \mathbf{\xi }\right) \text{ if }\
\left\vert \mathbf{\xi }\right\vert \leq 2\pi _{0},  \label{Y}
\end{equation}%
$\ $where $\pi _{0}$ is a small constant. \emph{It converts the dispersion
relation }$\omega _{n_{0}}\left( \mathbf{\mathbf{k}_{\ast }}+\mathbf{\eta }%
\right) $\emph{\ into its Taylor polynomial }$\gamma _{\left( \nu \right)
}\left( \mathbf{\mathbf{k}_{\ast }};\mathbf{\eta }\right) $\emph{\ at }$%
k_{\ast }$\emph{\ of the degree }$\nu $, i.e. 
\begin{equation}
\omega _{n_{0}}\left( \mathbf{k}_{\ast }+Y\left( \mathbf{\xi }\right)
\right) =\gamma _{\left( \nu \right) }\left( \mathbf{\mathbf{k}_{\ast }};%
\mathbf{\xi }\right) =\gamma _{\left( \nu \right) }\left( \mathbf{\xi }%
\right) .  \label{redY}
\end{equation}%
We will refer to coordinates $\mathbf{\xi }$ as to \emph{\ rectifying
coordinates}.

In the multidimensional case $d>1$ the rectifying change of variables is not
uniquely defined but it is not essential. We are interested primarily in $%
\nu =2$ or $\nu =3$ that are sufficient for approximations of nonlinear
interaction integrals up to the order $\beta ^{2}$. More accurate
approximation require larger values of $\nu $. The rectifying change of
varibale $Y\left( \mathbf{\xi }\right) $ satisfying (\ref{redY}) exists by
the Implicit function theorem and its power series expansions can be
explicitly found (see Subsection 8.4, in particular, the explicit formulas
for $\nu =1,2$). The rectifying change of varibales $Y\left( \mathbf{\xi }%
\right) $ is close to the identity, and if 
\begin{equation}
\omega _{n_{0}}^{\prime }\left( \mathbf{k}_{\ast }\right) \neq 0,
\label{om'ne0}
\end{equation}%
then 
\begin{equation}
Y\left( \mathbf{\xi }\right) =\mathbf{\xi }+O\left( \left\vert \mathbf{\xi }%
\right\vert ^{\nu +1}\right) ,\ Y^{-1}\left( \mathbf{\eta }\right) =\mathbf{%
\eta }+O\left( \left\vert \mathbf{\eta }\right\vert ^{\nu +1}\right) ,\ 
\text{if }\left\vert \mathbf{\eta }\right\vert \leq 2\pi _{0},\ \left\vert 
\mathbf{\xi }\right\vert \leq 2\pi _{0}.  \label{Ybet0}
\end{equation}%
If (\ref{om'ne0}) does not hold, but instead we have 
\begin{equation}
\omega _{n}^{\prime }\left( \mathbf{k}_{\ast }\right) =0,\ \det \omega _{%
\bar{n}}^{\prime \prime }\left( \mathbf{k}_{\ast }\right) \neq 0,
\label{om'0}
\end{equation}%
then $Y\left( \mathbf{\xi }\right) $ exists by the Morse lemma, \cite{Stein}%
, Section 8, Section 2.3.2, and 
\begin{equation}
Y\left( \mathbf{\xi }\right) =\mathbf{\xi }+O\left( \left\vert \mathbf{\xi }%
\right\vert ^{\nu }\right) ,\ Y^{-1}\left( \mathbf{\eta }\right) =\mathbf{%
\eta }+O\left( \left\vert \mathbf{\eta }\right\vert ^{\nu }\right) \text{ if 
}\left\vert \mathbf{\eta }\right\vert \leq 2\pi _{0},\ \left\vert \mathbf{%
\xi }\right\vert \leq 2\pi _{0}.  \label{Ybet00}
\end{equation}%
\emph{In this paper we assume that (\ref{om'ne0}) holds}. The case (\ref%
{om'0}) in many respects is similar, but requires somewhat different
treatment of higher order terms of asymptotic expansions. We will consider
this case in a separate paper.

By (\ref{invsym}), (\ref{invsym2}) and (\ref{Tayom}) we also have 
\begin{equation}
\omega _{n_{0}}\left( \zeta \mathbf{k}_{\ast }+\zeta Y\left( \zeta \mathbf{%
\xi }\right) \right) =\gamma _{\left( \nu \right) }\left( \mathbf{k}_{\ast
},\zeta \mathbf{\xi }\right) =\gamma _{\left( \nu \right) }\left( \zeta 
\mathbf{\xi }\right) ;\ Y\left( \zeta \mathbf{k}_{\ast },\mathbf{\xi }%
\right) =\zeta Y\left( \mathbf{k}_{\ast },\zeta \mathbf{\xi }\right) .
\label{redYz}
\end{equation}%
We will need the following \emph{scaled rectifying coordinates} $\mathbf{q}$
introduced in a vicinity of $\zeta \mathbf{k}_{\ast }$: 
\begin{gather}
\mathbf{q}=\frac{\mathbf{\xi }}{\beta }=\frac{\zeta }{\beta }Y^{-1}\left(
\zeta \mathbf{\eta }\right) =\frac{\zeta }{\beta }Y^{-1}\left( \zeta \beta 
\mathbf{s}\right) \ \text{or }\mathbf{s}=\frac{\zeta }{\beta }Y\left( \zeta
\beta \mathbf{q}\right) \text{ if }\left\vert \beta \mathbf{q}\right\vert
\leq 2\pi _{0},\ \left\vert \beta \mathbf{s}\right\vert \leq 2\pi _{0},
\label{Ys} \\
\text{where }\mathbf{\xi }\text{ is the rectifying coordinates, }\mathbf{s}=%
\frac{\mathbf{\eta }}{\beta }\text{\ is the scaled quasimomentum,}  \notag
\end{gather}%
and the notation 
\begin{equation}
\zeta Y\left( \zeta \beta \mathbf{q}\right) =Y_{\zeta }\left( \beta \mathbf{q%
}\right) .  \label{Yzet}
\end{equation}%
After this change of variables the linear response takes the form%
\begin{equation}
\tilde{U}_{\zeta ,n_{0}}^{\left( 0\right) }\left( \zeta \mathbf{k}_{\ast
}+Y_{\zeta }\left( \beta \mathbf{q}\right) ,t\right) =\psi \left( \tau
\right) \Psi _{0}\left( Y_{\zeta }\left( \beta \mathbf{q}\right) \right)
\beta ^{-d}\mathring{h}_{\zeta }\left( \frac{1}{\beta }Y_{\zeta }\left(
\beta \mathbf{q}\right) \right) \mathrm{e}^{-\mathrm{i}\zeta \gamma _{\left(
\nu \right) }\left( \zeta \beta \mathbf{q}\right) t},\ \tau =\varrho t
\label{u0sy}
\end{equation}%
with $\psi $, $\Psi _{0}$ and $\mathring{h}_{\zeta }$ defined by (\ref{psi0}%
), (\ref{j0}) and (\ref{jnn2}). Notice that for $\nu =2$ the phase function $%
\gamma _{\left( 2\right) }\left( \mathbf{\xi }\right) $ in (\ref{u0sy}) is a
quadratic polynomial which is identical to the phase function of the linear
response of the relevant linear Schrodinger equation (see (\ref{Slin}), (\ref%
{Slin1})). \emph{This formula shows that the linear Maxwell equation is
exactly equivalent in the quasimomentum domain to the linear Schrodinger
equation for a single doublet excitation localized around }$\mathbf{k}_{\ast
}$ (the relation between solutions written in the space domain is discussed
in Subsection 5.5).

Notice that the wavevectors (quasimomenta) $\mathbf{k}$ needed to compose a
solution to the NLS via its Fourier transform vary over the entire space $%
\mathbf{R}^{d}$ whereas for the NLM\ we use quasimomenta from the Brillouin
zone and in fact intend to use quasimomenta $\mathbf{k}$ from a small $\beta 
$-vicinity of $\zeta \mathbf{k}_{\ast }$. To deal with this difference we
introduce a cutoff function 
\begin{equation}
\Psi \left( \mathbf{\xi }\right) =\Psi \left( \zeta ,\mathbf{\xi }\right)
=\Psi _{0}\left( Y_{\zeta }\left( \mathbf{\xi }\right) \right) \text{ if }%
\left\vert \mathbf{\xi }\right\vert <\pi _{0};\ \Psi \left( \mathbf{\xi }%
\right) =0\text{ if }\left\vert \mathbf{\xi }\right\vert \geq \pi _{0},
\label{Psihat}
\end{equation}%
which is smooth for all $\mathbf{\xi }$, and has properties similar to $\Psi
_{0}\left( \mathbf{\xi }\right) $ in (\ref{j0}). We omit $\zeta $ in the
notation of $\Psi \left( \mathbf{\xi }\right) $ since it is not essential
for the analysis. Let us consider now the properties of the functions $%
\mathring{h}_{\zeta }\left( \mathbf{s}\right) $. We take two smooth function 
$\hat{h}_{\zeta }\left( \mathbf{q}\right) $, $\zeta =\pm $, defined for all 
\textbf{$q\ $}$\in \mathbf{R}^{d}$ that decay for large $\left\vert \mathbf{q%
}\right\vert $ faster than any negative power, i.e. 
\begin{equation}
\left\vert \hat{h}_{\zeta }\left( \mathbf{q}\right) \right\vert \leq
C_{N_{\Psi }}\left( 1+\left\vert \mathbf{q}\right\vert \right) ^{-N_{\Psi
}},\ \zeta =\pm ,\text{ with arbitrarily large }N_{\Psi }>0.  \label{hetap}
\end{equation}%
The condition (\ref{hetap}) is used primarily to show that as $\beta
\rightarrow 0$ the function $\Psi _{0}$ does not affect the asymptotic
expansions we derive below. The function $\mathring{h}_{\zeta }\left( 
\mathbf{s}\right) $ is defined by the equation 
\begin{equation}
\Psi \left( \beta \mathbf{q}\right) \hat{h}_{\zeta }\left( \mathbf{q}\right)
=\Psi \left( \beta \mathbf{q}\right) \mathring{h}_{\zeta }\left( \frac{%
Y_{\zeta }\left( \beta \mathbf{q}\right) }{\beta }\right) ,  \label{hcap}
\end{equation}%
or, equivalently%
\begin{equation}
\Psi _{0}\left( \beta \mathbf{s}\right) \mathring{h}_{\zeta }\left( \mathbf{s%
}\right) =\Psi _{0}\left( \beta \mathbf{s}\right) \hat{h}_{\zeta }\left( 
\frac{Y_{\zeta }^{-1}\left( \beta \mathbf{s}\right) }{\beta }\right) .
\end{equation}%
Obviously, this above equalities define $\mathring{h}\left( \mathbf{s}%
\right) $ only when $\left\vert \beta \mathbf{s}\right\vert <2\pi _{0}$,
yielding 
\begin{equation}
\mathring{h}_{\zeta }\left( \mathbf{s}\right) =\hat{h}_{\zeta }\left( \frac{1%
}{\beta }Y_{\zeta }^{-1}\left( \beta \mathbf{s}\right) \right) ,\ \hat{h}%
_{\zeta }\left( \mathbf{q}\right) =\mathring{h}_{\zeta }\left( \frac{%
Y_{\zeta }\left( \beta \mathbf{q}\right) }{\beta }\right) \text{ when }%
\left\vert \beta \mathbf{q}\right\vert <\pi _{0},\   \label{hcap1}
\end{equation}%
but it is sufficient since $\Psi _{0}\left( \beta \mathbf{s}\right) 
\mathring{h}_{\zeta }\left( \mathbf{s}\right) =0$ when $\left\vert \beta 
\mathbf{s}\right\vert \geq \pi _{0}$. To ensure (\ref{jreal}) and that the
excitation current is real-valued we assume 
\begin{equation}
\hat{h}_{-\zeta }\left( \mathbf{q}\right) =\hat{h}_{\zeta }\left( -\mathbf{q}%
\right) ^{\ast },  \label{hreal}
\end{equation}%
which is equivalent to (\ref{hhr}). Using the above notations we rewrite (%
\ref{u0sy}) in the form 
\begin{gather}
\tilde{U}_{\bar{n}}^{\left( 0\right) }\left( \zeta \mathbf{k}_{\ast
}+Y_{\zeta }\left( \beta \mathbf{q}\right) ,t\right) =\tilde{u}_{\bar{n}%
}^{\left( 0\right) }\left( \zeta \mathbf{k}_{\ast }+Y_{\zeta }\left( \beta 
\mathbf{q}\right) ,\tau \right) \mathrm{e}^{-\mathrm{i}\zeta \gamma _{\left(
\nu \right) }\left( \zeta \beta \mathbf{q}\right) t},\ \tau =\varrho t,
\label{U0cap} \\
\tilde{u}_{\bar{n}}^{\left( 0\right) }\left( \zeta \mathbf{k}_{\ast
}+Y_{\zeta }\left( \beta \mathbf{q}\right) ,\tau \right) =\psi \left( \tau
\right) \Psi \left( \beta \mathbf{q}\right) \beta ^{-d}\hat{h}_{\zeta
}\left( \mathbf{q}\right) \,,\ \bar{n}=\left( \zeta ,n_{0}\right) .  \notag
\end{gather}%
Notice that by (\ref{j0})%
\begin{equation}
\Psi _{0}\left( \beta \mathbf{s}\right) =1\text{ for }\left\vert \beta 
\mathbf{s}\right\vert \leq \pi _{0},  \label{ps01}
\end{equation}%
and it follows from (\ref{hetap}) that%
\begin{equation}
\left\vert \Psi _{0}\left( \beta \mathbf{q}\right) \hat{h}_{\zeta }\left( 
\mathbf{q}\right) -\hat{h}_{\zeta }\left( \mathbf{q}\right) \right\vert \leq
C_{N_{\Psi }}^{\prime }\beta ^{N_{\Psi }},\ \left\vert \mathbf{q}\right\vert
\geq \frac{\pi _{0}}{2\beta },\ \text{with arbitrarily large }N_{\Psi }>0.
\label{Psibet0}
\end{equation}%
Consequently, we have the representations%
\begin{equation}
\tilde{u}_{\bar{n}}^{\left( 0\right) }\left( \zeta \mathbf{k}_{\ast
}+Y_{\zeta }\left( \beta \mathbf{q}\right) ,\tau \right) =\psi \left( \tau
\right) \hat{h}_{\zeta }\left( \mathbf{q}\right) +O\left( \beta ^{N_{\Psi
}}\right) O\left( \left\vert \hat{h}\right\vert \right) ,  \label{U0cap1}
\end{equation}%
\begin{equation}
\tilde{u}_{\bar{n}}^{\left( 0\right) }\left( \mathbf{k},\tau \right) =\psi
\left( \tau \right) \hat{h}_{\zeta }\left( \frac{1}{\beta }Y_{\zeta
}^{-1}\left( \mathbf{k}-\zeta \mathbf{k}_{\ast }\right) \right) +O\left(
\beta ^{N_{\Psi }}\right) O\left( \left\vert \hat{h}\right\vert \right) ,
\label{U0cap2}
\end{equation}%
where $N_{\Psi }>0$ can be arbitrarily large.

\section{Asymptotic expansions for the first nonlinear response for the
Maxwell equations}

>From (\ref{Vn}), (\ref{U0n0}) and (\ref{vz}) it follows that the first
nonlinear response for every $\bar{n}=\left( \zeta ,n\right) $ is a sum of
only eight non-zero terms: 
\begin{equation}
\tilde{u}_{\bar{n}}^{\left( 1,0\right) }\left( \mathbf{k},\tau \right)
=\sum_{\zeta ^{\prime },\zeta ^{\prime \prime },\zeta ^{\prime \prime \prime
}}I_{\bar{n},\zeta ^{\prime },\zeta ^{\prime \prime },\zeta ^{\prime \prime
\prime }}\left( \tilde{u}^{\left( 0\right) }\right) \left( \mathbf{k},\tau
\right) -\tilde{u}_{\bar{n}}^{\left( 1\right) }\left( \mathbf{J}_{1};\mathbf{%
k},\tau \right) ,  \label{Vn00}
\end{equation}%
where the interaction integrals $I_{\bar{n},\zeta ^{\prime },\zeta ^{\prime
\prime },\zeta ^{\prime \prime \prime }}$ have the following representations 
\begin{gather}
I_{\bar{n},\zeta ^{\prime },\zeta ^{\prime \prime },\zeta ^{\prime \prime
\prime }}\left( \tilde{u}^{\left( 0\right) }\right) \left( \mathbf{k},\tau
\right) =I_{\bar{n},\zeta ^{\prime },\zeta ^{\prime \prime },\zeta ^{\prime
\prime \prime }}\left( \mathbf{k},\tau \right) =\frac{1}{\varrho }%
\int_{0}^{\tau }\int_{\substack{ \lbrack -\pi ,\pi ]^{2d}  \\ \mathbf{%
\mathbf{k}^{\prime }}+\mathbf{k}^{\prime \prime }+\mathbf{k}^{\prime \prime
\prime }=\mathbf{k}}}\exp \left\{ \mathrm{i}\phi _{\vec{n}}\left( \vec{k}%
\right) \frac{\tau _{1}}{\varrho }\right\}  \label{Vn0} \\
\breve{Q}_{\vec{n}}\left( \vec{k}\right) \tilde{u}_{\zeta ^{\prime
},n_{0}}^{\left( 0\right) }\left( \mathbf{k}^{\prime },\tau _{1}\right) 
\tilde{u}_{\zeta ^{\prime \prime },n_{0}}^{\left( 0\right) }\left( \mathbf{k}%
^{\prime \prime },\tau _{1}\right) \tilde{u}_{\zeta ^{\prime \prime \prime
},n_{0}}^{\left( 0\right) }\left( \mathbf{k}^{\prime \prime \prime },\tau
_{1}\right) \,\mathrm{d}\mathbf{k}^{\prime }\mathrm{d}\mathbf{k}^{\prime
\prime }\mathrm{d}\tau _{1},\ \vec{k}=\left( \mathbf{k},\mathbf{k}^{\prime },%
\mathbf{k}^{\prime \prime },\mathbf{k}^{\prime \prime \prime }\right) , 
\notag \\
\vec{n}=\left( \left( \zeta ,n\right) ,\left( \zeta ^{\prime },n_{0}\right)
,\left( \zeta ^{\prime \prime },n_{0}\right) ,\left( \zeta ^{\prime \prime
\prime },n_{0}\right) \right) .  \notag
\end{gather}%
The term $\tilde{u}_{\bar{n}}^{\left( 1\right) }\left( \mathbf{J}_{1};%
\mathbf{k},\tau \right) $ is defined by (\ref{un1J}). Note that indices $%
\vec{n}$ involved in the representation (\ref{Vn0}) satify the relation 
\begin{equation}
n^{\prime }=n^{\prime \prime }=n^{\prime \prime \prime }=n_{0}.
\label{ndiag}
\end{equation}%
Observe also that for $n\neq n_{0}$ the integral (\ref{Vn0}) describes the
nonlinear impact on the indirectly excited modes, it can be non-zero,
though, as we discussed in Subsection 1.2 and show later in Subsection 3.2,
it is small compared to $n=n_{0}$ since it is not frequency matched.

Since the tensors $\mathbf{R}_{D}^{\left( 3\right) }$ in (\ref{cd4}) are
symmetric, the coefficient $\breve{Q}_{\vec{n}}\left( \mathbf{k},\mathbf{k}%
^{\prime },\mathbf{k}^{\prime \prime },\mathbf{k}^{\prime \prime \prime
}\right) $ are symmetric with respect to the interchange of $\left( \zeta
^{\prime },\mathbf{\mathbf{k}^{\prime }}\right) $ and $\left( \zeta ^{\prime
\prime },\mathbf{\mathbf{k}^{\prime \prime }}\right) $, or $\left( \zeta
^{\prime },\mathbf{\mathbf{k}^{\prime }}\right) $ and $\left( \zeta ^{\prime
\prime \prime },\mathbf{\mathbf{k}^{\prime \prime \prime }}\right) $, or $%
\left( \zeta ^{\prime \prime },\mathbf{\mathbf{k}^{\prime \prime }}\right) $
and $\left( \zeta ^{\prime \prime \prime },\mathbf{\mathbf{k}^{\prime \prime
\prime }}\right) $ if the relations (\ref{ndiag}) hold. Consequently, we
have 
\begin{equation}
I_{\bar{n},\zeta ^{\prime },\zeta ^{\prime \prime },\zeta ^{\prime \prime
\prime }}=I_{\bar{n},\zeta ^{\prime \prime },\zeta ^{\prime },\zeta ^{\prime
\prime \prime }}=I_{\bar{n},\zeta ^{\prime \prime },\zeta ^{\prime \prime
\prime },\zeta ^{\prime }}.  \label{Izsym}
\end{equation}%
It follows from (\ref{un0}) that the integrands in the right-hand side of (%
\ref{Vn0}) are non-zero only when 
\begin{equation}
\left\vert \mathbf{\mathbf{k}^{\prime }-}\zeta ^{\prime }\mathbf{k}_{\ast
}\right\vert \leq \pi _{0},\ \left\vert \mathbf{\mathbf{k}^{\prime \prime }-}%
\zeta ^{\prime \prime }\mathbf{k}_{\ast }\right\vert \leq \pi _{0},\
\left\vert \mathbf{\mathbf{k}^{\prime \prime \prime }-}\zeta ^{\prime \prime
\prime }\mathbf{k}_{\ast }\right\vert \leq \pi _{0}.  \label{kbeta}
\end{equation}%
Observe that since the very form of the integral $I_{\bar{n},\zeta ^{\prime
},\zeta ^{\prime \prime },\zeta ^{\prime \prime \prime }}$ (\ref{Vn0}) obeys
the phase matching condition (\ref{PM0}) through its domain of integration ,
i.e. 
\begin{equation}
\mathbf{k}=\mathbf{\mathbf{k}^{\prime }}+\mathbf{k}^{\prime \prime }+\mathbf{%
k}^{\prime \prime \prime },  \label{PM}
\end{equation}%
the \emph{four-wave interactions may occur only if} 
\begin{equation}
\left\vert \zeta ^{\prime }\mathbf{k}_{\ast }+\zeta ^{\prime \prime }\mathbf{%
k}_{\ast }+\zeta ^{\prime \prime \prime }\mathbf{k}_{\ast }-\mathbf{k}%
\right\vert \leq 3\pi _{0}.  \label{kbeta3}
\end{equation}

We assume that $\mathbf{k}_{\ast }$ is a generic point in the following
sense.

\subparagraph{Genericity condition.}

A point $\mathbf{k}_{\ast }$ is called \emph{generic} if it satisfies the
relations%
\begin{equation}
3\mathbf{k}_{\ast }\neq \mathbf{k}_{\ast }(\func{mod}2\pi );  \label{modpi}
\end{equation}%
\begin{eqnarray}
\left\vert 3\omega _{n_{0}}\left( \mathbf{k}_{\ast }\right) -\omega
_{n}\left( 3\mathbf{k}_{\ast }\right) \right\vert &\neq &0,\ n=1,2,\ldots ;
\label{3om} \\
\left\vert \omega _{n_{0}}^{\prime }\left( \mathbf{k}_{\ast }\right) -\omega
_{n}^{\prime }\left( 3\mathbf{k}_{\ast }\right) \right\vert &\neq &0,\
\left\vert \omega _{n_{0}}^{\prime }\left( \mathbf{k}_{\ast }\right) +\omega
_{n}^{\prime }\left( 3\mathbf{k}_{\ast }\right) \right\vert \neq 0,\
n=1,2,\ldots ;  \notag \\
\left\vert \omega _{n_{0}}\left( \mathbf{k}_{\ast }\right) -\omega
_{n}\left( \mathbf{k}_{\ast }\right) \right\vert &\neq &0,\ \left\vert
\omega _{n_{0}}^{\prime }\left( \mathbf{k}_{\ast }\right) -\omega
_{n}^{\prime }\left( \pm \mathbf{k}_{\ast }\right) \right\vert \neq 0,\
n\neq n_{0}.  \notag
\end{eqnarray}%
We also assume in the strongly dispersive case (\ref{nonM}) that for a
generic $\mathbf{k}_{\ast }$ the relation (\ref{om'ne0}) holds together with%
\begin{equation}
\omega _{n_{0}}\left( \mathbf{k}_{\ast }\right) \neq 0,\ \det \omega
_{n_{0}}^{\prime \prime }\left( \mathbf{\mathbf{k}_{\ast }}\right) \neq 0.
\label{det}
\end{equation}%
Notice that by the inversion symmetry (\ref{invsym}) the relations (\ref{det}%
) readily imply%
\begin{equation}
\omega _{n_{0}}\left( -\mathbf{k}_{\ast }\right) \neq 0,\ \det \omega
_{n_{0}}^{\prime \prime }\left( -\mathbf{\mathbf{k}_{\ast }}\right) \neq 0.
\label{det1}
\end{equation}

According to (\ref{jnn1}), we compose the currents $\mathbf{J}$ from
eigenmodes $\left\{ \left( \zeta ,n_{0}\right) ,\mathbf{k}\right\} $
satisfying the following condition \emph{\ } 
\begin{equation}
\left\vert \mathbf{k}-\zeta \mathbf{k}_{\ast }\right\vert \leq \pi _{0},%
\text{ }\zeta =\pm 1,  \label{kkstar}
\end{equation}%
where $\pi _{0}$ is small constant. In fact when $\ \beta \rightarrow 0$ $%
\pi _{0}$ can be replaced for almost time-harmonic waves (\ref{jnn1}) with
even a smaller number $\beta \pi _{0}$.

We call a mode $\left( n,\mathbf{k}\right) $ \emph{indirectly excited} if 
\begin{equation}
\tilde{u}_{\bar{n}}^{\left( 0\right) }\left( \mathbf{k},\tau \right) =0\text{
for all }\tau \text{ and }\zeta =\pm .  \label{indir}
\end{equation}%
According to (\ref{vz}), the modes with 
\begin{equation}
n\neq n_{0}\text{ \ or }\left\vert \mathbf{k}-\zeta \mathbf{k}_{\ast
}\right\vert >\pi _{0}
\end{equation}%
are indirectly excited. All other modes are called \emph{directly excited},
obviously directly excited modes must satisfy (\ref{kkstar}).\ \emph{In
other words, directly excited modes are one excited through the linear
mechanism whereas indirectly excited ones are excited only through the
nonlinear mechanism}. Hence, based on the medium linear and the first
nonlinear responses all the eigenmodes labeled with $\left\{ \left( \zeta
,n\right) ,\mathbf{k}\right\} $ can be naturally partitioned into two
classes: the eigenmodes that are involved in the composition of the probing
excitation current and eignemodes that are not; the first class coincides
with the directly excited modes and the second with the indirectly excited.
The linear response of the medium obviously involves only the eigenmodes
presented in the source (current), i.e. ones satisfying the condition (\ref%
{kkstar}), and nothing else. If we look now at the first nonlinear response
we find that eigenmodes which don't satsify the condition (\ref{kkstar})
generically are also presented in its composition though with much smaller
amplitudes.

As it was shown in \cite{BF1}-\cite{BF3} stronger interactions must satisfy
the group velocity matching (\ref{GVM0}) and the frequency matching (\ref%
{FMC0}) conditions. It follows from (\ref{invsym}) that 
\begin{equation}
\left( \nabla \omega _{n}\right) \left( \zeta \mathbf{k}_{\ast }\right)
=\zeta \nabla \omega _{n}\left( \mathbf{k}_{\ast }\right) ,\ \zeta =\pm 1.
\label{invsym1}
\end{equation}%
Since (\ref{ndiag}) holds, the group velocity matching condition (\ref{GVM0}%
) at the points $\zeta ^{\left( i\right) }\mathbf{k}_{\ast }$ takes the form 
\begin{equation}
\nabla \left[ \zeta ^{\prime }\omega _{n_{0}}\left( \zeta ^{\prime }\mathbf{k%
}_{\ast }\right) \right] =\nabla \left[ \zeta ^{\prime \prime }\omega
_{n_{0}}\left( \zeta ^{\prime \prime }\mathbf{k}_{\ast }\right) \right]
,\nabla \left[ \zeta ^{\prime }\omega _{n_{0}}\left( \zeta ^{\prime }\mathbf{%
k}_{\ast }\right) \right] =\nabla \left[ \zeta ^{\prime \prime \prime
}\omega _{n_{0}}\left( \zeta ^{\prime \prime \prime }\mathbf{k}_{\ast
}\right) \right]  \label{GVMstar}
\end{equation}%
and by (\ref{invsym1}) it is always fulfilled. The \emph{frequency matching
(FM) condition }(\ref{FMC0}) can be written in the form 
\begin{gather}
\phi _{\vec{n}}\left( \vec{k}\right) =\phi _{\vec{n}}\left( \mathbf{k},%
\mathbf{k}^{\prime },\mathbf{k}^{\prime \prime },\mathbf{k}^{\prime \prime
\prime }\right) =\zeta \omega _{n}\left( \mathbf{k}\right) -\zeta ^{\prime
}\omega _{n_{0}}\left( \mathbf{k}^{\prime }\right) -\zeta ^{\prime \prime
}\omega _{n_{0}}\left( \mathbf{k}^{\prime \prime }\right) -\zeta ^{\prime
\prime \prime }\omega _{n_{0}}\left( \mathbf{k}^{\prime \prime \prime
}\right) =0,  \label{FMC} \\
\vec{n}=\left( \left( \zeta ,n\right) ,\left( \zeta ^{\prime },n_{0}\right)
,\left( \zeta ^{\prime \prime },n_{0}\right) ,\left( \zeta ^{\prime \prime
\prime },n_{0}\right) \right) .  \notag
\end{gather}%
Rather often the fulfillment of the equality (\ref{FMC}) is called \emph{%
phase matching} condtion, (see \cite{Mills}, \cite{BhatSipe}), but we prefer
to call it the \emph{frequency matching} condition and reserve the term
"phase matching condition" for the condition (\ref{PM0})).

At the points $\zeta ^{\left( i\right) }\mathbf{k}_{\ast }$ according to (%
\ref{PM0}) the relations (\ref{FMC0}), (\ref{FMC}) take the form 
\begin{equation}
\zeta \omega _{n}\left( \zeta ^{\prime }\mathbf{k}_{\ast }+\zeta ^{\prime
\prime }\mathbf{k}_{\ast }+\zeta ^{\prime \prime \prime }\mathbf{k}_{\ast
}\right) =\zeta ^{\prime }\omega _{n_{0}}\left( \zeta ^{\prime }\mathbf{k}%
_{\ast }\right) +\zeta ^{\prime \prime }\omega _{n_{0}}\left( \zeta ^{\prime
\prime }\mathbf{k}_{\ast }\right) +\zeta ^{\prime \prime \prime }\omega
_{n_{0}}\left( \zeta ^{\prime \prime \prime }\mathbf{k}_{\ast }\right) .
\label{FMCz}
\end{equation}%
Note now that the sum $\zeta ^{\prime }+\zeta ^{\prime \prime }+\zeta
^{\prime \prime \prime }$ equals either $\pm 1$ or $\pm 3$. From (\ref{3om})
we obtain that $\zeta ^{\prime }+\zeta ^{\prime \prime }+\zeta ^{\prime
\prime \prime }$ cannot be $\pm 3$, and, hence, 
\begin{equation}
n=n_{0},\ \zeta ^{\prime }\mathbf{k}_{\ast }+\zeta ^{\prime \prime }\mathbf{k%
}_{\ast }+\zeta ^{\prime \prime \prime }\mathbf{k}_{\ast }=\pm \mathbf{k}%
_{\ast },  \label{nzk}
\end{equation}%
The inequality (\ref{kbeta3}) implies that for frequency matched
interactions we have 
\begin{equation}
\left\vert \mathbf{k-}\zeta \mathbf{k}_{\ast }\right\vert \leq 3\pi _{0}.
\label{kbeta31}
\end{equation}%
Finally, the condition (\ref{FMCz}) together with (\ref{invsym}) and (\ref%
{3om}) imply that 
\begin{equation}
n=n^{\prime }=n^{\prime \prime }=n^{\prime \prime \prime }=n_{0},\ \zeta
^{\prime }+\zeta ^{\prime \prime }+\zeta ^{\prime \prime \prime }=\zeta .
\label{sumz}
\end{equation}%
We will refer to a situation when the multiindex $\vec{n}$ satisfies the
relation (\ref{sumz}) as the \emph{frequency-matched case (FM-case)}, and to
a situation when at least one of the relations (\ref{sumz}) does not hold as
to \emph{non-frequency-matched case (non-FM-case)}. \emph{Observe that for
the frequency-mathced interactions, i.e. for the FM-case, all significant
mode interactions are restricted to a single band }$n=n_{0}$\emph{.} Notice
also that in the FM-case the phase (\ref{FMC}) has the following more
special representation%
\begin{gather}
\;\phi _{n_{0},\vec{\zeta}}\left( \vec{k}\right) =\phi _{n_{0},\vec{\zeta}%
}\left( \mathbf{k},\mathbf{k}^{\prime },\mathbf{k}^{\prime \prime },\mathbf{k%
}^{\prime \prime \prime }\right) =\zeta \omega _{n_{0}}\left( \mathbf{k}%
\right) -\zeta ^{\prime }\omega _{n_{0}}\left( \mathbf{k}^{\prime }\right)
-\zeta ^{\prime \prime }\omega _{n_{0}}\left( \mathbf{k}^{\prime \prime
}\right) -\zeta ^{\prime \prime \prime }\omega _{n_{0}}\left( \mathbf{k}%
^{\prime \prime \prime }\right) ,  \label{FMC1} \\
\zeta =\zeta ^{\prime }+\zeta ^{\prime \prime }+\zeta ^{\prime \prime \prime
}.  \notag
\end{gather}%
It is convenient to introduce the interaction phase $\phi _{n_{0},\vec{\zeta}%
}$ for the special situation of the FM-case for which $\zeta ^{\prime \prime
\prime }=-\zeta $, $\zeta ^{\prime }=\zeta ^{\prime \prime }=\zeta $, and,
consequently, the phase function $\phi _{n_{0},\vec{\zeta}}\left( \vec{k}%
\right) $ in (\ref{FMC1}) takes the following form 
\begin{equation}
\phi _{n_{0},\zeta }\left( \vec{k}\right) =\zeta \left[ \omega
_{n_{0}}\left( \mathbf{k}\right) -\omega _{n_{0}}\left( \mathbf{k}^{\prime
}\right) -\omega _{n_{0}}\left( \mathbf{k}^{\prime \prime }\right) +\omega
_{n_{0}}\left( \mathbf{k}^{\prime \prime \prime }\right) \right] .
\label{FMC2}
\end{equation}%
We would like to remark that it turns out that the\ interaction integrals (%
\ref{Vn0}) in the non-FM case are much smaller than in the FM-case. \emph{%
Consequently, more significant nonlinear mode interactions are expected to
be frequency-matched}. The non-FM and FM cases will be discussed in detail
in the following two subsections.

\subsection{Frequency-matched interactions}

In this section we consider the interaction integrals $I_{\bar{n},\zeta
^{\prime },\zeta ^{\prime \prime },\zeta ^{\prime \prime \prime }}$ in the
frequency-matched (FM) case, i.e. if the relations (\ref{sumz}) are
fulfilled. In the next subsection we consider the same interaction integrals
in the non-FM case, i.e. when (\ref{sumz}) does not hold. Comparing the both
cases we will see in particular that the interactions in the FM-case are
stronger than the ones in the non-FM case.

Assuming that the relations (\ref{sumz}) hold we introduce the following
change of variables 
\begin{equation}
\mathbf{k-}\zeta \mathbf{k}_{\ast }=\beta \mathbf{s},\mathbf{\ k}^{\prime }%
\mathbf{-}\zeta ^{\prime }\mathbf{k}_{\ast }=\beta \mathbf{s}^{\prime },%
\mathbf{\ \mathbf{k}^{\prime \prime }\mathbf{-}}\zeta ^{\prime \prime }%
\mathbf{\mathbf{k}_{\ast }}=\beta \mathbf{\mathbf{s}}^{\prime \prime },%
\mathbf{\ k}^{\prime \prime \prime }-\zeta ^{\prime \prime \prime }\mathbf{k}%
_{\ast }=\beta \mathbf{\mathbf{s}}^{\prime \prime \prime },  \label{coords}
\end{equation}%
or in a shorter notation 
\begin{equation}
\vec{k}=\vec{\zeta}\vec{k}_{\ast }+\beta \vec{s},
\end{equation}%
where 
\begin{eqnarray}
\vec{\zeta} &=&\left( \zeta ,\zeta ^{\prime },\zeta ^{\prime \prime },\zeta
^{\prime \prime \prime }\right) ,\ \vec{k}=\left( \mathbf{\mathbf{k}},%
\mathbf{\mathbf{k}^{\prime }},\mathbf{\mathbf{k}^{\prime \prime }},\mathbf{%
\mathbf{k}^{\prime \prime \prime }}\right) ,\ \vec{s}=\left( \mathbf{\mathbf{%
s}},\mathbf{\mathbf{s}^{\prime }},\mathbf{\mathbf{s}^{\prime \prime }},%
\mathbf{\mathbf{s}^{\prime \prime \prime }}\right) ,  \label{zsy} \\
\vec{\zeta}\vec{s} &=&\left( \zeta \mathbf{\mathbf{s}},\zeta ^{\prime }%
\mathbf{\mathbf{s}}^{\prime },\zeta ^{\prime \prime }\mathbf{\mathbf{s}}%
^{\prime \prime },\zeta ^{\prime \prime \prime }\mathbf{\mathbf{s}}^{\prime
\prime \prime }\right) .  \notag
\end{eqnarray}%
Note that if (\ref{sumz}) holds the following two equalities are equivalent: 
\begin{equation}
\mathbf{\mathbf{k}^{\prime }}+\mathbf{k}^{\prime \prime }+\mathbf{k}^{\prime
\prime \prime }=\mathbf{k\ }\text{\ is equivalent to }\mathbf{s^{\prime }}+%
\mathbf{s}^{\prime \prime }+\mathbf{s}^{\prime \prime \prime }=\mathbf{s.}
\label{ks}
\end{equation}%
Obviously, there are three combinations of $\zeta ^{\prime },\zeta ^{\prime
\prime },\zeta ^{\prime \prime \prime }$ satisfying (\ref{sumz}) with $\zeta
=1$, and there are three more combinations with $\zeta =-1$. These
combinations correspond to the integrals in (\ref{Vn0}) yielding dominant
contributions. It follows from (\ref{sumz}) that two of the numbers $\zeta
^{\prime },\zeta ^{\prime \prime },\zeta ^{\prime \prime \prime }$ have to
coincide with $\zeta $, and the third one equals $-\zeta $. Let us fix $%
\zeta $ and assume that $\zeta ^{\prime \prime \prime }=-\zeta $, $\zeta
^{\prime }=\zeta ^{\prime \prime }=\zeta $, and denote 
\begin{equation}
\vec{\zeta}_{0}=\left( \zeta ,\zeta ,\zeta ,-\zeta \right) .  \label{z0}
\end{equation}%
Two other integrals with $\zeta ^{\prime }=-\zeta $, $\zeta ^{\prime \prime
}=-\zeta $ can be reduced to the above case with the help of the equalities (%
\ref{Izsym}), namely%
\begin{equation}
I_{\bar{n},\zeta ,\zeta ,-\zeta }=I_{\bar{n},\zeta ,-\zeta ,\zeta }=I_{\bar{n%
},-\zeta ,\zeta ,\zeta }.  \label{z01}
\end{equation}%
Using (\ref{Vn00}), (\ref{Vn0}) together with (\ref{z01}) and taking into
account the estimates of non-FM terms provided in the following subsection
we obtain the following important representation related to the first
nonlinear response 
\begin{equation}
\tilde{u}_{\bar{n}}^{\left( 1,0\right) }\left( \mathbf{k},\tau \right) =3I_{%
\bar{n},\zeta ,\zeta ,-\zeta }\left( \mathbf{k},\tau \right) -\tilde{u}_{%
\bar{n}}^{\left( 1\right) }\left( \mathbf{J}_{1};\mathbf{k},\tau \right)
+O\left( \varrho \right) O\left( \left\vert \mathbf{U}^{\left( 1\right)
}\right\vert \right)  \label{un13}
\end{equation}%
with the integral $I_{\bar{n},\zeta ,\zeta ,-\zeta }\left( \mathbf{k},\tau
\right) $ given by 
\begin{gather}
I_{\bar{n},\zeta ,\zeta ,-\zeta }\left( \mathbf{k},\tau \right) =\frac{1}{%
\varrho }\int_{0}^{\tau }\int_{\substack{ \lbrack -\pi ,\pi ]^{2d}  \\ 
\mathbf{\mathbf{k}^{\prime }}+\mathbf{k}^{\prime \prime }+\mathbf{k}^{\prime
\prime \prime }=\mathbf{k}}}\exp \left\{ \mathrm{i}\phi _{n_{0},\zeta
}\left( \vec{k}\right) \frac{\tau _{1}}{\varrho }\right\}  \label{Iz3} \\
\breve{Q}_{\vec{n}_{0}}\left( \vec{k}\right) \tilde{u}_{\zeta
,n_{0}}^{\left( 0\right) }\left( \mathbf{k}^{\prime },\tau _{1}\right) 
\tilde{u}_{\zeta ,n_{0}}^{\left( 0\right) }\left( \mathbf{k}^{\prime \prime
},\tau _{1}\right) \tilde{u}_{-\zeta ,n_{0}}^{\left( 0\right) }\left( 
\mathbf{k}^{\prime \prime \prime },\tau _{1}\right) \,\mathrm{d}\mathbf{k}%
^{\prime }\mathrm{d}\mathbf{k}^{\prime \prime }\mathrm{d}\tau _{1},  \notag
\\
\phi _{n_{0},\zeta }\left( \vec{k}\right) =\zeta \left[ \omega
_{n_{0}}\left( \mathbf{k}\right) -\omega _{n_{0}}\left( \mathbf{k}^{\prime
}\right) -\omega _{n_{0}}\left( \mathbf{k}^{\prime \prime }\right) +\omega
_{n_{0}}\left( \mathbf{k}^{\prime \prime \prime }\right) \right] ,  \notag
\end{gather}%
\begin{equation}
\vec{n}_{0}=\left( \left( \zeta ,n_{0}\right) ,\left( \zeta ,n_{0}\right)
,\left( \zeta ,n_{0}\right) ,\left( -\zeta ,n_{0}\right) \right) ,
\label{n0ar}
\end{equation}%
where $\breve{Q}_{\vec{n}_{0}}\ $and $\tilde{u}_{\zeta ,n_{0}}^{\left(
0\right) }$ are defined respectively by (\ref{Qn}) and (\ref{vz}).

Let us substitute for the factors $\tilde{u}_{\zeta ,n_{0}}^{\left( 0\right)
}$ in the (\ref{Iz3}) their expressions in terms of the currents $\tilde{j}%
_{n_{0},\zeta }^{\left( 0\right) }\left( \mathbf{k},\tau \right) $ and,
consequently, the initial data $\hat{h}_{\zeta }$. To this end, using (\ref%
{coords}) we obtain from (\ref{jnn1}), (\ref{vz}) respectively%
\begin{equation}
\tilde{j}_{\zeta ,n_{0}}^{\left( 0\right) }\left( \mathbf{k},\tau \right)
=-\varrho \psi _{0}\left( \tau \right) \Psi _{0}\left( \beta \mathbf{s}%
\right) \beta ^{-d}\mathring{h}_{\zeta }\left( \mathbf{s}\right) ,\ \tau
=\varrho t,  \label{jzs}
\end{equation}%
\begin{equation}
\tilde{u}_{\zeta ,n_{0}}^{\left( 0\right) }\left( \mathbf{k},\tau \right)
=\psi \left( \tau \right) \Psi _{0}\left( \mathbf{k-}\zeta \mathbf{k}_{\ast
}\right) \beta ^{-d}\mathring{h}_{\zeta }\left( \frac{\mathbf{k-}\zeta 
\mathbf{k}_{\ast }}{\beta }\right) =\psi \left( \tau \right) \Psi _{0}\left(
\beta \mathbf{s}\right) \beta ^{-d}\hat{h}_{\zeta }\left( \frac{1}{\beta }%
Y_{\zeta }^{-1}\left( \beta \mathbf{s}\right) \right) .  \label{u0z}
\end{equation}%
The equalities (\ref{Iz3}), (\ref{jzs}) and (\ref{u0z}) toghether with (\ref%
{zsy}), (\ref{z0}) yield the following expression for the interaction
integral in terms of $\mathring{h}_{\zeta }$ $\hat{h}_{\zeta }$ 
\begin{gather}
\beta ^{d}I_{\bar{n},\zeta ,\zeta ,-\zeta }\left( \zeta \mathbf{k}_{\ast
}+\beta \mathbf{s},\tau \right) =  \label{Inzbet} \\
\frac{1}{\varrho }\int_{0}^{\tau }\int_{\substack{ \mathbf{\mathbf{s}%
^{\prime }}+\mathbf{s}^{\prime \prime }+\mathbf{s}^{\prime \prime \prime }=%
\mathbf{s,}  \\ \mathbf{\mathbf{s}^{\prime }},\mathbf{s}^{\prime \prime }\in 
\mathbb{R}^{2d}}}\exp \left\{ \mathrm{i}\phi _{n_{0},\zeta }\left( \vec{\zeta%
}_{0}\mathbf{\mathbf{k}_{\ast }}+\beta \vec{s}\right) \frac{\tau _{1}}{%
\varrho }\right\} \breve{Q}_{\vec{n}_{0}}\left( \vec{\zeta}_{0}\mathbf{%
\mathbf{k}_{\ast }}+\beta \vec{s}\right) \Psi _{0}^{3}\left( \beta \vec{s}%
\right)  \notag \\
\psi ^{3}\left( \tau _{1}\right) \mathring{h}_{\zeta }\left( \mathbf{s}%
^{\prime }\right) \mathring{h}_{\zeta }\left( \mathbf{s}^{\prime \prime
}\right) \mathring{h}_{-\zeta }\left( \mathbf{s}^{\prime \prime \prime
}\right) \,\mathrm{d}\mathbf{s}^{\prime }\mathrm{d}\mathbf{s}^{\prime \prime
}\mathrm{d}\tau _{1},  \notag
\end{gather}%
where%
\begin{equation}
\Psi _{0}^{3}\left( \beta \vec{s}\right) =\Psi _{0}\left( \beta \mathbf{s}%
^{\prime }\right) \Psi _{0}\left( \beta \mathbf{s}^{\prime \prime }\right)
\Psi _{0}\left( \beta \mathbf{s}^{\prime \prime \prime }\right) .
\label{Psi3}
\end{equation}%
Notice that the domain of integration of the integral (\ref{Inzbet}) allow $%
\mathbf{s}^{\prime }$ and $\mathbf{s}^{\prime \prime }$ to vary over the
entire space $\mathbb{R}^{d}$ rather than restricting them to just $\left[
-\pi ,\pi \right] ^{d}$, that can be done since the function $\Psi
_{0}^{3}\left( \beta \vec{s}\right) $ by its definition (\ref{Psi3}), (\ref%
{j0}) is zero if either $\mathbf{s}^{\prime }$ or $\mathbf{s}^{\prime \prime
}$ is outside of $\left[ -\pi ,\pi \right] ^{d}$ for \ $\beta \leq 1$. We
remind that the functions $\Psi _{0}\left( \mathbf{k}\right) $ and $\Psi
_{0}^{3}\left( \beta \vec{s}\right) $ were introduced to do exactly that in
order to resolve the difference in setting of the NLS with the quasimomentum 
$\mathbf{k}$ varying in the entire space $\mathbb{R}^{d}$ and the NLM for
periodic medium with $\mathbf{k}$ varying in $\left[ -\pi ,\pi \right] ^{d}$.

\textbf{Remark.} The Floquet-Bloch representation of a single-mode function\ 
\begin{equation}
\mathbf{J}_{n}\left( \mathbf{r},t\right) =\frac{1}{\left( 2\pi \right) ^{d}}%
\int_{\left[ -\pi ,\pi \right] ^{d}}\tilde{J}_{n}\left( \mathbf{k}\right) 
\mathbf{\tilde{G}}_{n}\left( \mathbf{r},\mathbf{k}\right) \,\mathrm{d}%
\mathbf{k}
\end{equation}%
with the coefficient $\tilde{J}_{n}\left( \mathbf{k}\right) $ satisfying $%
\tilde{J}_{n}\left( \mathbf{k}\right) =0$ for $\left\vert \mathbf{k-k}_{\ast
}\right\vert \geq \pi _{0}$ can be rewritten in the form 
\begin{equation}
\mathbf{J}_{n}\left( \mathbf{r},t\right) =\beta ^{d}\frac{1}{\left( 2\pi
\right) ^{d}}\int_{\mathbb{R}^{d}}\tilde{J}_{n}\left( \mathbf{k}_{\ast
}+\beta \mathbf{s}\right) \mathbf{\tilde{G}}_{n}\left( \mathbf{r},\mathbf{k}%
_{\ast }+\beta \mathbf{s}\right) \,\mathrm{d}\mathbf{s}.  \label{Ubet}
\end{equation}%
This identity shows that the coefficient $\tilde{J}_{n}\left( \mathbf{k}%
_{\ast }+\beta \mathbf{s}\right) $ has to have the scaling factor $\beta
^{d} $ to determine a function $\mathbf{J}_{n}\left( \mathbf{r},t\right) $
which is bounded uniformly in $\beta $. Exactly this kind of an expression
is written in the left-hand side of (\ref{Inzbet}).$\blacklozenge $

\subsection{Non-frequency-matched interactions}

There are two different possibilities for the non-frequency-matched (non-FM)
case, i.e. when (\ref{ndiag}) holds and (\ref{FMCz}) does not, which are
described by the followng two alternatives:%
\begin{equation}
\zeta ^{\prime }+\zeta ^{\prime \prime }+\zeta ^{\prime \prime \prime
}=-\zeta ,\ n=n_{0},  \label{sumz-}
\end{equation}%
or%
\begin{equation}
\zeta ^{\prime }+\zeta ^{\prime \prime }+\zeta ^{\prime \prime \prime
}=3\zeta \text{ or }\zeta ^{\prime }+\zeta ^{\prime \prime }+\zeta ^{\prime
\prime \prime }=-3\zeta \text{ or }n\neq n_{0}.\text{\ }  \label{sumz3}
\end{equation}%
In the case (\ref{sumz3}) when $\mathbf{\mathbf{k}}=\pm 3\mathbf{k}_{\ast
}+\beta \mathbf{\mathbf{s}}$ or $n\neq n_{0}$ (\ref{nzk}) does not hold, the
FNLR is non-zero, but the linear response is zero. In the case when (\ref%
{sumz-}) holds according to (\ref{kbeta3}) the FNLR with $n=n_{0}$, $%
\left\vert -\zeta \mathbf{k}_{\ast }-\mathbf{k}\right\vert \leq 3\pi _{0}$
is non-zero, but by (\ref{un0}) the linear response $\tilde{u}_{\zeta
,n}^{\left( 0\right) }\left( \mathbf{k},\tau \right) $ is zero when the
linearly excited wave is uni-directional. Using (\ref{uur1}) we obtain 
\begin{equation}
\tilde{u}_{\bar{n}}^{\left( 1\right) }\left( \mathbf{k},t\right) =\tilde{u}_{%
\bar{n}}^{\left( 1,0\right) }\left( \mathbf{k},t\right) +O\left( \varrho
\right) O\left( \left\vert \mathbf{U}^{\left( 1\right) }\right\vert \right)
,\ \frac{\tau _{0}}{\varrho }\leq t\leq \frac{\tau _{\ast }}{\varrho },
\label{nonFM}
\end{equation}%
where $\tilde{u}_{\bar{n}}^{\left( 1,0\right) }$ is the time-harmonic
approximation for $\tilde{u}_{\bar{n}}^{\left( 1\right) }$ defined by (\ref%
{Vn00}), (\ref{Vn0}) and, in view of (\ref{FNR1}), we get 
\begin{equation}
\tilde{u}_{\bar{n}}^{\left( 1,0\right) }=\tilde{F}_{\bar{n}}\left[ \left( 
\mathbf{u}^{\left( 0\right) }\right) ^{3}\right] \left( \mathbf{k},\tau
\right) .  \label{un1Q}
\end{equation}

Let us show that for the both non-FM cases (\ref{sumz-}) and (\ref{sumz3})
the FNLR $\alpha \tilde{u}_{\bar{n}}^{\left( 1\right) }\left( \mathbf{k}%
,t\right) $ is much smaller than it its counterpart for the FM-case, and the
following estimation holds: 
\begin{equation}
\tilde{u}_{\bar{n}}^{\left( 1\right) }\left( \mathbf{k},t\right) =O\left(
\varrho \right) O\left( \left\vert \mathbf{U}^{\left( 1\right) }\right\vert
\right) ,\ \frac{\tau _{0}}{\varrho }\leq t\leq \frac{\tau _{\ast }}{\varrho 
}.  \label{un1nFM}
\end{equation}%
If the estimation (\ref{un1nFM}) holds then the FNLR $\tilde{u}_{\bar{n}%
}^{\left( 1\right) }\left( \mathbf{k},t\right) $ in the non-FM case
evidently is smaller by the factor $\varrho $ than the FNLR for the FM-case
in (\ref{Inzbet}) which is of order $O\left( \left\vert \mathbf{U}^{\left(
1\right) }\right\vert \right) $. Let us look first at how the non-FM
condition affects the magnitude of the interactions as described by the
interaction integral $I_{\bar{n},\zeta ^{\prime },\zeta ^{\prime \prime
},\zeta ^{\prime \prime \prime }}\left( \mathbf{k},\tau \right) $ in (\ref%
{Vn0}).\ Observe that if the relation (\ref{sumz}) does not hold, the
frequency matching condition (\ref{FMCz}) for the phase $\phi _{\vec{n}%
}\left( \vec{k}\right) $ does not hold too. Having this fact and integrating
by parts, as in \cite{BF1}, \cite{BF4}, it is convenient to recast the
interaction integral (\ref{Vn0}) as 
\begin{gather}
I_{\bar{n},\zeta ^{\prime },\zeta ^{\prime \prime },\zeta ^{\prime \prime
\prime }}\left( \mathbf{k},\tau \right) =\frac{1}{\varrho }\int_{0}^{\tau
}\int_{\substack{ \lbrack -\pi ,\pi ]^{2d}  \\ \mathbf{\mathbf{k}^{\prime }}+%
\mathbf{k}^{\prime \prime }+\mathbf{k}^{\prime \prime \prime }=\mathbf{k}}}%
\mathrm{e}^{\mathrm{i}\phi _{\vec{n}}\left( \vec{k}\right) \frac{\tau _{1}}{%
\varrho }}\mathcal{A}_{\vec{n}}\left( \vec{k},\tau _{1}\right) \,\mathrm{d}%
\mathbf{k}^{\prime }\mathrm{d}\mathbf{k}^{\prime \prime }\mathrm{d}\tau _{1}
\label{Inr} \\
=K_{1}\left( \mathbf{k},\tau \right) +K_{1,0}\left( \mathbf{k},\tau \right) ,
\notag
\end{gather}%
where the dominant term $K_{1}$ is given by 
\begin{gather}
K_{1}\left( \mathbf{k},\tau \right) =\int_{\substack{ \lbrack -\pi ,\pi
]^{2d}  \\ \mathbf{\mathbf{k}^{\prime }}+\mathbf{k}^{\prime \prime }+\mathbf{%
k}^{\prime \prime \prime }=\mathbf{k}}}\frac{\mathrm{e}^{\mathrm{i}\phi _{%
\vec{n}}\left( \vec{k}\right) \frac{\tau }{\varrho }}}{\mathrm{i}\phi _{\vec{%
n}}\left( \vec{k}\right) }\mathcal{A}_{\vec{n}}\left( \vec{k},\tau \right)
\,d\mathbf{k}^{\prime }d\mathbf{k}^{\prime \prime },  \label{K1} \\
\mathcal{A}_{\vec{n}}\left( \vec{k},\tau \right) =\breve{Q}_{\vec{n}}\left( 
\vec{k}\right) \tilde{u}_{\zeta ^{\prime },n_{0}}^{\left( 0\right) }\left( 
\mathbf{k}^{\prime },\tau \right) \tilde{u}_{\zeta ^{\prime \prime
},n_{0}}^{\left( 0\right) }\left( \mathbf{k}^{\prime \prime },\tau \right) 
\tilde{u}_{\zeta ^{\prime \prime \prime },n_{0}}^{\left( 0\right) }\left( 
\mathbf{k}^{\prime \prime \prime },\tau \right) ,  \notag \\
\vec{k}=\left( \mathbf{k},\mathbf{\mathbf{k}^{\prime }},\mathbf{\mathbf{k}%
^{\prime \prime }},\mathbf{\mathbf{k}^{\prime \prime \prime }}\right) . 
\notag
\end{gather}%
The integral with respect to $\mathrm{d}\mathbf{k}^{\prime }\mathrm{d}%
\mathbf{k}^{\prime \prime }$ is similar to $I_{\bar{n},\zeta ,\zeta ,-\zeta
} $ but without the factor $\frac{1}{\varrho }$, therefore $K_{1}\left( 
\mathbf{k},\tau \right) $ is of order $\varrho $ times (\ref{Inzbet}), that
is $K_{1}\left( \mathbf{k},\tau \right) =O\left( \varrho \right) O\left(
\left\vert \mathbf{U}^{\left( 1\right) }\right\vert \right) $. The higher
order term is 
\begin{equation}
K_{1,0}\left( \mathbf{k},\tau \right) =-\int_{0}^{\tau }\int_{\substack{ %
\lbrack -\pi ,\pi ]^{2d}  \\ \mathbf{\mathbf{k}^{\prime }}+\mathbf{k}%
^{\prime \prime }+\mathbf{k}^{\prime \prime \prime }=\mathbf{k}}}\frac{%
\mathrm{e}^{\mathrm{i}\phi _{\vec{n}}\left( \vec{k}\right) \frac{\tau _{1}}{%
\varrho }}}{\mathrm{i}\phi _{\vec{n}}\left( \vec{k}\right) }\partial _{\tau
_{1}}\mathcal{A}_{\vec{n}}\left( \vec{k},\tau _{1}\right) \,\mathrm{d}%
\mathbf{k}^{\prime }\mathrm{d}\mathbf{k}^{\prime \prime }\mathrm{d}\tau _{1}.
\label{K11}
\end{equation}%
To show that $K_{1,0}$ is of higher order in $\varrho $ than $K_{1}$ we can
integrate by parts one more time and obtain that%
\begin{gather}
K_{1,0}\left( \mathbf{k},\tau \right) =\varrho K_{2}\left( \mathbf{k},\tau
\right) +\varrho K_{2,0}\left( \mathbf{k},\tau \right)  \label{K10} \\
K_{2}\left( \mathbf{k},\tau \right) =-\int_{\substack{ \lbrack -\pi ,\pi
]^{2d}  \\ \mathbf{\mathbf{k}^{\prime }}+\mathbf{k}^{\prime \prime }+\mathbf{%
k}^{\prime \prime \prime }=\mathbf{k}}}\frac{\exp \left\{ \mathrm{i}\phi _{%
\vec{n}}\left( \vec{k}\right) \frac{\tau _{1}}{\varrho }\right\} }{\left( 
\mathrm{i}\phi _{\vec{n}}\left( \vec{k}\right) \right) ^{2}}\partial _{\tau
_{1}}^{2}\mathcal{A}_{\vec{n}}\left( \vec{k},\tau _{1}\right) \,\mathrm{d}%
\mathbf{k}^{\prime }\mathrm{d}\mathbf{k}^{\prime \prime },  \notag \\
K_{2,0}\left( \mathbf{k},\tau \right) =\int_{0}^{\tau }\int_{\substack{ %
\lbrack -\pi ,\pi ]^{2d}  \\ \mathbf{\mathbf{k}^{\prime }}+\mathbf{k}%
^{\prime \prime }+\mathbf{k}^{\prime \prime \prime }=\mathbf{k}}}\frac{\exp
\left\{ \mathrm{i}\phi _{\vec{n}}\left( \vec{k}\right) \frac{\tau _{1}}{%
\varrho }\right\} }{\left( \mathrm{i}\phi _{\vec{n}}\left( \vec{k}\right)
\right) ^{2}}\partial _{\tau _{1}}^{2}\mathcal{A}_{\vec{n}}\left( \vec{k}%
,\tau _{1}\right) \,\mathrm{d}\mathbf{k}^{\prime }\mathrm{d}\mathbf{k}%
^{\prime \prime }\mathrm{d}\tau _{1}.  \notag
\end{gather}%
Based on the above we get 
\begin{equation}
K_{1}\left( \mathbf{k},\tau \right) =O\left( \varrho \right) O\left(
\left\vert \mathbf{U}^{\left( 1\right) }\right\vert \right) ,\ K_{1,0}\left( 
\mathbf{k},\tau \right) =O\left( \varrho ^{2}\right) O\left\vert \mathbf{U}%
^{\left( 1\right) }\right\vert ,  \label{nonres1}
\end{equation}%
which implies that a greater contribution comes from $K_{1}$. Therefore, the
non-FM integral can be estimated using the principal term as follows: 
\begin{equation}
I_{\bar{n},\zeta ^{\prime },\zeta ^{\prime \prime },\zeta ^{\prime \prime
\prime }}\left( \mathbf{k},\tau \right) =O\left( \varrho \right) O\left(
\left\vert \mathbf{U}^{\left( 1\right) }\right\vert \right) .  \label{nonres}
\end{equation}%
which implies (\ref{un1nFM}). Now let us look at the higher order terms of
the expansion with respect to $\varrho $. Integrating (\ref{Inr}) by parts $%
m_{2}+1$ times we obtain the expansion 
\begin{equation}
I_{\bar{n},\zeta ^{\prime },\zeta ^{\prime \prime },\zeta ^{\prime \prime
\prime }}\left( \mathbf{k},\tau \right) =\frac{1}{\varrho }%
\sum_{m=1}^{m_{2}}\varrho ^{m}K_{m}\left( \mathbf{k},\tau \right) +O\left(
\varrho ^{m_{2}}\right) O\left( \left\vert \mathbf{U}^{\left( 1\right)
}\right\vert \right) .  \label{sumK}
\end{equation}%
Using the following change of variables 
\begin{equation}
\mathbf{k}^{\prime }\mathbf{-}\zeta ^{\prime }\mathbf{k}_{\ast }=\beta 
\mathbf{s}^{\prime },\mathbf{\ \mathbf{k}^{\prime \prime }\mathbf{-}}\zeta
^{\prime \prime }\mathbf{\mathbf{k}_{\ast }}=\beta \mathbf{\mathbf{s}}%
^{\prime \prime },\mathbf{\ k}^{\prime \prime \prime }-\zeta ^{\prime \prime
\prime }\mathbf{k}_{\ast }=\beta \mathbf{\mathbf{s}}^{\prime \prime \prime },%
\mathbf{\ k-}\left( \zeta ^{\prime }+\zeta ^{\prime \prime }+\zeta ^{\prime
\prime \prime }\right) \mathbf{k}_{\ast }=\beta \mathbf{s.}  \label{coords3}
\end{equation}%
we obtain 
\begin{equation}
K_{m}\left( \left( \zeta ^{\prime }+\zeta ^{\prime \prime }+\zeta ^{\prime
\prime \prime }\right) \mathbf{k}_{\ast }+\beta \mathbf{\mathbf{s}},\tau
\right) =\int_{\substack{ \lbrack -\pi ,\pi ]^{2d}  \\ \mathbf{\mathbf{k}%
^{\prime }}+\mathbf{k}^{\prime \prime }+\mathbf{k}^{\prime \prime \prime }=%
\mathbf{k}}}\exp \left\{ \mathrm{i}\phi _{\vec{n}}\left( \vec{k}\right) 
\frac{\tau }{\varrho }\right\} \mathcal{A}_{\vec{n}m}\left( \vec{k},\tau
\right) \,\mathrm{d}\mathbf{k}^{\prime }\mathrm{d}\mathbf{k}^{\prime \prime
}.  \label{I3k}
\end{equation}%
This integral is similar to the integral in (\ref{Inzbet}), but since it has
a factor $\varrho ^{m}$ in (\ref{sumK}) it may affect only the higher order
approximations. Note that since the operators $K_{m}$ in (\ref{sumK}) do not
involve integration with respect to $\tau $ they produce expressions in the
relevant extended NLS\ involving the time derivatives.

\subparagraph{Approximation of indirectly excited modes.}

For uni-directional excitation currents the indirecly excited modes $\left(
\left( \zeta ,n_{0}\right) ,\mathbf{k}\right) $, i.e. the ones not
satisfying the relation (\ref{kkstar}), have zero linear response, i.e. $%
\tilde{u}_{\bar{n}}^{\left( 0\right) }\left( \mathbf{k},t\right) =0$, and in
the both cases (\ref{sumz-}) or (\ref{sumz3}) the principal part of the
corresponding amplitudes $\tilde{u}_{\bar{n}}\left( \mathbf{k},t\right) $ is
given by formula (\ref{nonFM}). In particular, the relation (\ref{un1nFM})
holds, in view of (\ref{sumK}). For given $h_{\pm }$ for the indirectly
excited modes when $n\neq n_{0}$ or $n=n_{0}$ and $\left\vert \mathbf{k}-%
\mathbf{k}_{\ast }\right\vert \geq \pi _{0}$ we set 
\begin{equation}
\tilde{U}_{Z,\bar{n}}\left( \mathbf{k},t\right) =\alpha \mathrm{e}^{-\mathrm{%
i}\omega _{\bar{n}}\left( \mathbf{k}\right) t}\tilde{u}_{\bar{n}}^{\left(
1,0\right) }\left( \mathbf{k},t\right) ,  \label{UZnFM}
\end{equation}%
with the FNLR amplitudes $\tilde{u}_{\bar{n}}^{\left( 1,0\right) }$ being
defined by (\ref{Vn00}), (\ref{Vn0}). Then using (\ref{un1Q}) we can recast $%
\tilde{U}_{Z,\bar{n}}\left( \mathbf{k},t\right) $ in the form%
\begin{equation}
\tilde{U}_{Z,\bar{n}}\left( \mathbf{k},t\right) =\alpha \mathrm{e}^{-\mathrm{%
i}\omega _{\bar{n}}\left( \mathbf{k}\right) \left( \frac{\tau }{\varrho }%
t\right) }\tilde{u}_{Z,\bar{n}}\left( \mathbf{k},t\right) ,\text{ \ }\tilde{u%
}_{Z,\bar{n}}\left( \mathbf{k},t\right) =\tilde{F}_{\bar{n}}\left[ \left( 
\mathbf{u}^{\left( 0\right) }\right) ^{3}\right] \left( \mathbf{k},\tau
\right) ,  \label{UZnFMQ}
\end{equation}%
with formula (\ref{nonFM}) providing an estimate for the difference between
the modal coefficient of the exact solution of the NLM\ and the
approximation $\tilde{u}_{Z,\bar{n}}\left( \mathbf{k},t\right) $.

The above discussion shows that the amplitudes of indirectly excited modes
are determined mainly by the FNLR. The amplitudes of indirectly excited
modes are relatively small and are of the order $O\left( \varrho \alpha
\right) O\left( \mathbf{\mathbf{U}}^{\left( 1\right) }\right) $. \ 

\section{Asymptotic analysis of the FNLR}

In this section we write expansions of the FNLR of the NLM in $\beta $ up to
the order $\nu -2$ with given $\nu $ using Taylor expansions of the
integrands of the interaction integral (\ref{Inzbet}). More delicate are
expansions envolving the phase function. The corresponding asymptotic
expansions use the parameter $\theta =\frac{\varrho }{\beta ^{2}}$ in the
strongly dispersive case (\ref{nonM}) up to the order $N_{3}$ with
sufficiently large $N_{3}$, and, in the weakly dispersive case, the
expansions involve powers of $\frac{\beta ^{\nu +1}}{\varrho }$. Since for
waves under the study the linear effects of the medium are dominant for both
the NLM and the NLS when $\alpha $ is small, we establish first a
correspondence between the two by applying the rectifying change of
variables (\ref{Y}). That rectifying change of variables is designed to
transform the dispersion relation $\omega _{n_{0}}\left( \mathbf{\mathbf{k}}%
\right) $ related to the NLM with $\mathbf{\mathbf{k}}$ from the $\beta $%
-vicinity of a chosen $\mathbf{\mathbf{k}_{\ast }}$ to its $\nu $-th order
Taylor polynomial $\gamma _{\left( \nu \right) }\left( \mathbf{\mathbf{k}}-%
\mathbf{\mathbf{k}_{\ast }}\right) $ related to an NLS (in the classical
case $\nu =2$). Then we analyze the nonlinear interaction integrals and the
corresponding interaction phases $\phi _{\vec{n}}\left( \vec{k}\right) $ and 
$\phi _{n_{0},\zeta }\left( \vec{k}\right) $ defined by (\ref{FMC}), (\ref%
{FMC1}) and (\ref{FMC2}) in the rectifying variables, and obtain relatively
simple asymptotic expansions for the modal form of the FNLR. Those modal
expansions can be directly related to the FNLR of a proper NLS and, at the
same time, provide a basis for estimating the difference between a solution
to the NLM and its NLS approximation.

In the strongly dispersive case, when (\ref{nonM}) holds, we apply the
Stationary Phase Method (SPhM) to find the asymptotic expansion for a
solution to the NLM. The dominant term of that expansion happens to be
identical to its counterpart for the SPhM expansion of a corresponding
solution to the classical NLS with properly chosen coefficients. In the
weakly dispersive case, when (\ref{nonM1}) holds, the asymptotic analysis of
a solution to the NLM is carried straightforwardly based on the Taylor
expansion of the oscillating factor.

\subsection{Interaction integrals and phases in the scaled rectifying
coordinates}

The interaction integral $I_{\bar{n},\zeta ^{\prime },\zeta ^{\prime \prime
},\zeta ^{\prime \prime \prime }}$ defined by (\ref{Vn0})$\ $with%
\begin{equation}
\vec{\zeta}=\left( \zeta ,\zeta ^{\prime },\zeta ^{\prime \prime },\zeta
^{\prime \prime \prime }\right) =\vec{\zeta}_{0}=\left( \zeta ,\zeta ,\zeta
,-\zeta \right) \text{ and }n=n_{0}  \label{z00}
\end{equation}%
as in (\ref{z0}), (\ref{Iz3}) turns into the integral $I_{\bar{n},\zeta
,\zeta ,-\zeta }$ defined by (\ref{Inzbet}). To study the integral (\ref%
{Inzbet}) we use the scaled rectifying coordinates (\ref{Ys}) to get a
polynomial of degree $\ \nu $ phase function $\Phi ^{\left( \nu \right) }$
(for instance, quadratic when $\nu =2$): 
\begin{gather}
\Phi ^{\left( \nu \right) }\left( \vec{\zeta}_{0},\beta \vec{q}\right)
=\zeta \left[ \gamma _{\left( \nu \right) }\left( \zeta \beta \mathbf{q}%
\right) -\gamma _{\left( \nu \right) }\left( \zeta \beta \mathbf{q}^{\prime
}\right) -\gamma _{\left( \nu \right) }\left( \zeta \beta \mathbf{q}^{\prime
\prime }\right) +\gamma _{\left( \nu \right) }\left( -\zeta \beta \mathbf{q}%
^{\prime \prime \prime }\right) \right] ,  \label{phasefi2} \\
\vec{q}=\left( \mathbf{q},\mathbf{q}^{\prime },\mathbf{q}^{\prime \prime },%
\mathbf{q}^{\prime \prime \prime }\right) ,  \notag
\end{gather}%
in the place of the original interaction phase function $\phi _{n_{0},\zeta
}\left( \vec{k}\right) $ defined by (\ref{FMC2}), with the polynomial $%
\gamma _{\left( \nu \right) }$ being defined by (\ref{Tayom}) and (\ref%
{Tayom2}) based on $\omega _{n_{0}}\left( \mathbf{k}\right) $. When $\vec{%
\zeta}=\vec{\zeta}_{0}$ we denote 
\begin{gather}
\mathring{\Phi}\left( \mathbf{\mathbf{k}_{\ast }},\beta \vec{q}\right) =%
\mathring{\Phi}\left( \mathbf{\mathbf{k}_{\ast }},\beta \mathbf{q},\beta 
\mathbf{q}^{\prime },\beta \mathbf{q}^{\prime \prime },\beta \mathbf{q}%
^{\prime \prime \prime }\right) =\frac{1}{\beta ^{2}}\Phi ^{\left( \nu
\right) }\left( \vec{\zeta}_{0},\beta \vec{q}\right) =  \label{ficap} \\
=\frac{1}{\beta ^{2}}\phi _{n_{0},\zeta }\left( \zeta \mathbf{\mathbf{k}%
_{\ast }}+Y_{\zeta }\left( \beta \mathbf{q}\right) ,\zeta \mathbf{\mathbf{k}%
_{\ast }}+Y_{\zeta }\left( \beta \mathbf{q}^{\prime }\right) ,\zeta \mathbf{%
\mathbf{k}_{\ast }}+Y_{\zeta }\left( \beta \mathbf{q}^{\prime \prime
}\right) ,-\zeta \mathbf{\mathbf{k}_{\ast }}-Y_{\zeta }\left( -\beta \mathbf{%
q}^{\prime \prime \prime }\right) \right) =  \notag \\
\frac{\zeta }{\beta ^{2}}\left[ \omega _{n_{0}}\left( \zeta \mathbf{\mathbf{k%
}_{\ast }}+Y_{\zeta }\left( \beta \mathbf{q}\right) \right) +\omega
_{n_{0}}\left( \zeta \mathbf{\mathbf{k}_{\ast }}+Y_{\zeta }\left( -\beta 
\mathbf{q}^{\prime \prime \prime }\right) \right) \right]  \notag \\
-\frac{\zeta }{\beta ^{2}}\left[ \omega _{n_{0}}\left( \zeta \mathbf{\mathbf{%
k}_{\ast }}+Y_{\zeta }\left( \beta \mathbf{q}^{\prime }\right) \right)
+\omega _{n_{0}}\left( \zeta \mathbf{\mathbf{k}_{\ast }}+Y_{\zeta }\left(
\beta \mathbf{q}^{\prime \prime }\right) \right) \right]  \notag
\end{gather}%
where we used the notations (\ref{zsy}) and the changes of variables (\ref%
{eta}), (\ref{redYz}), (\ref{Ys}). The convenience of the rescaling in (\ref%
{ficap}) can be seen from (\ref{pfi2z}) below, since the principal part of
the phase function $\mathring{\Phi}\left( \mathbf{\mathbf{k}_{\ast }},\beta 
\vec{q}\right) $ does not depend on $\beta $.

Then the integral $I_{\bar{n},\zeta ,\zeta ,-\zeta }$ in (\ref{Inzbet})
written in the scaled rectifying variables takes the following form 
\begin{gather}
\beta ^{d}I_{\bar{n},\zeta ,\zeta ,-\zeta }\left( \zeta \mathbf{k}_{\ast
}+Y_{\zeta }\left( \beta \mathbf{q}\right) ,\tau \right) =\frac{1}{\varrho }%
\int_{0}^{\tau }\int_{Y_{\zeta }\left( \beta \mathbf{q}^{\prime }\right)
+Y_{\zeta }\left( \beta \mathbf{q}^{\prime \prime }\right) -Y_{\zeta }\left(
-\beta \mathbf{q}^{\prime \prime \prime }\right) =Y_{\zeta }\left( \beta 
\mathbf{q}\right) }  \label{IY} \\
\exp \left\{ \mathrm{i}\mathring{\Phi}\left( \mathbf{\mathbf{k}_{\ast }}%
,\beta \vec{q}\right) \frac{\tau _{1}}{\theta }\right\} \psi ^{3}\left( \tau
_{1}\right) A_{1}\left( \beta \vec{q}\right) \hat{h}_{\zeta }\left( \mathbf{q%
}^{\prime }\right) \hat{h}_{\zeta }\left( \mathbf{q}^{\prime \prime }\right) 
\hat{h}_{-\zeta }\left( \mathbf{q}^{\prime \prime \prime }\right) \,\mathrm{d%
}\mathbf{q}^{\prime }\mathrm{d}\mathbf{q}^{\prime \prime }\mathrm{d}\tau
_{1},\   \notag
\end{gather}%
\textbf{\ } 
\begin{equation}
A_{1}\left( \beta \vec{q}\right) =\breve{Q}_{\vec{n}_{0}}\left( \vec{\zeta}%
_{0}\left( \mathbf{\mathbf{k}_{\ast }}+Y\left( \beta \vec{\zeta}_{0}\vec{q}%
\right) \right) \right) \Psi _{0}^{3}\left( \vec{\zeta}_{0}Y\left( \beta 
\vec{\zeta}_{0}\vec{q}\right) \right) \det Y_{\zeta }^{\prime }\left( \beta 
\mathbf{q}^{\prime }\right) \det Y_{\zeta }^{\prime }\left( \beta \mathbf{q}%
^{\prime \prime }\right) ,  \label{A1y}
\end{equation}%
where $\hat{h}\left( \mathbf{q}\right) $ is the same as in (\ref{jnn2}) and (%
\ref{hcap}), $\breve{Q}_{\vec{n}_{0}}$, $\vec{\zeta}_{0}Y\left( \beta \vec{%
\zeta}_{0}\vec{q}\right) $ and $\Psi _{0}^{3}$ are defined respectively by (%
\ref{Qn}), (\ref{zsy}) and (\ref{Psi3}),$\ \theta =\frac{\varrho }{\beta ^{2}%
}$. Notice that the condition 
\begin{equation}
Y_{\zeta }\left( \beta \mathbf{q}^{\prime }\right) +Y_{\zeta }\left( \beta 
\mathbf{q}^{\prime \prime }\right) -Y_{\zeta }\left( -\beta \mathbf{q}%
^{\prime \prime \prime }\right) =Y_{\zeta }\left( \beta \mathbf{q}\right)
\label{Ysum}
\end{equation}%
describing the integration domain of the integral in (\ref{IY}) is the phase
matching condition in the rectifying coordinates which replaces the standard
phase matching condition (\ref{ks}) describing the integration domain in (%
\ref{Inzbet}). The condition (\ref{Ysum}) determines $\mathbf{q}^{\prime
\prime \prime }$ as a function of $\mathbf{q},\mathbf{q}^{\prime },\mathbf{q}%
^{\prime \prime }$, namely 
\begin{equation}
\mathbf{q}^{\prime \prime \prime }\left( \beta \right) =\mathbf{q}^{\prime
\prime \prime }\left( \beta ,\mathbf{q},\mathbf{q}^{\prime },\mathbf{q}%
^{\prime \prime }\right) =-\frac{1}{\beta }Y_{\zeta }^{-1}\left( -Y_{\zeta
}\left( \beta \mathbf{q}\right) +Y_{\zeta }\left( \beta \mathbf{q}^{\prime
}\right) +Y_{\zeta }\left( \beta \mathbf{q}^{\prime \prime }\right) \right) .
\label{Yy'}
\end{equation}%
We denote 
\begin{equation}
\ \vec{q}\left( \beta \right) =\left( \mathbf{q},\mathbf{q}^{\prime },%
\mathbf{q}^{\prime \prime },\mathbf{q}^{\prime \prime \prime }\left( \beta
\right) \right) ,  \label{qbet}
\end{equation}%
and notice that%
\begin{equation}
\mathbf{q}^{\prime \prime \prime }\left( \beta ,\mathbf{q},\mathbf{q}%
^{\prime },\mathbf{q}^{\prime \prime }\right) =-\mathbf{q}\text{ if }\mathbf{%
q}^{\prime }=\mathbf{q}^{\prime \prime }=\mathbf{q}.  \label{YY2}
\end{equation}%
Notice also that in view of (\ref{YY2}) and (\ref{Ybet0}) we have for small $%
\beta \mathbf{q},\ \beta \mathbf{q}^{\prime },\ \beta \mathbf{q}^{\prime
\prime }$ 
\begin{equation}
\mathbf{q}^{\prime \prime \prime }\left( \beta \right) =\left( \mathbf{q-q}%
^{\prime }-\mathbf{q}^{\prime \prime }\right) +O\left( \beta ^{\nu +1}\left(
\left\vert \mathbf{q}\right\vert ^{\nu +1}+\left\vert \mathbf{q}^{\prime
}\right\vert ^{\nu +1}+\left\vert \mathbf{q}^{\prime \prime }\right\vert
^{\nu +1}\right) \right) .  \label{ybet1}
\end{equation}%
>From (\ref{ybet1}) we infer that 
\begin{equation}
\mathbf{q}^{\prime \prime \prime }\left( 0\right) =\mathbf{q-q}^{\prime }-%
\mathbf{q}^{\prime \prime },\ \vec{q}\left( 0\right) =\left( \mathbf{q},%
\mathbf{q}^{\prime },\mathbf{q}^{\prime \prime },\mathbf{q-q}^{\prime }-%
\mathbf{q}^{\prime \prime }\right) .  \label{y0f0}
\end{equation}%
Let us consider now the phase function (\ref{phasefi2}), (\ref{ficap}) under
the constraint (\ref{Ysum}) that is 
\begin{equation}
\Phi ^{\left( \nu \right) }\left( \vec{\zeta}_{0},\beta \vec{q}\left( \beta
\right) \right) =\zeta \left[ \gamma _{\left( \nu \right) }\left( \zeta
\beta \mathbf{q}\right) -\gamma _{\left( \nu \right) }\left( \zeta \beta 
\mathbf{q}^{\prime }\right) -\gamma _{\left( \nu \right) }\left( \zeta \beta 
\mathbf{q}^{\prime \prime }\right) +\gamma _{\left( \nu \right) }\left(
-\zeta \beta \mathbf{q}^{\prime \prime \prime }\left( \beta \right) \right) %
\right] .  \label{Phz0}
\end{equation}%
Observe that (\ref{g2}) together with the inversion symmetry identities (\ref%
{invsym}), (\ref{invsym2}) imply that the constant and linear terms of the
phase $\Phi ^{\left( \nu \right) }\left( \zeta ,\beta \vec{q}\left( \beta
\right) \right) $ at $\mathbf{q}=\mathbf{q}^{\prime }=\mathbf{q}^{\prime
\prime }=0$ vanish, i.e.%
\begin{gather}
\Phi ^{\left( \nu \right) }\left( \vec{\zeta}_{0},0\right) =0,\nabla _{%
\mathbf{q}}\Phi ^{\left( \nu \right) }\left( \vec{\zeta}_{0},0\right)
=\nabla _{\mathbf{q}^{\prime }}\Phi ^{\left( \nu \right) }\left( \vec{\zeta}%
_{0},0\right) =\nabla _{\mathbf{q}^{\prime \prime }}\Phi ^{\left( \nu
\right) }\left( \vec{\zeta}_{0},0\right) =\mathbf{0},  \label{ybet2} \\
\mathring{\Phi}\left( \mathbf{\mathbf{k}_{\ast }},0\right) =0,\;\nabla _{%
\mathbf{q}}\mathring{\Phi}\left( \mathbf{\mathbf{k}_{\ast }},0\right)
=\nabla _{\mathbf{q}^{\prime }}\mathring{\Phi}\left( \mathbf{\mathbf{k}%
_{\ast }},0\right) =\nabla _{\mathbf{q}^{\prime \prime }}\mathring{\Phi}%
\left( \mathbf{\mathbf{k}_{\ast }},0\right) =\mathbf{0}.  \notag
\end{gather}%
In addition to that, the relations (\ref{ficap}), (\ref{YY2}) and (\ref{Phz0}%
) together with the inversion symmetry identities (\ref{invsym}), (\ref%
{invsym2}) imply that the constant and linear terms of the phase $\Phi
^{\left( \nu \right) }\left( \zeta ,\beta \vec{q}\left( \beta \right)
\right) $ vanish even in a more general situation, namely%
\begin{gather}
\mathring{\Phi}\left( \mathbf{\mathbf{k}_{\ast }},\beta \vec{q}^{\;\flat
}\right) =0,\;\nabla _{\mathbf{q}}\mathring{\Phi}\left( \mathbf{\mathbf{k}%
_{\ast }},\beta \vec{q}^{\;\flat }\right) =\nabla _{\mathbf{q}^{\prime }}%
\mathring{\Phi}\left( \mathbf{\mathbf{k}_{\ast }},\beta \vec{q}^{\;\flat
}\right) =\nabla _{\mathbf{q}^{\prime \prime }}\mathring{\Phi}\left( \mathbf{%
\mathbf{k}_{\ast }},\beta \vec{q}^{\;\flat }\right) =\mathbf{0},
\label{crity1} \\
\text{where }\vec{q}^{\;\flat }=\left. \vec{q}\left( 0\right) \right\vert _{%
\mathbf{q}^{\prime }=\mathbf{q}^{\prime \prime }=\mathbf{q}}=\left( \mathbf{q%
},\mathbf{q},\mathbf{q},\mathbf{-q}\right) .  \notag
\end{gather}%
As we will see in Subsection 4.1.3 the points $\vec{q}^{\;\flat }$ in the
formula (\ref{crity1}) describe the set of all critical points of the phase $%
\mathring{\Phi}\left( \mathbf{\mathbf{k}_{\ast }},\beta \vec{q}\left( \beta
\right) \right) $ located about $\mathbf{\mathbf{k}_{\ast }}$. Observe that
by (\ref{ybet1}) we have 
\begin{equation}
\mathring{\Phi}\left( \mathbf{\mathbf{k}_{\ast }},\beta \vec{q}\left( \beta
\right) \right) =\mathring{\Phi}\left( \mathbf{\mathbf{k}_{\ast }},\beta 
\vec{q}\left( 0\right) \right) +O\left( \beta ^{\nu -1}\left( \left\vert 
\mathbf{q}\right\vert ^{\nu +1}+\left\vert \mathbf{q}^{\prime }\right\vert
^{\nu +1}+\left\vert \mathbf{q}^{\prime \prime }\right\vert ^{\nu +1}\right)
\right) .  \label{fihatbet}
\end{equation}%
Using the phase $\mathring{\Phi}\left( \mathbf{\mathbf{k}_{\ast }},\beta 
\vec{q}\left( \beta \right) \right) $ we can rewrite the integral (\ref{IY})
in the form 
\begin{gather}
\beta ^{d}I_{\bar{n},\zeta ,\zeta ,-\zeta }\left( \zeta \mathbf{k}_{\ast
}+Y_{\zeta }\left( \beta \mathbf{q}\right) ,\tau \right) =\frac{1}{\varrho }%
\int_{0}^{\tau }\int_{\mathbb{R}^{2d}}\exp \left\{ \mathrm{i}\frac{\beta ^{2}%
}{\varrho }\mathring{\Phi}\left( \mathbf{\mathbf{k}_{\ast }},\beta \vec{q}%
\left( \beta \right) \right) \tau _{1}\right\} \psi ^{3}\left( \tau
_{1}\right)  \label{IY1} \\
A_{1}\left( \beta \vec{q}\right) \hat{h}_{\zeta }\left( \mathbf{q}^{\prime
}\right) \hat{h}_{\zeta }\left( \mathbf{q}^{\prime \prime }\right) \hat{h}%
_{-\zeta }\left( \mathbf{q}^{\prime \prime \prime }\left( \beta \right)
\right) \,\mathrm{d}\mathbf{q}^{\prime }\mathrm{d}\mathbf{q}^{\prime \prime }%
\mathrm{d}\tau _{1}.  \notag
\end{gather}%
Note that for $\nu =2$ only the quadratic part of $\Phi ^{\left( 2\right) }$
can be non-zero. This fact together with the identity $\omega _{n}^{\prime
\prime }\left( \zeta \mathbf{\mathbf{k}_{\ast }}\right) =\omega _{n}^{\prime
\prime }\left( \mathbf{\mathbf{k}_{\ast }}\right) $, which follows from (\ref%
{invsym2}), implies that 
\begin{gather}
\mathring{\Phi}\left( \mathbf{\mathbf{k}_{\ast }},\beta \vec{q}\left( \beta
\right) \right) =\frac{1}{\beta ^{2}}\Phi ^{\left( \nu \right) }\left( \vec{%
\zeta}_{0},\beta \vec{q}\left( \beta \right) \right) =  \label{pfi2z} \\
\frac{\zeta }{2}\left[ \mathbf{q}\cdot \omega _{n_{0}}^{\prime \prime }%
\mathbf{q}-\mathbf{q}^{\prime }\cdot \omega _{n_{0}}^{\prime \prime }\mathbf{%
q}^{\prime }-\mathbf{q}^{\prime \prime }\cdot \omega _{n_{0}}^{\prime \prime
}\mathbf{q}^{\prime \prime }-\mathbf{q}^{\prime \prime \prime }\left( \beta
\right) \cdot \omega _{n_{0}}^{\prime \prime }\mathbf{q}^{\prime \prime
\prime }\left( \beta \right) \right] +O\left( \beta \left\vert \vec{q}%
\mathbf{\mathbf{\mathbf{-}}}\vec{q}^{\;\flat }\right\vert ^{2}\left\vert 
\vec{q}\right\vert \right) ,  \notag \\
\omega _{n_{0}}^{\prime \prime }=\omega _{n_{0}}^{\prime \prime }\left( 
\mathbf{\mathbf{k}_{\ast }}+\mathbf{q}\right) ,\ \vec{q}^{\;\flat }=\left( 
\mathbf{q},\mathbf{q},\mathbf{q},\mathbf{-q}\right) .  \notag
\end{gather}

\subsubsection{Approximation of the modal susceptibility}

In this subsection we introduce the expansion producing powers $\beta
^{l_{2}}$ in the structured power series (\ref{asser}). To get an expansion
for the interaction integral in (\ref{IY}) we need to have an expansion for
function $A_{1}\left( \beta \vec{q}\right) $ as defined by (\ref{A1y}). To
have an expansion for $A_{1}\left( \beta \vec{q}\right) $ we need, in turn,
an expansion for one of its factors, namely the modal susceptibility $\breve{%
Q}_{\vec{n}_{0}}\left( \vec{\zeta}_{0}\mathbf{\mathbf{k}_{\ast }}+\beta \vec{%
s}\right) $ defined by (\ref{Qn}). In fact, the coefficients of the
expansion of the modal susceptibility determine the coefficients of a
corresponding NLS. The resulting approximation polynomials in $\vec{s}$ \
applied in the Fourier representation lead to differential operators which
are present in the NLS (see Subsection 8.7.1).

We use the Taylor expansion in $\beta $ for the modal susceptibility $\breve{%
Q}_{\vec{n}_{0}}\left( \vec{\zeta}_{0}\mathbf{\mathbf{k}_{\ast }}+\beta \vec{%
s}\right) $, namely%
\begin{gather}
\breve{Q}_{\vec{n}_{0}}\left( \vec{\zeta}_{0}\mathbf{\mathbf{k}_{\ast }}%
+\beta \vec{s}\right) =\breve{Q}_{\vec{n}_{0}}\left( \vec{\zeta}_{0}\mathbf{%
\mathbf{k}_{\ast }}\right) +  \label{Qnexp} \\
\beta \breve{Q}_{\vec{n}_{0}}^{\prime }\left( \vec{\zeta}_{0}\mathbf{\mathbf{%
k}_{\ast }},\vec{s}\right) +\frac{\beta }{2}^{2}\breve{Q}_{\vec{n}%
_{0}}^{\prime \prime }\left( \vec{\zeta}_{0}\mathbf{\mathbf{k}_{\ast }},\vec{%
s}\right) +\ldots +\frac{\beta ^{\nu -1}}{\left( \nu -1\right) !}\breve{Q}_{%
\vec{n}_{0}}^{\left( \sigma \right) }\left( \vec{\zeta}_{0}\mathbf{\mathbf{k}%
_{\ast }},\vec{s}\right) +O\left( \beta ^{\sigma +1}\right)  \notag
\end{gather}%
where $\breve{Q}_{\vec{n}_{0}}^{\left( j\right) }\left( \vec{\zeta}_{0}%
\mathbf{\mathbf{k}_{\ast }},\vec{s}\right) $ is a $j$-linear symmetric form
of $\mathbf{\vec{s}}$, in particular 
\begin{equation}
\breve{Q}_{\vec{n}}^{\prime }\left( \vec{\zeta}_{0}\mathbf{\mathbf{k}_{\ast }%
},\vec{s}\right) =\nabla _{\vec{s}}\breve{Q}_{\vec{n}_{0}}\left( \vec{\zeta}%
_{0}\mathbf{\mathbf{k}_{\ast }}\right) \cdot \vec{s},\ \breve{Q}_{\vec{n}%
}^{\prime \prime }\left( \vec{\zeta}_{0}\mathbf{\mathbf{k}_{\ast }},\vec{s}%
\right) =\nabla _{\vec{s}}^{2}\breve{Q}_{\vec{n}_{0}}\left( \vec{\zeta}_{0}%
\mathbf{\mathbf{k}_{\ast }}\right) \vdots \,\left( \vec{s}^{\ 2}\right)
,\ldots \ .  \label{Qn'}
\end{equation}%
We introduce the Taylor polynomial $p_{\text{T},\zeta }^{\left[ \sigma %
\right] }\left( \beta \vec{s}\right) $ of$\ \breve{Q}_{\vec{n}_{0}}$ of the
degree $\sigma $ by the formula 
\begin{gather}
p_{\text{T},\zeta }^{\left[ \sigma \right] }\left( \beta \vec{s}\right)
=\sum_{j=0}^{\sigma }\frac{1}{j!}\breve{Q}_{\vec{n}_{0}}^{\left( j\right)
}\left( \vec{\zeta}_{0}\mathbf{\mathbf{k}_{\ast }},\beta \vec{s}\right)
=\sum_{j=0}^{\sigma }\frac{\beta ^{j}}{j!}\breve{Q}_{\vec{n}_{0}}^{\left(
j\right) }\left( \vec{\zeta}_{0}\mathbf{\mathbf{k}_{\ast }},\vec{s}\right)
,\ \sigma \leq \nu -1,  \label{psig0} \\
\vec{\zeta}_{0}=\left( \zeta ,\zeta ,\zeta ,-\zeta \right) ,\ \vec{n}%
_{0}=\left( \left( \zeta ,n\right) ,\left( \zeta ,n_{0}\right) ,\left( \zeta
,n_{0}\right) ,\left( -\zeta ,n_{0}\right) \right) .  \notag
\end{gather}%
Now we consider vectors and polynomials with a smaller number of variables,
namely we eliminate $\mathbf{\mathbf{s}}$ using the relation $\mathbf{%
\mathbf{s=s}}^{\prime }+\mathbf{\mathbf{s}}^{\prime \prime }+\mathbf{\mathbf{%
s}}^{\prime \prime \prime }$. Given a vector $\vec{s}=\left( \mathbf{\mathbf{%
s}},\mathbf{\mathbf{s}}^{\prime },\mathbf{\mathbf{s}}^{\prime \prime },%
\mathbf{\mathbf{s}}^{\prime \prime \prime }\right) $ we introduce vectors 
\begin{equation}
\vec{s}^{\;\sharp }=\left( \mathbf{\mathbf{s}}^{\prime }+\mathbf{\mathbf{s}}%
^{\prime \prime }+\mathbf{\mathbf{s}}^{\prime \prime \prime },\mathbf{%
\mathbf{s}}^{\prime },\mathbf{\mathbf{s}}^{\prime \prime },\mathbf{\mathbf{s}%
}^{\prime \prime \prime }\right) ,\ \vec{s}^{\;\star }=\left( \mathbf{%
\mathbf{s}}^{\prime },\mathbf{\mathbf{s}}^{\prime \prime },\mathbf{\mathbf{s}%
}^{\prime \prime \prime }\right) ,  \label{sspr}
\end{equation}%
and the polynomial 
\begin{equation}
p_{\zeta }^{\left[ \sigma \right] }\left( \vec{s}^{\;\star }\right) =p_{%
\text{T},\zeta }^{\left[ \sigma \right] }\left( \vec{s}^{\;\sharp }\right)
,\ \zeta =\pm .  \label{psig}
\end{equation}%
Observe that the polynomial $p_{\zeta }^{\left[ \sigma \right] }\left( \beta 
\vec{s}^{\;\star }\right) $ defined by (\ref{psig}) has the following $m$%
-homogenious terms%
\begin{equation}
p_{\zeta }^{\left[ \sigma \right] }\left( \beta \vec{s}^{\;\star }\right)
=\sum_{j=0}^{\sigma }\beta ^{m}p_{m,\zeta }\left( \vec{s}^{\;\star }\right)
,\ p_{m,\zeta }\left( \vec{s}^{\;\star }\right) =\frac{1}{m!}\breve{Q}_{\vec{%
n}_{0}}^{\left( m\right) }\left( \vec{\zeta}_{0}\mathbf{\mathbf{k}_{\ast }},%
\vec{s}^{\;\sharp }\right)  \label{p0m}
\end{equation}%
which evidently depend only on $\mathbf{\mathbf{s}}^{\prime },\mathbf{%
\mathbf{s}}^{\prime \prime },\mathbf{\mathbf{s}}^{\prime \prime \prime }$.
In particular, for $\sigma =0$ we have \ 
\begin{equation}
p_{0,\zeta }=p_{\zeta }^{\left[ 0\right] }=\breve{Q}_{\vec{n}_{0}}\left( 
\vec{\zeta}_{0}\mathbf{\mathbf{k}_{\ast }}\right) =Q_{\zeta }=Q_{\pm },\
\zeta =\pm 1.  \label{Q0}
\end{equation}%
Formulas (\ref{p0m}) and (\ref{Qn'}) imply the following representation for
the linear form $p_{1,\zeta }\left( \vec{s}^{\;\star }\right) $: 
\begin{equation}
p_{1,\zeta }\left( \vec{s}^{\;\star }\right) =\nabla _{\vec{s}^{\;\star }}%
\breve{Q}_{\vec{n}_{0}}\left( \vec{\zeta}_{0}\mathbf{\mathbf{k}_{\ast }}%
\right) \cdot \vec{s}^{\;\star }+\nabla _{\mathbf{\mathbf{s}}}\breve{Q}_{%
\vec{n}_{0}}\left( \vec{\zeta}_{0}\mathbf{\mathbf{k}_{\ast }}\right) \cdot
\left( \mathbf{\mathbf{s}}^{\prime }+\mathbf{\mathbf{s}}^{\prime \prime }+%
\mathbf{\mathbf{s}}^{\prime \prime \prime }\right) .  \label{p01}
\end{equation}%
Recasting the formula (\ref{p01}) in terms of the gradients $\nabla _{%
\mathbf{\mathbf{s}}^{\prime }},\nabla _{\mathbf{\mathbf{s}}^{\prime \prime
}} $ we get 
\begin{gather*}
p_{1,\zeta }\left( \vec{s}^{\;\star }\right) =\nabla _{\mathbf{\mathbf{s}}%
^{\prime }}\breve{Q}_{\vec{n}_{0}}\left( \vec{\zeta}_{0}\mathbf{\mathbf{k}%
_{\ast }}\right) \cdot \mathbf{\mathbf{s}}^{\prime }+\nabla _{\mathbf{%
\mathbf{s}}^{\prime \prime }}\breve{Q}_{\vec{n}_{0}}\left( \vec{\zeta}_{0}%
\mathbf{\mathbf{k}_{\ast }}\right) \cdot \mathbf{\mathbf{s}}^{\prime \prime }
\\
+\nabla _{\mathbf{\mathbf{s}}^{\prime \prime \prime }}\breve{Q}_{\vec{n}%
_{0}}\left( \vec{\zeta}_{0}\mathbf{\mathbf{k}_{\ast }}\right) \cdot \mathbf{%
\mathbf{s}}^{\prime \prime \prime }+\nabla _{\mathbf{\mathbf{s}}}\breve{Q}_{%
\vec{n}_{0}}\left( \vec{\zeta}_{0}\mathbf{\mathbf{k}_{\ast }}\right) \cdot
\left( \mathbf{\mathbf{s}}^{\prime }+\mathbf{\mathbf{s}}^{\prime \prime }+%
\mathbf{\mathbf{s}}^{\prime \prime \prime }\right) ,
\end{gather*}%
implying%
\begin{equation}
p_{1,\zeta }\left( \vec{s}^{\;\star }\right) =a_{11,\zeta }\cdot \mathbf{%
\mathbf{s}}^{\prime }+a_{12,\zeta }\cdot \mathbf{\mathbf{s}}^{\prime \prime
}+a_{13,\zeta }\cdot \mathbf{\mathbf{s}}^{\prime \prime \prime },
\label{p1z}
\end{equation}%
with vectors $a_{11,\zeta }$, $a_{12,\zeta }$, $a_{13,\zeta }$ defined by 
\begin{gather}
a_{11,\zeta }=\nabla _{\mathbf{\mathbf{s}}^{\prime }}\breve{Q}_{\vec{n}%
_{0}}\left( \vec{\zeta}_{0}\mathbf{\mathbf{k}_{\ast }}\right) +\nabla _{%
\mathbf{\mathbf{s}}}\breve{Q}_{\vec{n}_{0}}\left( \vec{\zeta}_{0}\mathbf{%
\mathbf{k}_{\ast }}\right) ,\ a_{12,\zeta }=\nabla _{\mathbf{\mathbf{s}}%
^{\prime \prime }}\breve{Q}_{\vec{n}_{0}}\left( \vec{\zeta}_{0}\mathbf{%
\mathbf{k}_{\ast }}\right) +\nabla _{\mathbf{\mathbf{s}}}\breve{Q}_{\vec{n}%
_{0}}\left( \vec{\zeta}_{0}\mathbf{\mathbf{k}_{\ast }}\right) ,  \label{a11}
\\
a_{13,\zeta }=\nabla _{\mathbf{\mathbf{s}}^{\prime \prime \prime }}\breve{Q}%
_{\vec{n}_{0}}\left( \vec{\zeta}_{0}\mathbf{\mathbf{k}_{\ast }}\right)
+\nabla _{\mathbf{\mathbf{s}}}\breve{Q}_{\vec{n}_{0}}\left( \vec{\zeta}_{0}%
\mathbf{\mathbf{k}_{\ast }}\right) .  \notag
\end{gather}%
The quadratic polynomial $p_{2,\zeta }\left( \vec{s}^{\;\star }\right) $ has
the following representation%
\begin{gather}
p_{2,\zeta }\left( \vec{s}^{\;\star }\right) =\nabla _{\vec{s}^{\;\star
}}^{2}\breve{Q}_{\vec{n}_{0}}\left( \vec{\zeta}_{0}\mathbf{\mathbf{k}_{\ast }%
}\right) \vdots \,\left( \vec{s}^{\;\star }\right) ^{2}+2\nabla _{\vec{s}%
^{\;\star }}\nabla _{\mathbf{\mathbf{s}}}\breve{Q}_{\vec{n}_{0}}\left( \vec{%
\zeta}_{0}\mathbf{\mathbf{k}_{\ast }}\right) \vdots \,\left( \mathbf{\mathbf{%
s}}^{\prime }+\mathbf{\mathbf{s}}^{\prime \prime }+\mathbf{\mathbf{s}}%
^{\prime \prime \prime }\right) \left( \vec{s}^{\;\star }\right)  \label{p02}
\\
+\nabla _{\mathbf{\mathbf{s}}}^{2}\breve{Q}_{\vec{n}_{0}}\left( \vec{\zeta}%
_{0}\mathbf{\mathbf{k}_{\ast }}\right) \vdots \,\left( \mathbf{\mathbf{s}}%
^{\prime }+\mathbf{\mathbf{s}}^{\prime \prime }+\mathbf{\mathbf{s}}^{\prime
\prime \prime }\right) ^{2}.  \notag
\end{gather}

\subsubsection{Asymptotic expansion in the weakly dispersive case}

In this subsection we study the asymptotic expansions of the interaction
integrals (\ref{IY}), (\ref{IY1}) in the weakly dispersive case when (\ref%
{nonM1}) holds or, in other words, when the dispersion parameter $\theta $
satisfies the inequality 
\begin{equation}
\theta =\frac{\varrho }{\beta ^{2}}\geq \theta _{0}\text{ with a fixed }%
\theta _{0}>0.  \label{thet1}
\end{equation}%
In fact, the most interesting is the borderline case which corresponds to
the classical NLS scaling 
\begin{equation}
\alpha \sim \varrho \sim \beta ^{2}.  \label{border1}
\end{equation}%
To get the integral expansions in $\beta $ up to the order $\sigma =\nu -2$
we need the corresponding expansions for the involved integrands $%
A_{1}\left( \beta \vec{q}\left( \beta \right) \right) $ and $\mathring{h}%
_{-\zeta }\left( \mathbf{q}^{\prime \prime \prime }\left( \beta \right)
\right) $ which can be found as follows. First let us recall that the
dependence of $\mathbf{q}^{\prime \prime \prime }\left( \beta \right) $ on $%
\beta $ is described by (\ref{ybet1}) and it implies that 
\begin{equation}
\mathring{h}_{-\zeta }\left( \mathbf{q}^{\prime \prime \prime }\left( \beta
\right) \right) =\mathring{h}_{-\zeta }\left( \mathbf{q}^{\prime \prime
\prime }\left( 0\right) \right) +O\left( \beta ^{\nu }\right) =\hat{h}%
_{-\zeta }\left( \mathbf{q-q}^{\prime }-\mathbf{q}^{\prime \prime }\right)
+O\left( \beta ^{\nu }\right) .  \label{hybet1}
\end{equation}%
Note that also that by (\ref{Ybet0}) 
\begin{equation}
\det Y_{\zeta }^{\prime }\left( \beta \mathbf{q}^{\prime }\right) =1+O\left(
\beta ^{\nu }\left\vert \mathbf{q}^{\prime }\right\vert ^{\nu }\right) .
\label{detYbet}
\end{equation}%
Using (\ref{A1y}), (\ref{ybet1}) and (\ref{detYbet}) and taking into account
that for small $\beta $, in view of (\ref{j0}), $\Psi _{0}\left( Y\left(
\beta \mathbf{q}\right) \right) =1$ we obtain%
\begin{equation}
A_{1}\left( \beta \vec{q}\left( \beta \right) \right) =p^{\left[ \sigma %
\right] }\left( \beta \vec{q}\left( 0\right) \right) +O\left( \beta ^{\sigma
+1}\right) ,\ \sigma =0,\ldots ,\nu -2.  \label{A1exp}
\end{equation}%
\emph{Note that the above relations hold for the both weakly and strongly
dispersive cases.} In the interaction integral in (\ref{IY1}), according to (%
\ref{ybet1}) $\ $we can write taking into account (\ref{ficap})%
\begin{gather*}
\beta ^{2}\mathring{\Phi}\left( \mathbf{\mathbf{k}_{\ast }},\beta \vec{q}%
\left( \beta \right) \right) =\Phi ^{\left( \nu \right) }\left( \vec{\zeta}%
_{0},\beta \vec{q}\right) = \\
\zeta \left[ 
\begin{array}{c}
\omega _{n_{0}}\left( \zeta \mathbf{\mathbf{k}_{\ast }}+Y_{\zeta }\left(
\beta \mathbf{q}\right) \right) -\omega _{n_{0}}\left( \zeta \mathbf{\mathbf{%
k}_{\ast }}+Y_{\zeta }\left( \beta \mathbf{q}^{\prime }\right) \right) \\ 
-\omega _{n_{0}}\left( \zeta \mathbf{\mathbf{k}_{\ast }}+Y_{\zeta }\left(
\beta \mathbf{q}^{\prime \prime }\right) \right) +\omega _{n_{0}}\left(
\zeta \mathbf{\mathbf{k}_{\ast }}+Y_{\zeta }\left( -\beta \mathbf{q}^{\prime
\prime \prime }\right) \right)%
\end{array}%
\right] .
\end{gather*}%
According to (\ref{ybet1}), (\ref{redY}), (\ref{Tayom2}) and (\ref{YbetX}) 
\begin{equation}
\omega _{n_{0}}\left( \zeta \mathbf{\mathbf{k}_{\ast }}+Y_{\zeta }\left(
-\beta \mathbf{q}^{\prime \prime \prime }\left( \left( \beta \right) \right)
\right) \right) =\gamma _{\left( \nu \right) }\left( -\zeta \beta \mathbf{q}%
^{\prime \prime \prime }\left( \left( \beta \right) \right) \right) =\gamma
_{\left( \nu \right) }\left( -\zeta \beta \mathbf{q}^{\prime \prime \prime
}\left( 0\right) +O\left( \beta ^{\nu +1}\right) \right) .
\end{equation}%
Therefore 
\begin{equation}
\mathring{\Phi}\left( \mathbf{\mathbf{k}_{\ast }},\beta \vec{q}\left( \beta
\right) \right) \frac{\beta ^{2}\tau _{1}}{\varrho }=\Phi ^{\left( \nu
\right) }\left( \vec{\zeta}_{0},\beta \vec{q}\left( 0\right) \right) \frac{%
\tau _{1}}{\varrho }+\frac{\tau _{1}}{\varrho }O\left( \beta ^{\nu +1}\right)
\end{equation}%
and%
\begin{equation}
\exp \left\{ \mathrm{i}\mathring{\Phi}\left( \mathbf{\mathbf{k}_{\ast }}%
,\beta \vec{q}\left( \beta \right) \right) \frac{\beta ^{2}\tau _{1}}{%
\varrho }\right\} =\exp \left\{ \mathrm{i}\Phi ^{\left( \nu \right) }\left( 
\vec{\zeta}_{0},\beta \vec{q}\left( 0\right) \right) \frac{\tau _{1}}{%
\varrho }\right\} \exp \left\{ \mathrm{i}\frac{\tau _{1}}{\varrho }O\left(
\beta ^{\nu +1}\right) \right\} ,  \label{exp1}
\end{equation}%
where we have a standard series expansion 
\begin{equation}
\exp \left\{ \mathrm{i}\frac{\tau _{1}}{\varrho }O\left( \beta ^{\nu
+1}\right) \right\} =1+\frac{\tau _{1}}{\varrho }O\left( \beta ^{\nu
+1}\right) +\left[ \frac{\tau _{1}}{\varrho }O\left( \beta ^{\nu +1}\right) %
\right] ^{2}+\ldots .  \label{expbet}
\end{equation}%
This expansion leads in the case $\nu =4$ to (\ref{asserw}). We conclude
that in the interaction integral (\ref{IY1}) we can write 
\begin{equation}
\exp \left\{ \mathrm{i}\mathring{\Phi}\left( \mathbf{\mathbf{k}_{\ast }}%
,\beta \vec{q}\left( \beta \right) \right) \frac{\beta ^{2}\tau _{1}}{%
\varrho }\right\} =\exp \left\{ \mathrm{i}\Phi ^{\left( \nu \right) }\left( 
\vec{\zeta}_{0},\beta \vec{q}\left( 0\right) \right) \frac{\tau _{1}}{%
\varrho }\right\} \left\{ 1+\frac{\tau _{1}}{\varrho }O\left( \beta ^{\nu
+1}\right) \right\} ;  \label{e1bet}
\end{equation}%
recall that by (\ref{thet1}), in the weakly dispersive case $\frac{\beta
^{\nu +1}\tau _{1}}{\varrho }\ll 1$ when $\nu \geq 2$. Using (\ref{A1exp})
and (\ref{hybet1}) we infer from (\ref{IY1}) that%
\begin{gather}
\beta ^{d}I_{\bar{n},\zeta ,\zeta ,-\zeta }\left( \zeta \mathbf{k}_{\ast
}+Y_{\zeta }\left( \beta \mathbf{q}\right) ,\tau \right) =  \label{IY2} \\
\frac{1}{\varrho }\int_{0}^{\tau }\int_{\mathbb{R}^{2d}}\exp \left\{ \mathrm{%
i}\Phi ^{\left( \nu \right) }\left( \vec{\zeta}_{0},\beta \vec{q}\left(
0\right) \right) \frac{\tau _{1}}{\varrho }\right\} \left\{ 1+O\left( \frac{%
\beta ^{\nu +1}}{\varrho }\right) \right\} \psi ^{3}\left( \tau _{1}\right) 
\notag \\
\left( p^{\left[ \sigma \right] }\left( \beta \vec{q}\right) +O\left( \beta
^{\sigma +1}\right) \right) \left( \hat{h}_{\zeta }\left( \mathbf{q}^{\prime
}\right) \hat{h}_{\zeta }\left( \mathbf{q}^{\prime \prime }\right) \hat{h}%
_{-\zeta }\left( \mathbf{q}^{\prime \prime \prime }\left( 0\right) \right)
+O\left( \beta ^{\nu }\right) \right) \,\mathrm{d}\mathbf{q}^{\prime }%
\mathrm{d}\mathbf{q}^{\prime \prime }\mathrm{d}\tau _{1}  \notag \\
+O\left( \frac{\beta ^{N_{\Psi }-d}}{\varrho }\right) .  \notag
\end{gather}%
Note that in the case (\ref{border1}) $\frac{\beta ^{\nu +1}}{\varrho }$ is
of the same order $\ $as $\beta ^{\nu -1}$ and in the general weakly
dispersive case (\ref{nonM1})%
\begin{equation}
O\left( \frac{\beta ^{\nu +1}}{\varrho }\right) =O\left( \beta ^{\nu
-1}\right) .  \label{Obetw}
\end{equation}

\textbf{Remark. }The terms $O\left( \frac{\beta ^{N_{\Psi }-d}}{\varrho }%
\right) $ in (\ref{IY2}), (\ref{II1}) arise from replacing $\Psi _{0}$ \ by $%
1$. Indeed, when all $\left\vert \mathbf{s}^{\prime }\right\vert ,$ $%
\left\vert \mathbf{s}^{\prime \prime }\right\vert ,$ $\left\vert \mathbf{s}%
^{\prime \prime \prime }\right\vert $ are smaller than $\frac{\pi _{0}}{%
2\beta }$ we have%
\begin{equation}
\Psi _{0}\left( \beta \mathbf{s}^{\prime }\right) =\Psi _{0}\left( \beta 
\mathbf{s}^{\prime \prime }\right) =\Psi _{0}\left( \beta \mathbf{s}^{\prime
\prime \prime }\right) =1
\end{equation}%
and this replacement creates no error at all. Hence, it is sufficient to
consider the case when one of arguments, for example $\left\vert \mathbf{s}%
^{\prime \prime \prime }\right\vert $, is greater than $\frac{\pi _{0}}{%
2\beta }$. But then by (\ref{hetap}) $\hat{h}\left( \mathbf{s}^{\prime
}\right) $, $\hat{h}\left( \mathbf{s}^{\prime \prime }\right) $, $\hat{h}%
\left( \mathbf{s}^{\prime \prime \prime }\right) $ are very small for large
values of arguments. Subtracting from the integral \ (\ref{IY1}) the
integral obtained from (\ref{IY1}) through replacing $\Psi _{0}$ by $1$ we
obtain the integral over the domain where either $\left\vert \mathbf{s}%
^{\prime \prime \prime }\right\vert \geq \frac{\pi _{0}}{2\beta }$ or $%
\left\vert \mathbf{s}^{\prime \prime }\right\vert \geq \frac{\pi _{0}}{%
2\beta }$ or $\left\vert \mathbf{s}^{\prime }\right\vert \geq \frac{\pi _{0}%
}{2\beta }$.\ To estimate the integral over this domain $\left\vert \mathbf{s%
}^{\prime \prime \prime }\right\vert \geq \frac{\pi _{0}}{2\beta }$ we use (%
\ref{hetap}) and obtain 
\begin{eqnarray}
&&\left\vert \left[ \Psi _{0}\left( \beta \mathbf{s}^{\prime }\right) \Psi
_{0}\left( \beta \mathbf{s}^{\prime \prime }\right) \Psi _{0}\left( \beta 
\mathbf{s}^{\prime \prime \prime }\right) -1\right] \hat{h}_{\zeta }\left( 
\mathbf{s}^{\prime }\right) \hat{h}_{\zeta }\left( \mathbf{s}^{\prime \prime
}\right) \hat{h}_{-\zeta }\left( \mathbf{s}^{\prime \prime \prime }\right)
\right\vert  \notag \\
&\leq &C\left( 1+\left\vert \mathbf{s}^{\prime }\right\vert \right)
^{-d-1}\left( 1+\left\vert \mathbf{s}^{\prime \prime }\right\vert \right)
^{-d-1}\left\vert \mathbf{s}^{\prime \prime \prime }\right\vert ^{-N_{\Psi
}}.  \label{Psibet}
\end{eqnarray}%
After integration we obtain the term $O\left( \frac{\beta ^{N_{\Psi }-d}}{%
\varrho }\right) $. Similarly using (\ref{Psibet}) to estimate the
difference of (\ref{IY1}) and the integral obtained from (\ref{IY1}) by
replacing $\Psi _{0}$ with $1$ and looking at similar domains with $%
\left\vert \mathbf{s}^{\prime }\right\vert ,$ $\left\vert \mathbf{s}^{\prime
\prime }\right\vert \geq \frac{\pi _{0}}{2\beta }$ we obtain the term $%
O\left( \frac{\beta ^{N_{\Psi }-d}}{\varrho }\right) $ in (\ref{IY2}). We
assume that $N_{\Psi }$ is large enough to yield the following inequality 
\begin{equation}
O\left( \frac{\beta ^{N_{\Psi }-d}}{\varrho }\right) \ll \frac{\beta ^{2d}}{%
\varrho }O\left( \frac{\beta ^{3}}{\varrho }\right) .  \label{subPsi0}
\end{equation}%
Under this condition this term in (\ref{IY2}) is negligible. Hence, if one
replaces $\Psi _{0}$ by $1$ the error is negligible.$\blacklozenge $

\subsubsection{Asymptotic expansion in the strongly dispersive case and
critical points of the interaction phase}

In this and the following subsections we introduce an expansion yielding
powers $\left( \frac{\varrho }{\beta ^{2}}\right) ^{d+l_{3}}$ in the
structured power series (\ref{asser}). In the dispersive case when (\ref%
{nonM}) holds that is $\theta =\frac{\varrho }{\beta ^{2}}\ll 1$ we cannot
apply the elementary approach of the preceding subsection. Now we use the
Stationary Phase Method (SPhM)\ (see Subsection 8.1) to find an asymptotic
expansion for the interaction integral (\ref{IY}) with respect to the small
parameter $\theta $. According to the method, we need to find the critical
points of the phase $\mathring{\Phi}\left( \mathbf{\mathbf{k}_{\ast }},\beta 
\vec{q}\left( \beta \right) \right) $ defined by (\ref{ficap}) and (\ref%
{ybet1}) with respect to the variables $\mathbf{q}^{\prime }$, $\mathbf{q}%
^{\prime \prime }$. The critical points are the solutions to the following
system of equations 
\begin{equation}
\nabla _{\mathbf{q}^{\prime }}\mathring{\Phi}\left( \mathbf{\mathbf{k}_{\ast
}},\beta \vec{q}\left( \beta \right) \right) =0,\ \nabla _{\mathbf{q}%
^{\prime \prime }}\mathring{\Phi}\left( \mathbf{\mathbf{k}_{\ast }},\beta 
\vec{q}\left( \beta \right) \right) =0.  \label{GVMy}
\end{equation}%
Carrying out the differentiations in (\ref{GVMy}) of the phase $\mathring{%
\Phi}\left( \mathbf{\mathbf{k}_{\ast }},\beta \vec{q}\left( \beta \right)
\right) $, as defined by (\ref{ficap}), under the constraint (\ref{Ysum}) we
get the following equations for the critical points:%
\begin{eqnarray}
\left[ \omega _{n_{0}}^{\prime }\left( \zeta \mathbf{k}_{\ast }+Y_{\zeta
}\left( \beta \mathbf{q}^{\prime }\right) \right) +\omega _{n_{0}}^{\prime
}\left( -\zeta \mathbf{k}_{\ast }-Y_{\zeta }\left( -\beta \mathbf{q}^{\prime
\prime \prime }\right) \right) \right] Y_{\zeta }^{\prime }\left( \beta 
\mathbf{q}^{\prime }\right) &=&0,  \label{criteqy} \\
\left[ \omega _{n_{0}}^{\prime }\left( \zeta \mathbf{k}_{\ast }+Y_{\zeta
}\left( \beta \mathbf{q}^{\prime \prime }\right) \right) +\omega
_{n_{0}}^{\prime }\left( -\zeta \mathbf{k}_{\ast }-Y_{\zeta }\left( -\beta 
\mathbf{q}^{\prime \prime \prime }\right) \right) \right] Y_{\zeta }^{\prime
}\left( \beta \mathbf{q}^{\prime \prime }\right) &=&0.  \notag
\end{eqnarray}%
Taking into account the inversion symmetry identities (\ref{invsym2}) we
find that all small solutions to the system (\ref{criteqy}) together with (%
\ref{Ysum}) are exhausted by the following vectors%
\begin{equation}
\vec{q}^{\;\flat }=\left( \mathbf{q},\mathbf{q},\mathbf{q},-\mathbf{q}%
\right) \text{ or }\mathbf{q}^{\prime }=\mathbf{q}^{\prime \prime }=\mathbf{q%
},\ \mathbf{q}^{\prime \prime \prime }=-\mathbf{q}  \label{crity}
\end{equation}%
with $\mathbf{q}$ being arbitrary (but small). \emph{Remarkably, due the
inversion symmetry the set of critical points described by (\ref{crity})
does not depend on }$\beta $\emph{\ though the system (\ref{GVMy}) does.
Thus for any fixed and sufficiently small }$\mathbf{q}$\emph{\ the
interaction phase }$\mathring{\Phi}\left( \mathbf{\mathbf{k}_{\ast }},\beta 
\vec{q}\left( \beta \right) \right) $\emph{\ has a unique critical point }$%
\vec{q}^{\;\flat }$\emph{\ described by (\ref{crity})}. To prove rigorously
these statements notice first that the Jacobian of the system (\ref{GVMy})
coincides with the Hessian of $\mathring{\Phi}\left( \mathbf{\mathbf{k}%
_{\ast }},\beta \vec{q}\left( \beta \right) \right) $ defined by (\ref{Phz0})%
\begin{gather}
\mathring{\Phi}\left( \mathbf{\mathbf{k}_{\ast }},\beta \vec{q}\left( \beta
\right) \right) \beta ^{2}=\Phi ^{\left( \nu \right) }\left( \vec{\zeta}%
_{0},\beta \vec{q}\left( \beta \right) \right) \\
=\zeta \left[ \gamma _{\left( \nu \right) }\left( \zeta \beta \mathbf{q}%
\right) -\gamma _{\left( \nu \right) }\left( \zeta \beta \mathbf{q}^{\prime
}\right) -\gamma _{\left( \nu \right) }\left( \zeta \beta \mathbf{q}^{\prime
\prime }\right) +\gamma _{\left( \nu \right) }\left( -\zeta \beta \mathbf{q}%
^{\prime \prime \prime }\left( \beta \right) \right) \right] .  \notag
\end{gather}%
We evaluate the Hessian of $\Phi ^{\left( \nu \right) }\left( \vec{\zeta}%
_{0},\beta \vec{q}\left( \beta \right) \right) $ where $\mathbf{q}^{\prime
\prime \prime }\left( \beta \right) $ \ is given by (\ref{ybet1}). An
elementary computation shows that the matrix corresponding to the Hessian
(with respect to $\left( \mathbf{q}^{\prime },\mathbf{q}^{\prime \prime
}\right) $) \ has the following block-diagonal form: 
\begin{equation}
\mathring{\Phi}^{\prime \prime }\left( \mathbf{\mathbf{k}_{\ast }},\beta 
\vec{q}^{\;\flat }\right) =\zeta \left( 
\begin{array}{cc}
0 & \gamma _{\left( \nu \right) }^{\prime \prime }\left( \zeta \beta \mathbf{%
q}\right) \\ 
\gamma _{\left( \nu \right) }^{\prime \prime }\left( \zeta \beta \mathbf{q}%
\right) & 0%
\end{array}%
\right) +O\left( \beta ^{\nu -1}\right) ,  \label{fi''}
\end{equation}%
therefore%
\begin{equation}
\det \mathring{\Phi}^{\prime \prime }\left( \mathbf{\mathbf{k}_{\ast }}%
,\beta \vec{q}^{\;\flat }\right) =\left( -1\right) ^{d}\det \gamma _{\left(
\nu \right) }^{\prime \prime }\left( \zeta \beta \mathbf{q}\right)
^{2}+O\left( \beta ^{\nu -1}\right) .  \label{detF}
\end{equation}%
The relation (\ref{det}) implies that the determinant of the Hessian $%
\mathring{\Phi}^{\prime \prime }\left( \mathbf{\mathbf{k}_{\ast }},\beta 
\vec{q}^{\;\flat }\left( \beta \right) \right) $ does not vanish when $\beta 
$ is small, namely 
\begin{equation}
\det \mathring{\Phi}^{\prime \prime }\left( \mathbf{\mathbf{k}_{\ast }}%
,\beta \vec{q}^{\;\flat }\right) =\det \Phi ^{\left( \nu \right) \prime
\prime }\left( \vec{\zeta}_{0},0\right) +O\left( \left\vert \beta \mathbf{q}%
\right\vert \right) =\left( -1\right) ^{d}\det \omega _{n_{0}}^{\prime
\prime }\left( \mathbf{k}_{\ast }\right) ^{2}+O\left( \left\vert \beta 
\mathbf{q}\right\vert \right) \neq 0.  \label{detfi}
\end{equation}%
Hence for any sufficiently small $\mathbf{q}$ the system (\ref{criteqy}) has
a unique solution $\vec{q}=\vec{q}^{\;\flat }$, as defined by (\ref{crity}),
which evidently does not depend on $\beta $.

According to (\ref{detF}) 
\begin{equation}
\det \mathring{\Phi}^{\prime \prime }\left( \mathbf{\mathbf{k}_{\ast }}%
,\beta \vec{q}\left( \beta \right) \right) =\det \mathring{\Phi}^{\prime
\prime }\left( \mathbf{\mathbf{k}_{\ast }},\vec{\zeta}_{0},\beta \vec{q}%
\left( 0\right) \right) +O\left( \beta ^{\nu -1}\right)  \label{F-f}
\end{equation}%
and by (\ref{ybet1}) 
\begin{equation}
\hat{h}_{-\zeta }\left( \mathbf{q}^{\prime \prime \prime }\left( \beta
\right) \right) =\hat{h}_{-\zeta }\left( \mathbf{q}^{\prime \prime \prime
}\left( 0\right) \right) +O\left( \beta ^{\nu }\right) .  \label{hybet0}
\end{equation}

Let us turn now to the asymptotic expansion of (\ref{IY1}) for $\theta =%
\frac{\varrho }{\beta ^{2}}\rightarrow 0$. Using the SPhM (see Subsection
8.1) and the fact that the phase function has a single critical point $\vec{q%
}^{\;\flat }$ described by (\ref{crity}) we apply to (\ref{y0f0}) the
formula (\ref{I2d}) yielding 
\begin{gather}
\beta ^{d}I_{\bar{n},\zeta ,\zeta ,-\zeta }\left( \zeta \mathbf{k}_{\ast
}+Y_{\zeta }\left( \beta \mathbf{q}\right) ,\tau \right) =\frac{1}{\varrho }%
\int_{0}^{\tau }\frac{\left( 2\pi \right) ^{d}\psi ^{3}\left( \tau
_{1}\right) }{\left\vert \det \mathring{\Phi}^{\prime \prime }\left( \mathbf{%
\mathbf{k}_{\ast }},\beta \vec{q}^{\;\flat }\right) \right\vert ^{1/2}}
\label{Ias2} \\
\left( \frac{\theta }{\tau _{1}}\right) ^{d}\left\{
\sum_{m=0}^{N_{3}}b_{m}\left( \beta ,A\left( \beta ,\vec{q}^{\;\flat },\hat{h%
}\right) \right) \left( \frac{\theta }{\tau _{1}}\right) ^{m}+O\left( \theta
^{N_{3}+1}\right) \right\} \,\mathrm{d}\tau _{1},  \notag
\end{gather}%
where 
\begin{equation}
A\left( \beta ,\vec{q}^{\;\flat },\hat{h}\right) =A_{1}\left( \beta \vec{q}%
\right) \hat{h}^{3}\left( \vec{q}\left( \beta \right) \right) ,\ \hat{h}%
^{3}\left( \vec{q}\left( \beta \right) \right) =\hat{h}_{\zeta }\left( 
\mathbf{q}^{\prime }\right) \hat{h}_{\zeta }\left( \mathbf{q}^{\prime \prime
}\right) \hat{h}_{-\zeta }\left( \mathbf{q}^{\prime \prime \prime }\left(
\beta \right) \right) ,  \label{Ay}
\end{equation}%
with $A_{1}\left( \vec{q}\right) $ being defined by (\ref{A1y}). We use then
(\ref{Qnexp}) and \ (\ref{Ybet0}) to approximate $A_{1}\left( \beta \vec{q}%
\left( \beta \right) \right) \ $by $p^{\left[ \sigma \right] }\left( \beta 
\vec{q}\right) $ defined by (\ref{psig}), and based on the relations (\ref%
{hybet1}), (\ref{F-f}), (\ref{A1exp}) we get 
\begin{equation}
\frac{A_{1}\left( \beta ,\vec{q}^{\;\flat },\hat{h}\right) \hat{h}^{3}\left( 
\vec{q}^{\;\flat }\right) }{\left\vert \det \mathring{\Phi}^{\prime \prime
}\left( \mathbf{\mathbf{k}_{\ast }},\beta \vec{q}^{\;\flat }\right)
\right\vert ^{1/2}}=\frac{p^{\left[ \sigma \right] }\left( \beta \vec{q}%
^{\;\flat }\right) \hat{h}^{3}\left( \vec{q}^{\;\flat }\right) }{\left\vert
\det \mathring{\Phi}^{\prime \prime }\left( \mathbf{\mathbf{k}_{\ast }}%
,\beta \vec{q}^{\;\flat }\right) \right\vert ^{1/2}}+O\left( \beta ^{\sigma
+1}\right) ,\ \sigma \leq \nu -2.  \label{Aas10}
\end{equation}%
Note that $b_{m}\left( \beta ,A\right) $ in (\ref{Ias2}) are differential
operators with constant coefficients, \ determined by the phase function $%
\mathring{\Phi}\left( \mathbf{\mathbf{k}_{\ast }},\beta \vec{q}\left( \beta
\right) \right) $ and its derivatives up to the order $2m$ with respect to $%
\mathbf{q}^{\prime }$, $\mathbf{q}^{\prime \prime }$ at the critical point $%
\vec{q}^{\;\flat }$, see Subsection 8.1 for details. The following
expansions of the operators $b_{m}\left( \beta ,A\right) $ at $\beta =0$ are
obtained from the formula (\ref{fihatbet}): 
\begin{equation}
b_{m}\left( \beta ,A\right) =b_{m}\left( 0,A\right) +\beta ^{\nu
-1}b_{m}^{\prime }\left( \beta ,A\right) ,\ m=1,2,\ldots ,  \label{b1bet}
\end{equation}%
where $b_{m}^{\prime }$ is also a differential operator of the order $2m$.
Note that operator $b_{m}\left( 0,\cdot \right) $ is determined by the
polynomial phase $\Phi ^{\left( \nu \right) }$ \ and, ultimately, by
polynomials $\gamma _{\left( \nu \right) }$. \ Applying the above relations
we get 
\begin{equation}
\frac{b_{m}\left( \beta ,A\left( \beta ,\vec{q}^{\;\flat },\hat{h}\right)
\right) }{\left\vert \det \mathring{\Phi}^{\prime \prime }\left( \mathbf{%
\mathbf{k}_{\ast }},\beta \vec{q}^{\;\flat }\right) \right\vert ^{1/2}}=%
\frac{b_{m}\left( 0,p^{\left[ \sigma \right] }\left( \beta \vec{q}^{\;\flat
}\right) \hat{h}^{3}\left( \vec{q}^{\;\flat }\right) \right) }{\left\vert
\gamma _{\left( \nu \right) }^{\prime \prime }\left( \beta \mathbf{q}\right)
\right\vert }+O\left( \beta ^{\nu -1}\right) ,\ \sigma \leq \nu -2,\
m=1,2,\ldots .  \label{b1bet1}
\end{equation}%
For any $\sigma \leq \nu -2$ using (\ref{Ias2}), (\ref{Aas10}), (\ref{b1bet}%
) and (\ref{b1bet1}) we get the following expansion for the interaction
integral 
\begin{gather}
\beta ^{d}I_{\bar{n},\zeta ,\zeta ,-\zeta }\left( \zeta \mathbf{k}_{\ast
}+Y_{\zeta }\left( \beta \mathbf{q}\right) ,\tau \right) =  \label{Ias21} \\
\frac{1}{\varrho }\int_{0}^{\tau }\frac{\left( 2\pi \right) ^{d}\psi
^{3}\left( \tau _{1}\right) }{\left\vert \det \left( \omega _{n_{0}}^{\prime
\prime }\left( \mathbf{k}_{\ast }\right) \right) \right\vert }\left( \frac{%
\theta }{\tau _{1}}\right) ^{d}  \notag \\
\left\{ \sum_{m=0}^{N_{3}}b_{m}\left( 0,p^{\left[ \sigma \right] }\left(
\beta \vec{q}^{\;\flat }\right) \hat{h}^{3}\left( \vec{q}^{\;\flat }\right)
\right) \left( \frac{\theta }{\tau _{1}}\right) ^{m}+O\left( \theta
^{N_{3}+1}\right) +O\left( \beta ^{\nu -1}\right) \right\} \,\mathrm{d}\tau
_{1}.  \notag
\end{gather}%
Taking in (\ref{Ias21}) the number $N_{3}$ large enough to satisfy $O\left(
\theta ^{N_{3}+1}\right) \ll O\left( \beta ^{\nu -1}\right) $ we get the
final expansion%
\begin{gather}
\beta ^{d}I_{\bar{n},\zeta ,\zeta ,-\zeta }\left( \zeta \mathbf{k}_{\ast
}+Y_{\zeta }\left( \beta \mathbf{q}\right) ,\tau \right) =\frac{1}{\varrho }%
\int_{0}^{\tau }\frac{\left( 2\pi \right) ^{d}\psi ^{3}\left( \tau
_{1}\right) }{\left\vert \det \left( \omega _{n_{0}}^{\prime \prime }\left( 
\mathbf{k}_{\ast }\right) \right) \right\vert }\left( \frac{\theta }{\tau
_{1}}\right) ^{d}  \label{Ias21f} \\
\left\{ \sum_{m=0}^{N_{3}}b_{m}\left( 0,p^{\left[ \sigma \right] }\left(
\beta \vec{q}^{\;\flat }\right) \hat{h}^{3}\left( \vec{q}^{\;\flat }\right)
\right) \left( \frac{\theta }{\tau _{1}}\right) ^{m}+O\left( \beta ^{\nu
-1}\right) \right\} \,\mathrm{d}\tau _{1}.  \notag
\end{gather}%
Note that the significant terms of the expansion are determined by the
approximation $p^{\left[ \sigma \right] }$ of the modal susceptibility and,
through $\ b_{m}$, by the polynomial $\gamma _{\left( \nu \right) }$.

\textbf{Remark.} If $\tau _{1}$ is very small the factor $\frac{\beta
^{2}\tau _{1}}{\varrho }$ in (\ref{IY}) is not large and the SPhM cannot be
applied directly. To handle this case one has to consider small values of $%
\tau _{1}$ in (\ref{Ias2}) a bit differently. Namely, note first that
according to (\ref{psi0}) and (\ref{psi1}) $\psi \left( \tau \right) $ has
zero of infinite order at $\tau =0$, and $A\left( \vec{s}\right) $ is
bounded. Taking then sufficiently large compared with $N_{3}$ $\ $number $N$
and a number $\varkappa $ such that $2<\varkappa <\varkappa _{1}$, where $%
\varkappa _{1}$ is as in (\ref{kap1}), we get the following estimate 
\begin{gather}
\int_{0}^{\varrho /\beta ^{\varkappa }}\psi ^{3}\left( \tau _{1}\right) \,%
\mathrm{d}\tau _{1}\leq C_{N}\int_{0}^{\varrho /\beta ^{\varkappa }}\tau
_{1}^{3N}\,\mathrm{d}\tau _{1}=\left( \frac{\varrho }{\beta ^{\varkappa }}%
\right) ^{3N+1}\frac{1}{1+3N} \\
\leq C\beta ^{\left( 3N+1\right) \left( \varkappa _{1}-\varkappa \right)
}\ll \theta ^{N_{3}+1},  \notag
\end{gather}%
showing that the integral with respect to $\tau _{1}$ over the interval $%
\left[ 0,\varrho /\beta ^{\varkappa }\right] $ in view of (\ref{Ias21}) can
be neglected. For $\tau _{1}\geq \varrho /\beta ^{\varkappa }$ in view of (%
\ref{kap1}) we have $\frac{\beta ^{2}\tau _{1}}{\varrho }\gg 1$, and, hence,
we can apply the SPhM as we do above to get an asymptotic expansion for the
interaction integral (\ref{IY}) with the $m$-the term bounded by the
integral 
\begin{equation}
C_{j}\int_{\varrho /\beta ^{\varkappa }}^{\tau }\left( \frac{\theta }{\tau
_{1}}\right) ^{d+m}\tau _{1}^{3N}\,\mathrm{d}\tau _{1}\leq C_{j}^{\prime
}\theta ^{d+m},\ m=0,1,\ldots 3N-d.
\end{equation}%
Hence, the expansions (\ref{Ias2}) and (\ref{Ias21f}) hold in this case.$%
\blacklozenge $

\subsection{Approximation of the interaction integral}

Here we approximate the interaction integral $I_{\bar{n},\zeta ,\zeta
,-\zeta }$ defined by (\ref{IY}), (\ref{IY1}) \ by a simpler integral $I_{%
\bar{n},\zeta ,\zeta ,-\zeta }^{\left( \sigma \right) }$ which is introduced
below. Using this approximation we will be able to relate solutions to the
NLM and the NLS.

The approximation $I_{\bar{n},\zeta ,\zeta ,-\zeta }^{\left( \sigma \right)
} $ is constructed by the following alterations in the integral $I_{\bar{n}%
,\zeta ,\zeta ,-\zeta }$ represented by (\ref{IY1}) (i) integration domain (%
\ref{Ysum}) is replaced with $\mathbf{q}^{\prime \prime \prime }+\mathbf{q}%
^{\prime \prime }+\mathbf{q}^{\prime }=\mathbf{q}$ leading to a convolution
in the rectifying variables; (ii) the quantities $\mathbf{q}^{\prime \prime
\prime }\left( \beta \right) $ and $\vec{q}\left( \beta \right) $ defined by
(\ref{Yy'}), (\ref{qbet}) are replaced with respectively $\mathbf{q}^{\prime
\prime \prime }\left( 0\right) $ and $\vec{q}\left( 0\right) $ as in (\ref%
{y0f0}); (iii) in the expression $A_{1}\left( \beta \vec{q}\right) $ defined
by (\ref{A1y}) \ the modal susceptibility $\breve{Q}_{\vec{n}_{0}}\left( 
\vec{\zeta}_{0}\left( \mathbf{\mathbf{k}_{\ast }}+Y\left( \beta \vec{\zeta}%
_{0}\vec{q}\right) \right) \right) $ is replaced \ with the polynomial $%
p_{\zeta }^{\left[ \sigma \right] }\left( \beta \vec{q}\right) $ defined by (%
\ref{psig}), the Jacobians $\det Y_{\zeta }^{\prime }$ are replaced by $1$
and the cutoff function $\Psi _{0}$ defined by (\ref{j0}) is replaced by $1$
too. Thus, we introduce the integral $I_{\bar{n},\zeta ,\zeta ,-\zeta
}^{\left( \sigma \right) }$ by the following formula 
\begin{gather}
\beta ^{d}I_{\bar{n},\zeta ,\zeta ,-\zeta }^{\left( \sigma \right) }\left(
\zeta \mathbf{k}_{\ast }+Y_{\zeta }\left( \beta \mathbf{q}\right) ,\tau
\right) =\frac{1}{\varrho }\int_{0}^{\tau }d\tau _{1}\int_{\mathbb{R}%
^{2d}}\exp \left\{ \mathrm{i}\Phi ^{\left( \nu \right) }\left( \vec{\zeta}%
_{0},\beta \vec{q}\left( 0\right) \right) \frac{\tau _{1}}{\varrho }\right\}
p_{\zeta }^{\left[ \sigma \right] }\left( \beta \vec{q}^{\;\star }\right)
\label{I1} \\
\psi ^{3}\left( \tau _{1}\right) \hat{h}_{\zeta }\left( \mathbf{q}^{\prime
}\right) \hat{h}_{\zeta }\left( \mathbf{q}^{\prime \prime }\right) \hat{h}%
_{-\zeta }\left( \mathbf{q}-\mathbf{q}^{\prime }-\mathbf{q}^{\prime \prime
}\right) \,\mathrm{d}\mathbf{q}^{\prime }\mathrm{d}\mathbf{q}^{\prime \prime
},  \notag \\
\vec{q}\left( 0\right) =\left( \mathbf{q},\mathbf{q}^{\prime },\mathbf{q}%
^{\prime \prime },\mathbf{q}-\mathbf{q}^{\prime }-\mathbf{q}^{\prime \prime
}\right) ,  \notag
\end{gather}%
with the polynomials $p_{\zeta }^{\left[ \sigma \right] }\left( \beta \vec{q}%
^{\;\star }\right) $, $\sigma =0,1,2$, being defined by (\ref{psig}), (\ref%
{psig0}). In particular, for $\sigma =0$ we have 
\begin{gather}
\beta ^{d}I_{\bar{n},\zeta ,\zeta ,-\zeta }^{\left( 0\right) }\left( \zeta 
\mathbf{k}_{\ast }+Y_{\zeta }\left( \beta \mathbf{q}\right) ,\tau \right) =%
\frac{1}{\varrho }\int_{0}^{\tau }\mathrm{d}\tau _{1}\int_{\mathbb{R}%
^{2d}}\exp \left\{ \mathrm{i}\Phi ^{\left( \nu \right) }\left( \vec{\zeta}%
_{0},\vec{q}\left( 0\right) \right) \frac{\tau _{1}}{\theta }\right\} \breve{%
Q}_{\vec{n}_{0}}\left( \vec{\zeta}_{0}\mathbf{\mathbf{k}_{\ast }}\right)
\label{Inzbet2} \\
\psi ^{3}\left( \tau _{1}\right) \hat{h}_{\zeta }\left( \mathbf{q}^{\prime
}\right) \hat{h}_{\zeta }\left( \mathbf{q}^{\prime \prime }\right) \hat{h}%
_{-\zeta }\left( \mathbf{q}-\mathbf{q}^{\prime }-\mathbf{q}^{\prime \prime
}\right) \,\mathrm{d}\mathbf{q}^{\prime }\mathrm{d}\mathbf{q}^{\prime \prime
},\ \theta =\frac{\varrho }{\beta ^{2}}.  \notag
\end{gather}%
Let us show now that $I_{\bar{n},\zeta ,\zeta ,-\zeta }^{\left( \sigma
\right) }\left( \zeta \mathbf{k}_{\ast }+Y_{\zeta }\left( \beta \mathbf{q}%
\right) ,\tau \right) $ provides a good approximation to $I_{\bar{n},\zeta
,\zeta ,-\zeta }\left( \zeta \mathbf{k}_{\ast }+Y_{\zeta }\left( \beta 
\mathbf{q}\right) ,\tau \right) $ for the both weakly dispersive and
dispersive cases.

In the dispersive case as $\theta =\frac{\varrho }{\beta ^{2}}\rightarrow 0$
the principal term of the asymptotic expansion of (\ref{I1}) can be found by
the stationary phase method. Notice that the critical points with respect to 
$\mathbf{q}^{\prime }$, $\mathbf{q}^{\prime \prime }$ of the phase function $%
\mathring{\Phi}\left( \mathbf{\mathbf{k}_{\ast }},\beta \vec{q}\left( \beta
\right) \right) $, which is represented by (\ref{phasefi2}), (\ref{y0f0})
are determined by the equations 
\begin{equation}
-\gamma _{\left( \nu \right) }^{\prime }\left( \zeta \beta \mathbf{q}%
^{\prime }\right) +\gamma _{\left( \nu \right) }\left( -\zeta \beta \mathbf{q%
}^{\prime \prime \prime }\right) =0,\ -\gamma _{\left( \nu \right) }^{\prime
}\left( \zeta \beta \mathbf{q}^{\prime \prime }\right) +\gamma _{\left( \nu
\right) }\left( -\zeta \beta \mathbf{q}^{\prime \prime \prime }\right) =0,
\end{equation}%
the solution that satisfies (\ref{y0f0}) is given by (\ref{crity}).
Therefore the phase function $\mathring{\Phi}\left( \mathbf{\mathbf{k}_{\ast
}},\beta \vec{q}\right) $ \emph{has the same critical points} (\ref{crity})
under both restrictions (\ref{y0f0}), (\ref{Ysum}).

Similarly to the expansion of (\ref{IY}) given by (\ref{Ias2}) using SPhM\
we obtain for every $\mathbf{q}$ the following expansion of (\ref{I1}) at
the point determined by (\ref{crity}) as follows. Observe that the
differential operators $b_{m}\left( A\right) $ corresponding to the
polynomial phase function $\Phi ^{\left( \nu \right) }\left( \vec{\zeta}%
_{0},\beta \vec{q}\right) $, as in (\ref{Phz0}), coincide with $b_{m}\left(
0,A\right) $ in (\ref{b1bet}) implying that%
\begin{gather}
\beta ^{d}I_{\bar{n},\zeta ,\zeta ,-\zeta }^{\left( \sigma \right) }\left(
\zeta \mathbf{k}_{\ast }+Y_{\zeta }\left( \beta \mathbf{q}\right) ,\tau
\right) =\frac{1}{\varrho }\int_{0}^{\tau }\frac{\left( 2\pi \right)
^{d}\psi ^{3}\left( \tau _{1}\right) }{\left\vert \det \left( \omega
_{n_{0}}^{\prime \prime }\left( \mathbf{k}_{\ast }\right) \right)
\right\vert }  \label{I1as} \\
\left( \frac{\theta }{\tau _{1}}\right) ^{d}\left(
\sum_{m=0}^{N_{3}}b_{m}\left( 0,p_{\zeta }^{\left[ \sigma \right] }\left(
\beta \vec{q}^{\;\flat }\right) \hat{h}^{3}\left( \vec{q}^{\;\flat }\right)
\right) \left( \frac{\theta }{\tau _{1}}\right) ^{m}+O\left( \theta
^{N_{3}+1}\right) \right) \,\mathrm{d}\tau _{1},  \notag
\end{gather}%
where the critical point $\vec{q}^{\;\flat }$determined by (\ref{crity}) is
the same as in (\ref{Ias2}). This series gives the expansion with respect to 
$\beta $ and $\theta =\frac{\varrho }{\beta ^{2}}$ in (\ref{asser}) with $%
l_{1}=0,$ $N_{2}=\sigma $. Subtracting (\ref{I1as}) from (\ref{Ias21}) and
observing that the expansions (\ref{I1as}) and (\ref{Ias21}) coincide, we
obtain for $\sigma =0,1,2$, $\sigma +2\leq \nu $ the estimate 
\begin{gather}
\beta ^{d}\left[ I_{\bar{n},\zeta ,\zeta ,-\zeta }\left( \zeta \mathbf{k}%
_{\ast }+Y_{\zeta }\left( \beta \mathbf{q}\right) ,\tau \right) -I_{\bar{n}%
,\zeta ,\zeta ,-\zeta }^{\left( \sigma \right) }\left( \zeta \mathbf{k}%
_{\ast }+Y_{\zeta }\left( \beta \mathbf{q}\right) ,\tau \right) \right]
\label{II1} \\
=\left[ O\left( \theta ^{N_{3}+1}\right) +O\left( \beta ^{\nu -1}\right)
+O\left( \frac{\beta ^{N_{\Psi }-d}}{\varrho }\right) \right] O\left(
\left\vert \mathbf{U}^{\left( 1\right) }\right\vert \right) ,  \notag
\end{gather}%
when (\ref{nonM}) holds, and $\tau \geq \tau _{1}$. In particular, this
estimate holds when $\sigma =0$ for the integral $I_{\bar{n},\zeta ,\zeta
,-\zeta }^{\left( 0\right) }$ in (\ref{Inzbet2}). Note that we use (\ref%
{I1as}) and (\ref{Ias21}) when $\Psi _{0}=1$ and when $\Psi _{0}\neq 1$ we
obtain a negligible error $O\left( \frac{\beta ^{N_{\Psi }-d}}{\varrho }%
\right) $, see the Remark in the end of Subsection 4.1.2 for details. Note
that $O\left( \theta ^{N_{3}+1}\right) $ and $O\left( \frac{\beta ^{N_{\Psi
}-d}}{\varrho }\right) $ are negligible since $N_{3}$ and $N_{\Psi }$ are
arbitrary large.

Now we use the above results to estimate the leading term $\tilde{u}_{\bar{n}%
}^{\left( 1,0\right) }$ as in (\ref{Vn00}) of the expansion (\ref{u1rho}).
Note that 
\begin{equation}
U_{\bar{n}}^{\left( 1,0\right) }\left( \zeta \mathbf{\mathbf{k}_{\ast }}%
+Y_{\zeta }\left( \beta \mathbf{q}\right) ,t\right) =-\exp \left[ -\mathrm{i}%
\left( \zeta \gamma _{\left( 2\right) }\left( \zeta \beta \mathbf{q}\right)
\right) \frac{\tau }{\varrho }\right] \tilde{u}_{\bar{n}}^{\left( 1,0\right)
}\left( \zeta \mathbf{\mathbf{k}_{\ast }}+Y_{\zeta }\left( \beta \mathbf{q}%
\right) ,\tau \right) ,  \label{U1s}
\end{equation}%
where $\tilde{u}_{\bar{n}}^{\left( 1,0\right) }\left( \mathbf{k},\tau
\right) $ is given by (\ref{Vn0}). Applying (\ref{Vn00}), (\ref{Izsym}), (%
\ref{II1}), (\ref{nonres}) we get the following formula for $\sigma \leq \nu
-2$ 
\begin{gather}
\beta ^{d}\tilde{u}_{\bar{n}}^{\left( 1,0\right) }\left( \zeta \mathbf{k}%
_{\ast }+Y_{\zeta }\left( \beta \mathbf{q}\right) ,t\right) +\beta ^{d}%
\tilde{u}_{\bar{n}}^{\left( 1\right) }\left( \mathbf{J}_{1};\zeta \mathbf{k}%
_{\ast }+Y_{\zeta }\left( \beta \mathbf{q}\right) ,t\right) =  \label{U1z2}
\\
-3\beta ^{d}I_{\bar{n},\zeta ,\zeta ,-\zeta }^{\left( \sigma \right) }\left(
\zeta \mathbf{k}_{\ast }+Y_{\zeta }\left( \beta \mathbf{q}\right) ,\tau
\right) +\frac{1}{\varrho }\left[ O\left( \beta ^{\nu -1}\right) +O\left(
\varrho \right) \right] O\left( \left\vert \mathbf{U}^{\left( 1\right)
}\right\vert \right) .  \notag
\end{gather}%
In (\ref{U1z2}) the terms with $O\left( \theta ^{N_{3}+1}\right) $ and $%
O\left( \frac{\beta ^{N_{\Psi }-d}}{\varrho }\right) $ are neglected since $%
N_{\Psi }$ \ and $N_{3}+1$ can be chosen arbitrary large and we assume that 
\begin{equation}
O\left( \frac{\beta ^{N_{\Psi }-d}}{\varrho }\right) \ll \frac{1}{\varrho }%
O\left( \left\vert \mathbf{U}^{\left( 1\right) }\right\vert \right) O\left(
\beta ^{\nu -1}\right) ,O\left( \theta ^{N_{3}+1}\right) \ll O\left(
\left\vert \mathbf{U}^{\left( 1\right) }\right\vert \right) O\left( \beta
^{\nu -1}\right) ,  \label{subPsi}
\end{equation}%
since $N_{\Psi }$ is sufficiently large and the relations (\ref{kap1} or (%
\ref{kap11}) hold.

Now we consider the weakly dispersive case when (\ref{nonM1}) and (\ref%
{thet1}) hold. Comparing (\ref{IY2}) with (\ref{I1}) and using (\ref{subPsi}%
) we obtain for $\sigma +2\leq \nu \leq 4$ the following estimate 
\begin{gather}
I_{\bar{n},\zeta ,\zeta ,-\zeta }\left( \zeta \mathbf{k}_{\ast }+Y_{\zeta
}\left( \beta \mathbf{q}\right) ,\tau \right) =I_{\bar{n},\zeta ,\zeta
,-\zeta }^{\left( \sigma \right) }\left( \zeta \mathbf{k}_{\ast }+Y_{\zeta
}\left( \beta \mathbf{q}\right) ,\tau \right) +  \label{II2} \\
\left( O\left( \frac{\beta ^{\nu +1}}{\varrho }\right) +O\left( \beta ^{\nu
}\right) \right) O\left( \left\vert \mathbf{U}^{\left( 1\right) }\right\vert
\right) .
\end{gather}%
Taking into account (\ref{Obetw}) we see that (\ref{II2}) implies (\ref{II1}%
), therefore one can treat the weakly dispersive case similarly to the
dispersive case which we discuss in more detail.

\textbf{Remark.} Note that when we write in (\ref{II1}) and similar formulas
expressions of the form 
\begin{equation}
\left[ O\left( \theta ^{N_{3}+1}\right) +O\left( \beta ^{\nu -1}\right) %
\right] O\left( \left\vert \mathbf{U}^{\left( 1\right) }\right\vert \right)
\label{Inmul}
\end{equation}%
we assume that the principal term of the asymptotics of the integral $I_{%
\bar{n},\zeta ,\zeta ,-\zeta }^{\left( \sigma \right) }\left( \zeta \mathbf{k%
}_{\ast }+Y_{\zeta }\left( \beta \mathbf{q}\right) ,\tau \right) $ does not
vanish, and, hence, $O\left( \left\vert \mathbf{U}^{\left( 1\right)
}\right\vert \right) $ is of order $\varrho ^{d-1}$ in the strongly
dispersive case or $\varrho ^{-1}$ in the weakly dispersive case. Without
this kind of nondegeneracy assumption (\ref{II1}) is not equivalent to (\ref%
{I1as}). Writing error terms in the form (\ref{Inmul}) is more convenient
since it explicitly relates the magnitude of the approximation error to the
magnitude of the first nonlinear response. In addition to that, the
expression (\ref{Inmul}) has the same form in the both cases (\ref{nonM})
and (\ref{nonM1}), even when the asymptotic behavior of the principal term
may be different.$\blacklozenge $

\section{Tailoring the NLS to approximate the NLM}

In this section we introduce an NLS that is tailored to approximate the NLM.
NLS equations we are interested in are the two equations (\ref{Si}), (\ref%
{Si1}) or their $d$-dimensional analogs (\ref{GNLS+}), (\ref{GNLS-}). The
two equations correspond to the two values of $\zeta =\pm 1=\pm $. In this
section we mostly consider the case $\zeta =+$ with the understanding that
the case $\zeta =-$ can be treated similarly.

The NLS equation (\ref{Si}) involves some constants which are to be related
to the NLM. To do that let us construct first a linear NLS related to the
linear NLM. We begin with picking a $\nu =1,2,3$ or $4$ and then we
introduce the Taylor polynomial $\gamma _{\left( \nu \right) }\left( \mathbf{%
\eta }\right) $ of the order $\nu $ of the function $\omega _{n_{0}}\left( 
\mathbf{\mathbf{k}}\right) $ at the point $\mathbf{k}_{\ast }$ as defined by
(\ref{Tayom}). The case $\nu =2$, corresponds to the classical NLS, and $\nu
=3,4$ correspond to an extended NLS. For these two cases we have respectively%
\begin{gather}
\gamma _{\left( 2\right) }\left( \mathbf{\eta }\right) =\omega
_{n_{0}}\left( \mathbf{\mathbf{k}_{\ast }}\right) +\omega _{n_{0}}^{\prime
}\left( \mathbf{\mathbf{k}_{\ast }}\right) \left( \mathbf{\mathbf{\eta }}%
\right) +\frac{1}{2}\omega _{n_{0}}^{\prime \prime }\left( \mathbf{\mathbf{k}%
_{\ast }}\right) \left( \mathbf{\eta }^{2}\right) ,  \label{gam3} \\
\gamma _{\left( 3\right) }\left( \mathbf{\eta }\right) =\omega
_{n_{0}}\left( \mathbf{\mathbf{k}_{\ast }}\right) +\omega _{n_{0}}^{\prime
}\left( \mathbf{\mathbf{k}_{\ast }}\right) \left( \mathbf{\mathbf{\eta }}%
\right) +\frac{1}{2}\omega _{n_{0}}^{\prime \prime }\left( \mathbf{\mathbf{k}%
_{\ast }}\right) \left( \mathbf{\eta }^{2}\right) +\frac{1}{6}\omega
_{n_{0}}^{\prime \prime \prime }\left( \mathbf{\mathbf{k}_{\ast }}\right)
\left( \mathbf{\eta }\right) ^{3},  \notag \\
\gamma _{\left( 4\right) }\left( \mathbf{\eta }\right) =\omega
_{n_{0}}\left( \mathbf{\mathbf{k}_{\ast }}\right) +\omega _{n_{0}}^{\prime
}\left( \mathbf{\mathbf{k}_{\ast }}\right) \left( \mathbf{\mathbf{\eta }}%
\right) +\frac{1}{2}\omega _{n_{0}}^{\prime \prime }\left( \mathbf{\mathbf{k}%
_{\ast }}\right) \left( \mathbf{\eta }^{2}\right) +\frac{1}{6}\omega
_{n_{0}}^{\prime \prime \prime }\left( \mathbf{\mathbf{k}_{\ast }}\right)
\left( \mathbf{\eta }\right) ^{3}+  \notag \\
\frac{1}{24}\omega _{n_{0}}^{\prime \prime \prime \prime }\left( \mathbf{%
\mathbf{k}_{\ast }}\right) \left( \mathbf{\eta }\right) ^{4}.  \notag
\end{gather}%
Substituing $\eta _{j}=-\mathrm{i}\partial _{j}$ into the polynomial $\gamma
_{\left( \nu \right) }\left( \mathbf{\eta }\right) $ we obtain the
differential operator $\gamma _{\left( \nu \right) }\left[ -\mathrm{i}\vec{%
\nabla}_{\mathbf{r}}\right] $, in particular,%
\begin{equation}
\gamma _{\left( 3\right) }\left[ -\mathrm{i}\vec{\nabla}_{\mathbf{r}}\right]
V=\omega _{n_{0}}\left( \mathbf{\mathbf{k}_{\ast }}\right) V-\mathrm{i}%
\sum_{m}\gamma _{m}\partial _{m}V-\frac{1}{2}\sum_{m,l}\gamma _{ml}\partial
_{m}\partial _{l}V+\frac{\mathrm{i}}{6}\sum_{m,l,j}\gamma _{mlj}\partial
_{m}\partial _{l}\partial _{j}V  \label{Gam1}
\end{equation}%
where $\gamma _{m}$, $\gamma _{ml}$ and $\gamma _{mlj}$ are the real-valued
coefficients of the linear form $\omega _{n_{0}}^{\prime }\left( \mathbf{%
\mathbf{k}_{\ast }}\right) $, the quadratic form $\omega _{n_{0}}^{\prime
\prime }\left( \mathbf{\mathbf{k}_{\ast }}\right) $ and the cubic form $%
\omega _{n_{0}}^{\prime \prime \prime }\left( \mathbf{\mathbf{k}_{\ast }}%
\right) $ respectively. In the simplest classical case when $\nu =2$ and the
problem is one-dimensional, i.e. $d=1$, as in (\ref{Si}), the operator $%
\gamma _{\left( 2\right) }\left( -\mathrm{i}\partial _{x}\right) $ has the
form 
\begin{equation}
\gamma _{\left( 2\right) }\left( -\mathrm{i}\partial _{x}\right) V=\omega
_{n_{0}}\left( \mathbf{\mathbf{k}_{\ast }}\right) V-\mathrm{i}\omega
_{n_{0}}^{\prime }\left( \mathbf{\mathbf{k}_{\ast }}\right) \partial _{x}V-%
\frac{1}{2}\omega _{n_{0}}^{\prime \prime }\left( \mathbf{\mathbf{k}_{\ast }}%
\right) \partial _{x}^{2}V  \label{Gam11}
\end{equation}%
with the symbol $\gamma _{\left( 2\right) }\left( \mathbf{\eta }\right) $
being defined by (\ref{gam3}). The polynomial $\gamma _{\left( \nu \right)
}\left( \mathbf{\eta }\right) $ is called the \emph{symbol}\ of the operator 
$\gamma _{\left( \nu \right) }\left[ -\mathrm{i}\vec{\nabla}_{\mathbf{r}}%
\right] $.

Let us introduce a general \emph{linear Schrodinger equation} of the form 
\begin{equation}
\partial _{t}Z\left( \mathbf{\mathbf{r}},t\right) =-\mathrm{i}\gamma
_{\left( \nu \right) }\left[ -\mathrm{i}\vec{\nabla}_{\mathbf{r}}\right]
Z\left( \mathbf{\mathbf{r}},t\right) ,\ Z\left( \mathbf{\mathbf{r}},t\right)
|_{t=0}=h_{\beta }\left( \mathbf{\mathbf{r}}\right) ,\ h_{\beta }\left( 
\mathbf{\mathbf{r}}\right) =h\left( \beta \mathbf{\mathbf{r}}\right) .
\label{Slin}
\end{equation}%
It can be solved exactly in terms of the Fourier transform, namely%
\begin{equation}
\hat{Z}\left( \mathbf{\eta },t\right) =\hat{h}_{\beta }\left( \mathbf{\eta }%
\right) \exp \left\{ -\mathrm{i}\gamma _{\left( \nu \right) }\left( \mathbf{%
\eta }\right) t\right\} ,  \label{Slin1}
\end{equation}%
The properties of the Fourier transform are discussed in Subsection 8.6 (see
(\ref{Ftransform}) for its definition). Let us also consider the classical 
\emph{nonlinear Schrodinger equation} 
\begin{equation}
\partial _{t}Z_{+}=-\mathrm{i}\gamma _{\left( 2\right) }\left[ -\mathrm{i}%
\vec{\nabla}_{\mathbf{r}}\right] Z_{+}+\alpha _{\pi }Q_{+}\left\vert
Z_{+}\right\vert ^{2}Z_{+},\ Z_{+}\left( \mathbf{r},t\right)
|_{t=0}=h_{+}\left( \beta \mathbf{\mathbf{r}}\right) =h_{+,\beta }\left( 
\mathbf{\mathbf{r}}\right) ,  \label{schx}
\end{equation}%
where $Q_{+}$ is a complex constant, and the factor $\alpha _{\pi }=3\alpha
\left( 2\pi \right) ^{2d}$ is introduced for notational consistentcy with
the related NLM (we have used (\ref{conjug}) to simplify (\ref{Si}), (\ref%
{Si1})).

The simplest \emph{Extended Nonlinear Schrodinger equations (ENLS)} are
given in (\ref{Znu}), (\ref{Znu-}), they have the form 
\begin{gather}
\partial _{t}Z_{+}=-\mathrm{i}\gamma _{\left( \nu \right) }\left[ -\mathrm{i}%
\vec{\nabla}_{\mathbf{r}}\right] Z_{+}+\alpha _{\pi }p_{+}^{\left[ \nu -2%
\right] }\left[ -\mathrm{i}\vec{\nabla}_{\mathbf{r}}\right] \left(
Z_{+}^{2}Z_{-}\right) ,  \label{genS} \\
\ Z_{+}\left( \mathbf{r},t\right) |_{t=0}=h_{+}\left( \beta \mathbf{\mathbf{r%
}}\right) =h_{+,\beta }\left( \mathbf{\mathbf{r}}\right) ,  \notag
\end{gather}%
\begin{gather}
\partial _{t}Z_{-}=\mathrm{i}\gamma _{\left( \nu \right) }\left[ \mathrm{i}%
\vec{\nabla}_{\mathbf{r}}\right] Z_{-}+\alpha _{\pi }p_{-}^{\left[ \sigma %
\right] }\left[ -\mathrm{i}\vec{\nabla}_{\mathbf{r}}\right] \left(
Z_{-}^{2}Z_{+}\right) ,  \label{genSm} \\
\ Z_{-}\left( \mathbf{r},t\right) |_{t=0}=h_{-}\left( \beta \mathbf{\mathbf{r%
}}\right) =h_{-,\beta }\left( \mathbf{\mathbf{r}}\right) ,  \notag
\end{gather}%
where $p_{+}^{\left[ \sigma \right] }\left[ -\mathrm{i}\vec{\nabla}_{\mathbf{%
r}}\right] \left( Z_{+}^{2}Z_{-}\right) $ is a linear differential operator
with constant coefficients of the order $\sigma =\nu -2$ with its symbol $%
p_{\pm }^{\left[ \sigma \right] }\left( \vec{q}^{\;\star }\right) $ being
defined by (\ref{psig}). This operator acts on the product $Z_{+}^{2}Z_{-}$.
The action of such an operator on the factors of the product $Z_{+}^{2}Z_{-}$
is defined by (\ref{psigVVV}). Note that this operator acts on all factors
of the product $Z_{+}^{2}Z_{-}=Z_{+}Z_{+}Z_{-}$, and that the variables $%
\mathbf{q}^{\prime },\mathbf{q}^{\prime \prime },\mathbf{q}^{\prime \prime
\prime }$ of the symbol are replaced respectively by the differentiations of
the first, the second and the third factor. In particular, according to (\ref%
{Q0}) 
\begin{equation}
p_{+}^{\left[ 0\right] }\left[ -\mathrm{i}\vec{\nabla}_{\mathbf{r}}\right]
\left( Z_{+}^{2}Z_{-}\right) =Q_{+}Z_{+}^{2}Z_{-},  \label{P0P1}
\end{equation}%
where the coefficient $Q_{+}$ is given in (\ref{Q0}). The first order
operator $p_{+}^{\left[ 1\right] }=p_{+}^{\left[ 0\right] }+p_{1,+}$ where $%
p_{+}^{\left[ 0\right] }$ is given above and the symbol of $p_{1,+}$ is
defined by (\ref{p1z}), i.e.%
\begin{equation}
p_{1,\zeta }^{\left[ 1\right] }\left( \vec{q}^{\;\star }\right) =a_{11,\zeta
}\cdot \mathbf{q}^{\prime }+a_{12,\zeta }\cdot \mathbf{q}^{\prime \prime
}+a_{13,\zeta }\cdot \mathbf{q}^{\prime \prime \prime }.
\end{equation}%
The corresponding operator acts as follows 
\begin{gather}
p_{+}^{\left[ 1\right] }\left[ -\mathrm{i}\vec{\nabla}_{\mathbf{r}}\right]
\left( Z_{+}^{2}Z_{-}\right) =p_{+}^{\left[ 1\right] }\left[ \vec{\nabla}_{%
\mathbf{r}}\right] \left( Z_{+}Z_{+}Z_{-}\right) =  \label{P1} \\
Z_{+}Z_{-}\left( a_{11,+}+a_{12,+}\right) \cdot \nabla _{\mathbf{r}%
}Z_{+}+Z_{+}^{2}a_{13,+}\cdot \nabla _{\mathbf{r}}Z_{-},  \notag
\end{gather}%
where vectors $a_{12,+}\ $and $a_{13,+}$ are defined in (\ref{p01}) and (\ref%
{a11}) for $\zeta =+$, and the components of the vectors are complex.
Details concerning the values of the coefficients and the properties of the
equations will be considered in another paper. Observe that the extended NLS
equation (\ref{genS}) turns into the classical one (\ref{schx}) if we set $%
\nu =2,$ $\sigma =0$ \ and use (\ref{conjug}).

When comparing the solution $Z_{\pm }\left( \mathbf{r},t\right) $ to the NLS
(\ref{genS}) with a solution to the NLM we will need the following scaled
version of the function $Z_{\pm }\left( \mathbf{r},t\right) $ $\ $%
\begin{equation}
Z_{\beta ,\pm }\left( \mathbf{r},t\right) =Z_{\pm }\left( \frac{\mathbf{r}}{%
\beta },t\right) .  \label{ZZr}
\end{equation}%
The relation (\ref{ZZr}) between $Z_{\pm }\left( \mathbf{r},t\right) $ and
its scaled version $Z_{\beta ,\pm }\left( \mathbf{r},t\right) $ implies the
following relation between their Fourier transforms as defined by (\ref%
{Ftransform}): 
\begin{equation}
\widehat{Z}_{\pm }\left( \mathbf{\xi },t\right) =\int_{\mathbf{R}^{d}}%
\mathrm{e}^{-\mathrm{i}\beta \mathbf{r}\cdot \frac{1}{\beta }\mathbf{\xi }%
}Z_{\beta ,\pm }\left( \beta \mathbf{r},t\right) \,\mathrm{d}\mathbf{r}%
=\beta ^{-d}\widehat{Z}_{\beta ,\pm }\left( \frac{\mathbf{\xi }}{\beta }%
,t\right) .  \label{Fourbet}
\end{equation}%
It is convenient to recast the general NLS equation (\ref{genS}), (\ref%
{genSm}) as an equation for the quantity $Z_{\beta ,\pm }\left( \mathbf{r}%
,t\right) $, namely%
\begin{equation}
\partial _{t}Z_{\beta ,+}=-\mathrm{i}\gamma _{\left( \nu \right) }\left[ -%
\mathrm{i}\beta \nabla _{\mathbf{r}}\right] Z_{\beta ,+}+\alpha _{\pi
}p_{+}^{\left[ \sigma \right] }\left[ -\mathrm{i}\beta \nabla _{\mathbf{r}}%
\right] \left( Z_{\beta ,+}^{2}Z_{\beta ,-}\right) ,\ Z_{\beta ,+}\left( 
\mathbf{r},t\right) |_{t=0}=h_{+}\left( \mathbf{r}\right) ,  \label{genS1}
\end{equation}%
\begin{equation}
\partial _{t}Z_{\beta ,-}=\mathrm{i}\gamma _{\left( \nu \right) }\left[ 
\mathrm{i}\beta \nabla _{\mathbf{r}}\right] Z_{\beta ,-}+\alpha _{\pi
}p_{-}^{\left[ \sigma \right] }\left[ -\mathrm{i}\beta \vec{\nabla}_{\mathbf{%
r}}\right] \left( Z_{\beta ,-}^{2}Z_{\beta ,+}\right) ,\ Z_{\beta ,-}\left( 
\mathbf{r},t\right) |_{t=0}=h_{-}\left( \mathbf{r}\right) .  \label{genS1-}
\end{equation}%
Obviously, the initial data for the rescaled equation do not depend on $%
\beta $, but the coefficients explicitly depend on $\beta $.

\subsection{Total error of the approximation of the NLM by an NLS}

In this section we outline how we estimate the total error of the
approximation of the NLM with an NLS. For simplicity we discuss the case
when the NLM contains purely cubic nonlinearity, the weak dispersion case (%
\ref{nonM1}) and we use a second-order NLS (that is with the order of linear
part $\nu =2$) for the approximation.

An exact solution $\mathbf{U}\left( \mathbf{r},t\right) $ of the NLM
corresponding to an excitation current composed from a doublet of modes, as
in (\ref{Jn}), (\ref{jnn1}), splits naturally into two parts corresponding
to the directly and indirectly excited modes. The first part involves the
directly excited modes (excited through the linear medium response) with
modal amplitudes $U_{\zeta ,n}\left( \mathbf{k}_{\ast },t\right) $ with $%
n=n_{0}$,\ $\left\vert \mathbf{k-}\zeta \mathbf{k}_{\ast }\right\vert \leq
\pi _{0}$ as in (\ref{kkstar}). Their magnitude is $O\left( 1\right) $. The
second part consists of the indirectly excited modes, which are excited
solely through the nonlinear medium response, and this part involves modes
with either $n\neq n_{0}$ or \ $\left\vert \mathbf{k-}\zeta \mathbf{k}_{\ast
}\right\vert >\pi _{0}$. As was explained in Section 3.2 the magnitude of
the indirectly excited modes is estimated by $O\left( \varrho \right) $. The
approximate solution includes both directly and indirectly excited modes. We
take a solution $Z_{\zeta }\left( \mathbf{r},t\right) $ of the NLS (\ref%
{GNLS+}), (\ref{GNLS-}) and consider an exact solution $\mathbf{U}\left( 
\mathbf{r},t\right) $ related to $Z_{\zeta }\left( \mathbf{r},t\right) $
through properly chosen excitation currents. The currents are based on the
initial data of the NLS, see Section 5.2. We define the approximate solution 
$\mathbf{U}_{Z}\left( \mathbf{r},t\right) $ of the NLM by (\ref{UNLS}), (\ref%
{Uindn}), (\ref{Uindn0}).

The nonlinear interactions of the directly excited modes with themselves are
of order $O\left( 1\right) $ as in the case of the classical NLS scaling (%
\ref{thet1alph}) and they, of course, are taken into account. We approximate
the directly excited modes by appropriate solutions of \ the NLS. \emph{To
match/corresond the NLM and the NLS we use in concert the following two
options: (i) setting up the excitation currents; (i) choosing the
coefficients of the NLS.} The linear part of the NLS is obtained based on
the Taylor expansion of the dispersion relation, see (\ref{gam3}). We choose
the coefficients at the nonlinear terms of the NLS\ so that the first
nonlinear response of the NLM exactly matches with the first nonlinear
response of the NLS.

There are the following sources of the approximation error. First, we
replace the causal integral nonlinear operators which enter the operator $%
\mathcal{F}_{\text{NL}}$ by the instantaneous operators described by the
nonlinear susceptibilities, see Section 6 and \cite{BF5} for details. It is
necessary since the nonlinearity in the NLS is instantaneous. Second, we
neglect the impact of indirectly excited modes onto directly excited. More
precisely, we throw away all terms in (\ref{FNR1}) with $n^{\prime }\neq
n_{0}$, $n^{\prime \prime }\neq n_{0}$, $n^{\prime \prime \prime }\neq n_{0}$%
. This is necessary if we consider dynamics of modal amplitudes of the only
one band $n=n_{0}$ independently of all other bands. Third, we replace exact
dispersion relation $\omega _{n_{0}}\left( \mathbf{k}\right) $ by its Taylor
polynomial at $\mathbf{k}=\mathbf{k}_{\ast }$. Fourth, we replace
frequency-dependent susceptibilities by their values at $\omega =\omega
_{n_{0}}\left( \mathbf{k}_{\ast }\right) $; this is necessary since the
nonlinearity in the NLS is not frequency-dependent.

Note that all mentioned replacements and modifications affect the FNLR
exactly the same way as the exact solution. The only difference is that for
the FNLR the operators we mentioned above are applied to the linear
approximation $\mathbf{U}^{\left( 0\right) }$ whereas for the exact solution
they act on $\mathbf{U}$ itself. This explains why the choice of the
coefficients of the NLS based on matching FNLR of NLS and NLM\ gives a good
approximation of exact solutions of NLM\ even in the case of the classical
NLS scaling $\varrho \sim \alpha \sim \beta ^{2}$ for times $t\sim \frac{1}{%
\varrho }$. Since we match only the zero and the first order terms in $%
\alpha $ (the linear response and the FNLR), the higher order terms of order 
$\alpha ^{2}$ could create an additional discrepancy of order one, but the
effect of the higher order terms is effectively eliminated since \emph{we
approximate the solution of the NLM\ by the exact solution of the NLS}
rather than by the principal terms of the expansion in $\alpha $ of the
solution of the NLS, see Section 7 for details.

The total approximation error of the approximation of the exact solution $%
\mathbf{U}\left( \mathbf{r},t\right) $ by the approximate solution $\mathbf{U%
}_{Z}\left( \mathbf{r},t\right) $ defined by (\ref{UNLS}), (\ref{Uindn}), (%
\ref{Uindn0}) \ on the time interval (\ref{t0tt2}) in the case $\nu =2$
consists of the following components:

\begin{enumerate}
\item the error of the NLS\ approximation of the directly excited modes on
the interval $0\leq t\leq \frac{\tau _{\ast }}{\varrho }$ is estimated by $%
O\left( \beta \right) $;

\item the error of the zero-order time-harmonic approximation (\ref{suserror}%
) to the causal integral is estimated by $O\left( \varrho \right) $;

\item the error of\ the FNLR\ approximation (\ref{Uindn}), (\ref{Uindn0}) of
the indirectly excited modes, in particular through non-FM interactions is
estimated by $O\left( \varrho \right) $;

\item the error from the impact of indirectly excited modes onto the
directly excited modes (interband interactions) is estimated by $O\left(
\varrho \right) $;

\item the error of the polynomial approximation of the dispersion relation
in the weakly dispersive case $O\left( \frac{\beta ^{3}}{\varrho }\right) $
which gives $O\left( \beta \right) $ in the case of classical NLS scaling.
\end{enumerate}

The inequality $t\leq \frac{c_{0}}{\alpha }$ in (\ref{t0tt2}) ensures that
our analysis is applicable, see (\ref{difexp}). Consequently the total error
of approximation of a solution to the NLM by a solution to the classical NLS
when $\nu =2$ is of order 
\begin{equation}
O\left( \beta \right) +O\left( \varrho \right) .  \label{Toter}
\end{equation}%
The error estimate in the item 1 in above list is addressed below in this
section. The error estimate in the item 2 is discussed in Section 6.2. The
error estimate in the item 3 was discussed in Subsections 1.2 and 3.2. The
error in the item 4 caused by interband interactions includes higher order
terms of power expansions (\ref{uMv1}) and will be considered in a separate
paper. The error in the item 5 was discussed in Section 4.1. There are also
the negligible errors $O\left( \theta ^{N_{3}+1}\right) $ in the case (\ref%
{nonM}) when $\theta =\frac{\varrho }{\beta ^{2}}<1$ with arbitrary large $%
N_{3}$ and the error from the contribution of the cutoff function $\Psi $ of
the order $O\left( \beta ^{N_{\Psi }-d}\right) $ (see Remark in Subsection
4.1.2) where $N_{\Psi }$ is arbitrarily large; these errors are technical by
nature and are negligible at any order of accuracy.

Reduction of the errors by means of using extended NLS instead of classical
NLS\ is discussed in Subsection 1.2.\ 

\subsection{The linear response and the FNLR for an NLS}

To provide a basis for relating the NLM and an NLS using their linear and
the first nonlinear responses we need to construct for the general NLS (\ref%
{genS}) the linear and the first nonliner responses along the same lines as
we did for the NLM. For that we (i) single out the linear part of the
general NLS (\ref{genS}) and carry out its spectral analysis; (ii) introduce
the source term in the NLS which replaces the initial condition and study
the corresponding solution using the framework we developed for the NLM. The
source term is introduced based on the initial data of the NLS so that it:
(i) generates the same solution as the initial data; (ii) has the form of an
almost time-harmonic function consistent with (\ref{UU1}), (\ref{jjt2}), (%
\ref{jnn1}). The importance of the the relation between the excitation
current for the NLM\ and initial data for the NLS can be seen from the
following simple observation. When we compare solutions of two differential
equations the difference of two solutions originates from two sources: the
difference between the equations and the difference between the initial
data. Even when the equation is the same, the difference of solutions is
proportional to the difference of the initial data. Since we study
approximation of the solutions of the NLM by solutions of the NLS with a
high precision, and study effects of additional terms in the ENLS on the
accuracy of approximation, we want to eliminate completely the source of
differences which comes from the initial data. This is not trivial since the
initial data $h\left( \mathbf{r}\right) $ for the NLS are instantaneously
prescribed at $t=0$ and their counterpart--excitation currents $\mathbf{J}%
\left( \mathbf{r},t\right) $ for the NLM-- are defined on a time interval $%
0\leq t\leq \frac{\tau _{0}}{\varrho }$. That requires consideration of
technical issues, but the bottom line is that exact matching is possible for
arbitrary choice of \ $h$. We remind also that the constructed excitation
current $\mathbf{J}\left( \mathbf{r},t\right) $ vanishes for $t\geq \frac{%
\tau _{0}}{\varrho }$ \ and only after that time we compare solutions of the
NLM\ and the NLS.

In the subsequent treatment of the NLS we use the modal decomposition for
its analysis. Since the linear part of the NLS (\ref{genS}) is the
differential operator $-\mathrm{i}\gamma _{\left( \nu \right) }\left[ -%
\mathrm{i}\vec{\nabla}_{\mathbf{r}}\right] $ with constant coefficients, the
corresponding eigenmodes are just plane waves. Consequently, here we use the
plane waves and the standard Fourier transform (\ref{Ftransform}) instead of
the Bloch modes and the Floquet-Bloch transform.

\subparagraph{Source term for the NLS.}

First, let us show how the solution to the initial value problem (\ref{genS}%
) can be obtained as a solution to a similar differential equation with zero
initial data and a source term $f_{+}\left( \mathbf{r},t\right) $ based on $%
h_{+}\left( \beta \mathbf{\mathbf{r}}\right) $. This form of the solution
would be consistent with the form of the NLM (\ref{MXshort}). The general
form of such a nonlinear equation with a source is provided by (\ref{Veq}).
Hence, in the case of (\ref{genS}) the relevant evolution equation with a
source is%
\begin{equation}
\partial _{t}V_{+}=-\mathrm{i}\gamma _{\left( \nu \right) }\left[ -\mathrm{i}%
\vec{\nabla}_{\mathbf{r}}\right] V_{+}+\alpha _{\pi }p_{+}^{\left[ \sigma %
\right] }\left[ -\mathrm{i}\vec{\nabla}_{\mathbf{r}}\right] \left(
V_{+}^{2}V_{-}\right) -f_{+}\left( \mathbf{r},t\right) ,\ V_{+}=0\ \text{for 
}t\leq 0,  \label{Schgam}
\end{equation}%
and we want to find the source $f_{+}\left( \mathbf{r},t\right) $ so that
the solution $V_{+}\left( \mathbf{r},t\right) $ to (\ref{Schgam}) would be
equal to $Z_{+}\left( \mathbf{r},t\right) $ for $t\geq \tau _{0}/\varrho $.
The final form of the desired source $f_{+}\left( \mathbf{r},t\right) $ is
provided by the formula (\ref{frt}), and it is constructed as follows. We
begin with picking up a smooth real-valued function $\psi \left( \tau
\right) $ having the same properties as the function defined by (\ref{psi1}%
), (\ref{psi0}), namely 
\begin{equation}
0\leq \psi \left( \tau \right) \leq 1,\ \psi \left( \tau \right) =0,\ \tau
\leq 0;\ \psi \left( \tau \right) =1,\ \tau \geq \tau _{0}>0.  \label{psii}
\end{equation}%
Then taking the functions $Z_{\pm }\left( \mathbf{r},t\right) $ which solve
problem (\ref{genS}), (\ref{genSm}) we introduce 
\begin{equation}
V_{\pm }\left( \mathbf{r},t\right) =\psi \left( \varrho t\right) Z_{\pm
}\left( \mathbf{r},t\right) .  \label{VZ}
\end{equation}%
Multiplying the equation (\ref{genS}) by $\psi \left( t\right) $ we can
readily verify that $V_{+}$ is a solution of the equation 
\begin{equation}
\partial _{t}V_{+}-\varrho \psi ^{\prime }\left( \varrho t\right) Z_{+}=-%
\mathrm{i}\gamma _{\left( \nu \right) }\left[ -\mathrm{i}\vec{\nabla}_{%
\mathbf{r}}\right] V_{+}+\alpha _{\pi }\psi \left( \varrho t\right) p_{+}^{%
\left[ \sigma \right] }\left[ -\mathrm{i}\vec{\nabla}_{\mathbf{r}}\right]
\left( Z_{+}^{2}Z_{-}\right) .  \label{Schpsi}
\end{equation}%
Notice that in view of (\ref{VZ})%
\begin{equation}
\psi \left( \varrho t\right) Z_{+}^{2}Z_{-}-V_{+}^{2}V_{-}=\left( \psi
\left( \varrho t\right) -\psi ^{3}\left( \varrho t\right) \right)
Z_{+}^{2}Z_{-},
\end{equation}%
implying that the equation (\ref{Schpsi}) can be recast in the form of (\ref%
{Schgam}) with%
\begin{equation}
f_{+}=-\varrho \psi ^{\prime }\left( \varrho t\right) Z_{+}-\alpha _{\pi
}\left( \psi -\psi ^{3}\right) p_{+}^{\left[ \sigma \right] }\left[ -\mathrm{%
i}\vec{\nabla}_{\mathbf{r}}\right] \left( Z_{+}^{2}Z_{-}\right) .
\label{frt}
\end{equation}%
Evidently, in view of (\ref{psii}) and (\ref{VZ}) we have%
\begin{equation}
f_{\pm }\left( \mathbf{r},t\right) =0\text{ when }t\geq \tau _{0}/\varrho \ 
\text{or }t\leq 0,  \label{feq0}
\end{equation}%
\begin{equation}
V_{\pm }\left( \mathbf{r},t\right) =\psi \left( \varrho t\right) Z_{\pm
}\left( \mathbf{r},t\right) ,\ V_{\pm }\left( \mathbf{r},t\right) =Z_{\pm
}\left( \mathbf{r},t\right) \ \text{when }t\geq \tau _{0}/\varrho .
\label{veqz}
\end{equation}%
\emph{Notice that the equalities (\ref{veqz}) establish the relation between
the NLS as the initial value problem (\ref{genS}) and the NLS (\ref{Schgam})
with a source term}.

Using the equation (\ref{genSm}) and arguments similar to the above we get
the following equation for $V_{-}$: 
\begin{equation}
\partial _{t}V_{-}=\mathrm{i}\gamma _{\left( \nu \right) }\left( \mathrm{i}%
\beta \nabla _{\mathbf{r}}\right) Z_{-}+\alpha _{\pi }p_{-}^{\left[ \sigma %
\right] }\left[ -\mathrm{i}\vec{\nabla}_{\mathbf{r}}\right] \left(
Z_{-}^{2}Z_{-}^{\ast }\right) -f_{-}\left( \mathbf{r}\right) ,  \label{V-eq}
\end{equation}%
with the function $V_{-}\left( \mathbf{r},t\right) $ satisfying the identity%
\begin{equation}
V_{-}\left( \mathbf{r},t\right) =Z_{-}\left( \mathbf{r},t\right) ,\ t\geq
\tau _{0}/\varrho .  \label{VeqZ}
\end{equation}%
Notice that similarly to the NLM case the solution $Z_{\zeta }$, $\zeta =\pm 
$ $\ $to the NLS (\ref{genS}), (\ref{genSm}) admits the following expansion 
\begin{equation}
Z_{\zeta }=Z_{\zeta }^{\left( 0\right) }+\alpha Z_{\zeta }^{\left( 1\right)
}+\alpha ^{2}Z_{\zeta }^{\left( 2\right) }+\ldots ,\;\;\zeta =\pm .
\label{Zalpha}
\end{equation}%
In the expansion (\ref{Zalpha}) the term $Z_{+}^{\left( 0\right) }\left( 
\mathbf{r},t\right) $ evidently is a solution of the equation (\ref{genS})
with $\alpha =0$, namely 
\begin{equation}
\partial _{t}Z_{\zeta }^{\left( 0\right) }=-\mathrm{i}\zeta \gamma _{\left(
\nu \right) }\left[ -\mathrm{i}\beta \zeta \nabla _{\mathbf{r}}\right]
Z_{\zeta }^{\left( 0\right) },\ Z_{\zeta }^{\left( 0\right) }\left( 0\right)
=h_{\zeta ,\beta }\left( \mathbf{r}\right) ,\ h_{\zeta ,\beta }\left( 
\mathbf{r}\right) =h_{\zeta }\left( \beta \mathbf{r}\right) ,  \label{Z0bet}
\end{equation}%
and can be interpreted as the \emph{linear response} corresponding to (\ref%
{genS}). Its Fourier transform (\ref{Ftransform}) satisfies 
\begin{equation}
\hat{Z}_{\zeta }^{\left( 0\right) }\left( \mathbf{\xi },t\right) =\exp
\left( -\mathrm{i}\zeta \gamma _{\left( \nu \right) }\left( \zeta \mathbf{%
\xi }\right) t\right) \hat{h}_{\zeta ,\beta }\left( \mathbf{\xi }\right) ,\ 
\hat{h}_{\zeta ,\beta }\left( \mathbf{\xi }\right) =\beta ^{-d}\hat{h}%
_{\zeta }\left( \frac{1}{\beta }\mathbf{\xi }\right) ,\;\;\zeta =\pm ,
\label{Z0}
\end{equation}%
with the symbol $\gamma _{\left( \nu \right) }\left( \mathbf{\xi }\right) $
being defined by (\ref{Tayom}).

The function $Z_{\zeta }^{\left( 1\right) }$ in the expansion (\ref{Zalpha})
is the \emph{first nonlinear response} (FNLR) of (\ref{genS}), and based on (%
\ref{genS}) and (\ref{Zalpha}) one can verify that the FNLR $Z_{\zeta
}^{\left( 1\right) }$,$\zeta =\pm $, solves the following initial value
problem%
\begin{equation}
\partial _{t}Z_{\zeta }^{\left( 1\right) }=-\zeta \mathrm{i}\gamma _{\left(
\nu \right) }\left[ -\mathrm{i}\zeta \vec{\nabla}_{\mathbf{r}}\right]
Z_{\zeta }^{\left( 1\right) }+\alpha _{\pi }p_{\zeta }^{\left[ \sigma \right]
}\left[ -\mathrm{i}\vec{\nabla}_{\mathbf{r}}\right] \left( Z_{\zeta
}^{\left( 0\right) 2}Z_{\zeta }^{\left( 0\right) \ast }\right) ,\ Z_{\zeta
}^{\left( 1\right) }\left( 0\right) =0.  \label{Z1eq}
\end{equation}%
Using (\ref{veqz}) and (\ref{Zalpha}) we then obtain the following expansion
for the function $V_{+}\left( \mathbf{r},t\right) $ 
\begin{equation}
V_{+}\left( \mathbf{r},t\right) =\psi \left( \varrho t\right) Z_{+}\left( 
\mathbf{r},t\right) =\psi \left( \varrho t\right) Z_{+}^{\left( 0\right)
}\left( \mathbf{r},t\right) +\alpha \psi \left( \varrho t\right)
Z_{+}^{\left( 1\right) }\left( \mathbf{r},t\right) +\alpha ^{2}\psi \left(
\varrho t\right) Z_{+}^{\left( 2\right) }\left( \mathbf{r},t\right) +\ldots
\label{VZ01}
\end{equation}%
or, in other words%
\begin{eqnarray}
V_{+}\left( \mathbf{r},t\right) &=&V_{+}^{\left( 0\right) }\left( \mathbf{r}%
,t\right) +\alpha V_{+}^{\left( 1\right) }\left( \mathbf{r},t\right) +\alpha
^{2}V_{+}^{\left( 2\right) }\left( \mathbf{r},t\right) +\ldots  \label{V0V1}
\\
V_{+}^{\left( 0\right) }\left( \mathbf{r},t\right) &=&\psi \left( \varrho
t\right) Z_{+}^{\left( 0\right) }\left( \mathbf{r},t\right) ,  \label{V0psi}
\\
V_{+}^{\left( 1\right) }\left( \mathbf{r},t\right) &=&\psi \left( \varrho
t\right) Z_{+}^{\left( 1\right) }\left( \mathbf{r},t\right) ,  \label{V1psi}
\end{eqnarray}%
where $V_{+}^{\left( 0\right) }\left( \mathbf{r},t\right) $ and $%
V_{+}^{\left( 1\right) }\left( \mathbf{r},t\right) $ are respectively the
linear and the first nonlinear responses corresponding to the exact solution 
$V_{+}\left( \mathbf{r},t\right) $ of the NLS (\ref{Schgam}) with the sourse
(\ref{frt}). Similar statments hold for for $\zeta =-$.

One can verify that similar arguments hold for a general ENLS system
described by the equation 
\begin{equation}
\partial _{t}\vec{Z}=\mathcal{L}\vec{Z}+\alpha F^{\left( 3\right) }\left( 
\vec{Z}\right) ,\ \vec{Z}\left( 0\right) =\vec{h},  \label{ZLF}
\end{equation}%
with a\ cubic nonlinearity $F^{\left( 3\right) }\left( c\vec{Z}\right)
=c^{3}F^{\left( 3\right) }\left( \vec{Z}\right) $ where $c$ is a real
constant. Indeed, introducing $\vec{V}\left( t\right) =\psi \left( \varrho
t\right) \vec{Z}\left( t\right) $ we find that it satisfies the equation (%
\ref{Veq}), i.e.%
\begin{equation}
\partial _{t}\vec{V}=-\mathrm{i}\mathcal{L}\vec{V}+\alpha F^{\left( 3\right)
}\left( \vec{V}\right) -\vec{f}
\end{equation}%
with 
\begin{equation}
\vec{f}\left( t\right) =-\varrho \psi ^{\prime }\left( \varrho t\right) \vec{%
Z}-\alpha \left( \psi -\psi ^{3}\right) F^{\left( 3\right) }\left( \vec{Z}%
\right) .  \label{fF}
\end{equation}%
If the nonlinearity $F^{\left( 3\right) }\left( \vec{Z}\right) $ involves
the time derivatives of $\vec{Z}$ as in (\ref{GNLS1+}), (\ref{GNLS1-}) then
an analysis shows that the source term has the following form 
\begin{equation}
\vec{f}\left( t\right) =-\varrho \psi ^{\prime }\left( \varrho t\right) \vec{%
Z}-\alpha \left( \psi -\psi ^{3}\right) F^{\left( 3\right) }\left( \vec{Z}%
\right) +\alpha \varrho \psi ^{\prime }\left( \varrho t\right) F_{1}^{\left(
3\right) }\left( \varrho t,\vec{Z}\right) ,  \label{fFt}
\end{equation}%
where $F_{1}^{\left( 3\right) }\left( \varrho t,\vec{Z}\right) $ is a cubic
nonlinearity obtainable from $F^{\left( 3\right) }\left( \vec{Z}\right) $ by
a straightforward computation. An elementary examination shows that in the
both cases (\ref{fF}) and (\ref{fFt}) the source $\vec{f}\left( t\right) $
has the property (\ref{feq0}).

\textbf{Remark.} After the following change of variables 
\begin{equation}
Z_{+,\beta }\left( \mathbf{r}\right) =Z_{+}\left( \frac{\mathbf{r}}{\beta }%
\right) ,\;Z_{+,\beta }\left( \beta \mathbf{r}\right) =Z_{+}\left( \mathbf{r}%
\right) ,\ V_{+,\beta }\left( \mathbf{r}\right) =V_{+}\left( \frac{\mathbf{r}%
}{\beta }\right) ,V_{+,\beta }\left( \beta \mathbf{r}\right) =V_{+}\left( 
\mathbf{r}\right) ,
\end{equation}%
we obtain a solution $Z_{+,\beta }$ to the equation (\ref{genS1}) and a
solution $V_{+,\beta }$ to the following equation 
\begin{equation}
\partial _{t}V_{+,\beta }=-\mathrm{i}\gamma _{\left( \nu \right) }\left[ -%
\mathrm{i}\beta \nabla _{\mathbf{r}}\right] V_{+,\beta }+\alpha _{\pi
}p_{+}^{\left[ \sigma \right] }\left[ -\mathrm{i}\beta \nabla _{\mathbf{r}}%
\right] \left( V_{+,\beta }^{2}V_{-,\beta }\right) -f_{+}\left( \frac{%
\mathbf{r}}{\beta }\right) .  \label{V1eq}
\end{equation}%
The Fourier transforms $\hat{V}_{+,\beta }\left( \mathbf{\xi }\right) $ of $%
V_{+,\beta }\left( \mathbf{r}\right) $ and $\hat{V}_{+}\left( \mathbf{\xi }%
\right) $ of $V_{+}\left( \mathbf{r}\right) $ are related by the identity%
\begin{equation}
\hat{V}_{+}\left( \mathbf{\xi }\right) =\beta ^{-d}\hat{V}_{+,\beta }\left( 
\frac{\mathbf{\xi }}{\beta }\right) .  \label{vV1}
\end{equation}

\subparagraph{The linear response.}

To find the modal representation $\hat{V}_{+}^{\left( 0\right) }$ for the
linear response $V_{+}^{\left( 0\right) }$ as defined by (\ref{V0V1}) and (%
\ref{V0psi}) we use (\ref{Z0}) which implies 
\begin{equation}
\hat{V}_{+}^{\left( 0\right) }\left( \beta \mathbf{q},t\right) =\hat{v}%
_{+}^{\left( 0\right) }\left( \beta \mathbf{q},\tau \right) \mathrm{e}^{-%
\mathrm{i}\gamma _{\left( \nu \right) }\left( \beta \mathbf{q}\right)
t}=\psi \left( \tau \right) \beta ^{-d}\hat{h}_{+}\left( \mathbf{q}\right) 
\mathrm{e}^{-\mathrm{i}\gamma _{\left( \nu \right) }\left( \beta \mathbf{q}%
\right) t},\ \tau =\varrho t.  \label{V0s}
\end{equation}%
Observe now that the above expression for $\hat{V}_{+}^{\left( 0\right)
}\left( \beta \mathbf{q},t\right) $ coincides with the coefficient $\tilde{U}%
_{\bar{n}}^{\left( 0\right) }\left( \mathbf{k}_{\ast }+Y\left( \beta \mathbf{%
q}\right) ,t\right) $ for $\zeta =+$ as determined by (\ref{U0cap}), (\ref%
{u0s0}) and (\ref{jnn2}) implying%
\begin{equation}
\Psi \left( Y\left( \beta \mathbf{q}\right) \right) \hat{V}_{+}^{\left(
0\right) }\left( \beta \mathbf{q},t\right) =\tilde{U}_{\bar{n}}^{\left(
0\right) }\left( \mathbf{k}_{\ast }+Y\left( \beta \mathbf{q}\right)
,t\right) ,  \label{MSch0}
\end{equation}%
and for $\beta \rightarrow 0$ according to (\ref{U0cap1}) 
\begin{gather}
\hat{V}_{+}^{\left( 0\right) }\left( \beta \mathbf{q},t\right) =\tilde{U}_{%
\bar{n}}^{\left( 0\right) }\left( \mathbf{k}_{\ast }+Y\left( \beta \mathbf{q}%
\right) ,t\right) +O\left( \beta ^{N_{\Psi }}\right)  \label{vu0} \\
=\psi \left( \tau \right) \beta ^{-d}\hat{h}_{+}\left( \mathbf{q}\right) 
\mathrm{e}^{-\mathrm{i}\gamma _{\left( \nu \right) }\left( \beta \mathbf{q}%
\right) t}=\tilde{u}_{n_{0},+}^{\left( 0\right) }\left( \mathbf{k}_{\ast
}+Y\left( \beta \mathbf{q}\right) ,\tau \right) \mathrm{e}^{-\mathrm{i}%
\gamma _{\left( \nu \right) }\left( \beta \mathbf{q}\right) t}+O\left( \beta
^{N_{\Psi }}\right) ,  \notag
\end{gather}%
where $N_{\Psi }$ can be taken as large as we please, and the term $O\left(
\beta ^{N_{\Psi }}\right) $ comes from the cutoff function $\Psi $ (see the
Remark in Subsection 4.1.1).

\subparagraph{The first nonlinear response.}

The first nonlinear response (FNLR) $V_{+}^{\left( 1\right) }$ similarly to
its countepart $\mathbf{U}^{\left( 1\right) }$ is defined as the
proportional to $\alpha $ term (\ref{V1psi}) in the expansion (\ref{V0V1})
of the solution $V_{+}$ with respect to $\alpha $. Though the FNRL $%
V_{+}^{\left( 1\right) }$ is already described by (\ref{V1psi}), it is
useful to derive a differential equation with a source for $V_{+}^{\left(
1\right) }$ based on (\ref{Schgam}), (\ref{frt}) and to do the same for the
linear response $V_{+}^{\left( 0\right) }$ defined by (\ref{V0psi}). Notice
that (\ref{frt}) and (\ref{Zalpha}) imply 
\begin{equation}
f_{+}=-\varrho \psi ^{\prime }\left( \varrho t\right) \left[ Z_{+}^{\left(
0\right) }\left( \mathbf{r},t\right) +\alpha Z_{+}^{\left( 1\right) }\right]
-\alpha _{\pi }\left( \psi -\psi ^{3}\right) p_{+}^{\left[ \sigma \right] }%
\left[ -\mathrm{i}\vec{\nabla}_{\mathbf{r}}\right] \left( \left(
Z_{+}^{\left( 0\right) }\right) ^{2}Z_{-}^{\left( 0\right) }\right) +O\left(
\alpha ^{2}\right) .  \label{f0f1}
\end{equation}%
Substituting $\alpha =0$ in (\ref{Schgam}) we find that the linear response $%
V_{+}^{\left( 0\right) }$ solves the following problem%
\begin{equation}
\partial _{t}V_{+}^{\left( 0\right) }=-\mathrm{i}\gamma _{\left( \nu \right)
}\left[ -\mathrm{i}\vec{\nabla}_{\mathbf{r}}\right] V_{+}^{\left( 0\right)
}+\varrho \psi ^{\prime }\left( \varrho t\right) Z_{+}^{\left( 0\right)
}\left( \mathbf{r},t\right) ,\ V_{+}^{\left( 0\right) }=0.  \label{f0f2}
\end{equation}%
The equation for $V_{+}^{\left( 1\right) }$ can be similarly obtained from (%
\ref{Schgam}), (\ref{f0f1}) by collecting terms proportinal to $\alpha $
yielding%
\begin{equation}
\partial _{t}V_{+}^{\left( 1\right) }=-\mathrm{i}\gamma _{\left( \nu \right)
}\left[ -\mathrm{i}\vec{\nabla}_{\mathbf{r}}\right] V_{+}^{\left( 1\right) }+%
\frac{\alpha _{\pi }}{\alpha }p_{+}^{\left[ \sigma \right] }\left[ -\mathrm{i%
}\vec{\nabla}_{\mathbf{r}}\right] \left( V_{+}^{\left( 0\right)
2}V_{-}^{\left( 0\right) }\right) -f_{+}^{\left( 1\right) },  \label{V1eq1}
\end{equation}%
with the source%
\begin{equation}
f_{+}^{\left( 1\right) }\left( \mathbf{r},t\right) =-\varrho \psi ^{\prime
}\left( \varrho t\right) Z_{+}^{\left( 1\right) }-\frac{\alpha _{\pi }}{%
\alpha }\left( \psi \left( \varrho t\right) +\psi ^{3}\left( \varrho
t\right) \right) p_{+}^{\left[ \sigma \right] }\left[ -\mathrm{i}\vec{\nabla}%
_{\mathbf{r}}\right] \left( Z_{+}^{\left( 0\right) 2}Z_{-}^{\left( 0\right)
}\right) .  \label{f1}
\end{equation}%
Consequently, the Fourier transform $\hat{f}_{+}^{\left( 1\right) }\left( 
\mathbf{\xi },t\right) $ of $f_{+}^{\left( 1\right) }$ is given by the
formula 
\begin{equation}
\hat{f}_{+}^{\left( 1\right) }\left( \mathbf{\xi },t\right) =-\varrho \psi
^{\prime }\left( \varrho t\right) \hat{Z}_{+}^{\left( 1\right) }\left( 
\mathbf{\xi },t\right) -\frac{\alpha _{\pi }}{\alpha }\left( \psi -\psi
^{3}\right) \widehat{p_{+}^{\left[ \sigma \right] }\left( Z_{+}^{\left(
0\right) 2}Z_{-}^{\left( 0\right) }\right) }\left( \mathbf{\xi },t\right) ,
\label{f1F}
\end{equation}%
where $Z_{+}^{\left( 0\right) }\left( \mathbf{\xi },t\right) $ and $\hat{Z}%
_{+}^{\left( 1\right) }\left( \mathbf{\xi },t\right) $ are defined
respectively by (\ref{Z0}) and (\ref{Z1eq}).

Having described the source $\hat{f}_{+}^{\left( 1\right) }\left( \mathbf{%
\xi },t\right) $ by (\ref{f1F}) we can use now the equation (\ref{V1eq1})
together with (\ref{V0psi}) and find that 
\begin{gather}
\hat{V}_{+}^{\left( 1\right) }\left( \beta \mathbf{q},t\right) =\frac{\alpha
_{\pi }}{\alpha }\int_{0}^{t}\exp \left\{ -\mathrm{i}\gamma _{\left( \nu
\right) }\left( \beta \mathbf{q}\right) \left( t-t_{1}\right) \right\} \psi
^{3}\left( \varrho t\right) \widehat{p_{+}^{\left[ \sigma \right]
}Z_{+}^{\left( 0\right) 2}Z_{-}^{\left( 0\right) }}\left( \beta \mathbf{q}%
,t_{1}\right) \,\mathrm{d}t_{1}  \label{V1F} \\
-\int_{0}^{t}\exp \left\{ -\mathrm{i}\gamma _{\left( \nu \right) }\left(
\beta \mathbf{q}\right) \left( t-t_{1}\right) \right\} \hat{f}_{+}^{\left(
1\right) }\left( \beta \mathbf{q},t_{1}\right) \,\mathrm{d}t_{1}.  \notag
\end{gather}%
Since the Fourier transform of a product is given by the convolution and in
view of the representation (\ref{Z0}) for $\hat{Z}_{+}^{\left( 0\right)
}\left( \mathbf{\xi },t\right) $ and (\ref{z00}), (\ref{phasefi2}), (\ref%
{psigVVVF}) we have 
\begin{gather}
\widehat{p_{+}^{\left[ \sigma \right] }Z_{+}^{\left( 0\right)
2}Z_{-}^{\left( 0\right) }}\left( \mathbf{\xi },t_{1}\right) =  \label{Z03}
\\
\frac{1}{\left( 2\pi \right) ^{2d}}\int_{\mathbb{R}^{2d}}p_{+}^{\left[
\sigma \right] }\left( \vec{\xi}^{\;\star }\right) \hat{Z}_{+}^{\left(
0\right) }\left( \mathbf{\xi }^{\prime },t_{1}\right) \hat{Z}_{+}^{\left(
0\right) }\left( \mathbf{\xi }^{\prime \prime },t_{1}\right) \hat{Z}%
_{-}^{\left( 0\right) }\left( \mathbf{\xi -\xi }^{\prime }-\mathbf{\xi }%
^{\prime \prime },t_{1}\right) \,\mathrm{d}\mathbf{\xi }^{\prime }\mathrm{d}%
\mathbf{\xi }^{\prime \prime }  \notag \\
=\frac{1}{\left( 2\pi \right) ^{2d}}\int_{\mathbb{R}^{2d}}\exp \left\{ 
\mathrm{i}\left[ -\gamma _{\left( \nu \right) }\left( \beta \mathbf{\xi }%
^{\prime }\right) -\gamma _{\left( \nu \right) }\left( \beta \mathbf{\xi }%
^{\prime \prime }\right) +\gamma _{\left( \nu \right) }\left( -\beta \mathbf{%
\xi }^{\prime \prime \prime }\right) \right] t_{1}\right\}  \notag \\
p_{+}^{\left[ \sigma \right] }\left( \vec{\xi}^{\;\star }\right) \hat{h}%
_{+,\beta }\left( \mathbf{\xi }^{\prime }\right) \hat{h}_{+,\beta }\left( 
\mathbf{\xi }^{\prime \prime }\right) \hat{h}_{-,\beta }\left( \mathbf{\xi }%
^{\prime \prime \prime }\right) \,\mathrm{d}\mathbf{\xi }^{\prime }\mathrm{d}%
\mathbf{\xi }^{\prime \prime },  \notag
\end{gather}%
in the above formula $\mathbf{\xi }^{\prime \prime \prime }=\mathbf{\xi -\xi 
}^{\prime }-\mathbf{\xi }^{\prime \prime }.$ Analogously to (\ref{UFB1}) it
is convenient to single out a slow time factor $\hat{v}_{\zeta }\left( 
\mathbf{\xi },\tau \right) $ of the modal amplitude $\hat{V}_{\zeta }\left( 
\mathbf{\xi },t\right) $ defined by 
\begin{equation}
\hat{V}_{\zeta }\left( \mathbf{\xi },t\right) =\hat{v}_{\zeta }\left( 
\mathbf{\xi },\tau \right) e^{-\mathrm{i}\zeta \gamma _{\left( \nu \right)
}\left( \zeta \mathbf{\xi }\right) t},\ \tau =\varrho t,\ \hat{v}_{\zeta
}\left( \mathbf{\xi },\tau \right) =\hat{v}_{\zeta }^{\left( 0\right)
}\left( \mathbf{\xi },\tau \right) +\alpha \hat{v}_{\zeta }^{\left( 1\right)
}\left( \mathbf{\xi },\tau \right) +\ldots .  \label{Schlin}
\end{equation}%
Then (\ref{Z03}) together with (\ref{V1F}) imply after substitution $\mathbf{%
\xi =}\beta \mathbf{q}$ 
\begin{gather}
\hat{v}_{\zeta }^{\left( 1\right) }\left( \beta \mathbf{q},\tau \right)
=\exp \left\{ \mathrm{i}\zeta \gamma _{\left( \nu \right) }\left( \zeta
\beta \mathbf{q}\right) \frac{\tau }{\varrho }\right\} \hat{V}_{\zeta
}^{\left( 1\right) }\left( \beta \mathbf{q},\frac{\tau }{\varrho }\right) =
\label{convV1} \\
\frac{3\beta ^{-d}}{\varrho }\int_{0}^{\tau }\int_{\mathbf{q}^{\prime \prime
\prime }+\mathbf{q}^{\prime \prime }+\mathbf{q}^{\prime }=\mathbf{q}}\exp
\left\{ \mathrm{i}\Phi ^{\left( \nu \right) }\left( \vec{\zeta}_{0},\beta 
\vec{q}\right) \frac{\tau _{1}}{\varrho }\right\} \psi ^{3}\left( \tau
_{1}\right)  \notag \\
p_{\zeta }^{\left[ \sigma \right] }\left( \beta \vec{q}^{\;\star }\right) 
\hat{h}_{\zeta }\left( \mathbf{q}^{\prime }\right) \hat{h}_{\zeta }\left( 
\mathbf{q}^{\prime \prime }\right) \hat{h}_{-\zeta }\left( \mathbf{q}%
^{\prime \prime \prime }\right) \,\mathrm{d}\mathbf{q}^{\prime }\mathrm{d}%
\mathbf{q}^{\prime \prime }\mathrm{d}\tau _{1}  \notag \\
-\frac{1}{\varrho }\int_{0}^{\tau }\exp \left\{ \mathrm{i}\zeta \gamma
_{\left( \nu \right) }\left( \zeta \beta \mathbf{q}\right) \frac{\tau _{1}}{%
\varrho }\right\} \hat{f}_{\zeta }^{\left( 1\right) }\left( \beta \mathbf{q},%
\frac{\tau _{1}}{\varrho }\right) \,\mathrm{d}\tau _{1},\;\zeta =\pm , 
\notag
\end{gather}%
with $\vec{\zeta}_{0}$ and $\Phi ^{\left( \nu \right) }\left( \vec{\zeta}%
_{0},\beta \vec{q}\right) $ being defined respectively (\ref{z00}) and (\ref%
{phasefi2}). Analogously to (\ref{f1F}) in the equation (\ref{V-eq}) for $%
V_{-}$ we have%
\begin{equation}
f_{-}^{\left( 1\right) }\left( \mathbf{r},t\right) =f_{-1}^{\left( 1\right)
}\left( \mathbf{r},t\right) =-\varrho \psi ^{\prime }\left( \varrho t\right)
Z_{-}^{\left( 1\right) }-\frac{\alpha _{\pi }}{\alpha }\left( \psi \left(
\varrho t\right) +\psi ^{3}\left( \varrho t\right) \right) p_{+}^{\left[
\sigma \right] }\left[ -\mathrm{i}\vec{\nabla}_{\mathbf{r}}\right] \left(
Z_{-}^{\left( 0\right) 2}Z_{+}^{\left( 0\right) }\right) .  \label{f1m}
\end{equation}%
Consequently, the Fourier transform $\hat{f}_{-}^{\left( 1\right) }\left( 
\mathbf{\xi },t\right) $ of $f_{-}^{\left( 1\right) }$ is 
\begin{equation}
\hat{f}_{-}^{\left( 1\right) }\left( \mathbf{\xi },t\right) =-\varrho \psi
^{\prime }\left( \varrho t\right) \hat{Z}_{-}^{\left( 1\right) }\left( 
\mathbf{\xi },t\right) -\frac{\alpha _{\pi }}{\alpha }\left( \psi -\psi
^{3}\right) \widehat{\left( p_{-}^{\left[ \sigma \right] }Z_{-}^{\left(
0\right) 2}Z_{+}^{\left( 0\right) }\right) }\left( \mathbf{\xi },t\right) ,
\label{f1m2}
\end{equation}%
with $Z_{-}^{\left( 0\right) }\left( \mathbf{\xi },t\right) $ and $%
Z_{-}^{\left( 1\right) }\left( \mathbf{\xi },t\right) $ being defined
respectively by (\ref{Z0}) and (\ref{Z1eq}).

Comparing the equality (\ref{convV1}) for the NLS with the interaction
integral (\ref{I1}) for the NLM we establish the following relation between
them:%
\begin{gather}
\hat{v}_{\zeta }^{\left( 1\right) }\left( \beta \mathbf{q},\tau \right)
=-3I_{\bar{n},\zeta ,\zeta ,-\zeta }^{\left( \sigma \right) }\left( \zeta 
\mathbf{k}_{\ast }+Y_{\zeta }\left( \beta \mathbf{q}\right) ,\tau \right) -
\label{v1I} \\
\frac{1}{\varrho }\int_{0}^{\tau }\exp \left\{ -\mathrm{i}\zeta \gamma
_{\left( \nu \right) }\left( \zeta \beta \mathbf{q}\right) \frac{\tau _{1}}{%
\varrho }\right\} \hat{f}_{\zeta }^{\left( 1\right) }\left( \beta \mathbf{q},%
\frac{\tau _{1}}{\varrho }\right) \,\mathrm{d}\tau _{1}.  \notag
\end{gather}

\subsection{Relating the NLS and the NLM}

Using (\ref{un1J}), (\ref{jf}) and (\ref{redYz}) we obtain that the part of
the linear response $\tilde{u}_{\zeta ,n_{0}}^{\left( 1,0\right) }$ of the
NLM originating from $\mathbf{J}^{\left( 1\right) }$ \ equals 
\begin{equation}
\tilde{u}_{\zeta ,n_{0}}^{\left( 1\right) }\left( \mathbf{J}^{\left(
1\right) };\zeta \mathbf{k}_{\ast }+Y_{\zeta }\left( \beta \mathbf{q}\right)
,\tau \right) =\frac{1}{\varrho }\int_{0}^{\tau }\exp \left\{ -\mathrm{i}%
\zeta \gamma _{\left( \nu \right) }\left( \zeta \beta \mathbf{q}\right) 
\frac{\tau _{1}}{\varrho }\right\} \hat{f}_{\zeta }^{\left( 1\right) }\left(
\beta \mathbf{q},\frac{\tau _{1}}{\varrho }\right) \,\mathrm{d}\tau _{1}.
\label{u1f}
\end{equation}%
Therefore, from (\ref{v1I}) and (\ref{U1z2}) we readily obtain%
\begin{equation}
\beta ^{d}\tilde{u}_{\bar{n}_{0}}^{\left( 1,0\right) }\left( \zeta \mathbf{k}%
_{\ast }+Y_{\zeta }\left( \beta \mathbf{q}\right) ,t\right) =\beta ^{d}\hat{v%
}_{\zeta }^{\left( 1\right) }\left( \beta \mathbf{q},\tau \right) +\left[
O\left( \beta ^{\nu -1}\right) +O\left( \varrho \right) \right] O\left(
\left\vert \mathbf{U}^{\left( 1\right) }\right\vert \right) .  \label{u1v1}
\end{equation}%
Using (\ref{vu0}) we obtain that the first order response (it is the sum of
the linear response and the FNLR) has the form 
\begin{gather}
\hat{V}_{\zeta }^{\left( 0\right) }\left( \beta \mathbf{q},t\right) +\alpha 
\hat{V}_{\zeta }^{\left( 1\right) }\left( \beta \mathbf{q},t\right) =\beta
^{d}\tilde{U}_{\zeta ,n_{0}}^{\left( 0\right) }\left( \mathbf{k}_{\ast
}+Y_{\zeta }\left( \beta \mathbf{q}\right) ,t\right) +\alpha \beta ^{d}%
\tilde{U}_{\zeta ,n_{0}}^{\left( 1\right) }\left( \mathbf{k}_{\ast
}+Y_{\zeta }\left( \beta \mathbf{q}\right) ,t\right) +  \label{UVmatch} \\
\alpha \left[ O\left( \beta ^{\nu -1}\right) +O\left( \varrho \right) \right]
O\left( \left\vert \mathbf{U}^{\left( 1\right) }\right\vert \right) ,\ \zeta
=\pm 1.  \notag
\end{gather}%
Notice that in the equality (\ref{UVmatch}) and below we omit the error
terms which include $O\left( \beta ^{N_{\Psi }}\right) $ with a large $%
N_{\Psi }$ since such terms absorbed by larger terms in the relevant
expressions. Note that (\ref{UVmatch}) implies the fulfillment of (\ref%
{match}) with $N_{1}=0$, $N_{2}=\sigma $ \ and arbitrary large $N_{3}$. \ We
also used that the dominant term of\ $\mathbf{U}^{\left( 1\right) }$ is $%
\mathbf{U}_{n_{0}}^{\left( 1\right) }$ and the dominant term of the
asymptotic expansion of $V^{\left( 1\right) }$ is similar to the dominant
term of the asymptotic expansion of $\mathbf{U}_{n_{0}}^{\left( 1\right) },$
and consequently $V^{\left( 1\right) }=O\left( \left\vert \mathbf{U}^{\left(
1\right) }\right\vert \right) $.

\subparagraph{Some conclusions.}

Using (\ref{V0V1}), (\ref{Unexp}) and (\ref{UVmatch}) we find that the exact
solutions of NLM and NLS\ satisfy the following relations: 
\begin{equation}
\hat{V}_{\zeta }\left( \beta \mathbf{q},t\right) =\beta ^{d}\tilde{U}_{\bar{n%
}_{0}}\left( \zeta \mathbf{k}_{\ast }+Y_{\zeta }\left( \beta \mathbf{q}%
\right) ,t\right) +\left[ O\left( \alpha ^{2}\right) +O\left( \alpha \varrho
\right) +O\left( \alpha \beta ^{\nu -1}\right) \right] O\left( \left\vert 
\mathbf{U}^{\left( 1\right) }\right\vert \right) .  \label{VU0}
\end{equation}%
Conversly,%
\begin{equation}
\beta ^{d}\tilde{U}_{\zeta ,n_{0}}\left( \zeta \mathbf{k}_{\ast }+\beta 
\mathbf{s},t\right) =\hat{V}_{\zeta }\left( \frac{1}{\beta }Y_{\zeta
}^{-1}\left( \beta \mathbf{s}\right) ,t\right) +\left[ \ldots \right]
O\left( \alpha \left\vert \mathbf{U}^{\left( 1\right) }\right\vert \right) .
\label{UV}
\end{equation}%
According to (\ref{Zalpha}) $\hat{V}_{\zeta }^{\left( 0\right) }\left( \beta 
\mathbf{q},t\right) +\alpha \hat{V}_{\zeta }^{\left( 1\right) }\left( \beta 
\mathbf{q},t\right) $ is the Fourier transform of the first order
approximation for the solution of the equation (\ref{Schgam}). According to (%
\ref{VeqZ}) and (\ref{veqz}) for $t\geq \frac{\tau _{0}}{\varrho }$ the
definition (\ref{UNLS}) can be rewritten after the substitution $\mathbf{%
\eta }=\beta \mathbf{s}$ in the form%
\begin{gather}
\mathbf{U}_{Z,n_{0}}^{\text{dir}}\left( \mathbf{r},t\right) =\frac{1}{\left(
2\pi \right) ^{d}}\int_{\left[ -\pi ,\pi \right] ^{d}}\Psi _{0}\left( 
\mathbf{\eta }\right)  \label{UZV} \\
\left[ \hat{V}_{+}\left( \frac{Y_{+}^{-1}\left( \mathbf{\eta }\right) }{%
\beta },t\right) \mathbf{\tilde{G}}_{+,n_{0}}\left( \mathbf{r},\mathbf{k}%
_{\ast }+\mathbf{\eta }\right) +\hat{V}_{-}\left( Y_{-}^{-1}\left( \frac{%
\mathbf{\eta }}{\beta }\right) ,t\right) \mathbf{\tilde{G}}_{-,n_{0}}\left( 
\mathbf{r},-\mathbf{k}_{\ast }-\mathbf{\eta }\right) \right] \,\mathrm{d}%
\mathbf{\eta },  \notag
\end{gather}%
where $V_{\zeta }$, $\zeta =\pm $ is the solution of (\ref{Schgam}). Hence, (%
\ref{UNLS1}) follows from (\ref{VU0}).

In the following section we will explain what is the origin of the
additional terms in the ENLS with $\nu =2$ and $\nu =4$. After matching
these terms with the FNLR of the NLM \ the better error estimates are
derived similarly to the above derivation for the classical NLS with $\nu =2$%
.

\subsection{Bidirectional waves and four mode coupling}

In this subsection we assume that the condition (\ref{eer2}) for the
electric permittivity $\mathbf{\varepsilon }\left( \mathbf{r}\right) $ holds
implying, in particular, the property of complex conjugation (\ref{Gstar})
for the corresponding eigenmodes and allowing using (\ref{conjug}). Recall
now that for the excitation current $\mathbf{J}$ to be real-valued its modal
coefficients $\tilde{j}_{\zeta ,n_{0}}\left( \mathbf{k}_{\ast },t\right) $
must satisfy the relations (\ref{jstar}). In other words, if a mode $\left(
\zeta ,n_{0},\zeta \mathbf{k}_{\ast }\right) $ is in the modal composition
of $\mathbf{J}$ with an amplitude $j_{\zeta ,n_{0}}\left( \mathbf{k}_{\ast
},t\right) $ then the mode $\left( -\zeta ,n_{0},-\zeta \mathbf{k}_{\ast
}\right) $ is there as well with the amplitude $\left[ \tilde{j}_{\zeta
,n_{0}}\left( \mathbf{k}_{\ast },t\right) \right] ^{\ast }$. Evidently, for
real-valued currents the modes in their modal compositions are always
presented in pairs%
\begin{equation}
\left\uparrow n_{0},\mathbf{k}_{\ast }\right\downarrow =\left\{ \left(
1,n_{0},\mathbf{k}_{\ast }\right) ,\left( -1,n_{0},-\mathbf{k}_{\ast
}\right) \right\} =\cup _{\zeta =\pm 1}\left( \zeta ,n_{0},\zeta \mathbf{k}%
_{\ast }\right) ,  \label{ggk2}
\end{equation}%
and, in view of (\ref{invsym}), (\ref{invsym2}), every such a pair involves
modes $\mathbf{\tilde{G}}_{1,n}\left( \mathbf{r},\mathbf{k}_{\ast }\right) $
and $\mathbf{\tilde{G}}_{-1,n}\left( \mathbf{r},-\mathbf{k}_{\ast }\right) $
having the same frequency $\omega _{n_{0}}\left( \mathbf{k}_{\ast }\right)
=\omega _{n_{0}}\left( -\mathbf{k}_{\ast }\right) $, the same group velocity 
$\omega _{n_{0}}^{\prime }\left( \mathbf{k}_{\ast }\right) $ and complex
conjugate amplitudes. We refer to the modal pairs (\ref{ggk2}) as \emph{%
doublets}.

Observe that (\ref{invsym}), (\ref{invsym2}), (\ref{Gstar}) imply also that
modes involved in another modal pair $\left\{ \left( 1,n_{0},-\mathbf{k}%
_{\ast }\right) ,\left( -1,n_{0},\mathbf{k}_{\ast }\right) \right\} $ are%
\begin{equation}
\mathbf{\tilde{G}}_{1,n_{0}}\left( \mathbf{r},-\mathbf{k}_{\ast }\right) =%
\left[ \mathbf{\tilde{G}}_{-1,n_{0}}\left( \mathbf{r},\mathbf{k}_{\ast
}\right) \right] ^{\ast }\text{ and }\mathbf{\tilde{G}}_{-1,n_{0}}\left( 
\mathbf{r},\mathbf{k}_{\ast }\right) =\left[ \mathbf{\tilde{G}}%
_{1,n_{0}}\left( \mathbf{r},-\mathbf{k}_{\ast }\right) \right] ^{\ast },
\label{ggk1}
\end{equation}%
and that they have the frequency $\omega _{n_{0}}\left( \mathbf{k}_{\ast
}\right) $ and the group velocity $\omega _{n_{0}}^{\prime }\left( -\mathbf{k%
}_{\ast }\right) =-\omega _{n_{0}}^{\prime }\left( \mathbf{k}_{\ast }\right) 
$. Hence, the two doublets $\left\uparrow n_{0},\mathbf{k}_{\ast
}\right\downarrow $ and $\left\uparrow n_{0},-\mathbf{k}_{\ast
}\right\downarrow $ involve the complex conjugate eigenmodes (\ref{ggk1}) of
the same frequency $\omega _{n_{0}}\left( \mathbf{k}_{\ast }\right) $ and
they have opposite group velocities $\pm \omega _{n_{0}}^{\prime }\left( 
\mathbf{k}_{\ast }\right) $. Consequently, a wave composed of the modal
quadruplet%
\begin{gather}
\left\Uparrow n_{0},\mathbf{k}_{\ast }\right\Downarrow =\left\uparrow n_{0},%
\mathbf{k}_{\ast }\right\downarrow \cup \left\uparrow n_{0},-\mathbf{k}%
_{\ast }\right\downarrow =  \label{ggk4} \\
\left\{ \left( 1,n_{0},\mathbf{k}_{\ast }\right) ,\left( -1,n_{0},-\mathbf{k}%
_{\ast }\right) ,\left( 1,n_{0},-\mathbf{k}_{\ast }\right) ,\left( -1,n_{0},%
\mathbf{k}_{\ast }\right) \right\}  \notag
\end{gather}%
is \emph{bidirectional} since wavepackets corresponding to its two
constitutive doublets $\left\uparrow n_{0},\mathbf{k}_{\ast
}\right\downarrow $ and $\left\uparrow n_{0},-\mathbf{k}_{\ast
}\right\downarrow $ propagate with opposite group velocities $\pm \omega
_{n_{0}}^{\prime }\left( \mathbf{k}_{\ast }\right) $. Such a bidirectional
wave can be directly excited, i.e. excited through the linear mechanism, by
a current composed of the quadruplet $\left\Uparrow n_{0},\mathbf{k}_{\ast
}\right\Downarrow $ in (\ref{ggk4}). The four modes in the quadruplet $%
\left\Uparrow n_{0},\mathbf{k}_{\ast }\right\Downarrow $ are coupled through
a relatively strong nonlinear interactions and it is a subject of this
section. It turns out, that the quadruplet $\left\Uparrow n_{0},\mathbf{k}%
_{\ast }\right\Downarrow $ in (\ref{ggk4}) is the only generic with respect
to the excitation frequency variations modal quadruplet having relatively
strong (of the order $O\left( \varrho \right) $) \ nonlinear interaction
between its doublets, and we name and will refer to it as a \emph{%
bidirectional quadruplet}. Note that two doublets $\left\uparrow n_{0},%
\mathbf{k}_{\ast }\right\downarrow $ and $\left\uparrow n_{1},\mathbf{k}%
_{\ast }\right\downarrow $ corresponding to different bands, if directly
excited, also have the same order of \ interaction $\ O\left( \varrho
\right) $ between them, but one have to use a special pair of exactly
matched excitation carrier frequences $\omega _{1}=\omega _{n_{0}}\left( 
\mathbf{k}_{\ast }\right) $ and $\omega _{2}=\omega _{n_{1}}\left( \mathbf{k}%
_{\ast }\right) $ to directly excite them, whereas only one frequency $%
\omega _{n_{0}}\left( \mathbf{k}_{\ast }\right) $ is required to excite the
bi-directional quadruplet.

In the preceding sections we have considered the case where only one doublet 
$\left\uparrow n_{0},\mathbf{k}_{\ast }\right\downarrow $ was excited, this
case describes the situation when the directly excited wave propagates only
in one direction. When only the excitation frequences $\omega =\omega
_{n_{0}}\left( \mathbf{k}_{\ast }\right) $ and the wavenumbers $\pm \mathbf{k%
}_{\ast }$ are fixed, a general time-harmonic excitation produces waves
propagating in both directions. As we have shown in Subsection 3.2 the
magnitude of their non-FM interaction is $O\left( \alpha \varrho \right)
O\left( \left\vert \mathbf{U}^{\left( 1\right) }\right\vert \right) $, and
when we take into account corrections of the order $O\left( \alpha \varrho
\right) $ they have to be considered. For a discussion of multimode
interactions see Subsection 1.2.

Now let us introduce a bidirectional excitation current of the form (\ref%
{UU1}) where the amplitude $\tilde{j}_{\zeta ,n_{0}}^{\left( 0\right)
}\left( \mathbf{k},t\right) $ is defined by the following slightly more
general than (\ref{jnn1}) formula 
\begin{gather}
\tilde{j}_{\zeta ,n_{0}}^{\left( 0\right) }\left( \mathbf{k},t\right) =
\label{jbi} \\
-\varrho \psi _{0}\left( \varrho t\right) \beta ^{-d}\left[ \Psi _{0}\left( 
\mathbf{k-}\zeta \mathbf{k}_{\ast }\right) \mathring{h}_{\zeta }^{+}\left( 
\frac{\mathbf{k-}\zeta \mathbf{k}_{\ast }}{\beta }\right) +\Psi _{0}\left( 
\mathbf{k}+\zeta \mathbf{k}_{\ast }\right) \mathring{h}_{\zeta }^{-}\left( 
\frac{\mathbf{k}+\zeta \mathbf{k}_{\ast }}{\beta }\right) \right] ,  \notag
\end{gather}%
with functions $\mathring{h}_{\zeta }^{\pm }\left( \mathbf{s}\right) $
having the same properties as $\mathring{h}_{\zeta }\left( \mathbf{s}\right) 
$ in (\ref{jnn1}). The excitation current defined by (\ref{jbi}) directly
excites the four modes for the bidirectinal quadruplet $\left\Uparrow n_{0},%
\mathbf{k}_{\ast }\right\Downarrow $ (\ref{ggk4}) with the linear response
modal amplitudes $\tilde{U}_{\zeta ,n_{0}}^{\left( 0\right) }\left( \pm 
\mathbf{k}_{\ast }+\mathbf{\eta },t\right) $, $\zeta =\pm 1$. If we
introduce 
\begin{equation}
\mathbf{k}_{\ast }^{\pm }=\pm \mathbf{k}_{\ast },  \label{4mode}
\end{equation}%
then the bidirectional quadruplet $\left\Uparrow n_{0},\mathbf{k}_{\ast
}\right\Downarrow $ is composed of the two doublets $\left\uparrow n_{0},%
\mathbf{k}_{\ast }^{\pm }\right\downarrow $ with the group velocities $\pm
\omega _{n_{0}}^{\prime }\left( \mathbf{k}_{\ast }\right) $, and,
importantly, the nonlinear interactions betweeen those doublets are non-FM\
interactions satisfying the relation (\ref{sumz-}). Using the relation (\ref%
{sumz-}), the phase-matching relation (\ref{PM0}), with $\mathbf{\mathbf{k}%
^{\prime }}=\zeta ^{\prime }\mathbf{k}_{\ast }^{\pm }$, $\mathbf{\mathbf{k}%
^{\prime \prime }}=\zeta ^{\prime \prime }\mathbf{k}_{\ast }^{\pm }$, $%
\mathbf{\mathbf{k}^{\prime \prime \prime }}=\zeta ^{\prime \prime \prime }%
\mathbf{k}_{\ast }^{\pm }$, and arguments similar to ones used to derive (%
\ref{nzk}), (\ref{sumz}) and (\ref{sumz-}), we conclude that the triad of
modes from the quadruplet $\left\Uparrow n_{0},\mathbf{k}_{\ast
}\right\Downarrow $ affects only the mode from this quadruplet with the
quasimomentum 
\begin{equation}
\left( \zeta ^{\prime }+\zeta ^{\prime \prime }+\zeta ^{\prime \prime \prime
}\right) \mathbf{k}_{\ast }^{\pm }=-\zeta \mathbf{k}_{\ast }^{\pm }=\zeta 
\mathbf{k}_{\ast }^{\mp }.  \label{PM4}
\end{equation}%
This condition selects from formally possible $4^{4}$ interactions only a
few significant ones. Namely, the condition (\ref{PM4}) implies that a pair
of three numbers $\zeta ^{\prime },\zeta ^{\prime \prime },\zeta ^{\prime
\prime \prime }$ has the same sign as $-\zeta $, we set 
\begin{equation}
\zeta ^{\prime }=\zeta ^{\prime \prime }=-\zeta ,\ \zeta ^{\prime \prime
\prime }=\zeta ,\ \vec{\zeta}_{0,\times }=\left( \zeta ,-\zeta ,-\zeta
,\zeta \right) ,  \label{z0bi}
\end{equation}%
and two more cases are similar to the above. The corresponding interaction
wavevectors $\vec{k}$ are given by the formula: 
\begin{equation}
\vec{k}_{\ast ,\times ,-}=\left( -\mathbf{k}_{\ast },\mathbf{k}_{\ast },%
\mathbf{k}_{\ast },\mathbf{k}_{\ast }\right) ,\ \vec{k}_{\ast ,\times
,+}=\left( \mathbf{k}_{\ast },-\mathbf{k}_{\ast },-\mathbf{k}_{\ast },-%
\mathbf{k}_{\ast }\right) .  \label{kstar1}
\end{equation}

The interaction integral (\ref{Vn0}) with $\vec{\zeta},$ $\vec{k}$ \
satisfying (\ref{z0bi}) and (\ref{kstar1}) describes the impact of the triad
of waves from the forward-propagating doublet onto the modal coefficient of
the backward-propagating doublet, note that this interaction have three
modes (two of them forward propagating) with the same $\zeta $, that is in
the same band and one mode (forward-propagating) with the opposite $\zeta $,
that is in the opposite band.

The analysis of the interaction integral (\ref{Vn0}) with $\vec{\zeta}$
given by (\ref{z0bi}) and $\vec{k}$ in a vicinity of $\vec{k}_{\ast ,\times
} $ determined by (\ref{kstar1}) is similar to the analysis of the integral (%
\ref{Iz3}) in the FM case where $\vec{k}$ is in a vicinity of $\vec{k}_{\ast
}$ determined by (\ref{zsy}). We obtain that similarly to (\ref{I1}), (\ref%
{Inzbet2}) the principal part of the non-FM interaction integral is given by
the formula 
\begin{gather}
\beta ^{d}I_{\bar{n},-\zeta ,-\zeta ,\zeta }^{\left( 0\right) }\left( -\zeta 
\mathbf{k}_{\ast }+Y_{\zeta }\left( \beta \mathbf{q}\right) ,\tau \right) =
\label{I1bi} \\
\frac{1}{\varrho }\int_{0}^{\tau }\int_{\mathbf{q}^{\prime \prime \prime }+%
\mathbf{q}^{\prime \prime }+\mathbf{q}^{\prime }=\mathbf{q}}\exp \left\{ 
\mathrm{i}\Phi ^{\left( \nu \right) }\left( \vec{\zeta},\beta \vec{q}\right) 
\frac{\tau _{1}}{\varrho }\right\} \breve{Q}_{\vec{n}_{0}}\left( \vec{\zeta}%
_{0,\times }\vec{k}_{\ast ,\times ,-}\right)  \notag \\
\psi ^{3}\left( \tau _{1}\right) \hat{h}_{-\zeta }^{+}\left( \mathbf{q}%
^{\prime }\right) \hat{h}_{-\zeta }^{+}\left( \mathbf{q}^{\prime \prime
}\right) \hat{h}_{\zeta }^{+}\left( \mathbf{q}^{\prime \prime \prime }\left(
0\right) \right) \,\mathrm{d}\mathbf{q}^{\prime }\mathrm{d}\mathbf{q}%
^{\prime \prime }\mathrm{d}\tau _{1}  \notag
\end{gather}%
where we take $\sigma =0$ since we do not need to take into account the
higher approximations of this integral which is already $\varrho $ times
smaller than the FM\ interactions. Similarly to (\ref{II1}) the error of
approximation of the integral (\ref{Iz3}) is given by the formula 
\begin{gather}
\beta ^{d}\left[ I_{\bar{n},-\zeta ,-\zeta ,\zeta }\left( -\zeta \mathbf{k}%
_{\ast }+Y_{\zeta }\left( \beta \mathbf{q}\right) ,\tau \right) -I_{\bar{n}%
,-\zeta ,-\zeta ,\zeta }^{\left( 0\right) }\left( -\zeta \mathbf{k}_{\ast
}+Y_{\zeta }\left( \beta \mathbf{q}\right) ,\tau \right) \right] =
\label{II1bi} \\
\left[ O\left( \varrho \beta \right) \right] O\left( \left\vert \mathbf{U}%
^{\left( 1\right) }\right\vert \right) .  \notag
\end{gather}

The four modal amplitudes $\tilde{U}_{\zeta ,n_{0}}\left( \zeta \mathbf{k}%
_{\ast }^{\pm }+\mathbf{\eta },t\right) $ are approximated by the Fourier
transforms of the four NLS solutions $Z_{\zeta }^{\pm }\left( \mathbf{r}%
,t\right) $. We write the corresponding system in the most general case of
complex-valued excitation currents and then we will discuss the reduction of
the system when the excitation is real. To approximate the non-FM terms we
include the coupling terms $\delta _{\times ,\zeta }^{+}\left( \left(
Z_{-\zeta }^{-}\right) ^{2}Z_{\zeta }^{-}\right) $ in the equation for $%
Z_{\zeta }^{+}$ \ and $\delta _{\times ,\zeta }^{-}\left( \left( Z_{-\zeta
}^{+}\right) ^{2}Z_{\zeta }^{+}\right) $ in the equation for $Z_{\zeta }^{-}$
where 
\begin{equation}
\delta _{\times ,\zeta }^{-}=3\breve{Q}_{\vec{n}_{0}}\left( \vec{\zeta}%
_{0,\times }\vec{k}_{\ast ,\times ,-}\right) ,\ \delta _{\times ,\zeta
}^{+}=3\breve{Q}_{\vec{n}_{0}}\left( \vec{\zeta}_{0,\times }\vec{k}_{\ast
,\times ,+}\right) .  \label{dcross}
\end{equation}%
We take in (\ref{Znu}), (\ref{Znu-}) $\nu =4$ to take into account terms of
order $O\left( \alpha \beta ^{2}\right) $ which are comparable with the
terms coming from (\ref{I1bi}) in the dispersive case (\ref{nonM}). We
obtain two pairs of coupled equations for $Z_{\zeta }^{+}$ and $Z_{\zeta
}^{-}$, $\zeta =\pm $: 
\begin{equation}
\left\{ \partial _{t}+\zeta \mathrm{i}\gamma _{\left( 4\right) }\left( -%
\mathrm{i}\zeta \nabla _{\mathbf{r}}\right) \right\} Z_{\zeta }^{+}+\alpha
\delta _{\times ,\zeta }^{+}\left( \left( Z_{-\zeta }^{-}\right)
^{2}Z_{\zeta }^{-}\right) =+\alpha _{\pi }p_{\zeta }^{+,\left[ 2\right] }%
\left[ -\mathrm{i}\vec{\nabla}_{\mathbf{r}}\right] \left( Z_{-\zeta
}^{+}\left( Z_{\zeta }^{+}\right) ^{2}\right) ,  \label{EN+}
\end{equation}%
\begin{equation}
\left\{ \partial _{t}+\zeta \mathrm{i}\gamma _{\left( 4\right) }\left( 
\mathrm{i}\zeta \nabla _{\mathbf{r}}\right) \right\} Z_{\zeta }^{-}+\alpha
_{\pi }\delta _{\times ,\zeta }^{-}\left( \left( Z_{-\zeta }^{+}\right)
^{2}Z_{\zeta }^{+}\right) =\alpha _{\pi }p_{\zeta }^{-,\left[ 2\right] }%
\left[ -\mathrm{i}\vec{\nabla}_{\mathbf{r}}\right] \left( Z_{-\zeta
}^{-}\left( Z_{\zeta }^{-}\right) ^{2}\right) ,  \label{EN-}
\end{equation}%
with the initial data%
\begin{equation}
\left. Z_{\zeta }^{+}\left( \mathbf{r},t\right) \right\vert _{t=0}=h_{\zeta
}^{+}\left( \mathbf{r}\right) ,\ \left. Z_{\zeta }^{-}\left( \mathbf{r}%
,t\right) \right\vert _{t=0}=h_{\zeta }^{-}\left( \mathbf{r}\right) ,\ \zeta
=\pm .
\end{equation}%
\ We have similarly to (\ref{UZO}), (\ref{OFNLR1}) 
\begin{equation}
\tilde{U}_{\zeta ,n_{0}}\left( \zeta \mathbf{k}_{\ast }^{\pm }+\mathbf{\eta }%
,t\right) =\hat{Z}_{\zeta }^{\pm }\left( \pm Y_{\zeta }^{-1}\left( \pm 
\mathbf{\eta }\right) ,t\right) +\left[ O\left( \alpha \beta ^{3}\right)
+O\left( \alpha \varrho \right) +O\left( \alpha ^{2}\right) \right] O\left(
\left\vert \mathbf{U}^{\left( 1\right) }\right\vert \right)  \label{Bier}
\end{equation}%
If we take into account in ((\ref{ENc+}), (\ref{ENcx+}) the terms \ which
originate from the first order correction to the susceptibility, from (\ref%
{Ebir1}), (\ref{Ebir2}) we obtain the system 
\begin{gather}
\left\{ \partial _{t}+\zeta \mathrm{i}\gamma _{\left( 4\right) }\left( -%
\mathrm{i}\zeta \nabla _{\mathbf{r}}\right) \right\} Z_{\zeta }^{+}+\alpha
\delta _{\times ,\zeta }^{+}\left( \left( Z_{-\zeta }^{-}\right)
^{2}Z_{\zeta }^{-}\right) +  \label{ENc+} \\
\alpha _{\pi }\delta _{1,\zeta }^{+}Z_{\zeta }^{+}Z_{-\zeta }^{+}\left\{
\partial _{t}+\zeta \mathrm{i}\gamma _{\left( 4\right) }\left( -\mathrm{i}%
\zeta \nabla _{\mathbf{r}}\right) \right\} Z_{\zeta }^{+}=  \notag \\
-\alpha _{\pi }\delta _{2,\zeta }^{+}\left( Z_{\zeta }^{+}\right)
^{2}\left\{ \partial _{t}-\zeta \mathrm{i}\gamma _{\left( 3\right) }\left( 
\mathrm{i}\zeta \nabla _{\mathbf{r}}\right) \right\} Z_{-\zeta }^{+}+\alpha
_{\pi }p_{\zeta }^{+,\left[ 2\right] }\left[ -\mathrm{i}\vec{\nabla}_{%
\mathbf{r}}\right] \left( Z_{-\zeta }^{+}\left( Z_{\zeta }^{+}\right)
^{2}\right) ,  \notag
\end{gather}%
\begin{gather}
\left\{ \partial _{t}+\zeta \mathrm{i}\gamma _{\left( 4\right) }\left( 
\mathrm{i}\zeta \nabla _{\mathbf{r}}\right) \right\} Z_{\zeta }^{-}+\alpha
_{\pi }\delta _{\times ,\zeta }^{-}\left( \left( Z_{-\zeta }^{+}\right)
^{2}Z_{\zeta }^{+}\right) +  \label{ENcx+} \\
\alpha \delta _{1,\zeta }^{-}Z_{\zeta }^{-}Z_{-\zeta }^{-}\left\{ \partial
_{t}+\zeta \mathrm{i}\gamma _{\left( 4\right) }\left( \mathrm{i}\zeta \nabla
_{\mathbf{r}}\right) \right\} Z_{\zeta }^{-}=  \notag \\
-\alpha _{\pi }\delta _{2,\zeta }^{-}\left( Z_{\zeta }^{-}\right)
^{2}\left\{ \partial _{t}-\zeta \mathrm{i}\gamma _{\left( 4\right) }\left( -%
\mathrm{i}\zeta \nabla _{\mathbf{r}}\right) \right\} Z_{-\zeta }^{-}+\alpha
_{\pi }p_{\zeta }^{-,\left[ 2\right] }\left[ -\mathrm{i}\vec{\nabla}_{%
\mathbf{r}}\right] \left( Z_{-\zeta }^{-}\left( Z_{\zeta }^{-}\right)
^{2}\right) ,  \notag
\end{gather}%
These additional terms with coefficients $\delta _{1,\zeta }^{\pm }$ \ are
discussed in Section 6. Addition of these terms and taking into account
interband coupling (see Subsection 1.4.4) improves the error term in (\ref%
{Bier}) replacing $O\left( \alpha \varrho \right) $ by $O\left( \alpha
\varrho \beta \right) $. The term $O\left( \alpha ^{2}\right) $ in the the
approximation error term in (\ref{Bier}) can be replaced by $O\left( \alpha
^{2}\beta \right) $ if the fifth order terms of the nonlinearity are taken
into account, see Subsection 1.3.6 \ and Remark in the end of Section 7.
Note that if the initial data are real 
\begin{equation}
h_{\zeta }^{+}\left( \mathbf{r}\right) =h_{-\zeta }^{+}\left( \mathbf{r}%
\right) ^{\ast },\ h_{\zeta }^{-}\left( \mathbf{r}\right) =h_{-\zeta
}^{-}\left( \mathbf{r}\right) ^{\ast },
\end{equation}%
and the nonlinearity is real, then we have 
\begin{equation}
Z_{-\zeta }^{+}\left( \mathbf{r},t\right) =\left[ Z_{\zeta }^{+}\left( 
\mathbf{r},t\right) \right] ^{\ast },\ Z_{-\zeta }^{-}\left( \mathbf{r}%
,t\right) =\left[ Z_{\zeta }^{-}\left( \mathbf{r},t\right) \right] ^{\ast }.
\end{equation}%
and, consequently, we can apply (\ref{conjug}). Namely, we exclude $\zeta
=-1 $ and (\ref{ENc+}), (\ref{ENcx+}) is reduced to the following system of
two scalar equations (similar equations are known as \emph{coupled modes
equations}) for $Z_{+}^{+}$ and $\ Z_{+}^{-}$: 
\begin{gather}
\left\{ \partial _{t}+\mathrm{i}\gamma _{\left( 4\right) }\left[ -\mathrm{i}%
\vec{\nabla}_{\mathbf{r}}\right] \right\} Z_{+}^{+}+\alpha _{\pi }\delta
_{\times ,+}^{+}\left( \left\vert Z_{+}^{-}\right\vert ^{2}Z_{+}^{-\ast
}\right) +  \label{Ebir1} \\
\alpha _{\pi }\delta _{1,+}^{+}\left\vert \left( Z_{+}^{+}\right)
\right\vert ^{2}\left\{ \partial _{t}+\mathrm{i}\gamma _{\left( 4\right) }%
\left[ -\mathrm{i}\vec{\nabla}_{\mathbf{r}}\right] \right\} Z_{+}^{+}= 
\notag \\
-\alpha _{\pi }\delta _{2,+}^{+}\left( Z_{+}^{+}\right) ^{2}\left\{ \partial
_{t}-\mathrm{i}\gamma _{\left( 3\right) }\left[ \mathrm{i}\nabla _{\mathbf{r}%
}\right] \right\} \left( Z_{+}^{+}\right) ^{\ast }+\alpha _{\pi }p_{+}^{+,%
\left[ 2\right] }\left[ -\mathrm{i}\vec{\nabla}_{\mathbf{r}}\right] \left(
Z_{+}^{+}\left\vert Z_{+}^{+}\right\vert ^{2}\right) ,  \notag
\end{gather}%
\begin{gather}
\left\{ \partial _{t}+\mathrm{i}\gamma _{\left( 4\right) }\left[ \mathrm{i}%
\nabla _{\mathbf{r}}\right] \right\} Z_{+}^{-}+\alpha _{\pi }\delta _{\times
,+}^{-}\left( \left\vert Z_{+}^{+}\right\vert ^{2}Z_{+}^{+\ast }\right) +
\label{Ebir2} \\
\alpha _{\pi }\delta _{1,+}^{-}\left\vert \left( Z_{+}^{-}\right)
\right\vert ^{2}\left[ \partial _{t}+\mathrm{i}\gamma _{\left( 3\right) }%
\left[ \mathrm{i}\nabla _{\mathbf{r}}\right] \right] Z_{+}^{-}=  \notag \\
-\alpha _{\pi }\delta _{2,+}^{-}\left( Z_{+}^{-}\right) ^{2}\left\{ \partial
_{t}-\mathrm{i}\gamma _{\left( 4\right) }\left[ -\mathrm{i}\vec{\nabla}_{%
\mathbf{r}}\right] \right\} \left( Z_{+}^{-}\right) ^{\ast }+\alpha _{\pi
}p_{+}^{-,\left[ 2\right] }\left[ -\mathrm{i}\vec{\nabla}_{\mathbf{r}}\right]
\left( \left\vert Z_{+}^{-}\right\vert ^{2}Z_{+}^{-}\right) .  \notag
\end{gather}%
\ These equations imply (\ref{Bi1}), (\ref{Bi2}).

\subsection{Representation of solutions in the space domain}

The principal part of the approximate solutions $\mathbf{U}_{Z}\left( 
\mathbf{r},t\right) $ is determined by (\ref{UNLS}) in terms of their Bloch
modal coefficients and the Fourier transform $\hat{Z}\left( \mathbf{\xi }%
,t\right) $ of the solution of the NLS as follows: 
\begin{gather}
\mathbf{U}_{Z}^{\text{dir}}\left( \mathbf{r},t\right) =\frac{\beta ^{d}}{%
\left( 2\pi \right) ^{d}}\int_{\left[ -\pi /\beta ,\pi /\beta \right]
^{d}}\Psi _{0}\left( \beta \mathbf{s}\right)  \label{UZ} \\
\left[ \hat{Z}_{\beta ,+}\left( Y^{-1}\left( \beta \mathbf{s}\right)
,t\right) \mathbf{\tilde{G}}_{+,n_{0}}\left( \mathbf{r},\mathbf{k}_{\ast
}+\beta \mathbf{s}\right) +\hat{Z}_{\beta ,-}\left( -Y^{-1}\left( -\beta 
\mathbf{s}\right) ,t\right) \mathbf{\tilde{G}}_{-,n_{0}}\left( \mathbf{r},-%
\mathbf{k}_{\ast }+\beta \mathbf{s}\right) \right] \,\mathrm{d}\mathbf{s} 
\notag
\end{gather}%
where $\hat{Z}_{\zeta }\left( \mathbf{q},t\right) $ is the Fourier transform
of $Z_{\zeta }\left( \mathbf{r},t\right) $ of (\ref{Si}), (\ref{Si1}) or (%
\ref{Six}). According to (\ref{Fourbet}) $\beta ^{d}\hat{Z}_{\zeta }\left(
\beta \mathbf{q},t\right) =\hat{Z}_{\beta ,\zeta }\left( \mathbf{q},t\right) 
$ where $Z_{\beta ,\zeta }$ is a solution of (\ref{genS1}) which regularly
depends on $\beta $.In this subsection we derive the formula (\ref{Uphys})
providing a representation for $\mathbf{U}_{Z}^{\text{dir}}\left( \mathbf{r}%
,t\right) $ in the space domain in terms of $Z\left( \mathbf{r},t\right) $.
Note that (\ref{UZ}) \ has the form%
\begin{equation}
\mathbf{U}_{Z}^{\text{dir}}\left( \mathbf{r},t\right) =\mathbf{U}%
_{Z_{+}}\left( \mathbf{r},t\right) +\mathbf{U}_{Z_{-}}\left( \mathbf{r}%
,t\right) .  \label{UZ1}
\end{equation}

We begin with using change of variables $Y^{-1}\left( \beta \mathbf{s}%
\right) =\beta \mathbf{q}$ in (\ref{UZ}) and (\ref{Psihat}) to obtain%
\begin{equation}
\mathbf{U}_{Z_{+}}\left( \mathbf{r},t\right) =\frac{\beta ^{d}}{\left( 2\pi
\right) ^{d}}\int_{\mathbb{R}^{d}}\Psi \left( \beta \mathbf{q}\right) \hat{Z}%
_{+}\left( \beta \mathbf{q},t\right) \mathbf{\tilde{G}}_{+,n_{0}}\left( 
\mathbf{r},\mathbf{k}_{\ast }+Y\left( \beta \mathbf{q}\right) \right) \det
Y_{+}^{\prime }\left( \beta \mathbf{q}\right) \,\mathrm{d}\mathbf{q},
\label{UZ+}
\end{equation}%
\begin{equation}
\mathbf{U}_{Z_{-}}\left( \mathbf{r},t\right) =\frac{\beta ^{d}}{\left( 2\pi
\right) ^{d}}\int_{\mathbb{R}^{d}}\Psi \left( \beta \mathbf{q}\right) \hat{Z}%
_{-}\left( \beta \mathbf{q},t\right) \mathbf{\tilde{G}}_{-,n_{0}}\left( 
\mathbf{r},-\mathbf{k}_{\ast }+Y_{-}\left( \beta \mathbf{q}\right) \right)
\det Y_{-}^{\prime }\left( \beta \mathbf{q}\right) \,\mathrm{d}\mathbf{q}.
\label{UZ-}
\end{equation}%
Then taking into account (\ref{Ybet0}) and (\ref{detYbet}) we get 
\begin{equation}
\mathbf{U}_{Z_{+}}\left( \mathbf{r},t\right) =\frac{1}{\left( 2\pi \right)
^{d}}\int_{\mathbb{R}^{d}}\beta ^{d}\Psi \left( \beta \mathbf{q}\right) \hat{%
Z}_{+}\left( \beta \mathbf{q},t\right) \mathbf{\tilde{G}}_{1,n_{0}}\left( 
\mathbf{r},\mathbf{k}_{\ast }+\beta \mathbf{q}\right) \,\mathrm{d}\mathbf{q}%
+O\left( \beta ^{\nu }\right) .
\end{equation}%
The function $\mathbf{\tilde{G}}_{+,n_{0}}\left( \mathbf{r},\mathbf{k}%
\right) $ is $2\pi $ -periodic in $\mathbf{k}$\textbf{.} According to (\ref%
{nz2a}) the Bloch eigenfunctions can be presented in the form%
\begin{equation}
\mathbf{\tilde{G}}_{+,n_{0}}\left( \mathbf{r},\mathbf{k}\right) =\mathbf{%
\hat{G}}_{+,n_{0}}\left( \mathbf{r},\mathbf{k}\right) \mathrm{e}^{\mathrm{i}%
\mathbf{k}\cdot \mathbf{r}},
\end{equation}%
where $\mathbf{\hat{G}}_{+,n_{0}}\left( \mathbf{r},\mathbf{k}\right) $ is $1$%
-periodic function of $\mathbf{r}$\textbf{.} Hence,%
\begin{equation}
\mathbf{\tilde{G}}_{+,n_{0}}\left( \mathbf{r},\mathbf{k}_{\ast }+\beta 
\mathbf{q}\right) =\mathbf{\hat{G}}_{+,n_{0}}\left( \mathbf{r},\mathbf{k}%
_{\ast }+\beta \mathbf{q}\right) \mathrm{e}^{\mathrm{i}\left( \mathbf{k}%
_{\ast }+\beta \mathbf{q}\right) \cdot \mathbf{r}}.  \label{Ggy}
\end{equation}%
We approximate then $\mathbf{\hat{G}}_{+,n_{0}}\left( \mathbf{r},\mathbf{k}%
_{\ast }+\beta \mathbf{q}\right) \mathbf{\ }$by its Taylor polynomial of the
degree $\sigma $ 
\begin{equation}
\mathbf{\hat{G}}_{+,n_{0}}\left( \mathbf{r},\mathbf{k}_{\ast }+\beta \mathbf{%
q}\right) =\mathbf{p}_{g,\sigma }\left( \mathbf{r},\beta \mathbf{q}\right)
+O\left( \beta ^{\sigma +1}\right) ,\ \sigma +1\leq \nu ,  \label{GhatTay}
\end{equation}%
where, for $\sigma =2$, 
\begin{equation}
\mathbf{p}_{g,2}\left( \mathbf{r},\beta \mathbf{q}\right) =\mathbf{\hat{G}}%
_{1,n_{0}}\left( \mathbf{r},\mathbf{k}_{\ast }\right) +\beta \mathbf{\hat{G}}%
_{+,n_{0}}^{\prime }\left( \mathbf{r}\right) \left( \mathbf{q}\right) +\frac{%
1}{2}\beta ^{2}\mathbf{\hat{G}}_{+,n_{0}}^{\prime \prime }\left( \mathbf{r}%
\right) \left( \mathbf{q}^{2}\right) ,  \label{pgsig}
\end{equation}%
with tensors 
\begin{equation}
\mathbf{\hat{G}}_{+,n_{0}}^{\prime }\left( \mathbf{q}\right) =\nabla _{%
\mathbf{k}}\mathbf{\hat{G}}_{+,n_{0}}\left( \mathbf{r},\mathbf{k}_{\ast
}\right) \cdot \mathbf{q},\mathbf{\qquad \hat{G}}_{+,n_{0}}^{\prime \prime
}\left( \mathbf{q}^{2}\right) =\nabla _{\mathbf{k}}^{2}\mathbf{\hat{G}}%
_{+,n_{0}}\left( \mathbf{r},\mathbf{k}_{\ast }\right) \mathbf{q}\cdot 
\mathbf{q}.  \label{pgsig1}
\end{equation}%
Hence, 
\begin{equation}
\mathbf{U}_{Z_{+}}\left( \mathbf{r},t\right) =\frac{1}{\left( 2\pi \right)
^{d}}\int_{\mathbb{R}^{d}}\beta ^{d}\Psi \left( \beta \mathbf{q}\right) \hat{%
Z}_{+}\left( \beta \mathbf{q},t\right) \mathbf{p}_{g,\sigma }\left( \mathbf{r%
},\beta \mathbf{q}\right) \mathrm{e}^{\mathrm{i}\left( \mathbf{k}_{\ast
}+\beta \mathbf{q}\right) \cdot \mathbf{r}}\,\mathrm{d}\mathbf{q}+O\left(
\beta ^{\sigma +1}\right) .
\end{equation}%
Assuming that $\hat{Z}_{+}\left( \beta \mathbf{s},t\right) $ decays
sufficiently fast as $\left\vert \mathbf{s}\right\vert \rightarrow \infty $,
namely 
\begin{equation}
\beta ^{d}\left\vert \hat{Z}_{+}\left( \beta \mathbf{s},t\right) \right\vert
\leq C_{N}\left( 1+\left\vert \mathbf{s}\right\vert \right) ^{-N_{\Psi }}\ 
\end{equation}%
with large enough $N_{\Psi }$ (see (\ref{Fourbet}) on $\beta $-dependence)
we obtain%
\begin{equation}
\mathbf{U}_{Z_{+}}\left( \mathbf{r},t\right) =\frac{\mathrm{e}^{\mathrm{i}%
\mathbf{k}_{\ast }\cdot \mathbf{r}}}{\left( 2\pi \right) ^{d}}\int_{\mathbb{R%
}^{d}}\hat{Z}_{+}\left( \mathbf{q},t\right) \mathbf{p}_{g,\sigma }\left( 
\mathbf{r},\mathbf{q}\right) \mathrm{e}^{\mathrm{i}\mathbf{q}\cdot \mathbf{r}%
}\,\mathrm{d}\mathbf{q}+O\left( \beta ^{\sigma +1}\right) .  \label{UZr}
\end{equation}%
Note that 
\begin{equation}
\mathbf{p}_{g,\sigma }\left( \mathbf{r},\mathbf{q}\right) \hat{Z}_{+}\left( 
\mathbf{q},t\right) =\widehat{\mathbf{p}_{g,\sigma }\left( \mathbf{r},-%
\mathrm{i}\nabla _{\mathbf{r}}\right) Z_{+}\left( \mathbf{r},t\right) },
\end{equation}%
where $\mathbf{p}^{\left[ \sigma \right] }\left( \mathbf{r},-\mathrm{i}%
\nabla _{\mathbf{r}}\right) $ is a differential operator with the polynomial
symbol $\mathbf{p}^{\left[ \sigma \right] }\left( \mathbf{r},\mathbf{q}%
\right) $ with coefficients that depend on $\mathbf{r}$, see (\ref{GamFo}).
Hence, since the integral in (\ref{UZr}) is the inverse Fourier transform,
we obtain that 
\begin{equation}
\mathbf{U}_{Z_{+}}\left( \mathbf{r},t\right) =\mathrm{e}^{\mathrm{i}\mathbf{k%
}_{\ast }\cdot \mathbf{r}}\mathbf{p}_{g,\sigma }\left( \mathbf{r},-\mathrm{i}%
\nabla _{\mathbf{r}}\right) Z_{+}\left( \mathbf{r},t\right) +O\left( \beta
^{\sigma +1}\right) .
\end{equation}%
For $\sigma =2$ we obtain 
\begin{equation}
\mathbf{U}_{Z_{+}}\left( \mathbf{r},t\right) =\mathbf{U}_{Z_{+}}^{0}\left( 
\mathbf{r},t\right) +\mathbf{U}_{Z_{+}}^{1}\left( \mathbf{r},t\right) +%
\mathbf{U}_{Z_{+}}^{2}\left( \mathbf{r},t\right) +O\left( \beta ^{3}\right) .
\label{UZ+1}
\end{equation}%
According to (\ref{pgsig}), (\ref{pgsig1}) the dominant term is 
\begin{equation}
\mathbf{U}_{Z_{+}}^{0}\left( \mathbf{r},t\right) =\mathrm{e}^{\mathrm{i}%
\mathbf{k}_{\ast }\cdot \mathbf{r}}\mathbf{\hat{G}}_{+,n_{0}}\left( \mathbf{r%
},\mathbf{k}_{\ast }\right) Z_{+}\left( \mathbf{r},t\right) =\mathbf{\tilde{G%
}}_{+,n_{0}}\left( \mathbf{r},\mathbf{k}_{\ast }\right) Z_{+}\left( \mathbf{r%
},t\right) .  \label{UZr0}
\end{equation}%
Note that this term has the form which is used as an ansatz for the solution
of the NLM by \cite{BhatSipe}. The first order correction takes the form 
\begin{gather}
\mathbf{U}_{Z_{+}}^{1}\left( \mathbf{r},t\right) =-\mathrm{ie}^{\mathrm{i}%
\mathbf{k}_{\ast }\cdot \mathbf{r}}\nabla _{\mathbf{k}}\mathbf{\hat{G}}%
_{+,n_{0}}\left( \mathbf{r},\mathbf{k}_{\ast }\right) \cdot \nabla _{\mathbf{%
r}}Z_{+}\left( \mathbf{r},t\right)  \label{UZr1} \\
=-\mathrm{ie}^{\mathrm{i}\mathbf{k}_{\ast }\cdot \mathbf{r}}\left[ \partial
_{r_{1}}Z_{+}\left( \mathbf{r},t\right) \partial _{k_{1}}\mathbf{\hat{G}}%
_{+,n_{0}}\left( \mathbf{r},\mathbf{k}_{\ast }\right) +\ldots +\partial
_{r_{d}}Z_{+}\left( \mathbf{r},t\right) \partial _{k_{d}}\mathbf{\hat{G}}%
_{+,n_{0}}\left( \mathbf{r},\mathbf{k}_{\ast }\right) \right] .  \notag
\end{gather}%
The second order correction is 
\begin{equation}
\mathbf{U}_{Z_{+}}^{2}\left( \mathbf{r},t\right) =-\mathrm{e}^{\mathrm{i}%
\mathbf{k}_{\ast }\cdot \mathbf{r}}\frac{1}{2}\sum_{j,l=1}^{d}\partial
_{k_{j}}\partial _{k_{l}}\mathbf{\hat{G}}_{+,n_{0}}\left( \mathbf{r},\mathbf{%
k}_{\ast }\right) \partial _{r_{j}}\partial _{r_{l}}Z_{+}\left( \mathbf{r}%
,t\right) .  \label{UZr2}
\end{equation}%
Similarly, we can write higher terms of the expansion in $\beta $. Note that
since $Z_{+}\left( \mathbf{r},0\right) =h_{+}\left( \beta \mathbf{r}\right) $
we have 
\begin{equation}
\mathbf{U}_{Z_{+}}^{1}\left( \mathbf{r},t\right) =O\left( \beta \right) ,%
\mathbf{U}_{Z_{+}}^{2}\left( \mathbf{r},t\right) =O\left( \beta ^{2}\right) .
\end{equation}

\section{The first noninear response and its time-harmonic approximation \ }

To improve the accuracy of NLM-NLS\ approximation we have to improve the
term $O\left( \alpha \varrho \right) $ in (\ref{VU0}) originating from the
time-harmonic approximation. If we use a certain modification of the ENLS it
modifies $\widehat{v}_{\zeta }^{\left( 1\right) }\left( \mathbf{q},\tau
\right) $ so that the term $O\left( \alpha \varrho \right) $\ can be
replaced by $O\left( \alpha \varrho \beta \right) $. To this end we consider
in this Section an approximation of the first nonlinear response (FNLR) $%
\mathbf{U}^{\left( 1\right) }$, $\mathbf{\tilde{u}}_{\bar{n}}^{\left(
1\right) }$ determined by (\ref{UU1}), (\ref{uap1}), (\ref{UU3}) and (\ref%
{V01}) in terms of the susceptibility $\mathbf{\chi }_{D}^{\left( 3\right)
}\left( \mathbf{r};\omega _{1},\omega _{2},\omega _{3}\right) $ defined by (%
\ref{cd4ab}). This approximation is based on the following asymptotic
formulas for $\varrho \rightarrow 0$ 
\begin{gather}
\mathbf{\tilde{u}}_{\bar{n}}^{\left( 1\right) }\left( \mathbf{r},\mathbf{k}%
,\tau \right) =\mathbf{\tilde{u}}_{\bar{n}}^{\left( 1,0\right) }\left( 
\mathbf{r},\mathbf{k},\tau \right) +O\left( \varrho \right) O\left(
\left\vert \mathbf{\tilde{u}}_{\bar{n}}^{\left( 1,0\right) }\right\vert
\right) ,  \label{uur1} \\
\ \mathbf{\tilde{u}}_{\bar{n}}^{\left( 1,0\right) }\left( \mathbf{r},\mathbf{%
k},\tau \right) =\tilde{u}_{\bar{n}}^{\left( 1,0\right) }\left( \mathbf{k}%
,\tau \right) \mathbf{\tilde{G}}_{\bar{n}}\left( \mathbf{r},\mathbf{k}%
\right) ,  \notag
\end{gather}%
\begin{equation}
\mathbf{\tilde{u}}_{\bar{n}}^{\left( 1\right) }\left( \mathbf{r},\mathbf{k}%
,\tau \right) =\mathbf{\tilde{u}}_{\bar{n}}^{\left( 1,0\right) }\left( 
\mathbf{r},\mathbf{k},\tau \right) +\mathbf{\tilde{u}}_{\bar{n}}^{\left(
1,1\right) }\left( \mathbf{r},\mathbf{k},\tau \right) +O\left( \varrho
^{2}\right) O\left( \left\vert \mathbf{\tilde{u}}_{\bar{n}}^{\left(
1,0\right) }\right\vert \right) .  \label{uur2}
\end{equation}%
as well as higher order expansions which are derived below.

\subsection{The first noninear response as a causal integral}

In this subsection we recast the FNLR for an almost time-harmonic excitation
as a causal convolution integral. In next subsections we derive asymptotic
expansions without explicit convolution integration. Notice first that the
solution of (\ref{MXshort}) can be written in the form%
\begin{equation}
\mathbf{U}\left( t\right) =\int_{0}^{t}\mathrm{e}^{-\mathrm{i}\mathbf{M}%
\left( t-t^{\prime }\right) }\left[ \alpha \mathcal{F}_{\text{NL}}\left( 
\mathbf{U}\right) -\mathbf{J}\right] \,\mathrm{d}t^{\prime },  \label{MXint0}
\end{equation}%
which, after a change of variables $\tau =\varrho t$, yields the following
expression for the $\bar{n}$--th mode: 
\begin{equation}
\tilde{U}_{\bar{n}}\left( \mathbf{k},\frac{\tau }{\varrho }\right) =\frac{1}{%
\varrho }\int_{0}^{\tau /\varrho }\mathrm{e}^{-\mathrm{i}\omega _{\bar{n}%
}\left( \mathbf{k}\right) \frac{\left( \tau -\tau _{1}\right) }{\varrho }}%
\left[ \widetilde{\mathcal{F}_{\text{NL}}\left( \mathbf{U}\right) }_{\bar{n}%
}\left( \mathbf{k},\tau _{1}\right) -\tilde{J}_{\bar{n}}\left( \mathbf{k}%
,\tau _{1}\right) \right] \,\mathrm{d}\tau _{1}.  \label{MXint}
\end{equation}%
According to (\ref{FNL}) and (\ref{cd4})%
\begin{equation}
\mathcal{F}_{\text{NL}}\left( \mathbf{U}\right) =\mathcal{F}_{\text{NL}%
}^{\left( 3\right) }\left( \mathbf{U}\right) +\alpha \mathcal{F}_{\text{NL}%
}^{\left( 5\right) }\left( \mathbf{U}\right) +\alpha ^{2}\mathcal{F}_{\text{%
NL}}^{\left( 7\right) }\left( \mathbf{U}\right) +\ldots ,  \label{FNL7}
\end{equation}%
and as it follows from (\ref{FNL}) and (\ref{sigma}) the modal coefficient $%
\widetilde{\mathcal{F}_{\text{NL}}\left( \mathbf{U}\right) }_{\bar{n}}\left( 
\mathbf{k},\tau _{1}\right) $ is given by the formula 
\begin{gather}
\widetilde{\mathcal{F}_{\text{NL}}\left( \mathbf{U}\right) }_{\bar{n}}\left( 
\mathbf{k},\tau _{1}\right) =\left( \widetilde{\mathcal{F}_{\text{NL}}\left( 
\mathbf{U}\right) }\left( \cdot ,\tau _{1}\right) ,\mathbf{\tilde{G}}_{\bar{n%
}}\left( \cdot ,\mathbf{k}\right) \right) _{\mathcal{H}}  \label{FNLk} \\
=\int_{\left[ 0,1\right] ^{d}}\widetilde{\mathcal{F}_{\text{NL}}\left( 
\mathbf{U}\right) }\left( \mathbf{r},\tau _{1},\mathbf{k}\right) \cdot
\sigma _{\varepsilon }\left( \mathbf{r}\right) \mathbf{\tilde{G}}_{\bar{n}%
}^{\ast }\left( \mathbf{r},\mathbf{k}\right) \,\mathrm{d}\mathbf{r}.  \notag
\end{gather}%
In particular, using formula (217) of \cite{BF1} this coefficient can be
rewritten as 
\begin{gather}
\widetilde{\mathcal{F}_{\text{NL}}\left( \mathbf{U}\right) }_{\bar{n}}\left( 
\mathbf{k},\tau _{1}\right) =\int_{\left[ 0,1\right] ^{d}}\widetilde{\mathbf{%
S}_{D}\left( \mathbf{r},\tau _{1};\mathbf{D}\right) }\cdot \nabla \times 
\mathbf{\tilde{G}}_{B,\bar{n}}\left( \mathbf{r},\mathbf{k}\right) ^{\ast }\,%
\mathrm{d}\mathbf{r}  \label{FNLkD} \\
=-\mathrm{i}\zeta \omega _{\bar{n}}\left( \mathbf{k}\right) \int_{\left[ 0,1%
\right] ^{d}}\widetilde{\mathbf{S}_{D}\left( \mathbf{r},t;\mathbf{D}\right) }%
\cdot \mathbf{\tilde{G}}_{D,\bar{n}}\left( \mathbf{r},\mathbf{k}\right)
^{\ast }\,\mathrm{d}\mathbf{r}.  \notag
\end{gather}%
This form of coefficients may be useful in computations since it uses only
the $\mathbf{D}$-component of $\mathbf{\tilde{G}}_{\bar{n}}$. In fact, this
specific form is not important in our analysis. By (\ref{caus}) the cubic
part $\mathbf{S}_{D}^{\left( 3\right) }$ of $\mathbf{S}_{D}\left( \mathbf{r}%
,t;\mathbf{D}\right) $ can be written in the form of a causal integral which
involves a symmetric tensor $\mathbf{R}_{D}^{\left( 3\right) }$ 
\begin{gather}
\mathbf{S}_{D}^{\left( 3\right) }\left( \mathbf{r},t;\mathbf{U}\right) =
\label{S3} \\
\int_{-\infty }^{t}\int_{-\infty }^{t}\int_{-\infty }^{t}\mathbf{R}%
_{D}^{\left( 3\right) }\left( \mathbf{r};t-t_{1},t-t_{2},t-t_{3}\right)
\vdots \,\mathbf{D}\left( \mathbf{r},t_{1}\right) \mathbf{D}\left( \mathbf{r}%
,t_{2}\right) \mathbf{D}\left( \mathbf{r},t_{3}\right) \,\mathrm{d}t_{1}%
\mathrm{d}t_{2}\mathrm{d}t_{3},  \notag
\end{gather}%
where the tensors $\mathbf{R}_{D}^{\left( 3\right) }$ are smooth for $%
t_{1},t_{2},t_{3}\geq 0$ and satisfy the inequality 
\begin{equation}
\left\vert \mathbf{R}_{D}^{\left( 3\right) }\left( \mathbf{r}%
;t_{1},t_{2},t_{3}\right) \right\vert \leq C\exp \left[ -c_{0}\left(
t_{1}+t_{2}+t_{3}\right) \right] ,
\end{equation}%
with some $c_{0}>0.$ It is convenient to introduce similarly to (\ref{chiDB}%
) operators $\mathbf{S}_{D,B}^{\left( m\right) }$ and $\mathbf{R}%
_{D,B}^{\left( m\right) }$ that act in 6-dimensional $\left( \mathbf{D},%
\mathbf{B}\right) $-space, they act on the $\mathbf{D}$-components of $%
\mathbf{\tilde{u}}_{\bar{n}^{\prime }}^{\left( 0\right) }\ $and take values
in the $\mathbf{B}$-component. For example, when $m=3$ 
\begin{equation}
\mathbf{R}_{D,B}^{\left( 3\right) }\vdots \mathbf{U}_{1}\mathbf{U}_{2}%
\mathbf{U}_{3}=\left[ 
\begin{array}{c}
\mathbf{0} \\ 
\mathbf{R}_{D}^{\left( 3\right) }\vdots \mathbf{D}_{1}\mathbf{D}_{2}\mathbf{D%
}_{3}%
\end{array}%
\right] ,\quad \mathbf{U}_{j}=\left[ 
\begin{array}{c}
\mathbf{D}_{j} \\ 
\mathbf{B}_{j}%
\end{array}%
\right] .  \label{RDB}
\end{equation}%
Using the above notation we get%
\begin{gather}
\mathbf{S}_{D,B}^{\left( 3\right) }\left( \mathbf{r},t;\mathbf{U}\right) =
\label{SDB} \\
\int_{-\infty }^{t}\int_{-\infty }^{t}\int_{-\infty }^{t}\mathbf{R}%
_{D,B}^{\left( 3\right) }\left( \mathbf{r};t-t_{1},t-t_{2},t-t_{3}\right)
\vdots \,\mathbf{U}\left( \mathbf{r},t_{1}\right) \mathbf{U}\left( \mathbf{r}%
,t_{2}\right) \mathbf{U}\left( \mathbf{r},t_{3}\right) \,\mathrm{d}t_{1}%
\mathrm{d}t_{2}\mathrm{d}t_{3},  \notag
\end{gather}%
and from (\ref{FNL7}) together with (\ref{cd4}) we obtain the following
expansion 
\begin{equation}
\mathcal{F}_{\text{NL}}\left( \mathbf{U}\right) \left( \mathbf{r},t\right)
=\sum_{i=0}^{\infty }\alpha ^{i}\nabla \times \mathbf{S}_{D,B}^{\left(
2i+3\right) }\left( \mathbf{r},t;\mathbf{U}\right) .  \label{FNLDB}
\end{equation}%
To evaluate the integral in (\ref{MXint}) for $\varrho \ll 1$ we, as it is
commonly done in the nonlinear optics, represent the term $\widetilde{%
\mathcal{F}_{\text{NL}}\left( \mathbf{U}\right) }\left( \mathbf{k},\tau
_{1}\right) $ using the frequency-dependent susceptibilities. Using
expansion (\ref{Utild}) and the convolution formula (see \cite{BF1}) we
write the Floquet-Bloch transform of (\ref{SDB}) 
\begin{gather}
\widetilde{\mathbf{S}_{D,B}^{\left( 3\right) }\left( \cdot ;\mathbf{U}%
\right) }\left( \mathbf{r},\mathbf{k}\right) =  \label{R3U} \\
\sum_{\bar{n}^{\prime },\bar{n}^{\prime \prime },\bar{n}^{\prime \prime
\prime }}\int_{-\infty }^{t}\int_{-\infty }^{t}\int_{-\infty }^{t}\int 
_{\substack{ \lbrack -\pi ,\pi ]^{2d}  \\ \mathbf{\mathbf{k}^{\prime }}+%
\mathbf{k}^{\prime \prime }+\mathbf{k}^{\prime \prime \prime }=\mathbf{k}}}%
\frac{\mathrm{d}t_{1}\mathrm{d}t_{2}\mathrm{d}t_{3}}{\left( 2\pi \right)
^{2d}}\tilde{U}_{\bar{n}^{\prime }}\left( \mathbf{k}^{\prime },t_{1}\right) 
\tilde{U}_{\bar{n}^{\prime \prime }}\left( \mathbf{k}^{\prime \prime
},t_{2}\right) \tilde{U}_{\bar{n}^{\prime \prime \prime }}\left( \mathbf{k}%
^{\prime \prime \prime },t_{3}\right)  \notag \\
\mathbf{R}_{D,B}^{\left( 3\right) }\left( \mathbf{r};t-t_{1},t-t_{2},t-t_{3}%
\right) \vdots \mathbf{\tilde{G}}_{\bar{n}^{\prime }}\left( \mathbf{r},%
\mathbf{k}^{\prime }\right) \mathbf{\tilde{G}}_{\bar{n}^{\prime \prime
}}\left( \mathbf{r},\mathbf{k}^{\prime \prime }\right) \mathbf{\tilde{G}}_{%
\bar{n}^{\prime \prime \prime }}\left( \mathbf{r},\mathbf{k}^{\prime \prime
\prime }\right) \,\mathrm{d}\mathbf{k}^{\prime }\mathrm{d}\mathbf{k}^{\prime
\prime }.  \notag
\end{gather}%
We rewrite (\ref{R3U}) in terms of slowly varying coefficients $\tilde{u}_{%
\bar{n}}$ defined by (\ref{UFB1})

\begin{gather}
\widetilde{\mathbf{S}_{D,B}^{\left( 3\right) }\left( \cdot ;\mathbf{U}%
\right) }\left( \mathbf{r},\mathbf{k},t\right) =\sum_{\bar{n}^{\prime },\bar{%
n}^{\prime \prime },\bar{n}^{\prime \prime \prime }}\int_{\substack{ \lbrack
-\pi ,\pi ]^{2d}  \\ \mathbf{\mathbf{k}^{\prime }}+\mathbf{k}^{\prime \prime
}+\mathbf{k}^{\prime \prime \prime }=\mathbf{k}}}\mathrm{e}^{\left[ -\mathrm{%
i}\left( \omega _{\bar{n}^{\prime }}\left( \mathbf{k}^{\prime }\right)
+\omega _{\bar{n}^{\prime \prime }}\left( \mathbf{k}^{\prime \prime }\right)
+\omega _{\bar{n}^{\prime \prime \prime }}\left( \mathbf{k}^{\prime \prime
\prime }\right) \right) t\right] }  \label{Ru} \\
\int_{-\infty }^{t}\int_{-\infty }^{t}\int_{-\infty }^{t}\mathrm{e}^{\left[ 
\mathrm{i}\left( \omega _{\bar{n}^{\prime }}\left( \mathbf{k}^{\prime
}\right) \left( t-t_{1}\right) +\omega _{\bar{n}^{\prime \prime }}\left( 
\mathbf{k}^{\prime \prime }\right) \left( t-t_{2}\right) +\omega _{\bar{n}%
^{\prime \prime \prime }}\left( \mathbf{k}^{\prime \prime \prime }\right)
\left( t-t_{3}\right) \right) \right] }  \notag \\
\mathbf{R}_{D,B}^{\left( 3\right) }\left( \mathbf{r};t-t_{1},t-t_{2},t-t_{3}%
\right) \vdots \,\mathbf{\tilde{G}}_{\bar{n}^{\prime }}\left( \mathbf{r},%
\mathbf{k}^{\prime }\right) \mathbf{\tilde{G}}_{\bar{n}^{\prime \prime
}}\left( \mathbf{r},\mathbf{k}^{\prime \prime }\right) \mathbf{\tilde{G}}_{%
\bar{n}^{\prime \prime \prime }}\left( \mathbf{r},\mathbf{k}^{\prime \prime
\prime }\right)  \notag \\
\tilde{u}_{\bar{n}^{\prime }}\left( \mathbf{k}^{\prime },\tau _{1}\right) 
\tilde{u}_{\bar{n}^{\prime \prime }}\left( \mathbf{k}^{\prime \prime },\tau
_{2}\right) \tilde{u}_{\bar{n}^{\prime \prime \prime }}\left( \mathbf{k}%
^{\prime \prime \prime },\tau _{3}\right) \,\frac{\mathrm{d}t_{1}\mathrm{d}%
t_{2}\mathrm{d}t_{3}\mathrm{d}\mathbf{k}^{\prime }\mathrm{d}\mathbf{k}%
^{\prime \prime }}{\left( 2\pi \right) ^{2d}},  \notag \\
\tau _{j}=\varrho t_{j},\ j=1,2,3.  \notag
\end{gather}%
The FNLR has the form similar to (\ref{MXint0}) 
\begin{equation}
\tilde{U}_{\bar{n}}^{\left( 1\right) }\left( \mathbf{k},t^{\prime }\right)
=\int_{0}^{t^{\prime }}\mathrm{e}^{\mathrm{i}\omega _{\bar{n}}\left( \mathbf{%
k}\right) \left( t^{\prime }-t\right) }\left[ \widetilde{\mathcal{F}_{\text{%
NL}}\left( \mathbf{U}^{\left( 0\right) }\right) }_{\bar{n}}^{\left( 3\right)
}\left( \mathbf{k},t\right) -\tilde{J}_{\bar{n}}^{\left( 1\right) }\left( 
\mathbf{k},t\right) \right] \,\,\mathrm{d}t.  \label{FNRc}
\end{equation}%
By (\ref{FNL7}), (\ref{FNLk}), and (\ref{S3}) the modal coefficient of the
FNLR is given by%
\begin{equation}
\mathbf{\tilde{u}}_{\bar{n}}^{\left( 1\right) }\left( \mathbf{r},\mathbf{k}%
,\varrho t^{\prime }\right) =\mathrm{e}^{\mathrm{i}\omega _{\bar{n}}\left( 
\mathbf{k}\right) t^{\prime }}\widetilde{\mathbf{S}_{D,B}^{\left( 3\right)
}\left( \cdot ;\mathbf{D}^{\left( 0\right) }\right) }_{\bar{n}}\left( 
\mathbf{r},\mathbf{k},t^{\prime }\right) -\mathbf{\tilde{u}}_{\bar{n}%
}^{\left( 1\right) }\left( \mathbf{J}_{1};\mathbf{k},\varrho t^{\prime
}\right) ,  \label{ucS}
\end{equation}%
where $\mathbf{\tilde{u}}_{\bar{n}}^{\left( 1\right) }\left( \mathbf{J}_{1};%
\mathbf{k},\tau \right) $ is defined by (\ref{un1J}). The relations(\ref{vz}%
) and (\ref{Ru}) imply 
\begin{gather}
\widetilde{\mathbf{S}_{D,B}^{\left( 3\right) }\left( \mathbf{\cdot };\mathbf{%
U}^{\left( 0\right) }\right) }\left( \mathbf{r},\mathbf{k},t^{\prime
}\right) =  \label{intcaus} \\
=\sum_{\bar{n}^{\prime },\bar{n}^{\prime \prime },\bar{n}^{\prime \prime
\prime }}\frac{1}{\left( 2\pi \right) ^{2d}}\int_{0}^{t^{\prime }}\mathrm{e}%
^{\left\{ -\mathrm{i}\sum_{j=1}^{3}\omega _{\bar{n}^{\left( j\right)
}}\left( \mathbf{k}^{\left( j\right) }\right) t\right\} }\int_{\substack{ %
\lbrack -\pi ,\pi ]^{2d}  \\ \mathbf{k}^{\prime }+\mathbf{k}^{\prime \prime
}+\mathbf{k}^{\prime \prime \prime }=\mathbf{k}}}\int_{-\infty
}^{t}\int_{-\infty }^{t}\int_{-\infty }^{t}  \notag \\
\mathrm{e}^{\left\{ -\mathrm{i}\sum_{j=1}^{3}\omega _{\bar{n}^{\left(
j\right) }}(\mathbf{k}^{\left( j\right) })\left( t_{j}-t\right) \right\} }%
\mathbf{R}_{D,B}^{\left( 3\right) }\left( \mathbf{r};t-t_{1},t-t_{2},t-t_{3}%
\right) \mathbf{\vdots }  \notag \\
\prod_{j=1}^{3}\mathbf{\tilde{G}}_{\bar{n}^{\left( j\right) }}\left( \mathbf{%
k}^{\left( j\right) },\mathbf{r}\right) \Psi _{0}^{3}\left( \mathbf{k}%
\right) \psi \left( \varrho t_{j}\right) \mathring{h}_{\zeta ^{\left(
j\right) }}\left( \frac{\mathbf{k}^{\left( j\right) }\mathbf{-}\zeta
^{\left( j\right) }\mathbf{k}_{\ast }}{\beta }\right) \,\mathrm{d}t_{1}%
\mathrm{d}t_{2}\mathrm{d}t_{3}\mathrm{d}\mathbf{k}^{\prime }\mathrm{d}%
\mathbf{k}^{\prime \prime }\mathrm{d}t,  \notag
\end{gather}%
and the obtained integral is simplified in the next subsection.

\subsection{Time-harmonic approximation}

In this section we introduce an expansion yielding powers $\varrho ^{l_{1}}$
in the structured power series (\ref{asser}). In particular, we obtain (\ref%
{uur1}), (\ref{uur2}). If $\mathbf{U}_{n}^{\left( 0\right) }\left( \mathbf{r}%
,t\right) $ has the form (\ref{U0}) of a slowly modulated wavepacket with $%
\varrho \ll 1$ then the time-harmonic approximation can be applied. It
effectively substitutes the integration with respect to time in the causal
integral (\ref{intcaus}) with expressions invlolving frequency dependent
susceptibilities. This approximation is based on the Fourier transform $%
\mathbf{\chi }_{D}^{\left( 3\right) }$ of $\mathbf{R}^{\left( 3\right) }$
with respect to the time variables as in (\ref{cd4ab}), and it is
constructed as follows. Below we approximate $\tilde{u}_{\bar{n}}^{\left(
1\right) }\left( \mathbf{k},\tau \right) $ in (\ref{ucS}) by $\tilde{u}_{%
\bar{n}}^{\left( 1,0\right) }\left( \mathbf{k},t\right) $, which is defined
by (\ref{FNR1}), (\ref{uQ}), and then estimate the error providing the
higher order terms as well. Using in (\ref{intcaus}) the Taylor
approximation of $\psi \left( \varrho t_{j}\right) $ we get: 
\begin{equation}
\psi \left( \varrho \left( t_{j}-t\right) +\varrho t\right)
=\sum_{l=0}^{N_{1}}\frac{\left( -1\right) ^{l}\varrho ^{l}}{l!}\psi ^{\left(
l\right) }\left( \varrho t\right) \left( t-t_{j}\right) ^{l}+\psi _{\left(
N_{1}\right) }
\end{equation}%
with 
\begin{equation}
\left\vert \psi _{\left( N_{1}\right) }\right\vert \leq C_{N_{1}}\varrho
^{N_{1}+1}\left\vert t_{j}-t\right\vert ^{N_{1}+1}.  \label{psiN}
\end{equation}%
Substituting then the Taylor polynomial approximation for $\psi \left(
\varrho t_{j}\right) $ into (\ref{intcaus}) we obtain%
\begin{gather}
\int_{-\infty }^{t}\int_{-\infty }^{t}\int_{-\infty }^{t}\mathbf{R}%
_{D,B}^{\left( 3\right) }\left( \mathbf{r};t-t_{1},t-t_{2},t-t_{3}\right) 
\mathrm{e}^{\left\{ -\mathrm{i}\sum_{j=1}^{3}\omega _{\bar{n}^{\left(
j\right) }}(\mathbf{k}^{\left( j\right) })\left( t_{j}-t\right) \right\} }
\label{Rl} \\
\prod_{j=1}^{3}\frac{\left( -1\right) ^{l_{j}}\varrho ^{l_{j}}}{l_{j}!}%
\left( t-t_{j}\right) ^{l_{j}}\psi ^{\left( l_{j}\right) }\left( \varrho
t\right) \,\mathrm{d}t_{1}\mathrm{d}t_{2}\mathrm{d}t_{3}=  \notag \\
\int_{0}^{\infty }\int_{0}^{\infty }\int_{0}^{\infty }\mathbf{R}_{D}^{\left(
3\right) }\left( \mathbf{r};t_{1},t_{2},t_{3}\right) \mathrm{e}^{\left\{ 
\mathrm{i}\sum_{j=1}^{3}\omega _{\bar{n}^{\left( j\right) }}(\mathbf{k}%
^{\left( j\right) })t_{j}\right\} }\prod_{j=1}^{3}\frac{\left( -1\right)
^{l_{j}}\varrho ^{l_{j}}}{l_{j}!}t_{j}^{l_{j}}\psi ^{\left( l_{j}\right)
}\left( \varrho t\right) \,\mathrm{d}t_{1}\ldots \mathrm{d}t_{3}.  \notag
\end{gather}%
Let us introduce the following notation for the above integral with$\ \left(
l_{1},l_{2},l_{3}\right) =\bar{l}$: 
\begin{gather}
\mathbf{\chi }_{D,B,\bar{l}}^{\left( 3\right) }\left( \mathbf{r};\omega
_{1},\omega _{2},\omega _{3}\right) =  \label{chil} \\
\int_{0}^{\infty }\int_{0}^{\infty }\int_{0}^{\infty }\mathbf{R}%
_{D,B}^{\left( 3\right) }\left( \mathbf{r};t_{1},\ldots ,t_{3}\right) 
\mathrm{e}^{\mathrm{i}\left( \omega _{1}t_{1}+\omega _{2}t_{2}+\omega
_{3}t_{3}\right) }\prod_{j=1}^{3}\frac{\left( -1\right) ^{l_{j}}}{l_{j}!}%
t_{j}^{l_{j}}\,\mathrm{d}t_{1}\mathrm{d}t_{2}\mathrm{d}t_{3}.  \notag
\end{gather}%
Evidently the susceptibility $\mathbf{\chi }_{D,B}^{\left( 3\right) }$
defined by (\ref{chiDB}), (\ref{cd4ab}) equals $\mathbf{\chi }%
_{D,B,0}^{\left( 3\right) }$. A straightforward computation shows that the
quantity $\mathbf{\chi }_{D,B,\bar{l}}^{\left( 3\right) }\left( \mathbf{r}%
;\omega _{1},\omega _{2},\omega _{3}\right) $\ defined by (\ref{chil})
equals the partial derivative of $\mathbf{\chi }_{D,B}^{\left( 3\right) }$,
defined by (\ref{chiDB}), (\ref{cd4ab}), with respect to its frequency
arguments, namely 
\begin{equation}
\mathbf{\chi }_{D,B,\bar{l}}^{\left( 3\right) }\left( \mathbf{r};\omega
_{1},\omega _{2},\omega _{3}\right) =\frac{\mathrm{i}^{\left\vert \bar{l}%
\right\vert }}{l_{1}!l_{2}!l_{3}!}\frac{\partial ^{\left\vert \bar{l}%
\right\vert }\mathbf{\chi }_{D,B}^{\left( 3\right) }\left( \mathbf{r};\omega
_{1},\omega _{2},\omega _{3}\right) }{\partial \omega _{1}^{l_{1}}\partial
\omega _{2}^{l_{2}}\partial \omega _{3}^{l_{3}}}.  \label{child}
\end{equation}%
Substituting (\ref{Rl}) into (\ref{intcaus}) we obtain the following formula%
\begin{gather}
\widetilde{\nabla \times \mathbf{S}_{D,B}^{\left( 3\right) }\left( \cdot ;%
\mathbf{U}^{\left( 0\right) }\right) }_{\bar{n}}\left( \mathbf{k},t\right) =
\label{SDBsum} \\
\frac{1}{\left( 2\pi \right) ^{2d}}\sum_{\left\vert \bar{l}\right\vert
=0}^{N_{1}}\sum_{\bar{n}^{\prime },\bar{n}^{\prime \prime },\bar{n}^{\prime
\prime \prime }}\int_{0}^{t}\widetilde{\mathbf{Q}\left( \left( \mathbf{u}%
^{\left( 0\right) }\right) ^{3}\right) }\left( \vec{n},\bar{l},\mathbf{k}%
,\varrho t^{\prime }\right) \,\mathrm{d}t^{\prime }+O\left( \varrho
^{N_{1}+1}\right) ,  \notag
\end{gather}%
where%
\begin{gather}
\widetilde{\mathbf{Q}\left( \left( \mathbf{u}^{\left( 0\right) }\right)
^{3}\right) }\left( \vec{n},\bar{l},\mathbf{k},\varrho t\right) =\int_{0}^{t}%
\mathrm{e}^{\mathrm{i}\phi _{\vec{n}}\left( \vec{k}\right) t}\int_{\substack{
\lbrack -\pi ,\pi ]^{2d}  \\ \mathbf{k}^{\prime }+\mathbf{k}^{\prime \prime
}+\mathbf{k}^{\prime \prime \prime }=\mathbf{k}}}\int_{\left[ 0,1\right]
^{d}}  \label{Uc1} \\
\nabla \times \left[ \mathbf{\chi }_{D,B,\bar{l}}^{\left( 3\right) }\left( 
\mathbf{r};\omega _{\bar{n}^{\prime }}\left( \mathbf{k}^{\prime }\right)
,\omega _{\bar{n}^{\prime \prime }}\left( \mathbf{k}^{\prime \prime }\right)
,\omega _{\bar{n}^{\prime \prime \prime }}\left( \mathbf{k}^{\prime \prime
\prime }\right) \right) \mathbf{\vdots \,\,}\prod_{j=1}^{3}\mathbf{\tilde{G}}%
_{\bar{n}^{\left( j\right) }}\left( \mathbf{k}^{\left( j\right) },\mathbf{r}%
\right) \right] \cdot \mathbf{\tilde{G}}_{\bar{n}}\left( \mathbf{k},\mathbf{r%
}\right) \,\mathrm{d}\mathbf{r}  \notag \\
\Psi _{0}^{3}\left( \mathbf{\vec{s}}\right) \prod_{j=1}^{3}\psi ^{\left(
l_{j}\right) }\left( \varrho t^{\prime }\right) \mathring{h}_{\zeta ^{\left(
j\right) }}\left( \frac{\mathbf{k}^{\left( j\right) }-\zeta ^{\left(
j\right) }\mathbf{k}_{\ast }}{\beta }\right) \,\mathrm{d}\mathbf{k}^{\prime }%
\mathrm{d}\mathbf{k}^{\prime \prime }\mathrm{d}t^{\prime }.  \notag
\end{gather}%
Hence, we obtain from (\ref{intcaus}) the formula for the modal
coefficients: 
\begin{gather}
u_{\bar{n}}^{\left( 1,\bar{l}\right) }\left( \mathbf{k},\tau \right) =\frac{1%
}{\varrho }\sum_{\bar{n}^{\prime },\bar{n}^{\prime \prime },\bar{n}^{\prime
\prime \prime }}\int_{0}^{\tau }\mathrm{e}^{\mathrm{i}\phi _{\vec{n}}\left( 
\vec{k}\right) \frac{\tau _{1}}{\varrho }}\int_{\substack{ \lbrack -\pi ,\pi
]^{2d}  \\ \mathbf{k}^{\prime }+\mathbf{k}^{\prime \prime }+\mathbf{k}%
^{\prime \prime \prime }=\mathbf{k}}}  \label{Uc1m} \\
\breve{Q}_{\vec{n},\bar{l}}\left( \vec{k}\right) \Psi _{0}^{3}\left( \mathbf{%
\vec{s}}\right) \prod_{j=1}^{3}\psi ^{\left( l_{j}\right) }\left( \tau
_{1}\right) \beta ^{-d}\mathring{h}_{\zeta ^{\left( j\right) }}\left( \left( 
\mathbf{k}^{\left( j\right) }-\zeta ^{\left( j\right) }\mathbf{k}_{\ast
}\right) \mathbf{/}\beta \right) \,\mathrm{d}\mathbf{k}^{\prime }\mathrm{d}%
\mathbf{k}^{\prime \prime }\mathrm{d}\tau _{1},  \notag
\end{gather}%
where $\breve{Q}_{\vec{n},\bar{l}}\left( \vec{k}\right) $ is given by the
following formula similar to (\ref{Qn}) with $\mathbf{\chi }_{D}^{\left(
3\right) }$ being replaced with its frequency derivative $\mathbf{\chi }%
_{D,B,\bar{l}}^{\left( 3\right) }$%
\begin{gather}
\breve{Q}_{\vec{n},\bar{l}}\left( \vec{k}\right) =\frac{1}{(2\pi )^{2d}}%
\left( \left[ 
\begin{array}{c}
\mathbf{0} \\ 
\nabla \times \mathbf{\chi }_{D,B,\bar{l}}^{\left( 3\right) }%
\end{array}%
\right] ,\mathbf{\tilde{G}}_{\bar{n}}\left( \mathbf{r},\mathbf{k}\right)
\right) _{\mathcal{H}},  \label{Qnl} \\
\mathbf{\chi }_{D,B,\bar{l}}^{\left( 3\right) }=  \notag \\
\mathbf{\chi }_{D,B,\bar{l}}^{\left( 3\right) }\left( \omega _{\bar{n}%
^{\prime }}\left( \mathbf{k}^{\prime }\right) ,\omega _{\bar{n}^{\prime
\prime }}\left( \mathbf{k}^{\prime \prime }\right) ,\omega _{\bar{n}^{\prime
\prime \prime }}\left( \mathbf{k}^{\prime \prime \prime }\right) \right) 
\mathbf{\tilde{G}}_{D,\bar{n}^{\prime }}\left( \mathbf{r},\mathbf{k}^{\prime
}\right) \mathbf{\tilde{G}}_{D,\bar{n}^{\prime \prime }}\left( \mathbf{r},%
\mathbf{k}^{\prime \prime }\right) \mathbf{\tilde{G}}_{D,\bar{n}^{\prime
\prime \prime }}\left( \mathbf{r},\mathbf{k}^{\prime \prime \prime }\right) 
\notag
\end{gather}%
Observe that the modal susceptibility $\ \breve{Q}_{\vec{n}}\left( \vec{k}%
\right) =\breve{Q}_{\vec{n},0}\left( \vec{k}\right) $ defined by (\ref{Qn})
is symmetric with respect to permutations of $\zeta ^{\left( j\right) }$, $%
\mathbf{k}^{\left( j\right) }$ whereas $\breve{Q}_{\vec{n},\bar{l}}\left( 
\vec{k}\right) $ defined by (\ref{Qnl}) with non-symmetric $\bar{l}$ is not.

Hence, taking the term at $\alpha $ in (\ref{FNLDB}) and using (\ref{SDBsum}%
) we obtain that 
\begin{equation}
\tilde{u}_{\bar{n}}^{\left( 1\right) }\left( \mathbf{k},\tau \right) =\tilde{%
u}_{\bar{n}}^{\left( 1,0\right) }\left( \mathbf{k},\tau \right)
+\sum_{l=1}^{N_{1}}\varrho ^{l}\tilde{u}_{\bar{n}}^{\left( 1,l\right)
}\left( \mathbf{k},\tau \right) +O\left( \varrho ^{N_{1}+1}\right) .
\label{uc1}
\end{equation}%
Using (\ref{Uc1}), (\ref{Uc1m}) we express $\tilde{u}_{\bar{n}}^{\left(
1,l\right) }$ as an action of an operator $\tilde{\Xi}_{\bar{n},l}$ on $%
\mathbf{u}^{\left( 0\right) }$, namely 
\begin{equation}
\tilde{u}_{\bar{n}}^{\left( 1,l\right) }\left( \mathbf{k},\tau \right) =%
\tilde{\Xi}_{\bar{n},l}\left[ \mathbf{u}^{\left( 0\right) }\right] \left( 
\mathbf{k},\tau \right) ,\,\tilde{\Xi}_{\bar{n},l}\left[ \mathbf{u}^{\left(
0\right) }\right] =\sum_{l_{1}+l_{2}+l_{3}=l}\tilde{\Xi}_{\bar{n}}^{\left( 
\mathbf{l}\right) }\left[ \mathbf{u}^{\left( 0\right) }\right] ,\ \mathbf{l}%
=\left( l_{1},l_{2},l_{3}\right) ,  \label{FNRL}
\end{equation}%
\begin{gather}
\tilde{\Xi}_{\bar{n}}^{\left( \mathbf{l}\right) }\left[ \mathbf{u}^{\left(
0\right) }\right] \left( \mathbf{k},\tau \right) =\frac{1}{\varrho }\sum_{%
\bar{n}^{\prime },\bar{n}^{\prime \prime },\bar{n}^{\prime \prime \prime
}}\int_{0}^{\tau }\int_{\substack{ \lbrack -\pi ,\pi ]^{2d}  \\ \mathbf{%
\mathbf{k}^{\prime }}+\mathbf{k}^{\prime \prime }+\mathbf{k}^{\prime \prime
\prime }=\mathbf{k}}}\mathrm{e}^{\mathrm{i}\phi _{\vec{n}}\left( \vec{k}%
\right) \frac{\tau _{1}}{\varrho }}\,  \label{Ksi} \\
\left( \mathbf{\chi }_{D,\bar{l}}^{\left( 3\right) }\vdots \mathbf{\tilde{u}}%
_{\bar{n}^{\prime }}^{\left( 0\right) }\left( \mathbf{r},\mathbf{k}^{\prime
},\tau _{1}\right) \mathbf{\tilde{u}}_{\bar{n}^{\prime \prime }}^{\left(
0\right) }\left( \mathbf{r},\mathbf{k}^{\prime \prime },\tau _{1}\right) 
\mathbf{\tilde{u}}_{\bar{n}^{\prime \prime \prime }}^{\left( 0\right)
}\left( \mathbf{r},\mathbf{k}^{\prime \prime \prime },\tau _{1}\right) \,,%
\mathbf{\tilde{G}}_{\bar{n}}\left( \mathbf{r},\mathbf{k}\right) \right) _{%
\mathcal{H}}\mathrm{d}\mathbf{k}^{\prime }\mathrm{d}\mathbf{k}^{\prime
\prime }\mathrm{d}\tau _{1},  \notag \\
\mathbf{\chi }_{D,\bar{l}}^{\left( 3\right) }=\mathbf{\chi }_{D,\bar{l}%
}^{\left( 3\right) }\left( \mathbf{r};\omega _{1}\left( \mathbf{k}^{\prime
}\right) ,\omega _{2}\left( \mathbf{k}^{\prime \prime }\right) ,\omega
_{3}\left( \mathbf{k}^{\prime \prime \prime }\right) \right) .
\end{gather}%
Then we get from (\ref{uc1}) the following expansion for the modal
coefficients 
\begin{equation}
\tilde{u}_{\bar{n}}^{\left( 1\right) }\left( \mathbf{k},t\varrho \right) =%
\tilde{u}_{\bar{n}}^{\left( 1,0\right) }\left( \mathbf{k},t\varrho \right)
+\sum_{1\leq l_{1}+l_{2}+l_{3}\leq N_{1}}\varrho ^{l_{1}+l_{2}+l_{3}}\tilde{%
\Xi}_{\bar{n}}^{\left( \mathbf{l}\right) }\left[ \mathbf{u}^{\left( 0\right)
}\right] \left( \mathbf{k},t\varrho \right) +O\left( \varrho
^{N_{1}+1}\right) .  \label{VcV1}
\end{equation}%
This implies (\ref{uur1}), (\ref{uur2}) when $N_{1}=0,1$ respectively. The
dominant term $\tilde{u}_{\bar{n}}^{\left( 1,0\right) }\left( \mathbf{k}%
,t\varrho \right) $ corresponds to $\left\vert \mathbf{l}\right\vert
=l_{1}+l_{2}+l_{3}=0$, and it is given by (\ref{Vn}). In particular, using
rectifying variables about $\mathbf{k}_{\ast }$ we obtain the formula 
\begin{equation}
\beta ^{d}\tilde{u}_{\bar{n}}^{\left( 1\right) }\left( \zeta \mathbf{k}%
_{\ast }+Y_{\zeta }\left( \beta \mathbf{q}\right) ,\tau \right) =\beta ^{d}%
\tilde{u}_{\bar{n}}^{\left( 1,0\right) }\left( \zeta \mathbf{k}_{\ast
}+Y_{\zeta }\left( \beta \mathbf{q}\right) ,\tau \right) +O\left( \varrho
\right) O\left( \left\vert \mathbf{U}^{\left( 1\right) }\right\vert \right) .
\label{suserror}
\end{equation}

\textbf{An example.} A typical and rather common in optics example of the
response function is 
\begin{equation}
\mathbf{R}_{D}^{\left( 3\right) }\left( \mathbf{r};t_{1},t_{2},t_{3}\right)
\vdots \,\mathbf{\mathrm{e}}^{\prime }\mathbf{\mathrm{e}}^{\prime \prime }%
\mathbf{\mathrm{e}}^{\prime \prime \prime }=\left\{ 
\begin{tabular}{ll}
$\exp \left\{ -c\left[ t_{1}+t_{2}+t_{3}\right] \right\} \mathbf{R}%
_{0D}^{\left( 3\right) }\left( \mathbf{r}\right) \vdots \,$\textbf{$\mathrm{e%
}$}$^{\prime }$\textbf{$\mathrm{e}$}$^{\prime \prime }$\textbf{$\mathrm{e}$}$%
^{\prime \prime \prime }$ & if all $\tau _{j}\geq 0$ \\ 
$0$ & otherwise%
\end{tabular}%
\right.  \label{Rex}
\end{equation}%
where $\mathbf{R}_{0D}^{\left( 3\right) }\left( \mathbf{r}\right) \vdots
\,\, $\textbf{$\mathrm{e}$}$^{\prime }$\textbf{$\mathrm{e}$}$^{\prime \prime
}$\textbf{$\mathrm{e}$}$^{\prime \prime \prime }$ is a $3$-linear symmetric
form of vectors \textbf{$\mathrm{e}$}$^{\prime },$\textbf{$\mathrm{e}$}$%
^{\prime \prime },$\textbf{$\mathrm{e}$}$^{\prime \prime \prime }\in \mathbb{%
C}^{3}$ not depending on $t_{1},t_{2},t_{3}$, and $c>0$ is a constant. In
this case 
\begin{equation}
\mathbf{\chi }_{D}^{\left( 3\right) }\left( \mathbf{r};\omega _{1},\omega
_{2},\omega _{3}\right) =\frac{1}{\left( c-\mathrm{i}\omega _{1}\right)
\left( c-\mathrm{i}\omega _{2}\right) \left( c-\mathrm{i}\omega _{3}\right) }%
\mathbf{R}_{0D}^{\left( 3\right) }\left( \mathbf{r}\right)  \label{chiex}
\end{equation}%
and%
\begin{equation}
\mathbf{\chi }_{D,\bar{l}}^{\left( 3\right) }\left( \mathbf{r};\omega
_{1},\omega _{2},\omega _{3}\right) =\frac{\left( -1\right) ^{\left\vert 
\bar{l}\right\vert }}{\left( c-\mathrm{i}\omega _{1}\right) ^{l_{1}+1}\left(
c-\mathrm{i}\omega _{2}\right) ^{l_{2}+1}\left( c-\mathrm{i}\omega
_{3}\right) ^{l_{3}+1}}\mathbf{R}_{0D}^{\left( 3\right) }\left( \mathbf{r}%
\right) .  \label{chilom}
\end{equation}%
The modal susceptibility (\ref{Qn}) according to (\ref{FNLkD}) takes the form%
\begin{gather}
\breve{Q}_{\vec{n}}\left( \vec{k}\right) =\frac{1}{(2\pi )^{2d}}\frac{-%
\mathrm{i}\zeta \omega _{\zeta ,n}\left( \mathbf{k}\right) }{\left( c-%
\mathrm{i}\zeta ^{\prime }\omega _{n^{\prime }}\left( \mathbf{k}^{\prime
}\right) \right) \left( c-\mathrm{i}\zeta ^{\prime }\omega _{n^{\prime
\prime }}\left( \mathbf{k}^{\prime \prime }\right) \right) \left( c-\mathrm{i%
}\zeta ^{\prime }\omega _{n^{\prime \prime \prime }}\left( \mathbf{k}%
^{\prime \prime \prime }\right) \right) }  \label{QnD} \\
\int_{\left[ 0,1\right] ^{d}}\mathbf{R}_{0D}^{\left( 3\right) }\left( 
\mathbf{r}\right) \vdots \,\mathbf{\tilde{G}}_{D,\zeta ^{\prime },n^{\prime
}}\left( \mathbf{r},\mathbf{k}^{\prime }\right) \mathbf{\tilde{G}}_{D,\zeta
^{\prime \prime },n^{\prime \prime }}\left( \mathbf{r},\mathbf{k}^{\prime
\prime }\right) \mathbf{\tilde{G}}_{D,\zeta ^{\prime \prime \prime
},n^{\prime \prime \prime }}\left( \mathbf{r},\mathbf{k}^{\prime \prime
\prime }\right) \mathbf{\tilde{G}}_{D,\zeta ,n}^{\ast }\left( \mathbf{r},%
\mathbf{k}\right) \,\mathrm{d}\mathbf{r.}  \notag
\end{gather}%
If one would like to replace the function $\exp \left\{ -c\left[
t_{1}+t_{2}+t_{3}\right] \right\} $ in (\ref{Rex}) with a more general one
such as $\exp \left\{ -\left[ c_{1}t_{1}+c_{2}t_{2}+c_{3}t_{3}\right]
\right\} $ then, to provide the symmetry condition (\ref{Rsym}), one has to
apply the symmetrization operation with respect to permutations of indices $%
c_{j}$ after which the both $\mathbf{R}_{D}^{\left( 3\right) }$ and $\mathbf{%
\chi }_{D}^{\left( 3\right) }$ become the sums of 6 terms obtained by the
permutations.

\textbf{Remark.} Though here we consider the time-harmonic approximation of
the third order term $\nabla _{B}\times \mathbf{S}_{D,B}^{\left( 3\right)
}\left( \mathbf{r},t;\mathbf{U}^{\left( 0\right) }\right) $ in (\ref{FNLDB}%
), similar time-harmonic approximations are applicable to terms $\nabla
\times \mathbf{S}_{D,B}^{\left( m\right) }\left( \mathbf{r},t;\mathbf{U}%
^{\left( 0\right) }\right) $ of an arbitrary order $m$ of homogeneity.$%
\blacklozenge $

\subsection{The first order correction to the susceptibility}

For $\left\vert \bar{l}\right\vert =1$, the term $\tilde{u}_{\bar{n}%
}^{\left( 1,1\right) }$ involves three similar expressions of the form (\ref%
{Uc1m}) with $l_{1}+l_{2}+l_{3}=1$. Hence, 
\begin{gather}
\tilde{u}_{\bar{n}}^{\left( 1,1\right) }\left( \mathbf{r},\mathbf{k},\tau
\right) =\sum_{l_{1}+l_{2}+l_{3}=1}\tilde{u}_{\bar{n}}^{\left( 1,\bar{l}%
\right) }=\frac{1}{\varrho }\int_{0}^{\tau }\mathrm{e}^{\left\{ -\mathrm{i}%
\sum_{j=1}^{3}\omega _{\bar{n}^{\left( j\right) }}\left( \mathbf{k}^{\left(
j\right) }\right) \frac{\tau _{1}}{\varrho }\right\} }\int_{\substack{ %
\lbrack -\pi ,\pi ]^{2d}  \\ \mathbf{k}^{\prime }+\mathbf{k}^{\prime \prime
}+\mathbf{k}^{\prime \prime \prime }=\mathbf{k}}}  \label{unlt} \\
\breve{Q}_{\vec{n},\bar{l}}\left( \vec{k}\right) \Psi _{0}^{3}\left( \mathbf{%
\vec{s}}\right) \beta ^{-d}\psi ^{2}\left( \tau _{1}\right) \psi ^{\prime
}\left( \tau _{1}\right) \prod_{j=1}^{3}\mathring{h}_{\zeta ^{\left(
j\right) }}\left( \frac{1}{\beta }\left( \mathbf{k}^{\left( j\right) }-\zeta
^{\left( j\right) }\mathbf{k}_{\ast }\right) \right) \,\mathrm{d}\mathbf{k}%
^{\prime }\mathrm{d}\mathbf{k}^{\prime \prime }\mathrm{d}\tau _{1}.  \notag
\end{gather}%
The integral (\ref{Uc1m}) is similar to (\ref{Vn0}) and (\ref{Inzbet}).
Consequently, the principal contribution there is given by the FM terms such
that $\vec{n}$ \ satisfy (\ref{ndiag}) and (\ref{FMCz}). In the rectifying
variables\ the integral in (\ref{Uc1m}) with such $\vec{n}$ takes the form
similar to (\ref{IY}), i.e. 
\begin{gather}
\beta ^{d}\tilde{u}_{\bar{n}}^{\left( 1,\bar{l}\right) }\left( \zeta \mathbf{%
k}_{\ast }+Y_{\zeta }\left( \beta \mathbf{q}\right) ,\tau \right) =
\label{susy} \\
\frac{1}{\varrho }\int_{0}^{\tau }\int_{Y_{\zeta }\left( \beta \mathbf{q}%
^{\prime }\right) +Y_{\zeta }\left( \beta \mathbf{q}^{\prime \prime }\right)
-Y_{\zeta }\left( -\beta \mathbf{q}^{\prime \prime \prime }\right) =Y_{\zeta
}\left( \beta \mathbf{q}\right) }\mathrm{e}^{\mathrm{i}\mathring{\Phi}\left( 
\mathbf{\mathbf{k}_{\ast }},\beta \vec{q}\right) \frac{\beta ^{2}\tau _{1}}{%
\varrho }}  \notag \\
\partial _{\tau _{1}}\psi \left( \tau _{1}\right) \psi ^{2}\left( \tau
_{1}\right) \Psi _{0}^{3}\left( \vec{\zeta}_{0}Y\left( \beta \vec{\zeta}_{0}%
\vec{q}\right) \right) \breve{Q}_{\vec{n},\bar{l}}\left( \vec{\zeta}_{0}%
\mathbf{\mathbf{k}_{\ast }}+\vec{\zeta}Y\left( \beta \vec{\zeta}_{0}\vec{q}%
\right) \right)  \notag \\
\det \left[ Y_{\zeta }^{\prime }\left( \beta \mathbf{q}^{\prime }\right) %
\right] \det \left[ Y_{\zeta }^{\prime }\left( \beta \mathbf{q}^{\prime
\prime }\right) \right] \hat{h}_{\zeta }\left( \mathbf{q}^{\prime }\right) 
\hat{h}_{\zeta }\left( \mathbf{q}^{\prime \prime }\right) \hat{h}_{-\zeta
}\left( \mathbf{q}^{\prime \prime \prime }\right) \,\,\mathrm{d}\mathbf{q}%
^{\prime }\mathrm{d}\mathbf{q}^{\prime \prime }\mathrm{d}\tau _{1},  \notag
\end{gather}%
with the only difference that $\partial _{\tau _{1}}\psi \left( \tau
_{1}\right) \psi ^{2}\left( \tau _{1}\right) $ replaces $\psi ^{3}\left(
\tau _{1}\right) $ (see Subsection 8.7 where similar terms are derived from
ENLS). The integral in (\ref{susy}) can be treated similarly to (\ref{IY}).

The next, first order approximation is given by the formula 
\begin{gather}
\tilde{u}_{\bar{n}}^{\left( 1\right) }\left( \zeta \mathbf{k}_{\ast
}+Y_{\zeta }\left( \beta \mathbf{q}\right) ,\tau \right) =  \label{suserr1}
\\
\tilde{u}_{\bar{n}}^{\left( 1,0\right) }\left( \zeta \mathbf{k}_{\ast
}+Y_{\zeta }\left( \beta \mathbf{q}\right) ,\tau \right) +\varrho \tilde{u}_{%
\bar{n}}^{\left( 1,1\right) }\left( \zeta \mathbf{k}_{\ast }+Y_{\zeta
}\left( \beta \mathbf{q}\right) ,\tau \right) +O\left( \varrho ^{2}\right)
O\left( \left\vert \mathbf{U}^{\left( 1\right) }\right\vert \right) ,  \notag
\end{gather}%
where $\tilde{u}_{\bar{n}}^{\left( 1,1\right) }$is given by (\ref{unlt}), (%
\ref{susy}), which yields (\ref{uur2}).

\textbf{Remark.} Note that the above expansions for the FNLR\ integral (\ref%
{FNRc}) can be applied to the integral (\ref{MXint}) for the exact solution.
The formula (\ref{intcaus}) holds with $\mathbf{\tilde{U}}_{\bar{n}^{\left(
j\right) }}^{\left( 0\right) }$ replaced by $\mathbf{\tilde{U}}_{\bar{n}%
^{\left( j\right) }}$ and $\psi \left( \varrho t_{j}\right) \mathring{h}=%
\tilde{u}_{\bar{n}^{\left( j\right) }}^{\left( 0\right) }\left( \varrho
t_{j}\right) $ replaced by $\tilde{u}_{\bar{n}^{\left( j\right) }}\left(
\varrho t_{j}\right) .$ We have similarly to (\ref{uc1}) 
\begin{equation}
\mathbf{\tilde{u}}_{\bar{n}}\left( \mathbf{r},\mathbf{k},\tau \right) =%
\mathbf{\tilde{u}}_{\bar{n},0}\left( \mathbf{r},\mathbf{k},\tau \right)
+\sum_{l=1}^{N_{1}}\varrho ^{l}\mathbf{\tilde{u}}_{\bar{n},l}\left( \mathbf{r%
},\mathbf{k},\tau \right) +O\left( \varrho ^{N_{1}+1}\right) .
\end{equation}%
For the modal coefficients we get, similarly to (\ref{unlt}) 
\begin{gather}
\tilde{u}_{\bar{n}}\left( \mathbf{k},\tau \right) =\frac{1}{\varrho }%
\sum_{l=0}^{N_{1}}\sum_{l_{1}+l_{2}+l_{3}=l}\sum_{\bar{n}^{\prime },\bar{n}%
^{\prime \prime },\bar{n}^{\prime \prime \prime }}\varrho ^{l}\int_{0}^{\tau
}\int_{\substack{ \lbrack -\pi ,\pi ]^{2d}  \\ \mathbf{\mathbf{k}^{\prime }}+%
\mathbf{k}^{\prime \prime }+\mathbf{k}^{\prime \prime \prime }=\mathbf{k}}}%
\mathrm{e}^{\mathrm{i}\phi _{\vec{n}}\left( \vec{k}\right) \frac{\tau _{1}}{%
\varrho }}  \label{MXQ} \\
\breve{Q}_{\vec{n},\bar{l}}\left( \vec{k}\right) \partial _{\tau _{1}}^{l}%
\left[ \tilde{u}_{\bar{n}^{\prime }}\left( \mathbf{k}^{\prime },\tau
_{1}\right) \tilde{u}_{\bar{n}^{\prime \prime }}\left( \mathbf{k}^{\prime
\prime },\tau _{1}\right) \tilde{u}_{\bar{n}^{\prime \prime \prime }}\left( 
\mathbf{k}^{\prime \prime \prime },\tau _{1}\right) \right] \,\mathrm{d}%
\mathbf{k}^{\prime }\mathrm{d}\mathbf{k}^{\prime \prime }\mathrm{d}\tau _{1}-
\notag \\
\frac{1}{\varrho }\int_{0}^{\tau }\tilde{j}_{\bar{n}}\left( \mathbf{k},\tau
_{1}\right) \,\mathrm{d}\tau _{1}+O\left( \varrho ^{N_{1}}\right) ,  \notag
\end{gather}%
where $\breve{Q}_{\vec{n},\bar{l}}\left( \vec{k}\right) $ are the same as in
(\ref{Qnl}), $\tau \leq \tau _{\ast }$ with a fixed $\tau _{\ast }$. In the
above formula we assumed that the nonlinearity involves only the cubic term,
but in the general case a similar formula involves series with respect to
powers $\alpha ^{m}$ with coefficiens which are $\left( 2m+1\right) $-linear
tensors.$\blacklozenge $

\section{Beyond the FNLR\ }

In this section we discuss two subjects: (i) why do we use in\ repesentation
(\ref{UNLS}) the exact solution $Z\left( \mathbf{r},t\right) $ of the NLS
rather than its FNLR approximation; (ii) why do we impose a restriction $%
\frac{\tau _{\ast }}{\varrho }\leq \frac{\alpha _{0}}{\alpha }$ in (\ref%
{rhoalph}), and what can be expected on longer time intervals. These two
subjects happen to be related.

\subsubsection{Advantages of using an exact solution of the NLS}

So, why to use in (\ref{UNLS}) \ the exact solution $Z\left( \mathbf{r}%
,t\right) $ of the NLS rather than its FNLR approximation? \ There are at
least two advantages. First, the FNLR based on a solution of a linear
non-homogenious equation of the form (\ref{Z1eq}) may lead to functions
which grow linearly as $t\rightarrow \infty $, whereas an exact soliton-type
solution of the NLS is bounded for all $t$. Note though that on the time
interval $t\leq \frac{\tau _{\ast }}{\varrho }\leq \frac{\alpha _{0}}{\alpha 
}$ the both functions are bounded, therefore to see the difference one have
to consider longer time intervals. The second and more important advantage
is that using in (\ref{UNLS}) the exact solution $Z\left( \mathbf{r}%
,t\right) $ of the NLS rather than its FNLR approximation produces a smaller
approximation error. Now let us take a look at the above arguments trying to
avoid technical details.

We need to use some information on solutions of the NLS. In some cases the
NLS admit explicit solutions, which are regular. In many cases information
on the regularity of general solutions $Z\left( \mathbf{r},t\right) $ of NLS
and ENLS is available (see \cite{KenigPV93}, \cite{KenigPV98}, \cite{Sulem},
in particular, p.64 of \cite{Sulem} and references cited there concerning
non-elliptic NLS and ENLS). So, providing the error estimates we assume the
solutions of the NLS\ or ENLS\ to be sufficiently regular. Namely, we assume
that, in addition to (\ref{hetap}), we have the following estimate for
solutions of the NLS 
\begin{equation}
\left\vert \hat{Z}_{\beta ,\zeta }\left( \mathbf{q},t\right) \right\vert
+\sum_{\left\vert \bar{l}\right\vert \leq m_{0}}\left\vert \partial _{%
\mathbf{q}}^{\bar{l}}\hat{Z}_{\beta ,\zeta }\left( \mathbf{q},t\right)
\right\vert +\sum_{l\leq N_{0}}\left\vert \partial _{t}^{l}\hat{Z}_{\beta
,\zeta }\left( \mathbf{q},t\right) \right\vert \leq C_{N_{\Psi }}\left(
1+\left\vert \mathbf{q}\right\vert \right) ^{-N_{\Psi }},\ 0\leq t\leq \frac{%
\tau _{\ast }}{\varrho },  \label{Zin}
\end{equation}%
with a large enough $N_{\Psi }$. Note that the above condition includes
rescaled $\hat{Z}_{\beta ,\zeta }$ which is a solution of (\ref{genS1}).
Regular dependence of $\hat{Z}_{\beta ,\zeta }$ on $\beta $ is consistent
with the $\beta $- independent form of \ the above estimate. In fact, we
choose the value of $N_{\Psi }$ depending on the chosen order of
approximation, and if we take only a few lower order terms of the
approximation, the value of $N_{\Psi }$ does not have to be very large. The
value of $N_{\Psi }$ can be recovered from the Remark in Subsection 4.1.2.
Here, for simplicity, we primarily consider the case of the classical NLS
with $\nu =2$, $\sigma =0$ and the simplest ENLS with $\nu =3$, $\sigma =1$;
in the end of this section when we discuss fifth-order corrections we take $%
\nu =4$, $\sigma =2$.

Below we show that, in fact, the formula (\ref{UNLS}) gives a better
approximation than can be seen from the FNLR. To see that we first consider
a simpler case when the higher order terms in the expansion (\ref{cd4})
satisfy the estimate 
\begin{equation}
\mathbf{S}_{D}\left( \mathbf{r},t;\mathbf{D}\right) =\mathbf{S}_{D}^{\left(
3\right) }\left( \mathbf{r},t;\mathbf{D}\right) +O\left( \alpha _{5}\alpha
\right) ,  \label{alph1}
\end{equation}%
where the constant $\alpha _{5}\ll 1$ controls the magnitude of the next,
the fifth order term in the expansion of the nonlinearity (in particular, if
the nonlinearity in (\ref{MXshort}) is purely cubic, $\alpha _{5}=0$ ). Let 
\begin{equation}
Z_{\zeta }^{\left[ 1\right] }=Z_{\zeta }^{\left( 0\right) }+\alpha Z_{\zeta
}^{\left( 1\right) }
\end{equation}%
be the first-order approximation based on the linear and the first nonlinear
responses to the exact solution $Z_{\zeta }$ of the NLS. If in (\ref{UNLS})
we replaced $Z_{\zeta }$ by $Z_{\zeta }^{\left[ 1\right] }$ \ we would
obtain an approximate solution $\mathbf{U}_{Z^{\left[ 1\right] }}\left( 
\mathbf{r},t\right) $ of the NLM which satisfies \ the following equation\ 
\begin{gather}
\partial _{t}\mathbf{U}_{Z^{\left[ 1\right] }}=\mathbf{-}\mathrm{i}\mathbf{MU%
}_{Z^{\left[ 1\right] }}+\alpha \mathcal{F}_{\text{NL}}\left( \mathbf{U}_{Z^{%
\left[ 1\right] }}\right) -\mathbf{J}=  \label{UZ1a} \\
O\left( \alpha _{5}\alpha \right) +O\left( \frac{\alpha ^{2}}{\varrho }%
\right) O\left( \left\vert \mathbf{U}^{\left( 1\right) }\right\vert \right)
+O\left( \beta ^{\nu -1}\right) +O\left( \varrho \right) .  \notag
\end{gather}%
To see the origin of the leading term $O\left( \frac{\alpha ^{2}}{\varrho }%
\right) O\left( \left\vert \mathbf{U}^{\left( 1\right) }\right\vert \right) $
of the discrepancy in (\ref{UZ1a}) let us look at the second order term in
the expansion (\ref{UU1}). The next after the FNLR correction term $\alpha
^{2}\mathbf{U}^{\left( 2\right) }$ in the series (\ref{UU1}) for $\alpha
_{5}=0$ can be found by solving the equation%
\begin{equation}
\partial _{t}\mathbf{U}^{\left( 2\right) }=\mathbf{-}\mathrm{i}\mathbf{MU}%
^{\left( 2\right) }+3\mathcal{F}_{\text{NL}}^{\left( 1\right) }\left( 
\mathbf{U}^{\left( 0\right) },\mathbf{U}^{\left( 0\right) },\mathbf{U}%
^{\left( 1\right) }\right) -\mathbf{J}^{\left( 2\right) };\ \mathbf{U}%
^{\left( 2\right) }\left( t\right) =0\;\text{for }t\leq 0,  \label{FNL2}
\end{equation}%
where the form of the expression $\mathcal{F}_{\text{NL}}^{\left( 1\right)
}\left( \mathbf{U}^{\left( 0\right) },\mathbf{U}^{\left( 0\right) },\mathbf{U%
}^{\left( 1\right) }\right) $ is based on the fact that $\mathcal{F}_{\text{%
NL}}^{\left( 1\right) }$ is a tri-linear operator. Note that since the
expression for $\mathbf{U}^{\left( 1\right) }$ is itself a cubic with
respect to $\mathbf{U}^{\left( 0\right) }$, the next term \ $\mathbf{U}%
^{\left( 2\right) }$ is quintic. Note also that $\mathbf{U}^{\left( 1\right)
}$ is of order $\frac{1}{\varrho }$ and the time interval is of the same
order $\frac{1}{\varrho }$. Since $\mathcal{F}_{\text{NL}}^{\left( 1\right)
}\left( \mathbf{U}^{\left( 0\right) },\mathbf{U}^{\left( 0\right) },\mathbf{U%
}^{\left( 1\right) }\right) $ involves frequency matched terms, we can
conclude that that $\mathbf{U}^{\left( 2\right) }$ \ is of order $\frac{1}{%
\varrho ^{2}}$, or, equivalently, $\alpha ^{2}\mathbf{U}^{\left( 2\right) }$
is of order $O\left( \frac{\alpha ^{2}}{\varrho }\right) O\left( \left\vert 
\mathbf{U}^{\left( 1\right) }\right\vert \right) $. The deciding advantage
of using the exact solution $Z_{\zeta }$ of the NLS, as we do in (\ref{UNLS}%
), is making the discrepancy much smaller compared with (\ref{UZ1a}):%
\begin{equation}
\partial _{t}\mathbf{U}_{Z}=\mathbf{-}\mathrm{i}\mathbf{MU}_{Z}+\alpha 
\mathcal{F}_{\text{NL}}\left( \mathbf{U}_{Z^{\left[ 1\right] }}\right) -%
\mathbf{J}=\mathbf{J}_{Z},  \label{UZa}
\end{equation}%
\begin{equation}
\mathbf{J}_{Z}=O\left( \alpha _{5}\alpha ^{2}\right) +\left[ O\left( \alpha
\beta ^{\nu -1}\right) +O\left( \alpha \varrho \right) \right] O\left(
\left\vert \mathbf{U}^{\left( 1\right) }\right\vert \right) .  \label{JZ}
\end{equation}%
\emph{The terms of the order }$O\left( \frac{\alpha ^{2}}{\varrho ^{2}}%
\right) $\emph{, }$O\left( \frac{\alpha ^{3}}{\varrho ^{3}}\right) $\emph{\
and similar to them in the right-hand side of (\ref{FNL2}) and its higher
analogues disappear in (\ref{UZa}).} The reason is that those terms have
exactly the same form in the Floquet-Bloch expansion of solution of the NLM\
as the corresponding terms in the Fourier expansion of the solution of the
NLS. Since $Z_{\zeta }$ satisfy the NLS exactly, these terms completely
cancel in the expansion of the solution of the NLS, consequently
corresponding terms completely cancel in the expansion of the solution of
the NLM.

Now we provide some more details for the above considerations. We still use (%
\ref{UZdirind}), (\ref{UNLS}), (\ref{Uindsum}) to define the approximate
solution $\mathbf{U}_{Z}$. To make sure that the discrepancy does not
include terms coming from the mismatch in the initial data in all orders of
accuracy, the excitation current $\mathbf{J}$ in (\ref{MXshort}) is given by
(\ref{J01}) where $\mathbf{\tilde{J}}_{\bar{n}}$ has the form%
\begin{equation}
\mathbf{\tilde{J}}_{\zeta ,n}\left( \zeta \mathbf{k}_{\ast }+\zeta \mathbf{q}%
,t\right) =\Psi \left( \mathbf{q}\right) \hat{J}_{Z,\zeta }\left( \mathbf{q}%
,t\right) \mathbf{\tilde{G}}_{\bar{n}}\left( \zeta \mathbf{k}_{\ast }+\zeta 
\mathbf{q}\right) ,n=n_{0},\ \zeta =\pm 1,  \label{Jn11}
\end{equation}%
where $\ J_{Z,\zeta }\left( \mathbf{r},t\right) $ is given in (\ref{frt}).
Note that the difference between $\mathbf{\tilde{J}}_{\bar{n}}^{\left(
1\right) }\left( \mathbf{q},t\right) $ defined by (\ref{jf}), (\ref{jjt2})
and found by subtracting $\mathbf{J}^{\left( 0\right) }$ from (\ref{Jn11})
is of order $\alpha ^{2}$ and it does not affect the FNLR\ approximation.

Note that the term $\ O\left( \alpha \varrho \right) O\left( \left\vert 
\mathbf{U}^{\left( 1\right) }\right\vert \right) $ in (\ref{UZa}) comes from
the almost time-harmonic approximation of the nonlinearity in the NLM. We
consider then the NLS equations in the form (\ref{GNLS+}), (\ref{GNLS-})
with the initial conditions (\ref{inZ}). From the formula (\ref{UZdirind}), (%
\ref{UNLS}), (\ref{Uindsum}) we define the modal coefficients $\tilde{U}%
_{Z,\zeta ,n}\left( \zeta \mathbf{k}_{\ast }+\zeta \mathbf{\eta },t\right) $
of the approximate solution $\mathbf{U}_{Z}\left( \mathbf{r},t\right) $. To
show that $\mathbf{U}_{Z}\left( \mathbf{r},t\right) $ satisfies NLM with a
small discrepancy $\mathbf{J}_{Z}$ \ we consider equations for the
Floquet-Bloch modal coefficients $\tilde{U}_{\zeta ,n}\left( \zeta \mathbf{k}%
_{\ast }+\zeta \mathbf{\eta },t\right) $ of the exact solution $\mathbf{U}%
\left( \mathbf{r},t\right) $. We expand operators which enter the equations
with respect to $\varrho $ and $\beta $ as we did for the FNLR. The leading
part of the expansion of the equation which includes $n=n^{\prime
}=n^{\prime \prime }=n^{\prime \prime \prime }=n_{0}$ of the exact solution $%
\mathbf{U}\left( \mathbf{r},t\right) $ of the NLM has exactly the same form
as the NLS written in terms of Fourier transform. Therefore $\tilde{u}%
_{Z,\zeta ,n_{0}}\left( \zeta \mathbf{k}_{\ast }+\zeta \mathbf{\eta }%
,t\right) $ exactly satisfies this part of the equations. All remaining
terms of the expansion contribute to the discrepancy. The estimates of these
terms are completely similar to estimates for the FNLR. The only difference
is that instead of explicitly given $\tilde{u}_{\zeta ,n}^{\left( 0\right)
}\left( \zeta \mathbf{k}_{\ast }+\zeta \mathbf{\eta },t\right) $\ which was
in the FNLR \ we have have to consider the same formulas with $\tilde{u}%
_{Z,\zeta ,n}\left( \zeta \mathbf{k}_{\ast }+\zeta \mathbf{\eta },t\right) $%
. The analysis is the same, but now we have to use (\ref{Zin}) instead of (%
\ref{hetap}). The analysis implies that the discrepancy \ is small, namely $%
\mathbf{J}_{Z}$ satisfies (\ref{JZ}). From the estimate of the discrepancy
of the equations we derive the estimate for the difference of solutions 
\begin{equation}
\mathbf{U}\left( \mathbf{r},t\right) -\mathbf{U}_{Z}\left( \mathbf{r}%
,t\right) =\left[ O\left( \alpha _{5}\alpha ^{2}\right) +O\left( \alpha
\beta ^{\nu -1}\right) +O\left( \alpha \varrho \right) \right] O\left(
\left\vert \mathbf{U}^{\left( 1\right) }\right\vert \right)  \label{UUZ1}
\end{equation}%
on the interval $\frac{\tau _{0}}{\varrho }\leq t<\frac{\tau _{\ast }}{%
\varrho }$ (in the final part of this subsection we discuss the relevance of
the restriction on the length of the time interval). Estimate (\ref{UUZ1})
implies (\ref{UUZ}) and estimates in Subsection 1.3. Note that when $\tilde{u%
}_{Z}$ is defined by the FNLR, as in (\ref{UZ1a}), we would have much larger
term $O\left( \frac{\alpha ^{2}}{\varrho }\right) $ in addition to $O\left(
\alpha _{5}\alpha ^{2}\right) $. This is the main and very important
advantage of using the exact solution.

\subparagraph{The fifth order corrections.}

If the coefficient $\alpha _{5}$ in (\ref{alph1}) is not small, in order to
get the approximation by $\mathbf{U}_{Z}$ with error term $O\left( \alpha
_{5}\alpha ^{2}\right) $ replaced by $O\left( \beta \alpha ^{2}\right) $,
one has to take into account the fifth-order terms of $\mathcal{F}_{\text{NL}%
}$, and include into the NLS\ (\ref{genS}) a term similar to $\alpha _{\pi
}^{2}Q_{5,\pm }\left\vert Z_{\pm }\right\vert ^{4}Z_{\pm }$ as in (\ref{ENL+}%
): 
\begin{equation}
\partial _{t}Z_{\zeta }=-\mathrm{i}\zeta \gamma _{\left( 4\right) }\left[ -%
\mathrm{i}\zeta \vec{\nabla}_{\mathbf{r}}\right] Z_{\zeta }+\alpha _{\pi
}p_{\zeta }^{\left[ 2\right] }\left[ -\mathrm{i}\vec{\nabla}_{\mathbf{r}}%
\right] \left( Z_{\zeta }^{2}Z_{-\zeta }\right) +\alpha _{\pi
}^{2}Q_{5,\zeta }Z_{\zeta }^{3}Z_{-\zeta }^{2}.  \label{Z5}
\end{equation}%
The coefficient 
\begin{equation}
Q_{5,\zeta }=10\breve{Q}_{\vec{n},5}\left( \vec{\zeta}_{0}\vec{k}_{\ast
}\right)  \label{Q5}
\end{equation}%
\ is determined by the modal susceptibility of fifth order similar to (\ref%
{Qn}):%
\begin{equation}
\nabla \times \mathbf{\chi }_{D}^{\left( 5\right) }\left( \omega _{\bar{n}%
^{\prime }}\left( \mathbf{k}^{\prime }\right) ,\omega _{\bar{n}^{\prime
\prime }}\left( \mathbf{k}^{\prime \prime }\right) ,\ldots ,\omega _{\bar{n}%
^{\left( 5\right) }}\left( \mathbf{k}^{\left( 5\right) }\right) \right)
\vdots \,\mathbf{\tilde{G}}_{D,\bar{n}^{\prime }}\left( \cdot ,\mathbf{k}%
^{\prime }\right) \mathbf{\ldots \tilde{G}}_{D,\bar{n}^{\left( 5\right)
}}\left( \cdot ,\mathbf{k}^{\left( 5\right) }\right)
\end{equation}%
\begin{gather}
\breve{Q}_{\vec{n},5}\left( \vec{k}\right) =\frac{1}{(2\pi )^{4d}}\left( %
\left[ 
\begin{array}{c}
0 \\ 
\nabla \times \mathbf{\chi }_{D}^{\left( 5\right) }%
\end{array}%
\right] ,\mathbf{\tilde{G}}_{\bar{n}}\left( \cdot ,\mathbf{k}\right) \right)
_{\mathcal{H}},  \label{Qn5} \\
\mathbf{\chi }_{D}^{\left( 5\right) }=\mathbf{\chi }_{D}^{\left( 5\right)
}\left( \omega _{\bar{n}^{\prime }}\left( \mathbf{k}^{\prime }\right)
,\omega _{\bar{n}^{\prime \prime }}\left( \mathbf{k}^{\prime \prime }\right)
,\ldots ,\omega _{\bar{n}^{\left( 5\right) }}\left( \mathbf{k}^{\left(
5\right) }\right) \right) \vdots \,\mathbf{\tilde{G}}_{D,\bar{n}^{\prime
}}\left( \cdot ,\mathbf{k}^{\prime }\right) \mathbf{\ldots \tilde{G}}_{D,%
\bar{n}^{\left( 5\right) }}\left( \cdot ,\mathbf{k}^{\left( 5\right)
}\right) ,  \notag
\end{gather}%
with $n=n_{0}$, $\vec{\zeta}_{0}=\left( \zeta ,\zeta ,\zeta ,\zeta ,-\zeta
,-\zeta \right) $. The tensor $\mathbf{\chi }_{D}^{\left( 5\right) }$ is
defined by a formula similar to (\ref{cd4ab}) based on the kernel $\mathbf{R}%
_{D}^{\left( 5\right) }$ that corresponds to $\mathbf{S}_{D}^{\left(
5\right) }$ in (\ref{cd4}).

Note that to get high precision matching of initial data for the NLS and the
source term for the NLM one has to use there instead of (\ref{frt}) the
following modified source 
\begin{equation}
J_{Z,\zeta }=-\varrho \psi ^{\prime }\left( \varrho t\right) Z_{\zeta
}-\alpha _{\pi }\left( \psi -\psi ^{3}\right) p_{\zeta }^{\left[ \sigma %
\right] }\left[ -\mathrm{i}\vec{\nabla}_{\mathbf{r}}\right] \left( Z_{\zeta
}^{2}Z_{-\zeta }\right) -\alpha _{\pi }^{2}Q_{5,\zeta }\left( \psi -\psi
^{5}\right) Z_{\zeta }^{3}Z_{-\zeta }^{2}.  \label{frt5}
\end{equation}%
After the inclusion of $\ $the term $\alpha _{\pi }^{2}Q_{5,\zeta }Z_{\zeta
}^{3}Z_{-\zeta }^{2}$ \ the approximation error of the NLS-NLM\
approximation, that stems from the truncation of $\mathcal{F}_{\text{NL}}$,\
becomes $O\left( \beta \alpha ^{2}\right) $ instead of $O\left( \alpha
_{5}\alpha ^{2}\right) $ and the formula (\ref{UUZ1}) with $\sigma =2,\nu =4$
takes the form 
\begin{equation}
\mathbf{U}\left( \mathbf{r},t\right) -\mathbf{U}_{Z}\left( \mathbf{r}%
,t\right) =\left[ \alpha ^{2}\beta +O\left( \alpha \beta ^{3}\right)
+O\left( \alpha \varrho \right) \right] O\left( \left\vert \mathbf{U}%
^{\left( 1\right) }\right\vert \right) .  \label{UUZ2}
\end{equation}%
Similarly, a more elaborte analysis shows that if we take in the ENLS $\nu
=4 $, $\sigma =2$, and take into account the first order susceptibility
correction as in (\ref{GNLS1+}) \ or (\ref{ENL+}) with $Q_{5,\pm }$ defined
by (\ref{dl55}) we obtain the following improved error estimate 
\begin{equation}
\mathbf{U}\left( \mathbf{r},t\right) -\mathbf{U}_{Z}\left( \mathbf{r}%
,t\right) =\left[ \alpha ^{2}\beta +O\left( \alpha \beta ^{3}\right)
+O\left( \alpha \varrho \beta \right) \right] O\left( \left\vert \mathbf{U}%
^{\left( 1\right) }\right\vert \right) .  \label{UUZ4}
\end{equation}%
Note that in the above error estimates when $\nu =4$, $\sigma =2$ we assumed
that the ENLS are constructed so that they take into account effects of
interband interactions.

\subsubsection{Longer time intervals}

Here we we consider the case when (\ref{rhoalph}) does not hold, namely $%
\frac{1}{\varrho }\gg \frac{1}{\alpha }$, that is for time scales large
compared with the time scale $\frac{1}{\alpha }$ related with the magnitude
of the nonlinearity.

Still the approximate solution $\mathbf{U}_{Z}\left( \mathbf{r},t\right) $
which is constructed based on the ENLS (now we take $\nu =4$, $\sigma =2$ )
satisfies the Maxwell equation with a high precision on a long time
interval, namely 
\begin{gather}
\partial _{t}\mathbf{U}_{Z}\left( \mathbf{r},t\right) =\mathbf{-}\mathrm{i}%
\mathbf{MU}_{Z}\left( \mathbf{r},t\right) +\alpha \mathcal{F}_{\text{NL}%
}\left( \mathbf{U}_{Z}\left( \mathbf{r},t\right) \right) -\mathbf{J}+\mathbf{%
J}_{Z},\;t\leq \frac{\tau _{\ast }}{\varrho },  \label{UZeq} \\
\mathbf{J}_{Z}=O\left( \alpha ^{2}\beta \right) +O\left( \alpha \beta
^{3}\right) +O\left( \alpha \varrho \beta \right) ,  \notag
\end{gather}%
even when $\ \frac{1}{\varrho }\gg \frac{1}{\alpha }$. The only difference
between the equation (\ref{MXshort}) \ and the equation (\ref{UZeq}) is the
discrepancy term $\mathbf{J}_{Z}$. The discrepancy is small if 
\begin{equation}
\left[ O\left( \alpha ^{2}\beta \right) +O\left( \alpha \beta ^{3}\right)
+O\left( \alpha \varrho \beta \right) \right] \ll 1
\end{equation}%
in this case the equation (\ref{MXshort}) is satisfied by $\mathbf{U}_{Z}$
with a small error.

\textbf{\ }Smallness of the discrepancy $\mathbf{J}_{Z}$, generally
speaking, implies smallness of the approximation error only on time
intervals of order $\frac{1}{\alpha }$ or shorter. Without assumptions on
the stability of the the exact solution $\mathbf{U}$ of (\ref{MXshort}) and
the approximate solution $\mathbf{U}_{Z}$ the difference between $\mathbf{U}$
and $\mathbf{U}_{Z}$ can be estimated as follows%
\begin{equation}
\mathbf{U}\left( \mathbf{r},t\right) -\mathbf{U}_{Z}\left( \mathbf{r}%
,t\right) =O\left( \frac{1}{\alpha }\left[ e^{O\left( \frac{\alpha \tau
_{\ast }}{\varrho }\right) }-1\right] \right) \left[ O\left( \alpha
^{2}\beta \right) +O\left( \alpha \beta ^{3}\right) +O\left( \alpha \varrho
\beta \right) \right] .  \label{difexp}
\end{equation}%
Clearly, this estimate implies smallness of the difference between the
solutions of the equations (\ref{MXshort}) \ and the equation (\ref{UZeq})
if $\frac{\alpha }{\varrho }$ is bounded (or if it grows at most at a
logarithmic rate). If the discrepancy $\mathbf{J}_{Z}$ in (\ref{UZeq}) is
small and we want the exact solution $\mathbf{U}\left( \mathbf{r},t\right) $
to be close to the approximate solution $\mathbf{U}_{Z}\left( \mathbf{r}%
,t\right) $ for times much greater than $\frac{\alpha }{\varrho }$ then we
have to impose some kind of a stability condition on the nonlinearity $%
\mathcal{F}_{\text{NL}}$. More detailed analysis shows that it is sufficient
to impose a stability condition on the solution $Z$ of the NLS which serves
as a basis for $\mathbf{U}_{Z}$. For stability results for solutions of NLS
see \cite{Sulem}, section II.4, \cite{Sandstede02}, \cite{Weinstein85}, \cite%
{Weinstein86}. \ A detailed, mathematically rigorous analysis of the
validity of the approximation by a stable solution of the NLS\ on a long
time interval is done for some particular cases in \cite{Schneider03}.

\section{Some technical topics}

In this section for reader's convenience we discuss some technical topics
instrumental for the analysis of solutions to the NLM and their
approximations by the NLS.

\subsection{Stationary phase method}

To validate approximations for the dispersive case (\ref{nonM}) we use the
stationary phase method (SPhM). In this section we recall and review briefly
relevant concepts of the SPhM (see \cite{BF1}-\cite{BF3}, \cite{Stein}, \cite%
{Fedorjuk} for details). We consider oscillatory integrals of the form 
\begin{equation}
I\left( \theta \right) =\int_{\mathbf{R}^{d_{I}}}\mathrm{e}^{\mathrm{i}\frac{%
\Phi \left( \mathbf{\xi }\right) }{\theta }}\mathcal{A}\left( \mathbf{\xi }%
\right) \,\mathrm{d}\mathbf{\xi },\ \ \theta \rightarrow 0,  \label{osint0}
\end{equation}%
where $\mathcal{A}\left( \mathbf{\xi }\right) $ is assumed to be an
infinitely smooth function which vanish far from the origin. According to
the stationary phase method, the main contribution to $I\left( \theta
\right) $ as $\theta \rightarrow 0$ (up to $\theta ^{N}$ with arbitrary
large $N$) comes from small neighborhoods of critical points of the phase $%
\Phi \left( \mathbf{\xi }\right) $, that is the points $\mathbf{\xi }_{\ast
} $ which satisfy the equation 
\begin{equation}
\nabla _{\mathbf{\xi }}\Phi \left( \mathbf{\xi }\right) =\mathbf{0}.
\label{grFi}
\end{equation}%
Since (\ref{grFi}) is a system of $d_{I}$ equations for $d_{I}$ variables,
for a generic $\Phi \left( \mathbf{\xi }\right) $ there is a finite number
of such points. The integral over a small neighborhood of a critical point $%
\mathbf{\xi }_{\ast }$ expands into an asymptotic series in powers of $%
\theta $. The coefficients at the powers are written in terms of the values
of $\Phi \left( \mathbf{\xi }\right) $, $A\left( \mathbf{\xi }\right) $ and
their derivatives at the critical point $\mathbf{\xi }_{\ast }$. The most
important is the matrix of the second order derivatives, the so-called \emph{%
Hessian} defined by 
\begin{equation}
\Phi ^{\prime \prime }\left( \mathbf{\xi }_{\ast }\right) =\left\{ \frac{%
\partial ^{2}\Phi \left( \mathbf{\xi }_{\ast }\right) }{\partial \xi
_{i}\partial \xi _{j}}\right\} _{i,j=1}^{d_{I}}.  \label{FiHes}
\end{equation}%
The simplest case is the so-called non-degenerate one when $\det \Phi
^{\prime \prime }\left( \mathbf{\xi }_{\ast }\right) \neq 0$. \ At a
non-degenerate point the following classical \emph{asymptotic} formula holds
(see \cite{Stein}, \cite{Fedorjuk}):%
\begin{equation}
I\left( \theta \right) =b_{A_{1}}\theta ^{\frac{d_{I}}{2}}\left(
\sum_{m=0}^{\infty }b_{m}\left( \mathcal{A}\right) \left( \mathbf{\xi }%
_{\ast }\right) \theta ^{m}\right) ,\ \theta \rightarrow 0.  \label{Iclass}
\end{equation}%
The coefficient 
\begin{equation}
b_{A_{1}}=\frac{\left( 2\pi \right) ^{\frac{d_{I}}{2}}}{\sqrt{\left\vert
\det \Phi ^{\prime \prime }\left( \mathbf{\xi }_{\ast }\right) \right\vert }}%
\exp \left\{ \Phi \left( \mathbf{\xi }_{\ast }\right) +\frac{\mathrm{i}\pi }{%
4}\limfunc{sign}\left[ \Phi ^{\prime \prime }\left( \mathbf{\xi }_{\ast
}\right) \right] \right\} ,  \label{bA1}
\end{equation}%
with $\limfunc{sign}\left\{ \Phi ^{\prime \prime }\left( \mathbf{\xi }_{\ast
}\right) \right\} $ being the sum of signs of the eigenvalues of $\Phi
^{\prime \prime }\left( \mathbf{\xi }_{\ast }\right) ,$ \ and terms $%
b_{m}\left( \mathcal{A}\right) \left( \mathbf{\xi }_{\ast }\right) $ are
differential operators of order $2m$ applied to the function $\mathcal{A}%
\left( \mathbf{\xi }\right) $ at the point $\mathbf{\xi }_{\ast },$ in
particular the leading term with $m=0$ 
\begin{equation}
b_{0}\left( \mathcal{A}\right) \left( \mathbf{\xi }_{\ast }\right) =\mathcal{%
A}\left( \mathbf{\xi }_{\ast }\right) .
\end{equation}%
Let us consider now in more detail a special case when 
\begin{equation}
d_{I}=2d,\ \mathbf{R}^{d_{I}}=\mathbf{R}^{d}\times \mathbf{R}^{d},\ \mathbf{%
\xi }=\left( \xi ^{\prime },\xi ^{\prime \prime }\right) ,
\end{equation}%
and the matrix $\Phi ^{\prime \prime }\left( \mathbf{\xi }_{\ast }\right) $
has a special structure 
\begin{eqnarray}
\Phi ^{\prime \prime }\left( \mathbf{\xi }_{\ast }\right) &=&\left( 
\begin{array}{cc}
0 & \varphi \\ 
\varphi & 0%
\end{array}%
\right) \text{ where }\varphi \text{ is a symmetric matrix, and}
\label{fifi} \\
\mu _{1} &\neq &0,\ldots ,\mu _{d}\neq 0\text{ are the eigenvalus of }%
\varphi .  \notag
\end{eqnarray}%
This type of Hessian arises in nonlinear interaction integrals, see (\ref%
{fi''}). We set $\mathbf{\xi }_{\ast }$ to be the origin, that is now $%
\mathbf{\xi }_{\ast }=\left( 0,0\right) $. The representation (\ref{fifi})
implies that%
\begin{equation}
\limfunc{sign}\left\{ \Phi ^{\prime \prime }\left( \mathbf{\xi }_{\ast
}\right) \right\} =0,\text{ and}\det \Phi ^{\prime \prime }\left( \mathbf{%
\xi }_{\ast }\right) =-\det \varphi ^{2}
\end{equation}%
For $\Phi ^{\prime \prime }\left( \mathbf{\xi }_{\ast }\right) $ as (\ref%
{fifi}) the formulas (\ref{Iclass}), (\ref{bA1}) take the following form 
\begin{equation}
I\left( \theta \right) =\frac{\left( 2\pi \right) ^{d}}{\left\vert \det
\varphi \right\vert }\exp \left\{ \Phi \left( \mathbf{\xi }_{\ast }\right)
\right\} \theta ^{d}\left( \sum_{m=0}^{\infty }b_{m}\left( \mathcal{A}%
\right) \theta ^{m}\right) ,\ \theta \rightarrow 0.  \label{I2d}
\end{equation}%
Note that by the Morse lemma a function $\Phi \left( \mathbf{\xi }\right) $,
having at a critical point $\mathbf{\xi }_{\ast }$ a non-degenerate Hessian
of the form (\ref{fifi}), can be reduced by a smooth change of variables
with the unit Jacobian at $\mathbf{\xi }_{\ast }$ in a neighborhood of $%
\mathbf{\xi }_{\ast }$ to the form 
\begin{equation}
\Phi \left( \mathbf{\xi }_{\ast }+x\right) =2\mu _{1}x_{1}^{\prime
}x_{1}^{\prime \prime }+\ldots +2\mu _{d}x_{d}^{\prime }x_{d}^{\prime \prime
},\text{ where }\mu _{1}\neq 0,\ldots ,\mu _{d}\neq 0.  \label{x1x1}
\end{equation}%
If this change of variables is already made in (\ref{osint0}), the
coefficients $b_{m}\mathcal{A}\left( \mathbf{\xi }_{\ast }\right) $ in (\ref%
{I2d}) can be written explicitly, namely 
\begin{equation}
b_{m}\mathcal{A}\left( \mathbf{\xi }_{\ast }\right) =\frac{\mathrm{i}^{m}}{m!%
}\left. \left[ \frac{1}{\mu _{1}}\frac{\partial ^{2}}{\partial x_{1}^{\prime
}\partial x_{1}^{\prime \prime }}+\ldots +\frac{1}{\mu _{d}}\frac{\partial
^{2}}{\partial x_{d}^{\prime }\partial x_{d}^{\prime \prime }}\right] ^{m}%
\mathcal{A}\left( \mathbf{\xi }_{\ast }+x\right) \right\vert _{x=0},
\label{bi}
\end{equation}%
(see \cite{Stein} p. 355, \cite{EgorovKS} p. 80 for details).

\subsection{The Taylor formula}

Let us introduce notations related to the Taylor formula. For a function $h$
of variables $x_{1},\ldots ,x_{L}=\mathbf{x}$ we write the Taylor formula as
follows 
\begin{equation}
h\left( \mathbf{x}+\mathbf{y}\right) =h\left( x_{1}+y_{1},\ldots
,x_{L}+y_{L}\right) =h\left( x\right) +\sum_{\left\vert l\right\vert
=1}^{\nu }\frac{1}{\bar{l}!}h^{\left[ \bar{l}\right] }\left( x\right) y^{%
\bar{l}}+O\left( \left\vert y\right\vert ^{\nu +1}\right) ,  \label{Taylor}
\end{equation}%
where%
\begin{gather}
\bar{l}=\left( l_{1},\ldots ,l_{L}\right) ,\ \left\vert \bar{l}\right\vert
=l_{1}+\ldots +l_{L},\ y^{\bar{l}}=y_{1}^{l_{1}}\ldots y_{L}^{l_{L}},
\label{power} \\
\ \frac{1}{\bar{l}!}=\frac{1}{l_{1}!\ldots l_{L}!},\ h^{\left[ \bar{l}\right]
}\left( x\right) =\frac{\partial ^{\left\vert \bar{l}\right\vert }h\left(
x\right) }{\partial x_{1}^{l_{1}}\ldots \partial x_{L}^{l_{L}}}.  \notag
\end{gather}%
We often use a shorter notation 
\begin{equation}
\sum_{\left\vert \bar{l}\right\vert =l_{0}}\frac{1}{\bar{l}!}h^{\left[ \bar{l%
}\right] }\left( x\right) y^{\bar{l}}=\frac{1}{l_{0}!}H^{\left( l_{0}\right)
}\vdots \,\left( \mathbf{y}^{l_{0}}\right)  \label{power1}
\end{equation}%
where $l_{0}$ is an integer (not a integer vector) and $H^{\left(
l_{0}\right) }\left( \mathbf{y}^{l_{0}}\right) $ is a $l_{0}$-linear
symmetric form. For example, a symmetric cubic form can be written as
follows: 
\begin{equation}
H^{\left( 3\right) }\vdots \,\mathbf{uvw}%
=\sum_{j_{1},j_{2},j_{3}=1}^{L}H_{j_{1},j_{2},j_{3}}^{\left( 3\right)
}u_{j_{1}}v_{j_{2}}w_{j_{3}},  \label{H3}
\end{equation}%
with the following symmetry property satisfied by the coefficients: 
\begin{equation}
H_{j_{1},j_{2},j_{3}}^{\left( 3\right) }=H_{j_{2},j_{1},j_{3}}^{\left(
3\right) }=H_{j_{1},j_{3},j_{2}}^{\left( 3\right) }.
\end{equation}%
Using this notation we can rewrite (\ref{Taylor}) as 
\begin{equation}
h\left( \mathbf{x}+\mathbf{y}\right) =h\left( \mathbf{x}\right) +h^{\prime
}\left( \mathbf{x}\right) \left( \mathbf{y}\right) +\frac{1}{2}h^{\prime
\prime }\left( \mathbf{x}\right) \left( \mathbf{y}^{2}\right) +\ldots +\frac{%
1}{\nu !}h^{\left( \nu \right) }\left( \mathbf{x}\right) \left( \mathbf{y}%
^{\nu }\right) +O\left( \left\vert y\right\vert ^{\nu +1}\right) .
\label{Taylor1}
\end{equation}

\subsection{Almost time-harmonic waves and related expansions}

In this section we consider basic analytic properties of functions related
to almost time-harmonic excitations and a dispersive medium responses to
them as described by time convolution integrals which were considered in
Section 6. We define an \emph{almost time-harmonic} function $a\left(
t\right) $ as the one having the following form%
\begin{equation}
a\left( t\right) =a_{\varrho }\left( t\right) =\mathrm{e}^{-\mathrm{i}\omega
_{0}t}\psi \left( \varrho t\right) \text{ where }\psi \left( \tau \right) 
\text{ is smooth, }\psi \left( \tau \right) =0\text{ for }\tau \leq 0\text{
and }\tau \geq 1,  \label{aps1}
\end{equation}%
and $\varrho $ is a small positive parameter. We refer to the function $\psi
\left( \tau \right) $ in (\ref{aps1}) as the \emph{slow envelope} function
of the slow time $\tau $. The function $\psi \left( \tau \right) $ is
assumed to have the Taylor series satisfying$\ $ 
\begin{equation}
\psi \left( \tau +\xi \right) =\dsum\limits_{n=0}^{\infty }\frac{\psi
^{\left( n\right) }\left( \tau \right) }{n!}\xi ^{n},\ \left\vert \psi
\left( \tau +\xi \right) -\dsum\limits_{n=0}^{N_{1}}\frac{\psi ^{\left(
n\right) }\left( \tau \right) }{n!}\xi ^{n}\right\vert \leq C_{N_{1}}\xi
^{N_{1}+1}\text{ for any }\xi .  \label{aps2}
\end{equation}%
For $\varrho =0$ evidently $a_{0}\left( t\right) =\mathrm{e}^{-\mathrm{i}%
\omega _{0}t}\psi \left( 0\right) $ becomes a time harmonic function
justifying the term almost harmonic. For a small but finite $\varrho $ we
have the following formula for the Fourier transform $\hat{a}_{\varrho
}\left( \omega \right) $ for 
\begin{equation}
\hat{a}_{\varrho }\left( \omega \right) =\left[ \widehat{\mathrm{e}^{-%
\mathrm{i}\omega _{0}t}\psi \left( \varrho t\right) }\right] \left( \omega
\right) =\frac{1}{\varrho }\hat{\psi}\left( \frac{\omega -\omega _{0}}{%
\varrho }\right) ,  \label{aps3}
\end{equation}%
indicating that in the frequency domain the function $\hat{a}_{\varrho
}\left( \omega \right) $ has noticeble values only in the interval $%
\left\vert \omega -\omega _{0}\right\vert \leq \limfunc{Const}\varrho $.
Consequently, the frequency bandwidth of $a_{\varrho }\left( t\right) $ is
proportinal to $\varrho $.

Let us look what happens to an almost time-harmonic function if it is
convoluted with a smooth and exponentially decaying at infinity function $%
R\left( t\right) $: 
\begin{gather}
R\ast \left[ \mathrm{e}^{-\mathrm{i}\omega t_{1}}\psi \left( \varrho
t_{1}\right) \right] \left( t\right) =\dint\limits_{-\infty }^{\infty
}R\left( t-t_{1}\right) \mathrm{e}^{-\mathrm{i}\omega t_{1}}\psi \left(
\varrho t_{1}\right) \,\mathrm{d}t_{1}=  \label{aps4} \\
\mathrm{e}^{-\mathrm{i}\omega t}\dint\limits_{-\infty }^{\infty }R\left(
t_{1}\right) \mathrm{e}^{\mathrm{i}\omega t_{1}}\psi \left( \varrho \left(
t-t_{1}\right) \right) \,\mathrm{d}t_{1}.  \notag
\end{gather}%
Notice that we can write asymptotic (not convergent) expansion when $\varrho
\rightarrow 0$ 
\begin{gather}
\dint\limits_{-\infty }^{\infty }R\left( t_{1}\right) \mathrm{e}^{\mathrm{i}%
\omega t_{1}}\psi \left( \varrho \left( t-t_{1}\right) \right) \,\mathrm{d}%
t_{1}=  \label{aps5} \\
\dsum\limits_{n=0}^{\infty }\frac{\psi ^{\left( n\right) }\left( \varrho
t\right) }{n!}\left( -\varrho \right) ^{n}\dint\limits_{-\infty }^{\infty
}R\left( t_{1}\right) t_{1}^{n}\mathrm{e}^{\mathrm{i}\omega t_{1}}\,\mathrm{d%
}t_{1}=\dsum\limits_{n=0}^{\infty }\frac{\psi ^{\left( n\right) }\left(
\varrho t\right) }{n!}\left( \mathrm{i}\varrho \right) ^{n}\partial _{\omega
}^{n}\dint\limits_{-\infty }^{\infty }R\left( t_{1}\right) \mathrm{e}^{%
\mathrm{i}\omega t_{1}}\,\mathrm{d}t_{1}=  \notag \\
\dsum\limits_{n=0}^{\infty }\psi ^{\left( n\right) }\left( \varrho t\right) 
\frac{\left( \mathrm{i}\varrho \right) ^{n}}{n!}\hat{R}^{\left( n\right)
}\left( \omega \right) =\psi \left( \varrho t\right) \hat{R}\left( \omega
\right) +\psi ^{\prime }\left( \varrho t\right) \hat{R}^{\prime }\left(
\omega \right) \mathrm{i}\varrho +\ldots ,  \notag
\end{gather}%
where $\hat{R}^{\left( n\right) }\left( \omega \right) $ stands for the $n$%
-th derivative of the Fourier tranform $\hat{R}\left( \omega \right) $ of
the function $R\left( t\right) $. The relation (\ref{aps5}) implies%
\begin{equation}
R\ast \left[ \mathrm{e}^{-\mathrm{i}\omega t_{1}}\psi \left( \varrho
t_{1}\right) \right] =\mathrm{e}^{-\mathrm{i}\omega t}\mu _{R}\left[ \psi %
\right] \left( \varrho t\right) ,  \label{aps6}
\end{equation}%
where the transformation $\mu _{R}$ acts as follows 
\begin{gather}
\mu _{R}\left[ \psi _{0}\right] \left( t\right) =\dsum\limits_{n=0}^{\infty }%
\frac{\left( \mathrm{i}\varrho \right) ^{n}}{n!}\hat{R}^{\left( n\right)
}\left( \omega \right) \psi _{0}^{\left( n\right) }\left( t\right)
\label{aps7} \\
=\hat{R}\left( \omega \right) \psi _{0}\left( t\right) +\hat{R}^{\prime
}\left( \omega \right) \mathrm{i}\varrho \psi _{0}^{\prime }\left( t\right) -%
\frac{1}{2}\hat{R}^{\prime \prime }\left( \omega \right) \varrho ^{2}\psi
_{0}^{\prime \prime }\left( t\right) +\ldots  \notag
\end{gather}%
Note that equalities (\ref{aps6}) and (\ref{aps7}) has to be understood in
the asymptotic sense, namely when one truncates the series and takes $N_{1}$
terms the error in (\ref{aps6}) and (\ref{aps7}) is $O\left( \varrho
^{N_{1}+1}\right) $ but the series (\ref{aps7}) may not converge for a given 
$\varrho $. Observe that time convolution\ (\ref{aps6}) with any function $%
R\left( t\right) $ maps an almost time-harmonic function with a slow
envelope function $\psi _{0}\left( t\right) $ to an almost time-harmonic one
with the slow envelope function $\mu _{R}\left[ \psi _{0}\right] \left(
t\right) $ satisfying the relation (\ref{aps7}). We refer to expansion (\ref%
{aps7}) as to \emph{time-harmonic expansion}. Notice also that the expansion
(\ref{aps7}) for $\mu _{R}\left[ \psi _{0}\right] $ implies the following
approximate formlula%
\begin{equation}
\mu _{R}\left[ \psi \right] \left( t\right) \cong \hat{R}\left( \omega
\right) \psi \left( t\right) \text{ for }\varrho \ll 1.  \label{aps8}
\end{equation}

Mutlidimensional version of the time-harmonic expansion is as follows. The
multidimensional analog of the convolution mapping (\ref{aps4}) is defined
for a function $R\left( t_{1},\ldots ,t_{m}\right) $ and 
\begin{equation}
\mathrm{e}^{-\mathrm{i}\left\{ \omega _{1}t_{1}+\cdots +\omega
_{m}t_{m}\right\} }\psi \left( \varrho t_{1},\ldots ,\varrho t_{m}\right)
\label{aps9}
\end{equation}%
with $\psi \left( \tau _{1},\ldots ,\tau _{m}\right) $ being the slow
envelope funcitons of slow times $\tau _{1},\ldots ,\tau _{m}$, and it is
given by the formula 
\begin{gather}
R\ast \left[ \mathrm{e}^{-\mathrm{i}\left\{ \omega _{1}t_{1}+\cdots +\omega
_{m}t_{m}\right\} }\psi \left( \varrho t_{1},\ldots ,\varrho t_{m}\right) %
\right]  \label{aps10} \\
=\dint\limits_{-\infty }^{\infty }R\left( t-t_{1},\ldots ,t-t_{m}\right) 
\mathrm{e}^{-\mathrm{i}\left\{ \omega _{1}t_{1}+\cdots +\omega
_{m}t_{m}\right\} }\psi \left( \varrho t_{1},\ldots ,\varrho t_{m}\right) \,%
\mathrm{d}t_{1}\ldots \mathrm{d}t_{m}  \notag
\end{gather}%
Then%
\begin{equation}
R\ast \left[ \mathrm{e}^{-\mathrm{i}\left\{ \omega _{1}t_{1}+\cdots +\omega
_{m}t_{m}\right\} }\psi \left( \varrho t_{1},\ldots ,\varrho t_{m}\right) %
\right] =\mathrm{e}^{-\mathrm{i}\left\{ \omega _{1}t_{1}+\cdots +\omega
_{m}t_{m}\right\} }\mu _{R}\left[ \psi \right] \left( \varrho t\right) ,
\label{aps11}
\end{equation}%
where%
\begin{gather}
\mu _{R}\left[ \psi \right] \left( t_{1},\ldots ,t_{m}\right) =
\label{aps12} \\
\dsum\limits_{n=0}^{\infty }\frac{\left( \mathrm{i}\varrho \right)
^{l_{1}+\cdots +l_{m}}}{l_{1}!\cdots l_{m}!}\partial _{\omega
_{1}}^{l_{1}}\cdots \partial _{\omega _{m}}^{l_{m}}\hat{R}\left( \omega
_{1},\ldots ,\omega _{m}\right) \partial _{t_{1}}^{l_{1}}\cdots \partial
_{t_{m}}^{l_{m}}\psi \left( t_{1},\ldots ,t_{m}\right) .  \notag
\end{gather}

\subsection{Rectifying change of variables}

By a rectifying change of variables $\mathbf{\eta }=Y\left( \mathbf{\xi }%
\right) $ we call a solution of (\ref{Y}), (\ref{redY}) that reduces the
function $\omega _{n_{0}}\left( \mathbf{k}\right) $ to its Taylor polynomial 
$\gamma _{\left( \nu \right) }$ of the degree $\nu $: 
\begin{equation}
\omega _{n_{0}}\left( \mathbf{k}_{\ast }+Y\left( \mathbf{\xi }\right)
\right) =\gamma _{\left( \nu \right) }\left( \mathbf{\xi }\right)
\label{redY1}
\end{equation}%
and, equivalently,%
\begin{equation}
\omega _{n_{0}}\left( \mathbf{k}_{\ast }+\mathbf{\eta }\right) =\gamma
_{\left( \nu \right) }\left( Y^{-1}\left( \mathbf{\eta }\right) \right) .
\end{equation}%
We call the $\mathbf{\xi }$ the \emph{rectifying variable}. The Taylor
polynomial $\gamma _{\left( \nu \right) }\left( \mathbf{\eta }\right) $ of $%
\omega _{n_{0}}\left( \mathbf{\mathbf{k}_{\ast }}+\mathbf{\eta }\right) $ at 
$\mathbf{\mathbf{\eta }}=\mathbf{\mathbf{0}}$ of the degree $\nu $ is
defined by (\ref{Tayom}). For $\nu =2$ the polynomial takes the form (\ref%
{defom2}). Here we consider the case $\omega _{n_{0}}^{\prime }\left( 
\mathbf{\mathbf{k}_{\ast }}\right) \neq 0$. Let us discuss basic properties
of the rectifying change of variables and give some explicit formulas.

\subparagraph{One-dimensional case, $d=1$.}

Since the derivative $\omega _{n_{0}}^{\prime }\left( \mathbf{k}_{\ast
}\right) \neq 0$, the polynomial $\gamma _{\left( \nu \right) }\left( 
\mathbf{\xi }\right) $ is an invertible function in a vicinity $\mathbf{\xi }%
=\mathbf{0}$ implying that the function 
\begin{equation}
Y^{-1}\left( \mathbf{\eta }\right) =\gamma _{\left( \nu \right) }^{-1}\left(
\omega _{n_{0}}\left( \mathbf{k}_{\ast }+\mathbf{\eta }\right) \right)
\end{equation}%
is well defined. In particular,%
\begin{equation}
\gamma _{\left( 1\right) }\left( \mathbf{\xi }\right) =\omega _{n_{0}}\left( 
\mathbf{k}_{\ast }\right) +\omega _{n_{0}}^{\prime }\left( \mathbf{k}_{\ast
}\right) \mathbf{\xi }\text{ if }\nu =1  \label{gam1}
\end{equation}%
and for $d=1$ we obtain explicit expression 
\begin{equation}
Y^{-1}\left( \mathbf{\eta }\right) =\frac{\omega ^{0}\left( \mathbf{\eta }%
\right) }{\omega _{n_{0}}^{\prime }\left( \mathbf{k}_{\ast }\right) },\
\omega ^{0}\left( \mathbf{\eta }\right) =\omega _{n_{0}}\left( \mathbf{k}%
_{\ast }+\mathbf{\eta }\right) -\omega _{n_{0}}\left( \mathbf{k}_{\ast
}\right) .  \label{Y11}
\end{equation}%
If\ $\nu =2$ 
\begin{equation}
\gamma _{\left( 2\right) }\left( \mathbf{\eta }\right) =\omega
_{n_{0}}\left( \mathbf{\mathbf{k}_{\ast }}\right) +\omega _{n_{0}}^{\prime
}\left( \mathbf{\mathbf{k}_{\ast }}\right) \left( \mathbf{\mathbf{\eta }}%
\right) +\frac{1}{2}\omega _{n_{0}}^{\prime \prime }\left( \mathbf{\mathbf{k}%
_{\ast }}\right) \left( \mathbf{\eta }^{2}\right) \mathbf{.}  \label{gam2}
\end{equation}%
If $d=1$ we find that $\mathbf{\xi }=Y^{-1}\left( \mathbf{\eta }\right) $ is
a solution of the equation 
\begin{equation}
\omega _{n_{0}}\left( \mathbf{k}_{\ast }\right) +\omega _{n_{0}}^{\prime
}\left( \mathbf{k}_{\ast }\right) \mathbf{\xi }+\frac{1}{2}\omega
_{n_{0}}^{\prime \prime }\left( \mathbf{k}_{\ast }\right) \mathbf{\xi }%
^{2}=\omega _{n_{0}}\left( \mathbf{k}_{\ast }+\mathbf{\eta }\right) ,
\label{gam2a}
\end{equation}%
and for $\mathbf{\eta }=\mathbf{0}$ we have $\mathbf{\xi }=\mathbf{0}$.
Solving the equation (\ref{gam2a}) for $\mathbf{\xi }$ we get%
\begin{equation}
\mathbf{\xi }=Y^{-1}\left( \mathbf{\eta }\right) =\frac{\omega
_{n_{0}}^{\prime }\left( \mathbf{k}_{\ast }\right) }{2\omega
_{n_{0}}^{\prime \prime }\left( \mathbf{k}_{\ast }\right) }\left[ -1+\sqrt{1+%
\frac{2}{\omega _{n_{0}}^{\prime }\left( \mathbf{k}_{\ast }\right) ^{2}}%
\omega _{n_{0}}^{\prime \prime }\left( \mathbf{k}_{\ast }\right) \omega
^{0}\left( \mathbf{\eta }\right) }\right]  \label{Y21}
\end{equation}%
with $\omega ^{0}\left( \mathbf{\eta }\right) $ being defined by (\ref{Y11}%
). Note that that formula (\ref{Y21}) turns into the the formlula (\ref{Y11}%
) as $\omega _{n_{0}}^{\prime \prime }\left( \mathbf{k}_{\ast }\right)
\rightarrow 0$.

\subparagraph{Multidimensional case, $d>1$.}

In this case solution of (\ref{redY1}) is not unique. But under some
additional requirments on $Y\left( \mathbf{\xi }\right) $ it can become
unique. One way to do it is to set $Y^{-1}$ to be of the form 
\begin{equation}
Y^{-1}\left( \mathbf{\eta }\right) =\mathbf{\eta }+\vartheta _{Y}\left( 
\mathbf{\eta }\right) \omega _{n_{0}}^{\prime }\left( \mathbf{k}_{\ast
}\right) ,  \label{Yet}
\end{equation}%
where $\vartheta _{Y}\left( \mathbf{\eta }\right) $ is a scalar function.
Under the assumption (\ref{Yet}) the equation (\ref{redY1}) defining $Y$
turns into the following equation for the unknown scalar function $\vartheta
_{Y}\left( \mathbf{\eta }\right) $ 
\begin{equation}
\omega _{n_{0}}\left( \mathbf{k}_{\ast }+\mathbf{\eta }\right) =\gamma
_{\left( \nu \right) }\left( \mathbf{\eta }+\vartheta _{Y}\left( \mathbf{%
\eta }\right) \omega _{n_{0}}^{\prime }\left( \mathbf{k}_{\ast }\right)
\right) ,\;\vartheta _{Y}\left( \mathbf{0}\right) =0.  \label{ips}
\end{equation}%
If $\omega _{n_{0}}^{\prime }\left( \mathbf{k}_{\ast }\right) \neq 0$ this
equation has a unique small solution by the Implicit Function Theorem.

In particular, for $\nu =1$ we still have (\ref{gam1}) with $\omega
_{n_{0}}^{\prime }\left( \mathbf{k}_{\ast }\right) \mathbf{\xi }=\omega
_{n_{0}}^{\prime }\left( \mathbf{k}_{\ast }\right) \cdot \mathbf{\xi }$ \
and 
\begin{equation}
\vartheta _{Y}\left( \mathbf{\eta }\right) =\frac{1}{\left\vert \omega
_{n_{0}}^{\prime }\left( \mathbf{k}_{\ast }\right) \right\vert ^{2}}\left[
\omega ^{0}\left( \mathbf{\eta }\right) -\omega _{n_{0}}^{\prime }\left( 
\mathbf{k}_{\ast }\right) \cdot \mathbf{\eta }\right]  \label{ips1}
\end{equation}%
implying 
\begin{equation}
\vartheta _{Y}\left( \mathbf{0}\right) =0,\ \vartheta _{Y}^{\prime }\left( 
\mathbf{0}\right) =0,\ \vartheta _{Y}^{\prime \prime }\left( \mathbf{0}%
\right) =\frac{1}{\left\vert \omega _{n_{0}}^{\prime }\left( \mathbf{k}%
_{\ast }\right) \right\vert ^{2}}\omega _{n_{0}}^{\prime \prime }\left( 
\mathbf{\mathbf{k}_{\ast }}\right) .
\end{equation}

If $\nu =2$ the scalar function $\vartheta _{Y}\left( \mathbf{\eta }\right) $
solves the equation 
\begin{equation}
\omega _{n_{0}}^{\prime }\left( \mathbf{\mathbf{k}_{\ast }}\right) \left( 
\mathbf{\eta }+\vartheta \left( \mathbf{\eta }\right) \omega
_{n_{0}}^{\prime }\left( \mathbf{k}_{\ast }\right) \right) +\frac{1}{2}%
\omega _{n_{0}}^{\prime \prime }\left( \mathbf{\mathbf{k}_{\ast }}\right)
\left( \left( \mathbf{\eta }+\vartheta \left( \mathbf{\eta }\right) \omega
_{n_{0}}^{\prime }\left( \mathbf{k}_{\ast }\right) \right) ^{2}\right)
=\omega ^{0}\left( \mathbf{\eta }\right) ,  \label{ips2}
\end{equation}%
which is readily reduced to an elementary quadratic equation for $\vartheta
_{Y}\left( \mathbf{\eta }\right) =\vartheta $, namely 
\begin{gather}
\vartheta \left[ \left\vert \omega _{n_{0}}^{\prime }\left( \mathbf{\mathbf{k%
}_{\ast }}\right) \right\vert ^{2}+\omega _{n_{0}}^{\prime \prime }\left( 
\mathbf{\mathbf{k}_{\ast }}\right) \left( \left( \mathbf{\eta }\right)
\left( \omega _{n_{0}}^{\prime }\left( \mathbf{k}_{\ast }\right) \right)
\right) \right] +\frac{1}{2}\vartheta ^{2}\omega _{n_{0}}^{\prime \prime
}\left( \mathbf{\mathbf{k}_{\ast }}\right) \left( \left( \omega
_{n_{0}}^{\prime }\left( \mathbf{k}_{\ast }\right) \right) ^{2}\right) \\
=\omega ^{0}\left( \mathbf{\eta }\right) -\omega _{n_{0}}^{\prime }\left( 
\mathbf{\mathbf{k}_{\ast }}\right) \cdot \mathbf{\eta -}\frac{1}{2}\omega
_{n_{0}}^{\prime \prime }\left( \mathbf{\mathbf{k}_{\ast }}\right) \left( 
\mathbf{\eta }^{2}\right) ,  \notag
\end{gather}%
The coefficients of the Taylor expansion of $\vartheta _{Y}\left( \mathbf{%
\eta }\right) $ can be found recurrently, and, in particular,%
\begin{equation}
\vartheta _{Y}\left( \mathbf{0}\right) =0,\ \vartheta _{Y}^{\prime }\left( 
\mathbf{0}\right) =0,\ \vartheta _{Y}^{\prime \prime }\left( \mathbf{0}%
\right) =0,\ \vartheta _{Y}^{\prime \prime \prime }\left( \mathbf{0}\right)
\left( \mathbf{\eta }^{3}\right) =\frac{1}{\left\vert \omega
_{n_{0}}^{\prime }\left( \mathbf{k}_{\ast }\right) \right\vert ^{2}}\omega
_{n_{0}}^{\prime \prime \prime }\left( \mathbf{\mathbf{k}_{\ast }}\right)
\left( \mathbf{\eta }^{3}\right) .
\end{equation}%
Now we discuss some of general properties of \ $Y\left( \mathbf{q}\right) $
and related functions. According to (\ref{Ybet0})%
\begin{gather}
Y\left( \mathbf{q}\right) =\mathbf{q}+\mathbf{\Xi }\left( \mathbf{q}\right)
,\ \left\vert \mathbf{q}\right\vert \leq \pi _{0},  \label{YbetX} \\
\mathbf{\Xi }\left( \mathbf{q}\right) =\mathbf{\Xi }_{\nu +1}\left( \mathbf{q%
}\right) ^{\nu +1}+O\left( \left\vert \mathbf{q}\right\vert ^{\nu +2}\right)
,\;  \notag \\
\mathbf{\Xi }_{\nu +1}\left( \mathbf{q}\right) ^{\nu +1}=\frac{1}{\left( \nu
+1\right) !}Y^{\left( \nu +1\right) }\left( 0\right) \left( \mathbf{q}^{\nu
+1}\right) +O\left( \left\vert \mathbf{q}\right\vert ^{\nu +2}\right) . 
\notag
\end{gather}%
Therefore 
\begin{equation}
Y^{-1}\left( \mathbf{\eta }\right) =\mathbf{\eta }-\mathbf{\Xi }_{\nu
+1}\left( \mathbf{\eta }\right) +O\left( \left\vert \mathbf{\eta }%
\right\vert ^{\nu +2}\right) .  \label{Ym1eta}
\end{equation}%
The function $\mathbf{q}^{\prime \prime \prime }\left( \beta \right) $
described by (\ref{Yy'}) has the form 
\begin{equation}
\mathbf{q}^{\prime \prime \prime }\left( \beta \right) =\left( \mathbf{q-q}%
^{\prime }-\mathbf{q}^{\prime \prime }\right) +\beta ^{\nu }\delta _{\nu
+1}\left( \mathbf{q},\mathbf{q}^{\prime },\mathbf{q}^{\prime \prime }\right)
+O\left( \beta ^{\nu +1}\left( \left\vert \mathbf{q}\right\vert ^{\nu
+2}+\left\vert \mathbf{q}^{\prime }\right\vert ^{\nu +2}+\left\vert \mathbf{q%
}^{\prime \prime }\right\vert ^{\nu +2}\right) \right) ,
\end{equation}%
where $\delta _{\nu +1}\left( \mathbf{q,q}^{\prime },\mathbf{q}^{\prime
\prime }\right) $ is a $\nu +1$-linear form of $\mathbf{q,q}^{\prime },%
\mathbf{q}^{\prime \prime }$. From (\ref{Yy'}) using (\ref{YbetX}) and (\ref%
{Ym1eta}) we infer that 
\begin{equation}
\delta _{\nu +1}\left( \mathbf{q,q}^{\prime },\mathbf{q}^{\prime \prime
}\right) =\mathbf{\Xi }_{\nu +1}\left( \mathbf{q}\right) ^{\nu +1}-\mathbf{%
\Xi }_{\nu +1}\left( \mathbf{q}^{\prime }\right) ^{\nu +1}-\mathbf{\Xi }%
_{\nu +1}\left( \mathbf{q}^{\prime \prime }\right) ^{\nu +1}+\mathbf{\Xi }%
_{\nu +1}\left( \mathbf{q}^{\prime }+\mathbf{q}^{\prime \prime }-\mathbf{q}%
\right) ^{\nu +1}.  \label{delnu}
\end{equation}

We deduce an identity every rectifying change of variables must satisfy.
>From (\ref{redY1}) and (\ref{YbetX}) it follows that 
\begin{equation}
\omega _{n_{0}}^{\prime }\left( \mathbf{k}_{\ast }\right) \mathbf{\Xi }_{\nu
+1}\left( \mathbf{\xi }\right) ^{\nu +1}+\frac{1}{\left( \nu +1\right) !}%
\omega _{n_{0}}^{\left( \nu +1\right) }\left( \mathbf{k}_{\ast }\right) 
\mathbf{\xi }^{\nu +1}=0.  \label{omX}
\end{equation}%
Therefore 
\begin{gather}
\omega _{n_{0}}^{\prime }\left( \mathbf{k}_{\ast }\right) \delta _{\nu
+1}\left( \mathbf{q,q}^{\prime },\mathbf{q}^{\prime \prime }\right) =
\label{delnuX} \\
\frac{-1}{\left( \nu +1\right) !}\left[ \omega _{n_{0}}^{\left( \nu
+1\right) }\left( \mathbf{k}_{\ast }\right) \left( \mathbf{q}^{\prime }+%
\mathbf{q}^{\prime \prime }-\mathbf{q}\right) ^{\nu +1}+\omega
_{n_{0}}^{\left( \nu +1\right) }\left( \mathbf{k}_{\ast }\right) \mathbf{q}%
^{\nu +1}\right] +  \notag \\
\frac{1}{\left( \nu +1\right) !}\left[ \omega _{n_{0}}^{\left( \nu +1\right)
}\left( \mathbf{k}_{\ast }\right) \left( \mathbf{q}^{\prime }\right) ^{\nu
+1}+\omega _{n_{0}}^{\left( \nu +1\right) }\left( \mathbf{k}_{\ast }\right)
\left( \mathbf{q}^{\prime \prime }\right) ^{\nu +1}\right] .  \notag
\end{gather}

\subsection{ The Fourier-Bloch ansatz in the space domain}

Assume that solutions of two equations are exactly matched in the
quasimomentum domain with a help of the rectifying change of variables, as
in Subsection 5.5. What is the relation between the solutions in the spatial
domain? In this subsection we address this question.

Let $\bar{n}_{0}$ be the band number, $\eta =Y\left( \beta \mathbf{q}\right) 
$ be the rectifying change of variables in a vicinity of $\mathbf{k}_{\ast }$%
, which is given by (\ref{YbetX}). Let $v\left( \mathbf{r}\right) $ be a
given function, $v_{\beta }\left( \mathbf{r}\right) =v\left( \mathbf{r/}%
\beta \right) $ and $\hat{v}_{\beta }\left( \mathbf{q}\right) $ be its
Fourier transform, $\hat{v}_{\beta }\left( \mathbf{q}\right) =\beta ^{d}\hat{%
v}\left( \beta \mathbf{q}\right) .$ Then we write Fourier - Bloch ansatz as
follows: 
\begin{gather}
\mathbf{W}\left( v\right) \left( \beta ,\mathbf{r}\right) =\frac{1}{\left(
2\pi \right) ^{d}}\int_{\mathbb{R}^{d}}\Psi _{0}\left( \beta \mathbf{s}%
\right) \beta ^{d}\hat{v}\left( Y^{-1}\left( \beta \mathbf{s}\right) \right) 
\mathbf{\tilde{G}}_{\bar{n}_{0}}\left( \mathbf{r},\mathbf{k}_{\ast }+\beta 
\mathbf{s}\right) \,\mathrm{d}\mathbf{s}  \label{FBA} \\
=\frac{1}{\left( 2\pi \right) ^{d}}\int_{\mathbb{R}^{d}}\Psi \left( \beta 
\mathbf{q}\right) \beta ^{d}\hat{v}\left( \beta \mathbf{q}\right) \mathbf{%
\tilde{G}}_{\bar{n}_{0}}\left( \mathbf{r},\mathbf{k}_{\ast }+Y\left( \beta 
\mathbf{q}\right) \right) \det Y^{\prime }\left( \beta \mathbf{q}\right) \,%
\mathrm{d}\mathbf{q}  \notag
\end{gather}%
where we use (\ref{Psihat}). By (\ref{Ggy}) 
\begin{gather*}
\mathbf{\tilde{G}}_{1,n_{0}}\left( \mathbf{r},\mathbf{k}_{\ast }+Y\left(
\beta \mathbf{q}\right) \right) =\mathbf{\hat{G}}_{1,n_{0}}\left( \mathbf{r},%
\mathbf{k}_{\ast }+Y\left( \beta \mathbf{q}\right) \right) \mathrm{e}^{%
\mathrm{i}\left( \mathbf{k}_{\ast }+Y\left( \beta \mathbf{q}\right) \right)
\cdot \mathbf{r}} \\
=\mathbf{\hat{G}}_{1,n_{0}}\left( \mathbf{r},\mathbf{k}_{\ast }+Y\left(
\beta \mathbf{q}\right) \right) \mathrm{e}^{\mathrm{i}\left( \mathbf{k}%
_{\ast }+\beta \mathbf{q}\cdot \mathbf{r}\right) }\mathrm{e}^{\mathrm{i}%
\mathbf{\Xi }\left( \beta \mathbf{q}\right) \cdot \mathbf{r}}
\end{gather*}%
and%
\begin{gather}
\mathbf{W}\left( v\right) \left( \beta ,\mathbf{r}\right) =  \label{Uv0} \\
\frac{\mathrm{e}^{\mathrm{i}\left( \mathbf{k}_{\ast }\cdot \mathbf{r}\right)
}}{\left( 2\pi \right) ^{d}}\int_{R^{d}}\Psi \left( \beta \mathbf{q}\right)
\beta ^{d}\hat{v}\left( \beta \mathbf{q}\right) \mathbf{\hat{G}}%
_{1,n_{0}}\left( \mathbf{r},\mathbf{k}_{\ast }+Y\left( \beta \mathbf{q}%
\right) \right) \mathrm{e}^{\mathrm{i}\left( \beta \mathbf{q}\cdot \mathbf{r}%
\right) }\mathrm{e}^{\mathrm{i}\mathbf{\Xi }\left( \beta \mathbf{q}\right)
\cdot \mathbf{r}}\,\det Y^{\prime }\left( \beta \mathbf{q}\right) \,\mathrm{d%
}\mathbf{q}.  \notag
\end{gather}%
We introduce the Taylor polynomial $\mathbf{p}_{g,Y}^{\left[ \sigma _{g}%
\right] }$ of order $\sigma _{g},$ 
\begin{equation}
\mathbf{\hat{G}}_{1,n_{0}}\left( \mathbf{r},\mathbf{k}_{\ast }+Y\left( \beta 
\mathbf{q}\right) \right) \det Y^{\prime }\left( \beta \mathbf{q}\right) =%
\mathbf{p}_{g,Y}^{\left[ \sigma _{g}\right] }\left( \beta \mathbf{q}\right)
+O\left( \beta ^{\sigma _{g}+1}\right) .  \label{Tg}
\end{equation}%
>From (\ref{Uv0}), taking into account that $\Psi \left( \beta \mathbf{q}%
\right) =1$ for $\left\vert \beta \mathbf{q}\right\vert \leq \pi _{0},$ we
obtain 
\begin{equation}
\mathbf{W}\left( v\right) \left( \beta ,\mathbf{r}\right) =\frac{\mathrm{e}^{%
\mathrm{i}\left( \mathbf{k}_{\ast }\cdot \mathbf{r}\right) }}{\left( 2\pi
\right) ^{d}}\int_{\mathbb{R}^{d}}\beta ^{d}\hat{v}\left( \beta \mathbf{q}%
\right) \mathbf{p}_{g,Y}^{\left[ \sigma _{g}\right] }\left( \beta \mathbf{q}%
\right) \mathrm{e}^{\mathrm{i}\left( \beta \mathbf{q}\cdot \mathbf{r}\right)
}\mathrm{e}^{\mathrm{i}\mathbf{\Xi }\left( \beta \mathbf{q}\right) \cdot 
\mathbf{r}}\,\mathrm{d}\mathbf{q}+O\left( \beta ^{\sigma _{g}+1}\right) .
\label{Uv1}
\end{equation}%
We can simplify this formula when $\mathbf{r}$ is not too large, namely%
\begin{equation}
\left\vert \mathbf{\Xi }\left( \beta \mathbf{q}\right) \cdot \mathbf{r}%
\right\vert \ll 1.  \label{Uv1a}
\end{equation}%
Since $\hat{v}_{\beta }\left( \mathbf{q},t\right) $ decays fast for large $%
\left\vert \mathbf{q}\right\vert $ it is sufficient to have the above
inequality for 
\begin{equation}
\left\vert \mathbf{q}\right\vert \leq \beta ^{-\pi _{\Psi }}\text{ with a
small }\pi _{\Psi }>0.
\end{equation}%
Hence, (\ref{Uv1a}) is satisified if 
\begin{equation}
\left\vert \mathbf{r}\right\vert \left\vert \beta \mathbf{q}\right\vert
^{\nu +1}\ll 1\text{ for \ }\left\vert \mathbf{q}\right\vert \leq \beta
^{-\pi _{\Psi }}  \label{rbet}
\end{equation}%
or, equivalently, 
\begin{equation}
\left\vert \mathbf{r}\right\vert \left\vert \beta \right\vert ^{\left( \nu
+1\right) \left( 1-\pi _{\Psi }\right) }\ll 1.  \label{rbet1}
\end{equation}%
Under the condition (\ref{rbet1}) we can use the expansion 
\begin{gather*}
\mathrm{e}^{\mathrm{i}\mathbf{\Xi }\left( \beta \mathbf{q}\right) \cdot 
\mathbf{r}}=1+\mathrm{i}\mathbf{r}\cdot \mathbf{\Xi }\left( \beta \mathbf{q}%
\right) +O\left( \left\vert \mathbf{r}\right\vert ^{2}\beta ^{2\nu +2}\right)
\\
=1+\mathrm{i}\beta ^{\nu +1}\mathbf{r}\cdot \mathbf{\Xi }_{\nu +1}\left( 
\mathbf{q}\right) +O\left( \left\vert \mathbf{r}\right\vert \beta ^{\nu
+2}\right) +O\left( \left\vert \mathbf{r}\right\vert ^{2}\beta ^{2\nu
+2}\right) , \\
\mathbf{\Xi }_{\nu +1}\left( \beta \mathbf{q}\right) =\frac{\beta ^{\nu +1}}{%
\left( \nu +1\right) !}\mathbf{\Xi }^{\left( \nu +1\right) }\left( 0\right)
\left( \mathbf{q}^{\nu +1}\right) ,
\end{gather*}%
obtaining from (\ref{Uv1})%
\begin{gather}
\mathbf{W}\left( v\right) \left( \beta ,\mathbf{r}\right) =\frac{\mathrm{e}^{%
\mathrm{i}\left( \mathbf{k}_{\ast }\cdot \mathbf{r}\right) }}{\left( 2\pi
\right) ^{d}}\int_{\mathbb{R}^{d}}\mathrm{e}^{\mathrm{i}\left( \beta \mathbf{%
q}\cdot \mathbf{r}\right) }\left[ 1+\mathrm{i}\mathbf{r}\cdot \mathbf{\Xi }%
_{\nu +1}\left( \beta \mathbf{q}\right) \right] \mathbf{p}_{g,Y}^{\left[
\sigma _{g}\right] }\left( \beta \mathbf{q}\right) \beta ^{d}\hat{v}\left(
\beta \mathbf{q}\right) \,\mathrm{d}\mathbf{q}  \label{Uv2} \\
+O\left( \beta ^{\sigma _{g}+1}\right) +O\left( \left\vert \mathbf{r}%
\right\vert \beta ^{\nu +2}\right) +O\left( \left\vert \mathbf{r}\right\vert
^{2}\beta ^{2\nu +2}\right)  \notag \\
=\frac{\mathrm{e}^{\mathrm{i}\left( \mathbf{k}_{\ast }\cdot \mathbf{r}%
\right) }}{\left( 2\pi \right) ^{d}}\int_{\mathbb{R}^{d}}\mathrm{e}^{\mathrm{%
i}\left( \mathbf{\xi }\cdot \mathbf{r}\right) }\left[ 1+\mathrm{i}\mathbf{r}%
\cdot \mathbf{\Xi }_{\nu +1}\left( \mathbf{\xi }\right) \right] \mathbf{p}%
_{g,Y}^{\left[ \sigma _{g}\right] }\left( \mathbf{\xi }\right) \,\hat{v}%
\left( \mathbf{\xi }\right) \,\mathrm{d}\mathbf{\xi }  \notag \\
+O\left( \beta ^{\sigma _{g}+1}\right) +O\left( \left\vert \mathbf{r}%
\right\vert \beta ^{\nu +2}\right) +O\left( \left\vert \mathbf{r}\right\vert
^{2}\beta ^{2\nu +2}\right)  \notag
\end{gather}%
Consequently, we get the following relation between functions $\mathbf{W}%
\left( v\right) \left( \beta ,\mathbf{r}\right) $ and $v\left( \mathbf{r}%
\right) $ in the space domain: 
\begin{eqnarray}
\mathbf{W}\left( v\right) \left( \beta ,\mathbf{r}\right) &=&\mathrm{e}^{%
\mathrm{i}\left( \mathbf{k}_{\ast }\cdot \mathbf{r}\right) }\left[ 1+\mathrm{%
i}\mathbf{r}\cdot \mathbf{\Xi }_{\nu +1}\left[ -\mathrm{i}\vec{\nabla}_{%
\mathbf{r}}\right] \right] \mathbf{p}_{g,Y}^{\left[ \sigma _{g}\right] }%
\left[ -\mathrm{i}\vec{\nabla}_{\mathbf{r}}\right] v\left( \mathbf{r}\right)
\label{Uv} \\
&&+O\left( \left\vert \mathbf{r}\right\vert \beta ^{\nu +2}\right) +O\left(
\left\vert \mathbf{r}\right\vert ^{2}\beta ^{2\nu +2}\right) +O\left( \beta
^{\sigma _{g}+1}\right) .  \notag
\end{eqnarray}%
The principal part of (\ref{Uv}) takes the form 
\begin{gather}
\mathbf{W}\left( v\right) \left( \beta ,\mathbf{r}\right) =\mathrm{e}^{%
\mathrm{i}\left( \mathbf{k}_{\ast }\cdot \mathbf{r}\right) }\mathbf{\hat{G}}%
_{1,n_{0}}\left( \mathbf{r},\mathbf{k}_{\ast }\right) v\left( \mathbf{r}%
\right) -  \label{Uvp} \\
\mathrm{i}\beta \mathrm{e}^{\mathrm{i}\left( \mathbf{k}_{\ast }\cdot \mathbf{%
r}\right) }\nabla _{\mathbf{k}}\mathbf{\hat{G}}_{1,n_{0}}\left( \mathbf{r},%
\mathbf{k}_{\ast }\right) \cdot \nabla _{\mathbf{r}}v\left( \mathbf{r}%
\right) +O\left( \left( \left\vert \mathbf{r}\right\vert +1\right) \beta
^{2}\right) .  \notag
\end{gather}

\subsection{Fourier transform and linear differential operators}

The Fourier transform $\hat{u}\left( \mathbf{\xi }\right) $, $\mathbf{\xi }%
\in \mathbf{R}^{d}$, of a function $u\left( \mathbf{r}\right) $, $\mathbf{r}%
\in \mathbf{R}^{d}$, is defined by 
\begin{eqnarray}
u\left( \mathbf{r}\right) &=&\frac{1}{\left( 2\pi \right) ^{d}}\int_{\mathbb{%
R}^{d}}\hat{u}\left( \mathbf{\xi }\right) \mathrm{e}^{\mathrm{i}\mathbf{r}%
\cdot \mathbf{\xi }}\,\,\mathrm{d}\mathbf{\xi },\mathbf{\ }\mathrm{d}\mathbf{%
\xi }=\mathrm{d}\xi _{1}\ldots \mathrm{d}\xi _{d},  \label{Ftransform} \\
\hat{u}\left( \mathbf{\xi }\right) &=&\int_{\mathbf{R}^{d}}u\left( \mathbf{r}%
\right) \mathrm{e}^{-\mathrm{i}\mathbf{r}\cdot \mathbf{\xi }}\,\mathrm{d}%
\mathbf{r}.  \notag
\end{eqnarray}%
Evidently%
\begin{equation}
\left[ \hat{u}\left( -\mathbf{\xi }\right) \right] ^{\ast }=\widehat{\left[
u\left( \mathbf{\xi }\right) \right] ^{\ast }}.  \label{Fourstar0}
\end{equation}%
Let $\gamma \left( \mathbf{\xi }\right) $ be a polynomial of the variable $%
\mathbf{\xi }$ written in a form of Taylor polynomial similar to (\ref%
{Taylor1}) 
\begin{equation}
\gamma \left( \mathbf{\xi }\right) =\sum_{l=0}^{\nu }\frac{1}{\left\vert
l\right\vert !}\gamma ^{\left( l\right) }\vdots \left( \mathbf{\xi }%
^{l}\right)
\end{equation}%
where $\gamma ^{\left( l\right) }\vdots \left( \mathbf{\xi }^{l}\right) $ is
a $l$ -linear form similar to (\ref{H3}). Tnen the differential operator $%
\gamma \left[ -\mathrm{i}\vec{\nabla}_{\mathbf{r}}\right] $ is defined by
formally replacing variables $\xi _{j}$ by the differential operators $-%
\mathrm{i}\partial _{j}=-\mathrm{i}\frac{\partial }{\partial r_{j}}$ in the
polynomial $\gamma \left( \mathbf{\xi }\right) .$ The polynomial $\gamma
\left( \mathbf{\xi }\right) $ is called the \emph{symbol} of the operator $%
\gamma \left[ -\mathrm{i}\vec{\nabla}_{\mathbf{r}}\right] .$ In particular,
a general polynomial of third degree with $\nu =3$ takes the form 
\begin{equation}
\gamma _{\left( 3\right) }\left( \mathbf{\xi }\right) =\gamma
_{0}+\sum_{m}\gamma _{1m}\xi _{m}+\frac{1}{2}\sum_{m,l}\gamma _{ml}\xi
_{m}\xi _{l}+\frac{1}{6}\sum_{m,l,j}\gamma _{mlj}\xi _{m}\xi _{l}\xi _{j}.
\label{symbol}
\end{equation}%
Consequently, the operator $\gamma _{\left( 3\right) }\left[ -\mathrm{i}\vec{%
\nabla}_{\mathbf{r}}\right] $ with this symbol takes the form%
\begin{equation}
\gamma _{\left( 3\right) }\left[ -\mathrm{i}\vec{\nabla}_{\mathbf{r}}\right]
V=\omega _{n_{0}}\left( \mathbf{\mathbf{k}_{\ast }}\right) V-\mathrm{i}%
\sum_{m}\gamma _{1m}\partial _{m}V-\frac{1}{2}\sum_{m,l}\gamma _{ml}\partial
_{m}\partial _{l}V+\frac{\mathrm{i}}{6}\sum_{m,l,j}\gamma _{mlj}\partial
_{m}\partial _{l}\partial _{j}V  \label{opsym}
\end{equation}%
More exactly, the operator $\gamma \left[ -\mathrm{i}\beta \nabla _{\mathbf{r%
}}\right] $ is defined based on Fourier transform and the symbol $\gamma
\left( \beta \mathbf{\xi }\right) $ as follows%
\begin{equation}
\widehat{\gamma \left[ -\mathrm{i}\beta \nabla _{\mathbf{r}}\right] u}\left( 
\mathbf{\xi }\right) =\gamma \left( \beta \mathbf{\xi }\right) \hat{u}\left( 
\mathbf{\xi }\right) .  \label{GamFo}
\end{equation}

\subsection{Nonlinearity in ENLS}

Nonlinear terms in the ENLS involve spatial and time derivatives. We show
that their Fourier transforms have the same form as the convolution
integrals in (\ref{I1}) and similar convolution approximation of (\ref{VcV1}%
).

\subsubsection{Nonlinearity envolving spatial derivatives}

Here we biefly describe the Fourier transform of expressions that involve a
product of spatial derivatives. From (\ref{Ftransform}) we obtain for $\bar{l%
}=\left( l_{1},\ldots ,l_{d}\right) $ that 
\begin{equation}
\nabla _{\mathbf{r}}^{\bar{l}}V\left( \mathbf{r}\right) =\frac{\partial
^{\left\vert \bar{l}\right\vert }V}{\partial r_{1}^{l_{1}}\ldots \partial
r_{d}^{l_{d}}}=\frac{1}{\left( 2\pi \right) ^{d}}\int_{\mathbb{R}^{d}}%
\mathrm{i}^{\left\vert \bar{l}\right\vert }\mathbf{\xi }^{\bar{l}}\hat{V}%
\left( \mathbf{\xi }\right) \mathrm{e}^{\mathrm{i}\mathbf{r}\cdot \mathbf{%
\xi }}\,\mathrm{d}\mathbf{\xi .\ }  \label{dlF}
\end{equation}%
Consequently, 
\begin{equation}
\widehat{\left( -\mathrm{i}\right) ^{\left\vert \bar{l}\right\vert }\nabla ^{%
\bar{l}}V}\left( \mathbf{\xi }\right) =\mathbf{\xi }^{\bar{l}}\hat{V}\left( 
\mathbf{\xi }\right) .  \label{idl}
\end{equation}%
Now we introduce \emph{linear operators acting in a non-symmetric way on the
three factors of a product of three functions of }$\ d$\emph{\ variables as
in (\ref{p1})}. First, let us introduce a symbol of such an operator. If $p^{%
\left[ \sigma \right] }\left( \vec{s}^{\;\star }\right) $ is a polynomial of 
$\vec{s}^{\;\star }=\left( \mathbf{\mathbf{s}}^{\prime },\mathbf{\mathbf{s}}%
^{\prime \prime },\mathbf{\mathbf{s}}^{\prime \prime \prime }\right) $ of
the degree $\sigma $ it can be written as a sum of monomials in the form 
\begin{equation}
p^{\left[ \sigma \right] }\left( \vec{s}^{\;\star }\right) =\sum_{\left\vert 
\bar{l}^{\prime }\right\vert +\left\vert \bar{l}^{\prime \prime }\right\vert
+\left\vert \bar{l}^{\prime \prime \prime }\right\vert \leq \sigma }a_{\bar{l%
}^{\prime },\bar{l}^{\prime \prime },\bar{l}^{\prime \prime \prime }}\left( 
\mathbf{s}^{\prime }\right) ^{\bar{l}^{\prime }}\left( \mathbf{s}^{\prime
\prime }\right) ^{\bar{l}^{\prime \prime }}\left( \mathbf{s}^{\prime \prime
\prime }\right) ^{\bar{l}^{\prime \prime \prime }}  \label{psigexp}
\end{equation}%
where $a_{\bar{l}^{\prime },\bar{l}^{\prime \prime },\bar{l}^{\prime \prime
\prime }}$ are the coefficients of the polynomial, and mutiindices $\bar{l}%
^{\prime },\bar{l}^{\prime \prime },\bar{l}^{\prime \prime \prime }$ have
the form $\bar{l}^{\prime }=\left( l_{1}^{\prime },\ldots ,l_{d}^{\prime
}\right) $ etc.

We define the action of the differential operator $p_{m}\left[ -\mathrm{i}%
\vec{\nabla}_{\mathbf{r}}\right] $ on the product of three functions $%
V^{\prime },V^{\prime \prime },V^{\prime \prime \prime }$ \ by the following
formula 
\begin{gather}
p^{\left[ \sigma \right] }\left[ -\mathrm{i}\vec{\nabla}_{\mathbf{r}}\right]
\left( V^{\prime }V^{\prime \prime }V^{\prime \prime \prime }\right) =
\label{psigVVV} \\
\sum_{\left\vert \bar{l}^{\prime }\right\vert +\left\vert \bar{l}^{\prime
\prime }\right\vert +\left\vert \bar{l}^{\prime \prime \prime }\right\vert
\leq \sigma }a_{\bar{l}^{\prime },\bar{l}^{\prime \prime },\bar{l}^{\prime
\prime \prime }}\left( \left[ -\mathrm{i}\vec{\nabla}_{\mathbf{r}}\right] ^{%
\bar{l}^{\prime }}V^{\prime }\right) \left( \left[ -\mathrm{i}\vec{\nabla}_{%
\mathbf{r}}\right] ^{\bar{l}^{\prime \prime }}V^{\prime \prime }\right)
\left( \left[ -\mathrm{i}\vec{\nabla}_{\mathbf{r}}\right] ^{\bar{l}^{\prime
\prime \prime }}V^{\prime \prime \prime }\right) .  \notag
\end{gather}%
\emph{Notice that the order of factors in the product }$V^{\prime }V^{\prime
\prime }V^{\prime \prime \prime }$\emph{\ matters for the action of }$p^{%
\left[ \sigma \right] }\left[ -\mathrm{i}\vec{\nabla}_{\mathbf{r}}\right] $
and, generically,%
\begin{equation}
p^{\left[ \sigma \right] }\left[ -\mathrm{i}\vec{\nabla}_{\mathbf{r}}\right]
\left( V^{\prime }V^{\prime \prime }V^{\prime \prime \prime }\right) \neq p^{%
\left[ \sigma \right] }\left[ -\mathrm{i}\vec{\nabla}_{\mathbf{r}}\right]
\left( V^{\prime }V^{\prime \prime \prime }V^{\prime \prime }\right) \text{
if \ }V^{\prime \prime }\neq V^{\prime \prime \prime }.  \label{nonVVV}
\end{equation}%
For the Fourier transform we obtain the convolution formula%
\begin{gather}
\widehat{p^{\left[ \sigma \right] }\left[ -\mathrm{i}\vec{\nabla}_{\mathbf{r}%
}\right] \left( V^{\prime }V^{\prime \prime }V^{\prime \prime \prime
}\right) }\left( \mathbf{\xi }\right) =  \label{psigVVVF} \\
\frac{1}{\left( 2\pi \right) ^{2d}}\int_{\mathbf{\xi }^{\prime }+\mathbf{\xi 
}^{\prime \prime }+\mathbf{\xi }^{\prime \prime \prime }=\mathbf{\xi }}p^{%
\left[ \sigma \right] }\left( \vec{\xi}^{\;\star }\right) \hat{V}^{\prime
}\left( \mathbf{\xi }^{\prime }\right) \hat{V}^{\prime \prime }\left( 
\mathbf{\xi }^{\prime \prime }\right) \hat{V}^{\prime \prime \prime }\left( 
\mathbf{\xi }^{\prime \prime \prime }\right) \,\mathrm{d}\mathbf{\xi }%
^{\prime }\mathrm{d}\mathbf{\xi }^{\prime \prime }.  \notag
\end{gather}%
Multiplying this expression by an oscillatiing exponent we obtain
expressions completely similar to the integrands in (\ref{I1}) and (\ref%
{convV1}).

\subsubsection{Nonlinearity envolving time derivative}

The equations (\ref{GNLS1+}), (\ref{GNLS1-}), (\ref{Bi1}), (\ref{Bi2}) have
time derivatives in the nonlinear terms, for example $Z_{+}Z_{-}\left(
\partial _{t}+\mathrm{i}\mathcal{L}_{+}^{\left[ 4\right] }\right) Z_{+}$ in (%
\ref{GNLS1+}). We show here that the FNLR that corresponds to these terms
has the same form as the FNLR coming from the first order correction to the
susceptibility which is given by (\ref{unlt}). In the rectifying variables
the FM terms of the form (\ref{unlt}) lead to the FNLR\ of the following
form similar to (\ref{susy}): 
\begin{gather}
\beta ^{d}\tilde{u}_{\bar{n}}^{\left( 1,l\right) }\left( \zeta \mathbf{%
\mathbf{k}_{\ast }}+Y_{\zeta }\left( \beta \mathbf{q}\right) ,\frac{\tau }{%
\varrho }\right) =  \label{susy1} \\
\frac{1}{\varrho }\int_{0}^{\tau }\int_{Y_{\zeta }\left( \beta \mathbf{q}%
^{\prime }\right) +Y_{\zeta }\left( \beta \mathbf{q}^{\prime \prime }\right)
-Y_{\zeta }\left( -\beta \mathbf{q}^{\prime \prime \prime }\right) =Y_{\zeta
}\left( \beta \mathbf{q}\right) }\exp \left\{ \mathrm{i}\mathring{\Phi}%
\left( \mathbf{\mathbf{k}_{\ast }},\beta \vec{q}\right) \frac{\beta ^{2}\tau
_{1}}{\varrho }\right\} \partial _{\tau _{1}}\psi \left( \tau _{1}\right)
\psi ^{2}\left( \tau _{1}\right)  \notag \\
\Psi _{0}^{3}\left( \vec{\zeta}_{0}Y\left( \beta \vec{\zeta}_{0}\vec{q}%
\right) \right) \breve{Q}_{\vec{n},\bar{l}}\left( \vec{\zeta}_{0}\mathbf{%
\mathbf{k}_{\ast }}+\vec{\zeta}_{0}Y\left( \beta \vec{\zeta}_{0}\vec{q}%
\right) \right) \hat{h}_{\zeta }\left( \mathbf{q}^{\prime }\right) \hat{h}%
_{\zeta }\left( \mathbf{q}^{\prime \prime }\right) \hat{h}_{-\zeta }\left( 
\mathbf{q}^{\prime \prime \prime }\left( 0\right) \right) \,  \notag \\
\det Y_{\zeta }^{\prime }\left( \beta \mathbf{q}^{\prime }\right) \det
Y_{\zeta }^{\prime }\left( \beta \mathbf{q}^{\prime \prime }\right) \mathrm{d%
}\mathbf{q}^{\prime }\mathrm{d}\mathbf{q}^{\prime \prime }\mathrm{d}\tau
_{1}.  \notag
\end{gather}%
Using (\ref{hybet1}), (\ref{detYbet}), (\ref{Qnexp}), (\ref{psig}) we obtain
for the terms of the expansion (\ref{VcV1}) 
\begin{gather}
\varrho ^{\left\vert l\right\vert }\tilde{u}_{\bar{n}}^{\left( 1,l\right)
}\left( \zeta \mathbf{\mathbf{k}_{\ast }}+Y_{\zeta }\left( \beta \mathbf{q}%
\right) ,\frac{\tau }{\varrho }\right) =  \label{uc1ap} \\
\frac{\varrho ^{\left\vert l\right\vert }\beta ^{2d}}{\varrho }%
\int_{0}^{\tau }\int_{\mathbb{R}^{2d}}\exp \left\{ \mathrm{i}\mathring{\Phi}%
\left( \mathbf{\mathbf{k}_{\ast }},\beta \vec{q}\left( \beta \right) \right) 
\frac{\beta ^{2}\tau _{1}}{\varrho }\right\} \partial _{\tau _{1}}\psi
\left( \tau _{1}\right) \psi ^{2}\left( \tau _{1}\right) \left[ p_{\zeta ,%
\bar{l}}^{\left[ \sigma \right] }\left( \beta \vec{q}\right) +O\left( \beta
^{\sigma +1}\right) \right]  \notag \\
\left[ \hat{h}_{\zeta }\left( \mathbf{q}^{\prime }\right) \hat{h}_{\zeta
}\left( \mathbf{q}^{\prime \prime }\right) \hat{h}_{-\zeta }\left( \mathbf{q}%
^{\prime \prime \prime }\left( 0\right) \right) +O\left( \beta ^{\nu
}\right) \right] \left( 1+O\left( \beta ^{\nu }\right) \right) \,\mathrm{d}%
\mathbf{q}^{\prime }\mathrm{d}\mathbf{q}^{\prime \prime }\mathrm{d}\tau
_{1}+O\left( \frac{\beta ^{N_{\Psi }-2d}}{\varrho }\right) ,  \notag
\end{gather}%
where $p_{\zeta ,l}^{\left[ \sigma \right] }\left( \beta \vec{q}\right) $ is
the Taylor approximation for $\breve{Q}_{\vec{n},\bar{l}}\left( \vec{k}%
\right) $ calculated at $\vec{k}=\vec{\zeta}_{0}\mathbf{\mathbf{k}}_{\ast }$
by a formula similar to (\ref{psig}). We take in this formula $\sigma =0$, $%
\nu =2$. Note that $p_{\zeta ,\bar{l}}^{\left[ 0\right] }\left( \beta \vec{q}%
\right) =\breve{Q}_{\vec{n},\bar{l}}\left( \vec{\zeta}_{0}\mathbf{\mathbf{k}%
_{\ast }}\right) $ is \ the same for $l=\left( 1,0,0\right) $ and $l=\left(
0,1,0\right) $.

We introduce the additional terms in the NLS that approximate this integral.
>From (\ref{Schlin}) and (\ref{V0s}) we obtain the identity%
\begin{equation}
\varrho \exp \left\{ -\mathrm{i}\zeta \gamma _{\left( \nu \right) }\left(
\beta \zeta \mathbf{q}^{\prime }\right) t\right\} \partial _{t}\hat{v}%
_{\zeta }^{\left( 0\right) }\left( \mathbf{q},\varrho t\right) =\partial _{t}%
\left[ \hat{V}_{\zeta }^{\left( 0\right) }\left( \mathbf{q}^{\prime
},t\right) \right] +\mathrm{i}\zeta \gamma _{\left( \nu \right) }\left(
\beta \zeta \mathbf{q}^{\prime }\right) \hat{V}_{\zeta }^{\left( 0\right)
}\left( \mathbf{q}^{\prime },t\right) .  \label{expdt}
\end{equation}%
Since $\varrho t=\tau $, $\varrho \partial _{t}=$ $\partial _{\tau }$, (\ref%
{expdt}) implies 
\begin{gather*}
\varrho \exp \left\{ \mathrm{i}\mathring{\Phi}\left( \mathbf{\mathbf{k}%
_{\ast }},\beta \vec{q}\left( \beta \right) \right) \frac{\beta ^{2}\tau _{1}%
}{\varrho }\right\} \partial _{\tau _{1}}\psi \left( \tau _{1}\right) \psi
^{2}\left( \tau _{1}\right) \hat{h}_{\zeta }\left( \mathbf{q}^{\prime
}\right) \hat{h}_{\zeta }\left( \mathbf{q}^{\prime \prime }\right) \hat{h}%
_{-\zeta }\left( \mathbf{q}^{\prime \prime \prime }\left( 0\right) \right) \\
=\mathrm{e}^{\mathrm{i}\zeta \gamma _{\left( \nu \right) }\left( \beta \zeta 
\mathbf{q}\right) }\left[ \partial _{t}\left[ \hat{V}^{\left( 0\right)
}\left( \mathbf{q}^{\prime },t\right) \right] +\mathrm{i}\zeta \gamma
_{\left( \nu \right) }\left( \beta \zeta \mathbf{q}^{\prime }\right) \hat{V}%
_{\zeta }^{\left( 0\right) }\left( \mathbf{q}^{\prime },t\right) \right] 
\hat{V}_{\zeta }^{\left( 0\right) }\left( \zeta \mathbf{q}^{\prime \prime
},t\right) \hat{V}_{-\zeta }^{\left( 0\right) }\left( \mathbf{q}^{\prime
\prime \prime }\left( 0\right) ,t\right)
\end{gather*}%
\ where $\hat{V}_{\zeta }^{\left( 0\right) }\left( \mathbf{q},t\right) $ is
given by (\ref{V0s}). Hence, for $l=\left( 1,0,0\right) $ the principal part
of the integral (\ref{uc1ap}) coincides with the Fourier transform of the
term 
\begin{equation}
\frac{\left( 2\pi \right) ^{2d}}{\varrho }\int_{0}^{\tau }\breve{Q}_{\vec{n},%
\bar{l}}\left( \vec{\zeta}_{0}\mathbf{\mathbf{k}_{\ast }}\right) \left( \hat{%
V}_{\zeta }^{\left( 0\right) }\hat{V}_{-\zeta }^{\left( 0\right) }\left(
\partial _{t}V_{\zeta }^{\left( 0\right) }+\mathrm{i}\zeta \gamma _{\left(
\nu \right) }\left( -\mathrm{i}\zeta \beta \nabla _{\mathbf{r}}\right)
V_{\zeta }^{\left( 0\right) }\right) \right) \,\mathrm{d}\tau _{1}
\label{sus1}
\end{equation}%
\ where $V^{\left( 0\right) }$ is the linear response of the NLS given by (%
\ref{V0s}). Similar formula holds for $l=\left( 0,1,0\right) $. For $%
l=\left( 0,0,1\right) $ the principal part of the integral (\ref{uc1ap})
coincides with the Fourier transform of the term 
\begin{equation}
\frac{\left( 2\pi \right) ^{2d}}{\varrho }\int_{0}^{\tau }\breve{Q}_{\vec{n},%
\bar{l}}\left( \vec{\zeta}_{0}\mathbf{\mathbf{k}_{\ast }}\right) \left[
V^{\left( 0\right) 2}\left( \partial _{t}V_{\zeta ^{\prime \prime \prime
}}^{\left( 0\right) }+\frac{\mathrm{i}\zeta ^{\prime \prime \prime }}{%
\varrho }\gamma _{\left( \nu \right) }\left( -\mathrm{i}\beta \zeta ^{\prime
\prime \prime }\nabla _{\mathbf{r}}\right) V_{\zeta ^{\prime \prime \prime
}}^{\left( 0\right) }\right) \right] \,\mathrm{d}\tau _{1},\ \zeta ^{\prime
\prime \prime }=-\zeta .  \label{sus3}
\end{equation}%
Hence the part of the FNLR of the ENLS\ corresponding to the terms in (\ref%
{GNLS1+}), (\ref{GNLS1-}) with 
\begin{equation}
\delta _{1,\zeta }=2\breve{Q}_{\vec{n},1,0,0}\left( \vec{\zeta}_{0}\mathbf{%
\mathbf{k}_{\ast }}\right) ,\ \delta _{2,\zeta \ }=\breve{Q}_{\vec{n}%
,0,0,1}\left( \vec{\zeta}_{0}\mathbf{\mathbf{k}_{\ast }}\right) ,\ \zeta
=\pm .  \label{del+}
\end{equation}%
coincides with the principal part of (\ref{uc1ap}), here we use notation (%
\ref{Qnl}) where $\bar{l}=\left( 1,0,0\right) $ or $\bar{l}=\left(
0,0,1\right) $.

\section{Lattice Nonlinear Schrodinger equation}

In this section we show how the NLM can be approximated by a lattice
Nonlinear Schrodinger equation with the same precision as by the classical
NLS in the entire space. We consider for simplicity the case of real-valued
excitations and lower-order approximations.

In the one-dimensional case the lattice NLS replacing \ the NLS (\ref{Si}), (%
\ref{Si1}) have the form similar to (\ref{eqLatIntr}), namely 
\begin{gather}
\partial _{t}Z_{+}\left( m\right) =-\mathrm{i}\left( \gamma _{0}+\gamma
_{2}\right) Z_{+}\left( m\right) -\gamma _{1}\left( \frac{1}{2}\left[
Z_{+}\left( m+1\right) -Z_{+}\left( m-1\right) \right] \right)  \label{LSi}
\\
+\mathrm{i}\frac{\gamma _{2}}{2}\left[ Z_{+}\left( m+1\right) +Z_{+}\left(
m-1\right) \right] +\alpha _{\pi }Q_{+}Z_{-}\left( m\right) Z_{+}^{2}\left(
m\right) ,  \notag \\
Z_{+}\left( m\right) |_{t=0}=h_{+}\left( \beta m\right) ,\ \alpha _{\pi
}=3\alpha \left( 2\pi \right) ^{2},\ m=\ldots -1,0,1,2,\ldots .,  \notag
\end{gather}%
\begin{gather}
\partial _{t}Z_{-}\left( m\right) =\mathrm{i}\left( \gamma _{0}+\gamma
_{2}\right) Z_{-}\left( m\right) -\gamma _{1}\left( \frac{1}{2}\left[
Z_{-}\left( m+1\right) -Z_{-}\left( m-1\right) \right] \right)  \label{LSi1}
\\
-\mathrm{i}\frac{\gamma _{2}}{2}\left[ Z_{-}\left( m+1\right) +Z_{-}\left(
m-1\right) \right] +\alpha _{\pi }Q_{-}Z_{+}\left( m\right) Z_{-}^{2}\left(
m\right) ,  \notag \\
Z_{-}\left( m\right) |_{t=0}=h_{-}\left( \beta m\right) ,\ m=\ldots
-1,0,1,2,\ldots .  \notag
\end{gather}%
The equations do not involve the spatial derivatives and have the form of a
sequence of ODE describing coupled nonlinear oscillators.

The approximation of the NLM\ by the NLS is based on: (i) the approximation
of $\omega _{n_{0}}\left( \mathbf{k}_{\ast }+\mathbf{\eta }\right) $ by its
Taylor polynomial $\gamma _{\left( \nu \right) }\left( \mathbf{\eta }\right) 
$ in (\ref{Tayom}) and (ii) the approximation of the modal susceptibility $%
\breve{Q}_{\vec{n}_{0}}\left( \vec{\zeta}_{0}\mathbf{\mathbf{k}_{\ast }}+%
\vec{\eta}\right) $ defined by (\ref{Qn}) by its Taylor polynomial $p_{\text{%
T},\zeta }^{\left[ \sigma \right] }\left( \vec{\eta}\right) $ (see (\ref%
{psig0})) in a vicinity of $\zeta \mathbf{k}_{\ast }$. Here we consider the
case $\nu =2$,\ $\sigma =0$, with understanding that larger values of $\nu $
and\ $\sigma $ can be considered similarly. Using an orthogonal change of
variables $\mathbf{\mathbf{\eta }}=\mathbf{\mathbf{\Theta }\xi }$ we reduce
the quadratic form to the diagonal form $\omega _{n_{0}}^{\prime \prime
}\left( \mathbf{\mathbf{k}_{\ast }}\right) $ and obtain 
\begin{equation}
\gamma _{\left( 2\right) }\left( \mathbf{\mathbf{\Theta }\xi }\right)
=\gamma _{0}+\sum_{m}\Gamma _{1,m}\xi _{m}+\frac{1}{2}\sum_{m}\Gamma
_{2,m}\xi _{m}^{2},  \label{gam2xi}
\end{equation}%
where 
\begin{equation}
\omega _{n_{0}}\left( \mathbf{k}_{\ast }+\mathbf{\mathbf{\Theta }\xi }%
\right) =\gamma _{\left( 2\right) }\left( \mathbf{\mathbf{\Theta }\xi }%
\right) +O\left( \left\vert \xi \right\vert ^{3}\right) .
\end{equation}%
Now we use trigonometric polynomials instead of algebraic ones. Obviously%
\begin{eqnarray}
\xi _{m} &=&\sin \xi _{m}+O\left( \left\vert \xi \right\vert ^{3}\right) ,
\label{sin} \\
\xi _{m}^{2} &=&2-2\cos \xi _{m}+O\left( \left\vert \xi \right\vert
^{4}\right) .  \label{cos}
\end{eqnarray}%
We set 
\begin{eqnarray}
\Gamma _{\left( 2\right) }\left( \mathbf{\xi }\right) &=&\Gamma
_{0}+\sum_{m}\Gamma _{1,m}\sin \xi _{m}-\sum_{m}\Gamma _{2,m}\cos \xi _{m},
\label{Gam2} \\
\Gamma _{0} &=&\gamma _{0}+\sum_{m}\Gamma _{2,m},  \label{Gam0}
\end{eqnarray}%
which together with (\ref{gam2xi}) yield%
\begin{equation}
\omega _{n_{0}}\left( \mathbf{k}_{\ast }+\mathbf{\mathbf{\Theta }\xi }%
\right) =\Gamma _{\left( 2\right) }\left( \mathbf{\xi }\right) +O\left(
\left\vert \xi \right\vert ^{3}\right) .
\end{equation}%
In particular, for $d=1$ 
\begin{equation}
\Gamma _{\left( 2\right) }\left( \mathbf{\eta }\right) =\left[ \omega
_{n_{0}}\left( k_{\ast }\right) +\omega _{n_{0}}^{\prime \prime }\left(
k_{\ast }\right) \right] +\omega _{n_{0}}^{\prime }\left( k_{\ast }\right)
\sin \eta -\omega _{n_{0}}^{\prime \prime }\left( k_{\ast }\right) \cos \eta
,  \label{Gam21d}
\end{equation}%
\begin{gather*}
\omega _{n_{0}}\left( k_{\ast }\right) +\omega _{n_{0}}^{\prime }\left(
k_{\ast }\right) \eta \mathbf{\mathbf{+}}\frac{1}{2}\omega _{n_{0}}^{\prime
\prime }\left( k_{\ast }\right) \eta ^{2}= \\
\left[ \omega _{n_{0}}\left( k_{\ast }\right) +\omega _{n_{0}}^{\prime
\prime }\left( k_{\ast }\right) \right] +\omega _{n_{0}}^{\prime }\left(
k_{\ast }\right) \sin \eta -\omega _{n_{0}}^{\prime \prime }\left( k_{\ast
}\right) \cos \eta +O\left( \eta ^{3}\right) .
\end{gather*}%
An advantage of this representation compared with its algebraic counterpart
is that involves periodic function similar to s $\omega _{n}\left( \mathbf{k}%
\right) $, namely%
\begin{equation}
\Gamma _{\left( 2\right) }\left( \mathbf{\xi }+2\pi \mathbf{\zeta }\right)
=\Gamma _{\left( 2\right) }\left( \mathbf{\xi }\right) .
\end{equation}%
As before, in a vicinity $\mathbf{\xi =0}$ there is always a rectifying
change of variables 
\begin{equation}
\omega _{n_{0}}\left( \mathbf{k}_{\ast }+Y\left( \mathbf{\xi }\right)
\right) =\Gamma _{\left( \nu \right) }\left( \mathbf{\xi }\right)
,\;\left\vert \mathbf{\xi }\right\vert \leq \pi _{0}.
\end{equation}%
in the both cases when$\ \omega _{n_{0}}^{\prime }\left( \mathbf{k}_{\ast
}\right) \neq 0$ \ or $\det \omega _{n_{0}}^{\prime \prime }\left( \mathbf{k}%
_{\ast }\right) \neq 0$.

\textbf{Remark. }In the case $\nu \geq 3$ we can approximate functions $%
\omega _{n}\left( \mathbf{k}_{\ast }+\mathbf{\eta }\right) $ by $\sin
^{l}\left( \eta _{j}\right) ,$ $l=1,\ldots ,\nu $. Based on the Taylor
polynomial%
\begin{equation}
\gamma _{\left( \nu \right) }\left( \mathbf{\eta }\right) =\sum_{j=0}^{\nu }%
\frac{1}{j!}\omega _{n_{0}}^{\left( j\right) }\left( \mathbf{\mathbf{k}%
_{\ast }}\right) \left( \mathbf{\eta }^{j}\right) ,\ \mathbf{\eta }=\beta 
\mathbf{s}=\mathbf{k}-\zeta \mathbf{k}_{\ast }
\end{equation}%
we form a trigonometric polynomial 
\begin{equation}
\Gamma _{\left( \nu \right) }\left( \mathbf{\eta }\right) =\sum_{j=0}^{\nu }%
\frac{1}{j!}\Gamma _{\left( \nu \right) }^{\left( j\right) }\left( \sin 
\mathbf{\eta }\right) ^{j}  \label{Gamsin}
\end{equation}%
where%
\begin{equation}
\mathbf{\eta }=\left( \eta _{1},\ldots ,\eta _{d}\right) ,\;\sin \mathbf{%
\eta =}\left( \sin \eta _{1},\ldots ,\sin \eta _{d}\right) .
\end{equation}%
The coefficients $\Gamma _{\left( \nu \right) }^{\left( j\right) }$ are
uniquely determined by $\omega _{n_{0}}^{\left( i\right) }\left( \zeta 
\mathbf{\mathbf{k}_{\ast }}\right) $, $i=0,\ldots ,j$ since \ the change of
variables 
\begin{equation}
\sin \eta _{l}\leftrightarrow \eta _{l}
\end{equation}%
is invertible about the origin. We obviously have 
\begin{equation}
\left\vert \omega _{n_{0}}\left( \zeta \mathbf{\mathbf{k}_{\ast }}+\beta 
\mathbf{\mathbf{s}}\right) -\Gamma _{\left( \nu \right) }\left( \zeta \beta 
\mathbf{\mathbf{s}}\right) \right\vert \leq C\beta ^{\nu +1}\left\vert 
\mathbf{\mathbf{s}}\right\vert ^{\nu +1},\ \beta \mathbf{\mathbf{s}}\in %
\left[ -\pi _{0},\pi _{0}\right] ^{d},
\end{equation}%
and a rectifying change of variables exists is this case too.$\blacklozenge $

\subsubsection{Functions on a lattice and the discrete Fourier transform}

We consider the lattice of vectors with integer components 
\begin{equation}
\mathbf{\mathbf{m}}=\left( m_{1},m_{2},\cdots ,m_{d}\right) \in \mathbb{Z}%
^{d},
\end{equation}%
and functions $Z\left( \mathbf{\mathbf{m}}\right) $ on the lattice $\mathbb{Z%
}^{d}$. The shift operators are defined as follows 
\begin{equation}
\dot{\partial}_{+,j}Z=Z\left( \ldots ,m_{j}+1,\ldots \right) ,\ \dot{\partial%
}_{-,j}Z=Z\left( \ldots ,m_{j}-1,\ldots \right) .
\end{equation}%
The elementary difference operators then are defined as follows 
\begin{gather}
\dot{\Delta}_{+,j}Z=\frac{1}{2}\left[ Z\left( \ldots ,m_{j}+1,\ldots \right)
+Z\left( \ldots ,m_{j}-1,\ldots \right) \right] =\frac{1}{2}\left[ \dot{%
\partial}_{+,j}Z+\dot{\partial}_{-,j}Z\right] ,  \label{Del+} \\
\dot{\Delta}_{-,j}Z=\frac{1}{2\mathrm{i}}\left[ Z\left( \ldots
,m_{j}+1,\ldots \right) -Z\left( \ldots ,m_{j}-1,\ldots \right) \right] =%
\frac{1}{2\mathrm{i}}\left[ \dot{\partial}_{+,j}Z-\dot{\partial}_{-,j}Z%
\right] .  \notag
\end{gather}%
For every lattice function $Z\left( \mathbf{\mathbf{m}}\right) $ we define
its Fourier transform%
\begin{equation}
\bar{Z}\left( \mathbf{\mathbf{\xi }}\right) =\sum_{\mathbf{\mathbf{m}}%
}Z\left( \mathbf{\mathbf{m}}\right) e^{-i\mathbf{\mathbf{m\cdot \mathbf{%
\mathbf{\xi }}}}}  \label{LF}
\end{equation}%
with the inverse transform 
\begin{equation}
Z\left( \mathbf{\mathbf{m}}\right) =\frac{1}{\left( 2\pi \right) ^{d}}\int_{%
\left[ -\pi ,\pi \right] ^{d}}e^{i\mathbf{\mathbf{m\cdot \mathbf{\xi }}}}%
\bar{Z}\left( \mathbf{\mathbf{\xi }}\right) \,\mathrm{d}\mathbf{\mathbf{\xi }%
}.
\end{equation}%
Obviously $\bar{Z}\left( \mathbf{\mathbf{\xi }}\right) $ is a $2\pi $%
-periodic function of $\mathbf{\mathbf{\mathbf{\xi }\in }}\mathbb{R}^{d}$. \
The Fourier transform of the difference operators $\dot{\Delta}_{+,i}Z$ is
given by%
\begin{equation}
\overline{\left[ \dot{\Delta}_{+,j}Z\right] }\left( \mathbf{\mathbf{\mathbf{%
\xi }}}\right) =\cos \mathbf{\mathbf{\mathbf{\xi }}}_{j}\bar{Z}\left( 
\mathbf{\mathbf{\mathbf{\xi }}}\right) ,\;\overline{\left[ \dot{\Delta}%
_{-,j}Z\right] }\left( \mathbf{\mathbf{\mathbf{\xi }}}\right) =\sin \mathbf{%
\mathbf{\mathbf{\xi }}}_{j}\bar{Z}\left( \mathbf{\mathbf{\mathbf{\xi }}}%
\right) .  \label{Delsin}
\end{equation}%
When $d=1$ we omit $j$ and set 
\begin{equation}
\dot{\Delta}_{+}Z=\frac{1}{2}\left[ Z\left( m+1\right) +Z\left( m-1\right) %
\right] ,\ \dot{\Delta}_{-}Z=\frac{1}{2\mathrm{i}}\left[ Z\left( m+1\right)
-Z\left( m-1\right) \right] .  \label{Del1d}
\end{equation}%
Note that the Fourier transform of the product is given by the following
convolution formula%
\begin{equation}
\overline{XZ}\left( \mathbf{\mathbf{\xi }}\right) =\frac{1}{\left( 2\pi
\right) ^{d}}\int_{\left[ -\pi ,\pi \right] ^{d}}\overline{X}\left( \mathbf{%
\mathbf{s}}\right) \overline{Z}\left( \mathbf{\mathbf{\xi -s}}\right) \,%
\mathrm{d}\mathbf{s}
\end{equation}%
as in the case of the continuous Fourier transform.

\subsubsection{Lattice NLS (LNLS)}

When $\Gamma _{\left( 2\right) }\left( \mathbf{\mathbf{\mathbf{\mathbf{\xi }}%
}}\right) $ \ is given by (\ref{Gam2}), we define the difference operator on
the lattice by the formula 
\begin{equation}
\Gamma _{\left( 2\right) }\left( \zeta \dot{\nabla}\right) Z=\Gamma
_{0}Z+\zeta \sum_{m}\Gamma _{1,m}\dot{\Delta}_{-,m}Z-\sum_{m}\Gamma _{2,m}%
\dot{\Delta}_{+,m}Z.
\end{equation}%
Note that its Fourier transform is 
\begin{equation}
\overline{\Gamma _{\left( 2\right) }\left( \dot{\nabla}\right) Z}\left( 
\mathbf{\mathbf{\mathbf{\mathbf{\xi }}}}\right) =\Gamma _{\left( 2\right)
}\left( \mathbf{\mathbf{\mathbf{\mathbf{\xi }}}}\right) \overline{Z}\left( 
\mathbf{\mathbf{\mathbf{\mathbf{\xi }}}}\right) .
\end{equation}%
Let us introduce a \emph{linear lattice Schrodinger equation (LLS)} 
\begin{equation}
\partial _{t}Z\left( \mathbf{\mathbf{m}},t\right) =-\mathrm{i}\Gamma
_{\left( 2\right) }\left( \dot{\nabla}\right) Z\left( \mathbf{\mathbf{m}}%
,t\right) ,\ Z\left( \mathbf{\mathbf{m}},t\right) |_{t=0}=h\left( \mathbf{%
\mathbf{m}}\right) ,\;\mathbf{\mathbf{m\in }}\mathbb{Z}^{d}.
\end{equation}%
It can be solved exactly in terms of its lattice Fourier transform (\ref{LF}%
), namely%
\begin{equation}
\bar{Z}\left( \mathbf{\mathbf{\mathbf{\mathbf{\xi }}}},t\right) =\bar{h}%
\left( \mathbf{\mathbf{\mathbf{\mathbf{\xi }}}}\right) \exp \left\{ -\mathrm{%
i}\Gamma _{\left( 2\right) }\left( \mathbf{\mathbf{\mathbf{\mathbf{\xi }}}}%
\right) t\right\} .
\end{equation}%
Let us introduce now \emph{\ Lattice Nonlinear Schrodinger equation (LNLS)} 
\begin{equation}
\partial _{t}Z_{\zeta }=-\mathrm{i}\zeta \Gamma _{\left( 2\right) }\left(
\zeta \dot{\nabla}\right) Z_{\zeta }+\alpha _{\pi }Q_{\zeta }Z_{-\zeta
}Z_{+\zeta }^{2},\ Z_{\zeta }\left( \mathbf{\mathbf{m}},t\right)
|_{t=0}=h_{\beta ,\zeta }\left( \mathbf{\mathbf{m}}\right) ,\zeta =\pm ,
\label{LNLS}
\end{equation}%
where $Q_{\zeta }$ is a complex constant, and the factor $\alpha _{\pi
}=3\alpha \left( 2\pi \right) ^{2d}$ is introduced for notational
consistency with the related NLM. \ Here%
\begin{equation}
h_{\beta ,\zeta }\left( \mathbf{\mathbf{m}}\right) =h_{\zeta }\left( \beta 
\mathbf{\mathbf{m}}\right) ,\;\zeta =\pm
\end{equation}%
where $h_{\zeta }\left( \mathbf{\mathbf{r}}\right) $, $\mathbf{\mathbf{r\in }%
}\mathbb{R}^{d}$ is a given smooth function of continuous argument. Its
lattice Fourier transform is given by 
\begin{equation}
\bar{h}_{\beta ,\zeta }\left( \mathbf{\mathbf{\xi }}\right) =\sum_{\mathbf{%
\mathbf{m}}}h_{\zeta }\left( \beta \mathbf{\mathbf{m}}\right) e^{-i\mathbf{%
\mathbf{m\cdot \mathbf{\xi }}}},\;\zeta =\pm ,
\end{equation}%
with the inverse formula%
\begin{equation}
h_{\beta ,\zeta }\left( \mathbf{\mathbf{m}}\right) =\frac{1}{\left( 2\pi
\right) ^{d}}\int_{\left[ -\pi ,\pi \right] ^{d}}e^{i\mathbf{\mathbf{m\cdot 
\mathbf{\xi }}}}\bar{h}_{\beta ,\zeta }\left( \mathbf{\mathbf{\xi }}\right)
\,\mathrm{d}\mathbf{\mathbf{\xi }}.  \label{FBinv}
\end{equation}%
Note that this formula makes sense even for non-integer values of $\mathbf{%
\mathbf{m}}=\mathbf{\mathbf{r}}$ providing an interpolation to such values.
We can replace in (\ref{UNLS}) $\hat{Z}_{\zeta }$ based on a solution of the
NLS by by $\overline{Z}_{\zeta }$ based on the LNLS. \ Similarly to (\ref%
{UNLS1}) we obtain that the modal coefficient $\tilde{U}_{\zeta ,n_{0}}$ of
the solution of the NLM\ is well approximated in terms of the solution of
the LNLS, namely \ 
\begin{equation}
\tilde{U}_{\zeta ,n_{0}}\left( \zeta \mathbf{k}_{\ast }+\mathbf{\mathbf{%
\Theta }\xi },t\right) =\bar{Z}_{\zeta }\left( Y^{-1}\left( \mathbf{\xi }%
\right) ,t\right) +\left[ O\left( \alpha ^{2}\right) +O\left( \alpha \beta
\right) +O\left( \alpha \varrho \right) \right] O\left( \left\vert \mathbf{U}%
^{\left( 1\right) }\right\vert \right) ,  \label{UZlat}
\end{equation}%
Note that in the one-dimensional case $\mathbf{\mathbf{\Theta }\xi }=\mathbf{%
\xi }$. One can see that LNLS gives the same order of accuracy as the NLS.

\subparagraph{\textbf{Presentation in spatial domain.}}

The formula (\ref{UZ}) takes the form 
\begin{gather}
\mathbf{U}_{Z}\left( \mathbf{r},t\right) =\frac{\beta ^{d}}{\left( 2\pi
\right) ^{d}}\int_{\left[ -\pi /\beta ,\pi /\beta \right] ^{d}}\Psi
_{0}\left( \beta \mathbf{s}\right)  \\
\left[ \bar{Z}_{+}\left( Y^{-1}\left( \beta \mathbf{s}\right) ,t\right) 
\mathbf{\tilde{G}}_{+,n_{0}}\left( \mathbf{r},\mathbf{k}_{\ast }+\beta 
\mathbf{s}\right) +\bar{Z}_{-}\left( -Y^{-1}\left( -\beta \mathbf{s}\right)
,t\right) \mathbf{\tilde{G}}_{-,n_{0}}\left( \mathbf{r},-\mathbf{k}_{\ast
}+\beta \mathbf{s}\right) \right] \,\mathrm{d}\mathbf{s},  \notag
\end{gather}%
where%
\begin{equation}
\mathbf{\tilde{G}}_{+,n_{0}}\left( \mathbf{r},\mathbf{k}\right) =\mathbf{%
\hat{G}}_{+,n_{0}}\left( \mathbf{r},\mathbf{k}\right) \mathrm{e}^{\mathrm{i}%
\mathbf{k}\cdot \mathbf{r}},
\end{equation}%
where $\mathbf{\hat{G}}_{+,n_{0}}\left( \mathbf{r},\mathbf{k}\right) $ is $1$%
-periodic function of $\mathbf{r}$\textbf{. } Instead (\ref{GhatTay}) we
have similarly to (\ref{Gam2}) expansion into trigonometric functions 
\begin{equation}
\mathbf{\hat{G}}_{+,n_{0}}\left( \mathbf{r},\mathbf{k}_{\ast }+\beta \mathbf{%
q}\right) =\mathbf{\dot{p}}_{g,\sigma }\left( \mathbf{r},\beta \mathbf{q}%
\right) +O\left( \beta ^{\sigma +1}\right) ,\ \sigma +1\leq \nu ,
\end{equation}%
where, for $\sigma =2$, 
\begin{equation}
\mathbf{\dot{p}}_{g,\sigma }\left( \mathbf{r},\beta \mathbf{q}\right) =\left[
\mathbf{\hat{G}}_{+,n_{0}}\left( \mathbf{r},\mathbf{k}_{\ast }\right) +%
\mathbf{\hat{G}}_{+,n_{0}}^{\prime \prime }\left( \mathbf{r},\mathbf{k}%
_{\ast }\right) \right] +\mathbf{\hat{G}}_{+,n_{0}}^{\prime }\left( \mathbf{r%
}\right) \sin \left( \beta \mathbf{q}\right) -\mathbf{\hat{G}}%
_{+,n_{0}}^{\prime \prime }\left( \mathbf{r}\right) \cos \left( \beta 
\mathbf{q}\right) ,
\end{equation}%
We obtain (\ref{UZ1}) where, similarly to (\ref{UZ+1}),%
\begin{equation}
\mathbf{U}_{Z_{+}}\left( \mathbf{m},t\right) =\mathbf{U}_{Z_{+}}^{0}\left( 
\mathbf{m},t\right) +\mathbf{U}_{Z_{+}}^{1}\left( \mathbf{m},t\right) +%
\mathbf{U}_{Z_{+}}^{2}\left( \mathbf{m},t\right) +O\left( \beta ^{3}\right) ,
\label{UZLr}
\end{equation}%
where $\mathbf{U}_{Z_{+}}^{0}$ is given by 
\begin{equation}
\mathbf{U}_{Z_{+}}^{0}\left( \mathbf{m},t\right) =\mathrm{e}^{\mathrm{i}%
\mathbf{k}_{\ast }\cdot \mathbf{m}}\left[ \mathbf{\hat{G}}_{+,n_{0}}\left( 
\mathbf{m},\mathbf{k}_{\ast }\right) +\mathbf{\hat{G}}_{+,n_{0}}^{\prime
\prime }\left( \mathbf{m},\mathbf{k}_{\ast }\right) \right] Z_{+}\left( 
\mathbf{m},t\right) .  \label{UZLr0}
\end{equation}%
and, similarly to (\ref{UZr0}),%
\begin{equation}
\mathbf{U}_{Z_{+}}^{1}\left( \mathbf{m},t\right) =\mathrm{e}^{\mathrm{i}%
\mathbf{k}_{\ast }\cdot \mathbf{m}}\mathbf{\hat{G}}_{+,n_{0}}^{\prime
}\left( \mathbf{m},\mathbf{k}_{\ast }\right) \cdot \dot{\Delta}%
_{-}Z_{+}\left( \mathbf{m},t\right) ,  \label{UZLr1}
\end{equation}%
\begin{equation}
\mathbf{U}_{Z_{+}}^{2}\left( \mathbf{m},t\right) =-\mathrm{e}^{\mathrm{i}%
\mathbf{k}_{\ast }\cdot \mathbf{m}}\mathbf{\hat{G}}_{+,n_{0}}^{\prime \prime
}\left( \mathbf{m},\mathbf{k}_{\ast }\right) \cdot \dot{\Delta}%
_{+}Z_{+}\left( \mathbf{m},t\right) .  \label{UZLr2}
\end{equation}%
\ Note that using (\ref{FBinv}) we can interpolate $\mathbf{U}_{Z_{\pm
}}\left( \mathbf{m},t\right) $ to non-integer $\mathbf{m}=\mathbf{r}$.

\subparagraph{\textbf{Comparision of the Lattice NLS with the NLS.}}

The LNLS approximation, compared with the classical NLS. has the following
properties:

\begin{itemize}
\item The accuracy of approximation by the lattice NLS (\ref{LNLS}) is the
same as by the NLS (\ref{GNLS+}).

\item The right-hand side of \ (\ref{LNLS}) is a bounded operator which is
an advantate compared with the NLS.

\item The lattice system (\ref{LNLS}) is already in a spatially discretized
form which can be advantageous for numerical simulations.

\item The form of LNLS (\ref{LNLS}) suggests that small scale (compared with
the cell size) features of the wave dynamics are effectively eliminated.
Note that the derivation of the NLS also assumes the elimination of the
small scale, but the differential form of the NLS\ still allows small scale
perturbations to be of importance for large scale wave dynamics.
\end{itemize}

There is extensive literature on coupled nonlinear oscillators on lattices,
see for example \cite{PhysD}, \cite{MacKay} and references therein. For
photonic crystals such equations were used in \cite{EisenbergS}.

\textbf{Remark.} In the case when we use (\ref{Gamsin}) the
difference-differential equation (\ref{LNLS}) takes the form 
\begin{equation}
\partial _{t}Z=-\mathrm{i}\Gamma _{\left( 2\right) }\left( \dot{\Delta}%
_{-}\right) Z+\alpha _{\pi }Q_{+}\left\vert Z\right\vert ^{2}Z,\ Z\left( 
\mathbf{\mathbf{m}},t\right) |_{t=0}=h_{\beta }\left( \mathbf{\mathbf{m}}%
\right) .
\end{equation}%
where the operator $\ \Gamma _{\left( 2\right) }\left( \dot{\Delta}%
_{-}\right) $ $\ $is obtained by substituting $\dot{\Delta}_{-,j}$ in place
of $\sin \eta _{j}$ in the polynomial (\ref{Gamsin}).$\blacklozenge $

\textbf{Remark.} Note that usually the terms with operators $\dot{\Delta}%
_{-,m}$ are not involved into the lattice Schrodinger equations considered
in the literature. The reason is that the influence of these terms on
solutions with the initial data $h\left( \beta \mathbf{\mathbf{m}}\right) $
for $\beta \ll 1$ can be taken into account by choosing a coordinate frame
moving with the group velocity. To give a simple explanation, we use another
approximation for $\omega _{n_{0}}\left( \zeta \mathbf{k}_{\ast }+\mathbf{%
\mathbf{\Theta }\xi }\right) $, namely 
\begin{equation}
\omega _{n_{0}}\left( \zeta \mathbf{k}_{\ast }+\mathbf{\mathbf{\Theta }\xi }%
\right) =\Gamma _{\left( 2\right) }\left( \mathbf{\xi }\right) =\Gamma
_{0}+\sum_{m}\Gamma _{1,m}\xi _{m}-\sum_{m}\Gamma _{2,m}\cos \xi
_{m}+O\left( \left\vert \xi \right\vert ^{3}\right) ,
\end{equation}%
which combines linear functions with trigonometric. Corresponding
difference-differential equation in $\mathbb{R}^{d}$ has the form 
\begin{gather}
\partial _{t}Z\left( \mathbf{\mathbf{x}},t\right) =-\mathrm{i}\Gamma
_{0}Z\left( \mathbf{\mathbf{x}},t\right) +\sum_{m}\ \Gamma _{1,m}\frac{%
\partial }{\partial x_{m}}Z\left( \mathbf{\mathbf{x}},t\right)  \label{fdNLS}
\\
+\mathrm{i}\sum_{m}\Gamma _{2,m}\dot{\Delta}_{+,m}Z\left( \mathbf{\mathbf{x}}%
,t\right) +\alpha _{\pi }Q_{+}\left\vert Z\right\vert ^{2}Z\left( \mathbf{%
\mathbf{x}},t\right) ,\;\mathbf{\mathbf{x}}\in \mathbb{R}^{d},  \notag
\end{gather}%
involving the both differential and finite difference operators. The
standard change of variables 
\begin{equation}
Z\left( \mathbf{\mathbf{x}},t\right) =z\left( \left( \mathbf{\mathbf{x+}}%
\Gamma t,t\right) \right) ,\ \Gamma =\left( \Gamma _{1,1},\ldots ,\Gamma
_{1,d}\right)
\end{equation}%
reduces this equation to the following NLS difference equation 
\begin{equation}
\partial _{t}z\left( \mathbf{\mathbf{x}},t\right) =-\mathrm{i}\Gamma
_{0}z\left( \mathbf{\mathbf{x}},t\right) +\mathrm{i}\sum_{m}\Gamma _{2,m}%
\dot{\Delta}_{+,m}z\left( \mathbf{\mathbf{x}},t\right) +\alpha _{\pi
}Q_{+}\left\vert z\right\vert ^{2}z\left( \mathbf{\mathbf{x}},t\right) .
\end{equation}%
Obviously this equation is equivalent to a family of independent equations
on the lattice $\mathbb{Z}^{d}$ of the same form as (\ref{LNLS}) but without
the terms $\dot{\Delta}_{-,m}$.$\blacklozenge $

\section{Conclusions}

The basic conditions on a periodic dielectric medium to support NLS regimes
of electromagnetic wave propagation are (i) the inversion symmetry $\omega
_{n}\left( -\mathbf{k}\right) =\omega _{n}\left( \mathbf{k}\right) $ of the
dispersion relations, and\ (ii) the leading term in the nonlinearity is
cubic. NLS regimes are generated by almost time harmonic excitation currents
with localized quasimomenta and their most essential properties are as
follows.

\begin{itemize}
\item The asymptotic nature of NLS regimes is determined by three small
parameters $\alpha ,\varrho ,\beta $ . The parameter $\alpha $ scales the
magnitude of the nonlinearity, it is proportional to the square of the
amplitude of the excitation. The parameter $\frac{1}{\varrho }$ is
proportional to time extention of the initial current excitation. The
parameter $\beta $ describes the range of quasimomenta $\mathbf{k}$ about a
fixed $\mathbf{k}_{\ast }$ in the modal composition of\ the excitation
current. The NLS regimes arise when $\alpha \sim \varrho $ and 
\begin{equation*}
\alpha \sim \varrho \text{ and }\varrho \sim \beta ^{\varkappa _{1}},\ \text{%
for some }\varkappa _{1}>0.
\end{equation*}%
In particular, the classical NLS regime is characterized by the following
relations between the tree small parameters%
\begin{equation*}
\alpha \sim \varrho \sim \beta ^{2}.
\end{equation*}

\item The NLS and as well their extended versions describe approximately the
evolution of the Floquet-Bloch modal coefficients $\ \tilde{U}_{\pm
,n_{0}}\left( \mathbf{k}_{\ast }+\mathbf{\eta },t\right) $ of the
propagating wave.

\item Multi-modal excitation currents about several $\mathbf{k}_{\ast ,j}$
generate NLS regimes which satisfy the principle of approximate
superposition when with a very high accuracy $O\left( \beta ^{\infty
}\right) $ the modal components about different $\mathbf{k}_{\ast ,j}$
evolve essentially indepently according to NLS or extended NLS equations.

\item Higher accuracy approximations for longer time intervals are achieved
by the analysis of the modal decomposition of the wave and proper rectifying
change of variables for the phase of noninear modal interactions.
\end{itemize}

The accuracy of NLS/ENLS approximation by developed methods can be
characterized as follows.

\begin{itemize}
\item Classical NLS gives aproximation with the error $O\left( \beta \right) 
$ on the time interval $O\left( \frac{1}{\beta ^{2}}\right) $.

\item To improve the accuracy to $O\left( \beta ^{2}\right) $ on the time
interval $O\left( \frac{1}{\beta ^{2}}\right) $ it is sufficient to take
into account the frequency dependence of the sussectibility tensor (in terms
of first order derivatives with respect to $\mathbf{k}$ of the tensor at $%
\mathbf{k}=\mathbf{k}_{\ast }$ ) and the third-order derivatives of the
dispersion relation $\omega _{n_{0}}\left( \mathbf{k}\right) $ at $\mathbf{k}%
=\mathbf{k}_{\ast }$, leading to the third order extended NLS (see (\ref{G3+}%
), (\ref{G3-})).

\item To improve the accuracy furher to $O\left( \beta ^{3}\right) $ on the
time interval $O\left( \frac{1}{\beta ^{2}}\right) $ the following
characteristics of the media have to be taken into account: (i) fourth-order
derivatives of the dispersion relation $\omega _{n_{0}}\left( \mathbf{k}%
\right) $ at $\mathbf{k}=\mathbf{k}_{\ast }$; (ii) second order derivatives
of the susceptibility tensor at $\mathbf{k}=\mathbf{k}_{\ast }$; (iii)
nonlinear interactions between the forward and backward propagating waves;
(iv) \ finer effects of the susceptibility approximation expressed in terms
of the first order frequency derivatives of the susceptibility; (v) fifth
order terms in the nonlinearity; (vi) nonlinear interactions between
different spectral bands. The above effects are taken into account in the
fourth-order extended NLS (see Subsection 1.4.3).

\item The lattice NLS provides the same accuracy of approximation as the
regular NLS with evident advantages for numerically efficient analysis.
\end{itemize}

\section{Notations and abbreviations}

For reader's convenience we provide below a list of notations and
abbreviations used in this paper.

\textbf{almost single-mode excitation} see (\ref{Jzin})

\textbf{bidirectional quadruplet }(\ref{ggk4})

\textbf{directly excited modes}, see \ (\ref{kkstar}), (\ref{UNLS})

\textbf{doublet} (\ref{ggk2})

\textbf{ENLS} Extended Nonlinear Schrodinger equation, see (\ref{GNLS+}), (%
\ref{GNLS-}) or (\ref{GNLS1+}), (\ref{GNLS1-}) and Section 1.3

\textbf{FNLR} First nonlinear response, see (\ref{FNLR}), (\ref{Vn})

\textbf{Floquet-Bloch modal decomposition }see (\ref{UFB})

\textbf{Fourier transform}, see (\ref{Ftransform})

\textbf{FM} frequency matching condition (\ref{FMC}), frequency-matched, see
(\ref{sumz})

\textbf{indirectly excited modes}, see (\ref{indir}), (\ref{Uindn}), (\ref%
{Uindn0})

\textbf{interaction quadruplet} see (\ref{triad})

\textbf{GVM} group velocity matching condition (\ref{GVM0}), (\ref{GVMstar})

\textbf{non-FM}\ non-frequency-matched

\textbf{NLM} Nonlinear Maxwell equation, see (\ref{MXshort})

\textbf{NLS} Nonlinear Schrodinger equation, see (\ref{Si}, (\ref{Si1}), and
(\ref{genS1}), (\ref{genS1-})

\textbf{NLS regime} a situation when the evolultion of an electromagnetic
(EM) wave is governed by the NLM equations and it can be approximated by an
NLS equation or, may be, by a slightly more general extended NLS equation

\textbf{linear response}, see (\ref{MXlin})

\textbf{rectifying coordinates} (\ref{Y}), (\ref{redY})

\textbf{scaled rectifying coordinates} (\ref{Ys})

\textbf{susceptibility} $\mathbf{\chi }_{D}^{\left( 3\right) }$ (\ref{cd4ab})

\textbf{susceptibility} $\mathbf{\chi }^{\left( 3\right) }$ (\ref{cd5b})

\textbf{uni-directional excitation} (\ref{jnn1})

$\alpha _{\pi }=3\alpha \left( 2\pi \right) ^{2d}$

$\gamma _{\left( \nu \right) }\left( \mathbf{\eta }\right) $ - the Taylor
polynomial of $\omega _{\bar{n}}\left( \mathbf{k}_{\ast }+\mathbf{\eta }%
\right) $ of order $\nu $ (\ref{Tayom})

$\gamma _{\left( 2\right) }\left( \mathbf{\eta }\right) =\gamma _{2}\left( 
\mathbf{\eta }^{2}\right) +\gamma _{1}\left( \mathbf{\eta }\right) +\gamma
_{0}\left( \mathbf{\eta }\right) $ - the second order Taylor polynomial of $%
\omega _{\bar{n}}\left( \mathbf{k}_{\ast }+\mathbf{\eta }\right) $ (\ref%
{defom2})

$\delta _{\times ,\zeta }^{\pm }$ - coefficients defined by (\ref{dcross}), (%
\ref{z0bi}), (\ref{kstar1})

$\zeta =\pm 1$ or $\zeta =\pm $ - band binary number, when used in indices
is abbriviated to $\zeta =\pm $, namely 
\begin{equation}
V_{\zeta }=V_{+}\text{ if\ \ }\zeta =+1,\ V_{\zeta }=V_{-}\text{ if\ \ }%
\zeta =-1.
\end{equation}

$\vec{\zeta}_{0,\times }$ - vectors defined by (\ref{z0bi})

$\theta =\frac{\varrho }{\beta ^{2}}$ - inverse dispersion parameter, see (%
\ref{theta})

$\xi =\left( \xi _{1},\ldots ,\xi _{d}\right) $ - Fourier wavevector variable

$\omega _{\bar{n}}\left( \mathbf{k}\right) =\zeta \omega _{n}\left( \mathbf{k%
}\right) $ - dispersion relation of the band $\left( \zeta ,n\right) $, see (%
\ref{omn})

$\omega _{n_{0}}^{\prime }\left( \mathbf{k}\right) =\nabla _{\mathbf{k}%
}\omega _{n_{0}}\left( \mathbf{k}\right) $ - group velocity vector.

$\omega _{n_{0}}^{\prime \prime }\left( \mathbf{k}_{\ast }\right) =\nabla _{%
\mathbf{k}}\nabla _{\mathbf{k}}\omega _{n_{0}}\left( \mathbf{k}_{\ast
}\right) $ - Hessian matrix of $\omega _{n_{0}}\left( \mathbf{k}\right) $ at 
$\mathbf{k}_{\ast }$.

$\mathbf{\tilde{G}}_{\bar{n}}\left( \mathbf{r},\mathbf{k}\right) $ -the
eigenfunction (eigenmode) corresponding to band index $\bar{n}$ and
quasimomentum $\mathbf{k}$, see (\ref{eigen}), (\ref{nz2}).

$h_{\zeta }\left( \mathbf{r}\right) ,\ \zeta =\pm $ - initial data for the
NLS\ equation (\ref{Si}), (\ref{Si1}), (\ref{schx}), (\ref{genS1}), (\ref%
{genS1-}).

$\mathcal{F}_{\text{NL}}^{\left( 0\right) }$ \ see (\ref{UU3})

$\hat{h}_{\zeta }\left( \frac{1}{\beta }\mathbf{\xi }\right) ,\ \zeta =\pm $
- Fourier transform of the initial data $h_{\zeta }\left( \beta \mathbf{r}%
\right) $ for the NLS (\ref{jnn2}), (\ref{Ftransform}).

$\mathring{h}_{\zeta }\left( \frac{1}{\beta }Y\left( \mathbf{\xi }\right)
\right) =\hat{h}_{\zeta }\left( \frac{1}{\beta }\mathbf{\xi }\right) $ -
function used to form excitation curreents for the NLM, (\ref{hcap}) and (%
\ref{hcap1}).

$I_{\bar{n},\zeta ^{\prime },\zeta ^{\prime \prime },\zeta ^{\prime \prime
\prime }}\left( \mathbf{k},\tau \right) $ - interaction integral (\ref{Vn0}).

$\mathbf{J}$ - excitation current see (\ref{J01}).

$\mathbf{k}=\left( k_{1},\ldots ,k_{d}\right) $ - quasimomentum (wavevector)
variable.

$\mathbf{k}_{\ast }=\left( k_{\ast 1},\ldots ,k_{\ast d}\right) $ - center
of the wavepacket, directly excited mode.

$\vec{k}=\left( \mathbf{k},\mathbf{k}^{\prime },\mathbf{k}^{\prime \prime },%
\mathbf{k}^{\prime \prime \prime }\right) $, $\vec{q}=\left( \mathbf{q},%
\mathbf{q}^{\prime },\mathbf{q}^{\prime \prime },\mathbf{q}^{\prime \prime
\prime }\right) $ - four-wave interaction wavevector, (\ref{karrow}), (\ref%
{zsy}).

$\vec{k}_{\ast ,\times ,\pm }$ - vectors defined by (\ref{kstar1})

$\bar{n}=\left( \zeta ,n\right) $ - band index, see (\ref{omn}).

$n$ - band number.

$n_{0}$ - band number of a chosen band.

$\vec{n}=\left( \overline{n},\overline{n}^{\prime },\overline{n}^{\prime
\prime },\bar{n}^{\prime \prime \prime }\right) $ - four-wave interaction
band index.

$\vec{n}_{0}=\left( \left( \zeta ,n\right) ,\left( \zeta ,n_{0}\right)
,\left( \zeta ,n_{0}\right) ,\left( -\zeta ,n_{0}\right) \right) $ - (\ref%
{n0ar}).

$\left\uparrow n_{0},\mathbf{k}_{\ast }\right\downarrow =\left\{ \left(
1,n_{0},\mathbf{k}_{\ast }\right) ,\left( -1,n_{0},-\mathbf{k}_{\ast
}\right) \right\} $ - modal doublet, doublet, (\ref{ggk2}).

$\left\Uparrow n_{0},\mathbf{k}_{\ast }\right\Downarrow =\left\uparrow n_{0},%
\mathbf{k}_{\ast }\right\downarrow \cup \left\uparrow n_{0},-\mathbf{k}%
_{\ast }\right\downarrow $ - bidirectional quadruplet, (\ref{ggk4}).

$\nabla _{\mathbf{r}}=\left( \frac{\partial }{\partial r_{1}},\frac{\partial 
}{\partial r_{2}},\cdots ,\frac{\partial }{\partial r_{d}}\right) $.

$O\left( \mu \right) $ - any quantity having the property that $\frac{%
O\left( \mu \right) }{\mu }$ is bounded as $\mu \rightarrow 0$.

$O\left( \left\vert \mathbf{U}^{\left( 1\right) }\right\vert \right) $
magnitude of the FNLR, estimated by (\ref{OU1})

$\breve{Q}_{\vec{n}}\left( \vec{k}\right) $ - modal susceptibility defined
by (\ref{Qn}).

$\breve{Q}_{\vec{n},\bar{l}}\left( \vec{k}\right) $ - a component of the
modal susceptibility susceptibility, (\ref{Qnl}).

$\vec{q}=\left( \mathbf{q},\mathbf{q}^{\prime },\mathbf{q}^{\prime \prime },%
\mathbf{q}^{\prime \prime \prime }\right) $ - (\ref{karrow}), (\ref{zsy}).

$\vec{q}^{\;\flat }=\left( \mathbf{q},\mathbf{q},\mathbf{q},-\mathbf{q}%
\right) $ - see (\ref{crity1}).

$\mathbf{q}^{\prime \prime \prime }\left( \beta \right) $ expression of $%
\mathbf{q}^{\prime \prime \prime }$ from the phase-matching condition, (\ref%
{Yy'})

$\vec{q}\left( \beta \right) =\left( \mathbf{q},\mathbf{q}^{\prime },\mathbf{%
q}^{\prime \prime },\mathbf{q}^{\prime \prime \prime }\left( \beta \right)
\right) $ - rectifying variables subjected to the phase-matching condition, (%
\ref{qbet}).

$\mathbf{q}^{\prime \prime \prime }\left( 0\right) =\mathbf{0}$, $\vec{q}%
\left( 0\right) =\left( \mathbf{q},\mathbf{q}^{\prime },\mathbf{q}^{\prime
\prime },\mathbf{q-q}^{\prime }-\mathbf{q}^{\prime \prime }\right) $ - see (%
\ref{y0f0}).

$\mathbf{r}=\left( r_{1},\ldots ,r_{d}\right) $ - spatial variable.

$\mathbf{s}=\left( s_{1},\ldots ,s_{d}\right) $ - local quasimomentum
variable, see (\ref{eta}), (\ref{coords}).

$\vec{s}=\left( \mathbf{\mathbf{s}},\mathbf{\mathbf{s}}^{\prime },\mathbf{%
\mathbf{s}}^{\prime \prime },\mathbf{\mathbf{s}}^{\prime \prime \prime
}\right) $ - see (\ref{zsy}).

$\vec{s}^{\;\star }=\left( \mathbf{\mathbf{s}}^{\prime },\mathbf{\mathbf{s}}%
^{\prime \prime },\mathbf{\mathbf{s}}^{\prime \prime \prime }\right) $ - see
(\ref{sspr}).

$\vec{s}^{\;\sharp }=\left( \mathbf{\mathbf{s}}^{\prime }+\mathbf{\mathbf{s}}%
^{\prime \prime }+\mathbf{\mathbf{s}}^{\prime \prime \prime },\mathbf{%
\mathbf{s}}^{\prime },\mathbf{\mathbf{s}}^{\prime \prime },\mathbf{\mathbf{s}%
}^{\prime \prime \prime }\right) $ - see (\ref{sspr}).

$\mathbf{U}$ - solution of the NLM, see (\ref{ML1}).

$\mathbf{\tilde{U}}\left( \mathbf{k},\mathbf{r},t\right) $ Floquet-Bloch
transform of $\mathbf{U}\left( \mathbf{r},t\right) $, (\ref{Utild})

$\mathbf{\tilde{U}}_{\bar{n}}^{\left( 0\right) }$ - modal component of the
linear response, (\ref{UlinBloch}).

$\mathbf{\tilde{U}}_{\bar{n}}^{\left( 1\right) }$ - modal component of the
first nonlinear response, (\ref{UlinBloch}).

$\tilde{U}_{\bar{n}}\left( \mathbf{k},\tau \right) =\tilde{u}_{\bar{n}%
}\left( \mathbf{k},\tau \right) \mathrm{e}^{-\mathrm{i}\omega _{\bar{n}%
}\left( \mathbf{k}\right) t}$ - modal amplitudes, (\ref{UG}), (\ref{UFB1}).

$u^{\left( m\right) },u^{\left( m_{1},m_{2}\right) }$ - coefficients of the
power series expansions of $u=u\left( \alpha ,\beta ,\varrho \right) $ - see
(\ref{UU1}), (\ref{uap1}), (\ref{U1rho}), (\ref{u1rho}).

$\tilde{u}_{\bar{n}}^{\left( 1\right) }\left( \mathbf{k},\tau \right) $ -
modal amplitude depending on the slow time $\tau $ of the first nonlinear
response in written causal form, (\ref{ucS}), (\ref{uc1}).

$\tilde{u}_{\bar{n}}^{\left( 1,0\right) }\left( \mathbf{k},\tau \right) $ -
modal amplitude depending on the slow time $\tau $ of the first nonlinear
response in the time-harmonic approximation, (\ref{Vn}).

$\hat{V}\left( \mathbf{q}\right) $ - Fourier transform of $V\left( \mathbf{r}%
\right) $, see (\ref{Ftransform}).

$\hat{V}_{\zeta }\left( \mathbf{\xi },t\right) =\hat{v}_{\zeta }\left( 
\mathbf{\xi },\tau \right) e^{-\mathrm{i}\zeta \gamma _{\left( \nu \right)
}\left( \zeta \mathbf{\xi }\right) t}$ - amplitudes, see (\ref{Schlin}).

$\psi _{0}\left( \tau \right) $ - slowly time cutoff function, see (\ref%
{psi0}).

$\psi \left( \tau \right) =\int_{0}^{\tau }\psi _{0}\left( \tau \right) \,%
\mathrm{d}\tau _{1}$ - auxiliary function with $\psi _{0}\left( \tau \right) 
$ satisfying (\ref{psi0}), see also (\ref{psii}).

$\Psi _{0}$ - cutoff function in quasimomentum domain, see (\ref{j0}).

$\Psi \left( \mathbf{\xi }\right) =\Psi \left( \zeta ,\mathbf{\xi }\right)
=\Psi _{0}\left( \zeta Y\left( \zeta \mathbf{\xi }\right) \right) $ - cutoff
function in the local rectifying coordinates, see (\ref{Psihat}).

$\phi _{\vec{n}}\left( \vec{k}\right) =\zeta \omega _{n}\left( \mathbf{k}%
\right) -\zeta ^{\prime }\omega _{n^{\prime }}\left( \mathbf{k}^{\prime
}\right) -\zeta ^{\prime \prime }\omega _{n^{\prime \prime }}\left( \mathbf{k%
}^{\prime \prime }\right) -\zeta ^{\prime \prime \prime }\omega _{n^{\prime
\prime \prime }}\left( \mathbf{k}^{\prime \prime \prime }\right) $ -
four-wave interaction phase function, (\ref{phink}).

$\mathring{\Phi}\left( \mathbf{\mathbf{k}_{\ast }},\beta \vec{q}\right) $ -
four-wave interaction phase function in rectifying variables (\ref{ficap}).

$\Phi ^{\left( \nu \right) }\left( \vec{\zeta}_{0},\beta \vec{s}\right) $ -
polynomial phase function (\ref{phasefi2}).

$Y\left( \mathbf{\xi }\right) $ - rectifying change of variables, (\ref{Y}),
(\ref{redY}).

$Y_{\zeta }\left( \mathbf{\xi }\right) =\zeta Y\left( \zeta \mathbf{\xi }%
\right) $, (\ref{Yzet})

$\zeta =\pm $ binary index.

$\vec{\zeta}=\left( \zeta ,\zeta ^{\prime },\zeta ^{\prime \prime },\zeta
^{\prime \prime \prime }\right) $ - four-wave interaction binary band index.

$\vec{\zeta}_{0}=\left( \zeta ,\zeta ,\zeta ,-\zeta \right) $ - see (\ref{z0}%
).

$Z_{\pm }$ - solution of the NLS or ENLS.

$Z^{\ast }$ - complex conjugate to $Z$.\newline

\textbf{Acknowledgment and Disclaimer:} Effort of A. Babin and A. Figotin is
sponsored by the Air Force Office of Scientific Research, Air Force
Materials Command, USAF, under grant number F49620-01-1-0567. The US
Government is authorized to reproduce and distribute reprints for
governmental purposes notwithstanding any copyright notation thereon. The
views and conclusions contained herein are those of the authors and should
not be interpreted as necessarily representing the official policies or
endorsements, either expressed or implied, of the Air Force Office of
Scientific Research or the US Government.\newline

\end{document}